École doctorale IAEM Lorraine

# Formalization and Implementation of Safe Destination Passing in Pure Functional Programming Settings

## THÈSE

présentée et soutenue publiquement le 14 novembre 2025

pour l'obtention du

### Doctorat de l'Université de Lorraine

**(mention informatique)**

par

Thomas Bagrel

**Composition du jury**

| | | |
|---|---|---|
| *Président :* | Laurent Vigneron | |
| *Rapporteurs :* | Dominic Orchard | University of Kent |
| | Gabriele Keller | Utrecht University |
| *Examinateurs :* | Delia Kesner | IRIF – Université Paris Cité |
| | Laurent Vigneron | LORIA – Université de Lorraine |
| | Gabriel Scherer | IRIF – Université Paris Cité |
| *Encadrants :* | Horatiu Cirstea | LORIA – Université de Lorraine (*directeur de thèse*) |
| | Arnaud Spiwack | Tweag (*encadrant industriel*) |

**Laboratoire Lorrain de Recherche en Informatique et ses Applications — UMR 7503**

*Je dédie cette thèse à Pascal et Gérard, que j'espère rendre fiers aujourd'hui.*

# Résumé

La programmation par passage de destination introduit le concept de *destination*, qui représente l'adresse d'une cellule mémoire encore vierge sur laquelle on ne peut écrire qu'une fois. Ces destinations peuvent être passées en tant que paramètres de fonction, permettant à l'appelant de garder le contrôle de la gestion mémoire : la fonction appelée se contente de remplir la cellule au lieu d'allouer de l'espace pour une valeur de retour. Bien que principalement utilisé en programmation système, le passage de destination trouve aussi des applications en programmation fonctionnelle pure, où il permet d'écrire des programmes auparavant inexpressibles avec les structures de données immuables usuelles.

Dans cette thèse, nous développons un $\lambda$-calcul avec destinations, $\lambda_d$. Ce nouveau système théorique est plus expressif que les travaux similaires existants, le passage de destination y étant conçu pour être aussi flexible que possible. Cette expressivité est rendue possible par un système de types modaux combinant types linéaires et un système d'âges pour gérer le contrôle lexical des ressources, afin de garantir la sûreté du passage de destination. Nous avons prouvé la sûreté de notre système via les théorèmes habituels de progression et de préservation des types, de façon mécanisée, avec l'assistant de preuve Coq.

Nous montrons ensuite comment le passage de destination, formalisé dans ce calcul théorique, peut être intégré à un langage fonctionnel pur existant, Haskell, dont le système de types est moins puissant que notre système théorique. Préserver la sûreté nécessite alors de restreindre la flexibilité dans la gestion des destinations. Nous affinons par la suite l'implémentation pour retrouver une grande partie de cette flexibilité, au prix d'une complexité accrue pour l'utilisateur.

L'implémentation prototype en Haskell montre des résultats encourageants pour l'adoption du passage de destination pour des parcours ou du mapping de grandes structures de données, telles que les listes ou les arbres.

# Abstract

Destination-passing style programming introduces destinations, which represent the address of a write-once memory cell. These destinations can be passed as function parameters, allowing the caller to control memory management: the callee simply fills the cell instead of allocating space for a return value. While typically used in systems programming, destination passing also has applications in pure functional programming, where it enables programs that were previously unexpressible using usual immutable data structures.

In this thesis, we develop a core $\lambda$-calculus with destinations, $\lambda_d$. Our new calculus is more expressive than similar existing systems, with destination passing designed to be as flexible as possible. This is achieved through a modal type system combining linear types with a system of ages to manage scopes, in order to make destination-passing safe. Type safety of our core calculus was proved formally with the Coq proof assistant.

Then, we see how this core calculus can be adapted into an existing pure functional language, Haskell, whose type system is less powerful than our custom theoretical one. Retaining safety comes at the cost of removing some flexibility in the handling of destinations. We later refine the implementation to recover much of this flexibility, at the cost of increased user complexity.

The prototype implementation in Haskell shows encouraging results for adopting destination-passing style programming when traversing or mapping over large data structures such as lists or data trees.

# Remerciements

C'est avec beaucoup d'émotions que j'arrive au terme de cette aventure qui aura occupé trois années et demi de ma vie. Ce parcours a été difficile et exigeant, autant vis-à-vis de moi-même que du sujet à traiter. Heureusement, j'ai eu l'honneur d'être encadré et soutenu par un très riche entourage, où chacun a contribué, à sa façon, pour que je puisse aujourd'hui déposer le point final sur ce manuscrit. Un entourage plus riche que ce que je peux exprimer sur cette page d'ailleurs; aussi, je dédie mes plus sincères remerciements à ceux que je n'ai pas eu la place de mentionner ici.

Je suis immensément reconnaissant envers mes parents, qui, bien avant même le début de cette thèse, m'ont toujours donné guidé, soutenu, et donné les encouragements et la reconnaissance nécessaire pour que je suive la voie de mon choix malgré les difficultés pouvant se présenter. Mon coeur se tourne également vers Marinette, qui suit aujourd'hui une voie proche de la mienne, avec une détermination et une excellence que j'admire, et dont la réussite ne fait aucun doute. Mes pensées vont aussi immédiatement à Julien et Ocelline, qui ont supporté mes états d'âme, mes soucis, et m'auront toujours ouvert leur porte pour repartir plus fort et plus motivé que la veille. À Luna également, qui aura été un rayon de soleil pour égayer les moments plus sombres au quotidien, et qui n'aura cessé de croire en moi, même quand l'espoir me quittait. À Tim aussi, dont la vision du monde me redonne un peu d'optimisme, et m'aide à trouver un sens à mon travail.

Je dois aussi tous mes remerciements à Arnaud, qui malgré ses nombreuses responsabilités, aura été présent chaque jour, sans faute, pour répondre à mes interrogations, à mes doutes, et pour suivre avec attention mes élucubrations. Sa passion scientifique rend chacune des interactions avec lui captivante et instructive, et je le remercie pour toutes les connaissances qu'il a partagé avec moi au cours de cette aventure, ainsi que pour sa patience face à mon entêtement régulier. Je remercie également Horatiu, pour son soutien, son aide pour naviguer la bureaucratie académique, et ses conseils avisés pour améliorer mes travaux. Je les remercie tous deux de m'avoir fait confiance et d'avoir accepté d'embarquer dans ce projet avec moi; projet que j'ai entrepris suite aux conseils avisés de Dominique Mery et Clément Hurlin.

Mais de nombreux autres visages m'ont permis, au quotidien, de rester à flot. D'abord, mes joyeux collègues doctorants, Sarah, Vincent, Ghislain, Alessio, dont l'estime que j'ai pour eux est bien supérieure à ce que les remarques qu'on s'envoyait dans le bureau pouvait faire croire. Je leur souhaite la plus sincère réussite dans leurs propres aventures. L'équipe Mosel-Veridis plus généralement, et nos discussions qui alternaient souvent entre sciences et vêtements. Mes *fellow Tweagers* également. À Clément (à nouveau), qui m'a pris sous son aile, et avec lequel j'ai la chance de partager une passion pour Warhammer, en plus d'un certain regard sur notre profession. À Camilla, dont la joie de vivre et la bienveillance inonde le bureau parisien, et qui s'est toujours débrouillée pour que je ne manque de rien d'un point de vue logistique sur ces années. À Florent, mon partenaire de cynisme préféré; à Alex, Nicolas et Simeon, et nos lunchs très sympas; à Yann, qui m'a aidé à sortir de l'impasse du *scope escape*; à Guillaume, avec son smile et ses pas de danse; à Benjamin, qui supervise ce coin de Paris où il fait bon travailler (et vivre).

Dans mon cercle personnel, je me dois également de remercier Julien, Florian et Alexandre, qui ont suivi mes avancées avec attention, et dont j'ai malheureusement souvent dû refuser les propositions de sorties ou d'activités. Ysaline aussi, qui m'a aidé à traverser de nombreuses épreuves, et dont j'admire la détermination et la pertinence. J'ai aussi une pensée toute particulière pour Marion, qui m'a montré la voie dans le monde académique, en suivant de près mon parcours. C'est aussi grâce à elle que j'ai très tôt commencé à utiliser LaTeX; que j'utilise aujourd'hui encore pour produire ce document. Enfin, je dédie mes derniers mots à mon grand-père Gérard, qui m'aura transmis sa rigueur et sa passion pour les mathématiques, et à mon oncle Pascal, avec qui j'ai très tôt exploré l'ingénieurie. Même s'ils ne sont plus là pour lires ces lignes, ils ont toujours eu très à coeur de suivre mes avancées et réussites, et c'est en partie grâce à eux que j'en suis là aujourd'hui.

# Table of Contents

# List of Figures

# List of Listings

# Introduction

## 1 Memory management: a core design axis for programming languages

Over the last fifty years, programming languages have evolved into indispensable tools, profoundly shaping how we interact with computers and process information across almost every domain. Much like human languages, the vast variety of programming languages reflects the diversity of computing needs, design philosophies, and objectives that developers and researchers have pursued. This diversity is a response to specialized applications, user preferences, and the continuous search for improvements in either speed or expressiveness. Today, hundreds of programming languages exist, each with unique features and tailored strengths, making language selection a nuanced process for developers.

At a high level, programming languages differ in how they handle core aspects of data representation and program execution, and they can be classified along several key dimensions. These dimensions often reveal the underlying principles of the language and its suitability for different types of applications. Some of the main characteristics that distinguish programming languages include:

- how the user communicates intent to the computer—whether through explicit step-by-step instructions in procedural languages or through a more functional/declarative style that emphasizes what to compute rather than how;
- the organization and manipulation of data—using either object-oriented paradigms that encapsulate domain data within objects than can interact with each other or using simpler data structures that can be operated on and dissected by functions and procedures;
- the level of abstraction—a higher-level language abstracts technical details to enable more complex functionality in fewer lines of code, while lower-level languages provide more control over the environment and the detailed execution of a task;
- the management of memory—whether memory allocation and deallocation are handled automatically by the language, or require explicit intervention from the programmer;
- side effects and exceptions—whether they can be represented, caught, and/or manipulated with the language.

Among these, memory management is one of the most critical aspects of programming language design. Every program requires memory to store data and instructions, and this memory must be managed judiciously to prevent errors and maintain performance. Typically, programs operate with input sizes and data structures that can vary greatly, requiring dynamic memory allocation to ensure smooth execution. Therefore, the chosen memory management scheme will shape very deeply how programmers write and structure their code but also which features can be included in the language with reasonable effort.

High-level languages, such as Python or Java, often manage memory automatically, through garbage collection. With automatic garbage collection (thanks to a tracing or reference-counting garbage collector), memory allocation and deallocation happen behind the scenes, freeing developers from the complexities of manual memory control. This automatic memory management simplifies programming, allowing developers to focus on functionality and making the language more accessible for rapid development. Although garbage collection is decently fast overall, it can be slow for some specific use-cases, and is also dreaded for its unpredictable overhead or pauses in the program execution, which makes it unsuited for real-time applications.

In contrast, low-level languages like C or Zig tend to provide developers with direct control over memory allocation and deallocation. It indeed allows for greater optimization and efficient resource usage, particularly useful in systems programming or performance-sensitive applications. This control, however, comes with increased risks; errors like memory leaks, buffer overflows, and dangling pointers can lead to instability and security vulnerabilities if not carefully detected and addressed.

Interestingly, some languages defy these typical categorizations by providing precise memory control in the form of safe, high-level abstractions. Such languages let the user manage resource lifetimes explicitly while taking the responsibility of allocating and deallocating memory at the start and end of each resource's lifetime. The most well-known examples are smart pointers in C++ and the ownership model in Rust, whose founding principles are also known as *Scope-Bound Resource Management* (SBRM) or *Resource Acquisition is Initialization* (RAII). Initially applicable only to stack-allocated objects in early C++, SBRM evolved with the introduction of smart pointers for heap-allocated objects in the early 2000s. These tools have since become fundamental to modern C++ and Rust, significantly improving memory safety. In particular, Rust provides safety guarantees comparable to those of garbage-collected languages but without the garbage collection overhead, at the cost of a steep learning curve. This makes Rust suitable for both high-level application programming and low-level systems development, bridging a gap that was traditionally divided by memory management style.

## 2 Destination passing: taking roots in the old imperative world

In systems programming, where fine control over memory is essential, the same memory buffer is often (re)used several times, in order to save up memory space and decrease the number of costly calls to the memory allocator. As a result, we do not want intermediary functions to allocate memory for the result of their computations. Instead, we want to provide them with a memory address in which to write their result, so that we can decide to reuse an already allocated memory slot or maybe write to a memory-mapped file, or RDMA buffer. Similarly, we can provide a function with multiple output slots, allowing it to write several results.

In practice, this means that the parent scope manages not only the allocation and deallocation of the function inputs, but also those of *the function outputs*. Output slots are passed to functions as pointers (or mutable references in higher-level languages), allowing the function to write directly into memory managed by the caller. These pointers, called out parameters or *destinations*, specify the exact memory location where the function should store its output.

This idea forms the foundation of destination-passing style programming (DPS): instead of letting a function allocate memory for its outputs, the caller provides it with write access to a pre-allocated memory space that can hold the function's results. Assigning memory responsibility to the caller, rather than the callee, makes functions more flexible to use: this is the core advantage of destination-passing style programming (DPS).

In a low-level, system-programming setting, DPS also offers direct performance benefits. For example, a large memory block can be allocated in advance and divided into smaller segments, each designated as an output slot for a specific function call. This minimizes the number of separate allocations required, which is often a performance bottleneck, especially in contexts where memory allocation is intensive.

In the functional interpretation of DPS that will follow, the memory saving will be more subtle, and we will instead leverage instead the flexibility and expressivity benefits of DPS to lift some usual limitations of functional programming.

# 3 Functional programming languages

Functional programming languages are often seen as the opposite end of the spectrum from systems languages.

Functional programming lacks a single definition, but most agree that a functional language:

- supports lambda abstractions, a.k.a. functions as first-class values, that can an capture parts of their parent environment (closures), can be stored, and can passed as parameters;
- emphasizes expressions over statements, where each instruction always produces a value;
- builds complex expressions by applying and composing functions rather than chaining statements.

From these principles, functional languages tend to favor immutable data structures and a declarative style, where new data structures are defined by specifying their differences from existing ones, often relying on structural sharing and persistent data structures instead of direct mutation.

Since "everything is an expression" in functional languages, they are particularly amenable to mathematical formalization. This is no accident: many core concepts in functional programming originate in mathematics, such as the $\lambda$-calculus—a minimal yet fully expressive model of computation, that also connects closely with formal proofs through the Curry-Howard isomorphism.

Despite this, many functional languages still permit expressions with side effects. Side effects encompass all the observable actions a program can take on its environment, e.g. writing to a file, or to the console, or altering memory. Side effects are hard to reason about, especially if any expression of the language is likely to emit one, but they are in fact a very crucial piece of any programming language: without them, there is no way for a program to communicate its results to the "outside".

**Pure functional programming languages** Some functional languages enforce a stricter approach by disallowing side effects except at the boundaries of the main program, a concept known as *pure functional programming*. Here, all programmer-defined expressions compute and return values without side effects. Instead, the *intention* of performing side effects (like printing to the console) can be represented as a value, which is only turned into the intended side effect by the language runtime.

This restriction provides substantial benefits. Programs can be wholly modeled as mathematical functions, making reasoning about behavior easier and allowing for more powerful analysis of a program's behavior.

A key property of pure functional languages is *referential transparency*: any function call can be substituted with its result without altering the program's behavior. This property is obviously absent in languages like C; for example, replacing `write(myfile, string, 12)` with the value `12` (its return if the write is successful) would not produce the intended behavior: the latter would not produce any write effect at all.

This ability to reason about programs as pure functions greatly improves the predictability of programs, especially across library boundaries, and overall improves software safety.

**Memory management in functional languages** Explicit memory management is very often *impure*, as modifying memory in a way that can later be inspected *is* a side effect. Consequently, most functional languages (even those that aren't fully *pure* like OCaml or Scala) tend to rely on a garbage collector to abstract away memory management and memory access for the user.

However, this doesn't mean that functional languages must entirely give up any aspect of memory control; it simply means that memory control cannot involve explicit effectful statements like `malloc` or `free`. In practice, it requires that memory management should not affect the value of an expression being computed within the language. One way to achieve this is by using annotations or pragmas attached to specific functions or blocks of code, indicating how or where memory should be allocated. Another approach is to use a custom type system with *modes*, to track at type level how values are managed in memory, as in the recent modal development for OCaml ⟦Lorenzen, White, et al., 2024⟧. Additionally, some languages use special built-in functions, like `oneShot` in Haskell, which do not affect the result of the expression they wrap around, but carry special significance for the compiler in managing memory.

There is also another possible solution. We previously mentioned that side effects—prohibited in pure functional languages—are any modifications of the environment or memory that are *observable* by the user. What if we allow explicit memory management expressions while ensuring that these (temporary) changes to memory or the environment remain unobservable?

This is the path we will adopt—one that others have taken before—allowing for finer-grained memory control while upholding the principles of purity.

## 4 Functional structures with holes: Minamide's foundational work

In most contexts, functional languages with garbage collectors efficiently manage memory without requiring programmer intervention. However, a major limitation arises from the *immutability* characteristic common to most functional languages, which restricts us to constructing data structures directly in their final form. That means that the programmer has to give an initial value to every field of the structure, even if no meaningful value for them has been computed yet. And later, any update beyond simply expanding the original structure requires creating a (partial) copy of that structure. This incurs a big load on the allocator and garbage collector.

As a result, algorithms that generate large structures—whether by reading from an external source or transforming an existing structure—might need many intermediate allocations, each representing a temporary processing state. While immutability has advantages, in this case, it can become an obstacle to optimizing performance.

To lift this limitation, Minamide ⟦1998⟧ introduced an extension for functional languages that allows to represent *yet incomplete* data structures, that can be extended and completed later. Incomplete structures are not allowed to be read until they are completed—this is enforced by the type system—so that the underlying mutations that occur while updating the structure from its incomplete state to completion are hidden from the user, who can only observe the final result. This way, it preserves the feel of an immutable functional language and does not break purity per se. This method of making imperative or impure computations opaque to the user by making sure that their effects remain unobservable to the user (thanks to type-level guarantees) is central to the approach developed in this document.

In Minamide's work, incomplete structures i.e. *structures with a hole* are represented by hole abstractions—essentially pure linear functions that take the missing component of the structure as an argument and return the completed structure. In other terms, it represents the pending construction of a structure, than can theoretically only be carried out when all its missing pieces have been supplied. Hole abstractions can also be composed, like functions: $(\lambda x \mapsto 1 :: x) \circ (\lambda y \mapsto 2 :: y) \rightsquigarrow \lambda y \mapsto 1 :: 2 :: y$. Behind the scenes however, each hole abstraction is not represented in memory by a function object, but rather by a pair of pointers, one read pointer to the root of the data structure that the hole abstraction describes, and one write pointer to the yet unspecified field in the data structure. Composition of two holes abstractions can be carried out immediately, with no need to wait for the missing value of the second hole abstraction, and results, memory-wise, in the mutation of the previously missing field of the first one, to point to the root of the second one.

In Minamide's system, the (pointer to the) hole and the structure containing it are indissociable. One interacts with the hole in the structure by interacting with the structure itself, which limits an incomplete structure to having only a single hole. For these reasons, we would argue that Minamide's approach does not yet qualify as destination-passing style programming. Nonetheless, it is one of the earliest examples of a pure functional language allowing data structures to be passed with write permissions while preserving key language guarantees such as memory safety and purity.

This idea has been refreshed and extended more recently with ⟦Leijen and Lorenzen, 2025⟧ and ⟦Lorenzen, Leijen, Swierstra, and Lindley, 2024⟧, where structures with holes are referred to as *first-class contexts*.

To reach true destination-passing style programming however, we need a way to refer to the hole(s) of the incomplete structure without having to pass around the structure that has the hole(s). This in fact has not been explored much in functional settings so far. ⟦Bour, Clément, and Scherer, 2021⟧ is probably the closest existing work on this matter, but destination-passing style is only used at the intermediary language level, in the compiler, to do optimizations, so the user cannot make use of DPS in their code.

# 5 Formalizing functional destination passing

What we will develop in this thesis is a functional language in which *structures with holes* are a first-class type in the language, as in ⟦Minamide, 1998⟧, but that also integrate first-class *pointers to these holes*, a.k.a. *destinations*, as part of the user-facing language.

We will see that combining these two features gives us performance improvement for some functional algorithms like in ⟦Bour, Clément, and Scherer, 2021⟧ or ⟦Leijen and Lorenzen, 2025⟧, but we also get some extra expressiveness that is usually a privilege of imperative languages only. It should be indeed the first instance of a functional language that supports write pointers in the frontend (and not only in intermediate representation like ⟦Bour, Clément, and Scherer, 2021; Leijen and Lorenzen, 2025⟧) to make flexible and efficient data structure building!

In fact, we will design a formal language, based on a linear types approach, in which *structures with holes* serve as the core abstraction for constructing data structures, departing from the traditional use of data constructors in functional programming. Later, we will demonstrate how many of the ideas and features from this theoretical calculus can be implemented in a practical functional language—namely Haskell—while retaining as much expressivity as possible. We will also show that the predicted benefits of the approach are effectively observed in the prototype Haskell implementation.

The formal language will also aim at providing a theoretical framework that encompasses most existing work on functional DPS and can be used to reason about safety and correctness properties of earlier works.

## 6 Reconciling write pointers and immutability

Introducing unrestricted write pointers in a functional setting would not play well as it would break many properties that make functional languages easy to reason about. Instead, we need tools to ensure that only controlled mutations occur. The goal is to allow a structure to remain partially uninitialized temporarily, with write pointers referencing these uninitialized portions. However, once the structure has been fully populated, with a value assigned to each field, it should become immutable to maintain the integrity of the functional paradigm. Put simply, we need a write-once discipline for fields within a structure, which can be mutated through their associated write pointers a.k.a. *destinations*. While enforcing a single-use discipline for destinations at runtime would be unwieldy, we can instead deploy static guarantees through the type system to ensure that every destination will be used exactly once.

In programming language theory, when new guarantees are required, it's common to design a type system to enforce them, ensuring the program remains well-typed. This is precisely the approach we'll take here, facilitated by an established type system—*linear types*—which can monitor resource usage, especially the number of times a resource is used. Leveraging a linear type system, we can design a language in which structures with holes are first-class citizens, and write pointers to these holes are only usable once. This ensures that each hole is filled exactly once, preserving immutability and making sure that structure are indeed completed before being read. This is also the approach taken in ⟦Minamide, 1998⟧ and subsequent work.

# Chapter 1

# Linear $\lambda$-calculus and Linear Haskell

In the following we assume that the reader is familiar with the simply-typed $\lambda$-calculus, its typing rules, and usual semantics. We also assume some degree of familiarity with the intuitionistic fragment of natural deduction. We kindly refer the reader to ⟦Pierce, 2002⟧ and ⟦Sørensen and Urzyczyn, 2006⟧ for a proper introduction to these notions.

## 1.1 From the $\lambda$-calculus to the linear $\lambda$-calculus

At the end of the 1930s, Alonzo Church introduced the *untyped $\lambda$-calculus* as a formal mathematical model of computation. Untyped $\lambda$-calculus is based on the concept of function abstraction and application, and is Turing-complete: in other terms, it has the same expressive power as the rather imperative model of *Turing machines*, introduced in the same decade. The only available objects in the untyped $\lambda$-calculus are pure functions; but they can be used smartly to encode many other kind of data, and as a result they are enough to form a solid model of computation.

In 1940, Church defined a typed variant of its original calculus, denoted the *simply-typed $\lambda$-calculus*, or STLC, that gives up Turing-completeness but becomes strongly-normalizing: every well-typed term eventually reduces to a normal form. STLC assign types to terms, and restricts the application of functions to terms of the right type. This restriction is enforced by the typing rules of the calculus.

It was observed by Howard in 1969 ⟦Howard, 1969⟧ that the simply-typed $\lambda$-calculus is actually isomorphic to the logical framework known as *natural deduction*. This observation relates to earlier work by Curry, who had also noted similar symmetries between systems on the computational side and those on the logical side. These insights were later formalized into the *Curry-Howard isomorphism*, which states that types in a typed $\lambda$-calculus correspond to formulae in a proof system, and that terms correspond to proofs of these formulae. The Curry-Howard isomorphism has since been extended to other logics and calculi and has been a rich source of inspiration for research on both the logical and computational sides. For instance, building on an idea from his work on models of the $\lambda$-calculus, Jean-Yves Girard first introduced Linear Logic ⟦Jean-Yves Girard, 1987⟧, and later the corresponding calculus, known as the *linear $\lambda$-calculus*.

**What is linear logic?** In some logical presentations, notably the sequent calculus[1], hypotheses are duplicated or discarded explicitly using *structural* rules of the system, named respectively *contraction* and *weakening* (in contrast to natural deduction where all that happens implicitly, as it is part of the meta-theory). As a result, it is possible to track the number of times a formula is used by counting the use of these structural rules. In Linear Logic, contraction and weakening are deliberately restricted, so by default, every hypothesis must be used exactly once. Consequently, logical implication $\mathtt{T} \to \mathtt{U}$ is not part of linear logic, but is replaced by linear implication $\mathtt{T} \multimap \mathtt{U}$, where $\mathtt{T}$ must be used exactly once to prove $\mathtt{U}$. Linear logic also introduces a modality $!$, pronounced *of course* or *bang*, to allow weakening and contraction on specific formulae: $!\mathtt{T}$ denotes that $\mathtt{T}$ can be used an arbitrary number of times (we say it is *unrestricted*). We say that linear logic is a *substructural* logic because it restricts the use of structural rules of usual logic.

We present in Figure 1.1 a natural deduction formulation[2] of intuitionistic linear logic (ILL), as it lends itself well to a computational interpretation as a linear λ-calculus with usual syntax.

All the rules of ILL, except the ones related to the $!$ modality, are directly taken (and slightly adapted) from natural deduction. $\mathtt{1}$ denotes the truth constant, $\oplus$ denotes disjunction, and $\otimes$ denotes conjunction[3]. Hypotheses are represented by multisets $\Gamma$. These multisets keep track of how many times each formula appear in them. The comma operator in $\Gamma_1,\ \Gamma_2$ is multiset union, so it sums the number of occurrences of each formula in $\Gamma_1$ and $\Gamma_2$. We denote the empty multiset by $\bullet$.

Let's focus on the four rules for the $!$ modality now. The promotion rule ILL/!P states that a formula $\mathtt{T}$ can become an unrestricted formula $!\mathtt{T}$ if it only depends on formulae that are themselves unrestricted. For this purpose, we introduce the notation $!\Gamma$ to denote a context which contains the same hypotheses as $\Gamma$ but all wrapped in the $!$ modality in other terms, every hypothesis in the (potentially empty) multiset $!\Gamma$ is unrestricted. The dereliction rule ILL/!D states that an unrestricted formula $!\mathtt{T}$ can be used in a place expecting a linear formula $\mathtt{T}$. The contraction rule ILL/!C and weakening rule ILL/!W state respectively that an unrestricted formula $!\mathtt{T}$ can be cloned or discarded at any time.

The linear logic system might appear very restrictive, but we can always simulate the usual non-linear natural deduction in ILL, through the $!$ modality. Girard gives precise rules for such a translation in Section 2.2.6 of [[J.-Y. Girard, 1995]], whose main idea is to prefix most formulae with $!$ and encode the non-linear implication $\mathtt{T} \to \mathtt{U}$ as $!\mathtt{T} \multimap \mathtt{U}$.

To become more familiar with the system, let us explore how linear logic effectively restricts the duplication of linear hypotheses through its rules. For example, let's try (and fail) to derive the implication $\mathtt{T} \multimap (\mathtt{T} \otimes \mathtt{T})$ from an empty environment:

$$\dfrac{\dfrac{\dfrac{}{\mathtt{T} \vdash \mathtt{T}}\ \text{ID} \qquad \dfrac{}{\mathtt{T} \vdash \mathtt{T}}\ \text{ID}}{\mathtt{T},\ \mathtt{T} \vdash \mathtt{T} \otimes \mathtt{T}}\ \otimes\text{I}}{\mathtt{T} \vdash \mathtt{T} \multimap (\mathtt{T} \otimes \mathtt{T})}\ \multimap\text{I}$$

[1] Sequent calculus is a very popular logic framework that is an alternative to natural deduction and that has also been introduced by Gentzen in the 30s

[2] Though we borrow the *sequent style* notation of sequent calculus for easier transition into typing rules of terms later. However, as we are in an intuitionistic setting, rules only derive a single conclusion from a multiset of formulae.

[3] Linear logic distinguishes between multiplicative and additive conjunctions and disjunctions, which differ in how they interact with negation and De Morgan duality: multiplicative connectives (e.g., $\otimes$, $\parr$) are dual to each other, as are additive ones (e.g., $\oplus$, $\&$). In our setting, we focus on $\otimes$ (multiplicative conjunction) and $\oplus$ (additive disjunction), as they correspond more closely to the standard logical connectives $\wedge$ and $\vee$.

$\boxed{\Gamma \vdash \mathtt{T}}$ *(Deduction rules)*

$$\frac{}{\mathtt{T} \vdash \mathtt{T}}\ \textsc{id} \qquad \frac{\Gamma,\ \mathtt{T} \vdash \mathtt{U}}{\Gamma \vdash \mathtt{T} \multimap \mathtt{U}}\ {\multimap}\mathrm{I} \qquad \frac{}{\bullet \vdash \mathtt{1}}\ \mathrm{1I} \qquad \frac{\Gamma \vdash \mathtt{T}_1}{\Gamma \vdash \mathtt{T}_1 \oplus \mathtt{T}_2}\ \oplus\mathrm{I}_1 \qquad \frac{\Gamma \vdash \mathtt{T}_2}{\Gamma \vdash \mathtt{T}_1 \oplus \mathtt{T}_2}\ \oplus\mathrm{I}_2$$

$$\frac{\Gamma_1 \vdash \mathtt{T}_1 \qquad \Gamma_2 \vdash \mathtt{T}_2}{\Gamma_1,\ \Gamma_2 \vdash \mathtt{T}_1 \otimes \mathtt{T}_2}\ \otimes\mathrm{I} \qquad \frac{\Gamma_1 \vdash \mathtt{T} \qquad \Gamma_2 \vdash \mathtt{T} \multimap \mathtt{U}}{\Gamma_1,\ \Gamma_2 \vdash \mathtt{U}}\ {\multimap}\mathrm{E} \qquad \frac{\Gamma_1 \vdash \mathtt{1} \qquad \Gamma_2 \vdash \mathtt{U}}{\Gamma_1,\ \Gamma_2 \vdash \mathtt{U}}\ \mathrm{1E}$$

$$\frac{\begin{array}{c}\Gamma_1 \vdash \mathtt{T}_1 \oplus \mathtt{T}_2 \\ \Gamma_2,\ \mathtt{T}_1 \vdash \mathtt{U} \\ \Gamma_2,\ \mathtt{T}_2 \vdash \mathtt{U}\end{array}}{\Gamma_1,\ \Gamma_2 \vdash \mathtt{U}}\ \oplus\mathrm{E} \qquad \frac{\begin{array}{c}\Gamma_1 \vdash \mathtt{T}_1 \otimes \mathtt{T}_2 \\ \Gamma_2,\ \mathtt{T}_1,\ \mathtt{T}_2 \vdash \mathtt{U}\end{array}}{\Gamma_1,\ \Gamma_2 \vdash \mathtt{U}}\ \otimes\mathrm{E} \qquad \frac{!\Gamma \vdash \mathtt{T}}{!\Gamma \vdash\ !\mathtt{T}}\ !\mathrm{P} \qquad \frac{\Gamma \vdash\ !\mathtt{T}}{\Gamma \vdash \mathtt{T}}\ !\mathrm{D} \qquad \frac{\begin{array}{c}\Gamma_1 \vdash\ !\mathtt{T} \\ \Gamma_2,\ !\mathtt{T},\ !\mathtt{T} \vdash \mathtt{U}\end{array}}{\Gamma_1,\ \Gamma_2 \vdash \mathtt{U}}\ !\mathrm{C}$$

$$\frac{\Gamma_1 \vdash\ !\mathtt{T} \qquad \Gamma_2 \vdash \mathtt{U}}{\Gamma_1,\ \Gamma_2 \vdash \mathtt{U}}\ !\mathrm{W}$$

Figure 1.1: Intuitionistic linear logic in natural deduction presentation, sequent-style (ILL)

In sequent calculus, we would insert a contraction step between $\otimes$I and $\multimap$I to deduce $\mathtt{T} \vdash \mathtt{T} \otimes \mathtt{T}$ from $\mathtt{T},\ \mathtt{T} \vdash \mathtt{T} \otimes \mathtt{T}$. But in linear logic, contraction is not allowed for linear hypotheses. Each application of the ILL$\multimap$I rule consumes exactly one copy of $\mathtt{T}$, so we still need an extra $\mathtt{T}$ in the context to make this implication work.

But all is not lost; we *can* duplicate a value if it comes wrapped in the ! modality. That's precisely what $!\mathtt{T}$ is for: an unrestricted version of $\mathtt{T}$ that allows copying. So let's see how we can derive the implication $!\mathtt{T} \multimap (\mathtt{T} \otimes \mathtt{T})$ from an empty environment:

$$\dfrac{\dfrac{\dfrac{}{!\mathtt{T} \vdash\ !\mathtt{T}}\ \textsc{id} \qquad \dfrac{\dfrac{\dfrac{}{!\mathtt{T} \vdash\ !\mathtt{T}}\ \textsc{id}}{!\mathtt{T} \vdash \mathtt{T}}\ !\mathrm{D} \qquad \dfrac{\dfrac{}{!\mathtt{T} \vdash\ !\mathtt{T}}\ \textsc{id}}{!\mathtt{T} \vdash \mathtt{T}}\ !\mathrm{D}}{!\mathtt{T},\ !\mathtt{T} \vdash \mathtt{T} \otimes \mathtt{T}}\ \otimes\mathrm{I}}{!\mathtt{T} \vdash \mathtt{T} \otimes \mathtt{T}}\ !\mathrm{C}}{\bullet \vdash\ !\mathtt{T} \multimap (\mathtt{T} \otimes \mathtt{T})}\ {\multimap}\mathrm{I}$$

Here, we use identity and dereliction to extract linear $\mathtt{T}$s from the unrestricted $!\mathtt{T}$. The product $\mathtt{T} \otimes \mathtt{T}$ requires two copies, so we need two $!\mathtt{T}$s in the context; and that corresponds exactly to the setting required to apply the contraction rule ILL/!C. In the next section, we'll see how this derivation corresponds to a well-typed program.

## 1.2 Linear λ-calculus: a computational interpretation of linear logic

There are several possible interpretations of linear logic as a linear λ-calculus. The first—referred to as the *monadic* presentation, following the terminology of ⟦Andreoli, 1992⟧—arises from a direct term assignment to the natural deduction rules of ILL shown in Figure 1.1. The syntax and typing rules for this monadic formulation, denoted $\lambda_{L1}$, are presented in Figures 1.2 and 1.3, and are heavily inspired by the work of Bierman ⟦1994⟧.

$$
\begin{array}{rcl}
v & ::= & \lambda x \mapsto u \mid () \mid \mathsf{Inl}\, v \mid \mathsf{Inr}\, v \mid (v_1, v_2) \mid \mathsf{Many}\, v \\
t, u & ::= & v \mid x \mid \mathsf{Inl}\, t \mid \mathsf{Inr}\, t \mid (t_1, t_2) \mid \mathsf{Many}\, t \mid t\ t' \mid t \mathbin{;} t' \\
& \mid & \mathbf{case}\ t\ \mathbf{of}\ \{\mathsf{Inl}\, x_1 \mapsto u_1 ,\ \mathsf{Inr}\, x_2 \mapsto u_2\} \mid \mathbf{case}\ t\ \mathbf{of}\ (x_1, x_2) \mapsto u \\
& \mid & \mathbf{dup}\ t\ \mathbf{as}\ x_1 ,\ x_2\ \mathbf{in}\ u \mid \mathbf{drop}\ t\ \mathbf{in}\ u \mid \mathbf{derelict}\ t \\
\\
\mathsf{T}, \mathsf{U} & ::= & \mathsf{T} \multimap \mathsf{U} \mid \mathsf{1} \mid \mathsf{T}_1 \oplus \mathsf{T}_2 \mid \mathsf{T}_1 \otimes \mathsf{T}_2 \mid\ !\mathsf{T} \\
\\
\Gamma & ::= & \bullet \mid x : \mathsf{T} \mid \Gamma_1, \Gamma_2
\end{array}
$$

Figure 1.2: Grammar of linear λ-calculus in monadic presentation ($\lambda_{L1}$)

$\boxed{\Gamma \vdash t : \mathsf{T}}$ *(Typing judgment for terms)*

$$
\frac{}{x : \mathsf{T} \vdash x : \mathsf{T}}\ \textsc{id}
\qquad
\frac{\Gamma,\ x : \mathsf{T} \vdash u : \mathsf{U}}{\Gamma \vdash \lambda x \mapsto u : \mathsf{T} \multimap \mathsf{U}}\ {\multimap}\mathrm{I}
\qquad
\frac{}{\bullet \vdash () : \mathsf{1}}\ \mathrm{1I}
\qquad
\frac{\Gamma \vdash t_1 : \mathsf{T}_1}{\Gamma \vdash \mathsf{Inl}\, t_1 : \mathsf{T}_1 \oplus \mathsf{T}_2}\ \oplus\mathrm{I}_1
$$

$$
\frac{\Gamma \vdash t_2 : \mathsf{T}_2}{\Gamma \vdash \mathsf{Inr}\, t_2 : \mathsf{T}_1 \oplus \mathsf{T}_2}\ \oplus\mathrm{I}_2
\qquad
\frac{\Gamma_1 \vdash t_1 : \mathsf{T}_1 \quad \Gamma_2 \vdash t_2 : \mathsf{T}_2}{\Gamma_1, \Gamma_2 \vdash (t_1, t_2) : \mathsf{T}_1 \otimes \mathsf{T}_2}\ \otimes\mathrm{I}
\qquad
\frac{\Gamma_1 \vdash t : \mathsf{T} \quad \Gamma_2 \vdash t' : \mathsf{T} \multimap \mathsf{U}}{\Gamma_1, \Gamma_2 \vdash t'\ t : \mathsf{U}}\ {\multimap}\mathrm{E}
$$

$$
\frac{\Gamma_1 \vdash t : \mathsf{1} \qquad \Gamma_2 \vdash u : \mathsf{U}}{\Gamma_1, \Gamma_2 \vdash t \mathbin{;} u : \mathsf{U}}\ \mathrm{1E}
\qquad
\frac{\Gamma_1 \vdash t : \mathsf{T}_1 \oplus \mathsf{T}_2 \quad \Gamma_2,\ x_1 : \mathsf{T}_1 \vdash u_1 : \mathsf{U} \quad \Gamma_2,\ x_2 : \mathsf{T}_2 \vdash u_2 : \mathsf{U}}{\Gamma_1, \Gamma_2 \vdash \mathbf{case}\ t\ \mathbf{of}\ \{\mathsf{Inl}\, x_1 \mapsto u_1 ,\ \mathsf{Inr}\, x_2 \mapsto u_2\} : \mathsf{U}}\ \oplus\mathrm{E}
$$

$$
\frac{\Gamma_1 \vdash t : \mathsf{T}_1 \otimes \mathsf{T}_2 \quad \Gamma_2,\ x_1 : \mathsf{T}_1,\ x_2 : \mathsf{T}_2 \vdash u : \mathsf{U}}{\Gamma_1, \Gamma_2 \vdash \mathbf{case}\ t\ \mathbf{of}\ (x_1, x_2) \mapsto u : \mathsf{U}}\ \otimes\mathrm{E}
\qquad
\frac{!\Gamma \vdash t : \mathsf{T}}{!\Gamma \vdash \mathsf{Many}\, t : !\mathsf{T}}\ !\mathrm{P}
\qquad
\frac{\Gamma \vdash t : !\mathsf{T}}{\Gamma \vdash \mathbf{derelict}\ t : \mathsf{T}}\ !\mathrm{D}
$$

$$
\frac{\Gamma_1 \vdash t : !\mathsf{T} \quad \Gamma_2,\ x_1 : !\mathsf{T},\ x_2 : !\mathsf{T} \vdash u : \mathsf{U}}{\Gamma_1, \Gamma_2 \vdash \mathbf{dup}\ t\ \mathbf{as}\ x_1 ,\ x_2\ \mathbf{in}\ u : \mathsf{U}}\ !\mathrm{C}
\qquad
\frac{\Gamma_1 \vdash t : !\mathsf{T} \quad \Gamma_2 \vdash u : \mathsf{U}}{\Gamma_1, \Gamma_2 \vdash \mathbf{drop}\ t\ \mathbf{in}\ u : \mathsf{U}}\ !\mathrm{W}
$$

Figure 1.3: Typing rules for linear λ-calculus in monadic presentation ($\lambda_{L1}$-TY)

In $\lambda_{L1}$, $\Gamma$ is now a finite map from variables to types that can be represented as a set of variable bindings $x : \mathtt{T}$. As usual, duplicated variable names are not allowed in a context $\Gamma$. The comma operator now denotes disjoint union for finite maps.

Term grammar for $\lambda_{L1}$ borrows most of simply-typed $\lambda$-calculus grammar. Elimination of unit type $\mathtt{1}$ is made with the $\fatsemi$ operator. Pattern-matching on sum and product types is made with the **case** keyword. Finally, we have new operators **dup**, **drop** and **derelict** for, respectively, contraction, weakening and dereliction of unrestricted terms with type $!\mathtt{T}$. In addition, the language includes a data constructor for unrestricted terms and values, denoted by $\mathtt{Many}\ t$ and $\mathtt{Many}\ v$. Promotion of a term to an unrestricted form is made by direct application of constructor $\mathtt{Many}$. For easier exposition, we allow ourselves to use syntactic sugar $\mathbf{let}\ x := t\ \mathbf{in}\ u$ and encode it as $(\boldsymbol{\lambda} x \mapsto u)\ (t)$.

The derivation of the linear implication $!\mathtt{T} \multimap (\mathtt{T} \otimes \mathtt{T})$ from previous section now translates into a typing derivation for a well-typed term of type $!\mathtt{T} \multimap (\mathtt{T} \otimes \mathtt{T})$. Let's see how it goes:

$$\dfrac{\dfrac{\dfrac{}{x : !\mathtt{T} \vdash x : !\mathtt{T}}\ \textsc{id} \qquad \dfrac{\dfrac{\dfrac{}{x_1 : !\mathtt{T} \vdash x_1 : !\mathtt{T}}\ \textsc{id}}{x_1 : !\mathtt{T} \vdash \mathbf{derelict}\ x_1 : \mathtt{T}}\ !\mathrm{D} \qquad \dfrac{\dfrac{}{x_2 : !\mathtt{T} \vdash x_2 : !\mathtt{T}}\ \textsc{id}}{x_2 : !\mathtt{T} \vdash \mathbf{derelict}\ x_2 : \mathtt{T}}\ !\mathrm{D}}{x_1 : !\mathtt{T},\ x_2 : !\mathtt{T} \vdash (\mathbf{derelict}\ x_1\ ,\ \mathbf{derelict}\ x_2) : \mathtt{T} \otimes \mathtt{T}}\ \otimes\mathrm{I}}{x : !\mathtt{T} \vdash \mathbf{dup}\ x\ \mathbf{as}\ x_1\ ,\ x_2\ \mathbf{in}\ (\mathbf{derelict}\ x_1\ ,\ \mathbf{derelict}\ x_2) : \mathtt{T} \otimes \mathtt{T}}\ !\mathrm{C}}{\cdot \vdash \boldsymbol{\lambda} x \mapsto \mathbf{dup}\ x\ \mathbf{as}\ x_1\ ,\ x_2\ \mathbf{in}\ (\mathbf{derelict}\ x_1\ ,\ \mathbf{derelict}\ x_2) : !\mathtt{T} \multimap (\mathtt{T} \otimes \mathtt{T})}\ \multimap\mathrm{I}$$

This translation is very direct. The only change is that we have to give distinct variable names to hypotheses of the same type inside the typing context, while there was no concept of named copies of a same formula in ILL. We also see that dealing with unrestricted hypotheses can be rather heavyweight because of the need of manual duplication and dereliction. Finally, in this system, substitution is not admissible, as detailled in ⟦Wadler, 1992⟧. Next section present a way to circumvent these issues.

## 1.3 Ergonomic handling of unrestricted resources

In the monadic presentation $\lambda_{L1}$, the use of unrestricted terms can become very verbose and cumbersome because of the need for explicit contraction, weakening and dereliction. A second and more powerful presentation of the linear $\lambda$-calculus, named *dyadic* presentation or $\lambda_{L2}$, tends to alleviate this burden by using two typing contexts on each judgment, one for linear variables and one for unrestricted variables. The syntax and typing rules of $\lambda_{L2}$ are given in Figures 1.4 and 1.5.

In $\lambda_{L2}$ there are no longer rules for contraction, weakening, and dereliction of unrestricted resources. Instead, each judgment is equipped with a second context $\mho$ that holds variable bindings that can be used in an unrestricted fashion. The $\mathbf{case}\ t\ \mathbf{of}\ \mathtt{Many}\ x \mapsto u$ construct let us access a term of type $!\mathtt{T}$ as a variable $x$ of type $\mathtt{T}$ that lives in the unrestricted context $\mho$. It's important to note that $\mathbf{case}\ t\ \mathbf{of}\ \mathtt{Many}\ x \mapsto u$ is not dereliction: one can still use $x$ several times within body $u$, or recreate $t$ by wrapping $x$ back as $\mathtt{Many}\ x$; while that would not be possible with $\mathbf{let}\ x := \mathbf{derelict}\ t\ \mathbf{in}\ u$ of $\lambda_{L1}$. Morally, we can understand the pair of contexts $\Gamma; \mho$ of $\lambda_{L2}$ as the single context $\Gamma,\ !\mho$ of $\lambda_{L1}$, where $!\mho$ is the context with the same variable bindings as $\mho$, except that all types are prefixed by $!$. This presentation has the advantage of making substitution admissible for $\lambda_{L2}$, which was not the case for $\lambda_{L1}$.

In $\lambda_{L2}$, contraction for unrestricted resources happens implicitly every time a rule has two subterms as premises. Indeed, the unrestricted context $\mho$ is duplicated in both premises, as we observe in particular in rule $\lambda_{L2}$–TY/$\otimes$I, unlike the linear context $\Gamma$ that must be split into two disjoint parts. All unrestricted

$$
\begin{array}{rcl}
v & ::= & \lambda x \mapsto u \mid () \mid \mathsf{Inl}\, v \mid \mathsf{Inr}\, v \mid (v_1\,,\, v_2) \mid \mathsf{Many}\, v \\
t, u & ::= & v \mid x \mid \mathsf{Inl}\, t \mid \mathsf{Inr}\, t \mid (t_1\,,\, t_2) \mid \mathsf{Many}\, t \mid t\; t' \mid t \,\fatsemi\, t' \\
& \mid & \mathbf{case}\; t\; \mathbf{of}\; \{\mathsf{Inl}\, x_1 \mapsto u_1\,,\; \mathsf{Inr}\, x_2 \mapsto u_2\} \mid \mathbf{case}\; t\; \mathbf{of}\; (x_1\,,\, x_2) \mapsto u \mid \mathbf{case}\; t\; \mathbf{of}\; \mathsf{Many}\, x \mapsto u \\
\\
\mathsf{T}, \mathsf{U} & ::= & \mathsf{T} \multimap \mathsf{U} \mid \mathsf{1} \mid \mathsf{T}_1 \oplus \mathsf{T}_2 \mid \mathsf{T}_1 \otimes \mathsf{T}_2 \mid\ !\mathsf{T} \\
\\
\Gamma & ::= & \bullet \mid x : \mathsf{T} \mid \Gamma_1,\ \Gamma_2 \\
\mho & ::= & \bullet \mid x : \mathsf{T} \mid \mho_1,\ \mho_2
\end{array}
$$

Figure 1.4: Grammar of linear λ-calculus in dyadic presentation ($\lambda_{L2}$)

$\boxed{\Gamma\,;\mho \vdash t : \mathsf{T}}$ *(Typing judgment for terms)*

$$
\frac{}{x : \mathsf{T}\,;\mho \vdash x : \mathsf{T}}\ \textsc{IdLin}
\qquad
\frac{}{\bullet\,;\mho,\ x : \mathsf{T} \vdash x : \mathsf{T}}\ \textsc{IdUr}
\qquad
\frac{\Gamma,\ x : \mathsf{T}\,;\mho \vdash u : \mathsf{U}}{\Gamma\,;\mho \vdash \lambda x \mapsto u : \mathsf{T} \multimap \mathsf{U}}\ {\multimap}\mathrm{I}
$$

$$
\frac{}{\bullet\,;\mho \vdash () : \mathsf{1}}\ \mathsf{1}\mathrm{I}
\qquad
\frac{\Gamma\,;\mho \vdash t_1 : \mathsf{T}_1}{\Gamma\,;\mho \vdash \mathsf{Inl}\, t_1 : \mathsf{T}_1 \oplus \mathsf{T}_2}\ \oplus\mathrm{I}_1
\qquad
\frac{\Gamma\,;\mho \vdash t_2 : \mathsf{T}_2}{\Gamma\,;\mho \vdash \mathsf{Inr}\, t_2 : \mathsf{T}_1 \oplus \mathsf{T}_2}\ \oplus\mathrm{I}_2
$$

$$
\frac{\Gamma_1\,;\mho \vdash t_1 : \mathsf{T}_1 \quad \Gamma_2\,;\mho \vdash t_2 : \mathsf{T}_2}{\Gamma_1,\ \Gamma_2\,;\mho \vdash (t_1\,,\, t_2) : \mathsf{T}_1 \otimes \mathsf{T}_2}\ \otimes\mathrm{I}
\qquad
\frac{\bullet\,;\mho \vdash t : \mathsf{T}}{\bullet\,;\mho \vdash \mathsf{Many}\, t : !\mathsf{T}}\ !\mathrm{I}
\qquad
\frac{\Gamma_1\,;\mho \vdash t : \mathsf{T} \quad \Gamma_2\,;\mho \vdash t' : \mathsf{T} \multimap \mathsf{U}}{\Gamma_1,\ \Gamma_2\,;\mho \vdash t'\; t : \mathsf{U}}\ {\multimap}\mathrm{E}
$$

$$
\frac{\Gamma_1\,;\mho \vdash t : \mathsf{1} \quad \Gamma_2\,;\mho \vdash u : \mathsf{U}}{\Gamma_1,\ \Gamma_2\,;\mho \vdash t \,\fatsemi\, u : \mathsf{U}}\ \mathsf{1}\mathrm{E}
\qquad
\frac{\Gamma_1\,;\mho \vdash t : \mathsf{T}_1 \oplus \mathsf{T}_2 \quad \Gamma_2,\ x_1 : \mathsf{T}_1\,;\mho \vdash u_1 : \mathsf{U} \quad \Gamma_2,\ x_2 : \mathsf{T}_2\,;\mho \vdash u_2 : \mathsf{U}}{\Gamma_1,\ \Gamma_2\,;\mho \vdash \mathbf{case}\; t\; \mathbf{of}\; \{\mathsf{Inl}\, x_1 \mapsto u_1\,,\; \mathsf{Inr}\, x_2 \mapsto u_2\} : \mathsf{U}}\ \oplus\mathrm{E}
$$

$$
\frac{\Gamma_1\,;\mho \vdash t : \mathsf{T}_1 \otimes \mathsf{T}_2 \quad \Gamma_2,\ x_1 : \mathsf{T}_1,\ x_2 : \mathsf{T}_2\,;\mho \vdash u : \mathsf{U}}{\Gamma_1,\ \Gamma_2\,;\mho \vdash \mathbf{case}\; t\; \mathbf{of}\; (x_1\,,\, x_2) \mapsto u : \mathsf{U}}\ \otimes\mathrm{E}
\qquad
\frac{\Gamma_1\,;\mho \vdash t : !\mathsf{T} \quad \Gamma_2\,;\mho,\ x : \mathsf{T} \vdash u : \mathsf{U}}{\Gamma_1,\ \Gamma_2\,;\mho \vdash \mathbf{case}\; t\; \mathbf{of}\; \mathsf{Many}\, x \mapsto u : \mathsf{U}}\ !\mathrm{E}
$$

Figure 1.5: Typing rules for linear λ-calculus in dyadic presentation ($\lambda_{L2}$–ty)

variable bindings are thus propagated to the leaves of the typing tree, that is, the rules with no premises $\lambda_{L2}$–TY/ID$_{\mathrm{LIN}}$, $\lambda_{L2}$–TY/ID$_{\mathrm{UR}}$, and $\lambda_{L2}$–TY/1I. These three rules discard all bindings of the unrestricted context $\mho$ that are not used, performing several implicit weakening steps.

Finally, this system has two identity rules. The first one, $\lambda_{L2}$–TY/ID$_{\mathrm{LIN}}$, is the usual linear identity: it consumes the variable $x$ present in the linear typing context. Because the linear typing context must be split into disjoint parts between sibling subterms (when a rule has several premises, like the product rule), it means $x$ can only be used in one of them. The second identity rule, $\lambda_{L2}$–TY/ID$_{\mathrm{UR}}$, is the unrestricted identity: it lets us use the variable $x$ from the unrestricted typing context in a place where a linear variable is expected, performing a sort of implicit dereliction. Because the unrestricted typing context is duplicated between sibling subterms, an unrestricted variable can be used in several of them. For instance, the following derivation is *not valid* because we do not respect the disjointness condition of the comma operator for the resulting typing context of the pair:

$$\dfrac{\dfrac{}{x' : \mathsf{T} \,;\, \bullet \vdash x' : \mathsf{T}}\ \mathrm{ID_{LIN}} \qquad \dfrac{}{x' : \mathsf{T} \,;\, \bullet \vdash x' : \mathsf{T}}\ \mathrm{ID_{LIN}}}{x' : \mathsf{T},\ x' : \mathsf{T} \,;\, \bullet \vdash (x' , x') : \mathsf{T} \otimes \mathsf{T}}\ \otimes\mathrm{I}$$

But the following derivation is valid, as the unrestricted typing context is duplicated between sibling subterms:

$$\dfrac{\dfrac{}{\bullet \,;\, x' : \mathsf{T} \vdash x' : \mathsf{T}}\ \mathrm{ID_{UR}} \qquad \dfrac{}{\bullet \,;\, x' : \mathsf{T} \vdash x' : \mathsf{T}}\ \mathrm{ID_{UR}}}{\bullet \,;\, x' : \mathsf{T} \vdash (x' , x') : \mathsf{T} \otimes \mathsf{T}}\ \otimes\mathrm{I}$$

At this point, we are very close to recreating a term of type $!\mathsf{T} \multimap (\mathsf{T} \otimes \mathsf{T})$ as we did in previous section. Let's finish the example:

$$\dfrac{\dfrac{\dfrac{}{x : !\mathsf{T} \,;\, \bullet \vdash x : !\mathsf{T}}\ \mathrm{ID_{LIN}} \qquad \dfrac{\dfrac{}{\bullet \,;\, x' : \mathsf{T} \vdash x' : \mathsf{T}}\ \mathrm{ID_{UR}} \qquad \dfrac{}{\bullet \,;\, x' : \mathsf{T} \vdash x' : \mathsf{T}}\ \mathrm{ID_{UR}}}{\bullet \,;\, x' : \mathsf{T} \vdash (x' , x') : \mathsf{T} \otimes \mathsf{T}}\ \otimes\mathrm{I}}{x : !\mathsf{T} \,;\, \bullet \vdash \mathtt{case}\ x\ \mathtt{of}\ \mathtt{Many}\ x' \mapsto (x' , x') : \mathsf{T} \otimes \mathsf{T}}\ !\mathrm{E}}{\bullet \,;\, \bullet \vdash \lambda x \mapsto \mathtt{case}\ x\ \mathtt{of}\ \mathtt{Many}\ x' \mapsto (x' , x') : !\mathsf{T} \multimap \mathsf{T} \otimes \mathsf{T}}\ \multimap\mathrm{I}$$

If we had any useless variable binding polluting our unrestricted typing context (let's say $\mho = y : !\mathsf{U}$), the derivation would still hold without any change at term level; the useless bindings would just be carried over throughout the typing tree, and eliminated at the leaves (via implicit weakening happening in leaf rules):

$$\dfrac{\dfrac{\dfrac{}{x : !\mathsf{T} \,;\, y : !\mathsf{U} \vdash x : !\mathsf{T}}\ \mathrm{ID_{LIN}} \qquad \dfrac{\dfrac{}{\bullet \,;\, y : !\mathsf{U},\ x' : \mathsf{T} \vdash x' : \mathsf{T}}\ \mathrm{ID_{UR}} \qquad \dfrac{}{\bullet \,;\, y : !\mathsf{U},\ x' : \mathsf{T} \vdash x' : \mathsf{T}}\ \mathrm{ID_{UR}}}{\bullet \,;\, y : !\mathsf{U},\ x' : \mathsf{T} \vdash (x' , x') : \mathsf{T} \otimes \mathsf{T}}\ \otimes\mathrm{I}}{x : !\mathsf{T} \,;\, y : !\mathsf{U} \vdash \mathtt{case}\ x\ \mathtt{of}\ \mathtt{Many}\ x' \mapsto (x' , x') : \mathsf{T} \otimes \mathsf{T}}\ !\mathrm{E}}{\bullet \,;\, y : !\mathsf{U} \vdash \lambda x \mapsto \mathtt{case}\ x\ \mathtt{of}\ \mathtt{Many}\ x' \mapsto (x' , x') : !\mathsf{T} \multimap \mathsf{T} \otimes \mathsf{T}}\ \multimap\mathrm{I}$$

Managing unrestricted hypotheses is, as we just demonstrated, much easier in the dyadic system $\lambda_{L2}$. Typing trees are shorter, and probably easier to read too.

Also, we do not loose anything by going from $\lambda_{L1}$ presentation to $\lambda_{L2}$: Andreoli ⟦1992⟧ has a detailed proof that $\lambda_{L1}$–TY and $\lambda_{L2}$–TY are equivalent, in other terms, that a program $t$ types in the pair of contexts $\Gamma ; \mho$ in $\lambda_{L2}$–TY if and only if it types in context $\Gamma,\ !\mho$ in $\lambda_{L1}$–TY.

## 1.4 Back to a single context with the graded modal approach

So far, we only considered type systems that are linear, but that are otherwise fairly standard with respect to usual simply-typed λ-calculus. Anticipating on our future needs, further in this document we will need to carry more information and restrictions throughout the type system than just linearity alone.

Naively, we could just multiply the typing contexts, for each new modality that we need. The problem is, that approach is not really scalable: at the end we need one context per possible combination of modalities (if modalities represent orthogonal principles), so it grows really quickly.

In fact, without thinking too far away, we already have a problem if we want a finer control over linearity. What if we want to allow variables to be used a fixed number of times that is not just one or unlimited? In ⟦Jean-Yves Girard, Scedrov, and Scott, 1992⟧, Girard considers the *bounded* extension of linear logic, where the number of uses of hypotheses can be restricted to any value $\mathtt{m}$—named *multiplicity*—instead of being just linear or unrestricted. For that he extends the $!$ modality with an index $\mathtt{m}$ that specifies how many times an hypothesis should be used. We say that $!_{\mathtt{m}}$ is a *graded modality*.

Having a family of modalities $(!_{\mathtt{m}})_{\mathtt{m} \in \mathtt{M}}$ with an arbitrary number of elements instead of the single $!$ modality of original linear logic means that we cannot really have a distinct context for each of them as in the dyadic presentation. One solution is to go back to the monadic presentation, where variables carry their modality on their type until the very end when they get used. We would also need a new operator to go from $!_{\mathtt{m}}\mathtt{T}$ to a pair $!_{\mathtt{m}-1}\mathtt{T} \otimes \mathtt{T}$ so that we can extract a single use from a value wrapped in a modality allowing several uses. As one might imagine, this approach becomes unpractical rather quickly.

Fortunately, there's a way out of this. Instead of having $\mathtt{m}$ be part of the modality and thus of the type of terms, we can bake it in as an annotation on variable bindings. In place of $x : \mathtt{S}, y : !_{\mathtt{m}}\mathtt{T}, z : !_{\mathtt{n}}\mathtt{U} \vdash \ldots$, we can have $x :_{\mathtt{1}} \mathtt{S}, y :_{\mathtt{m}} \mathtt{T}, z :_{\mathtt{n}} \mathtt{U} \vdash$. We will call the new annotations on bindings *modes*. Note that every binding will be equipped with a mode $\mathtt{m}$, even linear bindings that previously did not have modalities on their types[4]. In addition to adding *modes* to bindings, we will also define a few algebraic operations on modes (and later by extension on bindings and typing contexts) that aim to encode the (sub)structural rules of our system, in a simpler and more extensive manner, without needing explicit syntax for them like **derelict** or **dup** anymore.

With the modal approach, we recover a system, like the dyadic one, in which contraction, weakening, and dereliction can be made conveniently implicit, without loosing any control power over resource use, and with no explosion of the number of contexts!

Furthermore, we build on the key insight—originally proposed in ⟦Brunel et al., 2014⟧ and ⟦Ghica and Smith, 2014⟧—that equipping the set of modes with a semiring structure is sufficient to express, algebraically, all the typing context manipulations we need. The idea is to define two main operations on modes: *multiplication* $\cdot$, which models how modes are compounded across nested expressions or function calls (for example, if a function $f$ uses its argument with mode $2$, and $g$ also uses its argument with mode $2$, then $f\ (g\ x)$ uses $x$ with mode $2 \cdot 2$); and *addition* $+$, which accounts for what happens when a variable is used across multiple subterms (for example, if subterm $t$ uses $x$ with mode $2$ and $u$ uses $x$ with mode $3$, then $t \fatsemi u$ uses $x$ with mode $2 + 3$). By abstracting away the structural restrictions of the type system as a mode system with a semiring structure, we make the type system much more scalable: if we were to enrich the type system, and if we are able to express the extension as a new semiring, then we almost do not have to modify the typing rules; just the underlying mode semiring. We then lift the mode operators to variable bindings and typing contexts in the following way:

[4] This uniform approach where every binding receives a mode seems to originate in ⟦Ghica and Smith, 2014⟧ and ⟦Petricek, Orchard, and Mycroft, 2014⟧.

$$
\begin{array}{rcl}
v & ::= & \lambda x\,{}_{\mathsf{m}}\!\mapsto u \mid () \mid \mathsf{Inl}\,v \mid \mathsf{Inr}\,v \mid (v_1\,,\,v_2) \mid \mathsf{Mod}_{\mathsf{n}}\,v \\
t, u & ::= & v \mid x \mid \mathsf{Inl}\,t \mid \mathsf{Inr}\,t \mid (t_1\,,\,t_2) \mid \mathsf{Mod}_{\mathsf{n}}\,t \mid t\ t' \mid t \mathbin{;} t' \\
& \mid & \mathsf{case}\ t\ \mathsf{of}\ \{\mathsf{Inl}\,x_1 \mapsto u_1\,,\ \mathsf{Inr}\,x_2 \mapsto u_2\} \mid \mathsf{case}\ t\ \mathsf{of}\ (x_1\,,\,x_2) \mapsto u \mid \mathsf{case}\ t\ \mathsf{of}\ \mathsf{Mod}_{\mathsf{n}}\,x \mapsto u \\
\\
\mathsf{T}, \mathsf{U} & ::= & \mathsf{T}\ {}_{\mathsf{m}}\!\multimap \mathsf{U} \mid \mathsf{1} \mid \mathsf{T}_1 \oplus \mathsf{T}_2 \mid \mathsf{T}_1 \otimes \mathsf{T}_2 \mid !_{\mathsf{n}}\mathsf{T} \\
\mathsf{m}, \mathsf{n} & ::= & 1 \mid \omega \\
\\
\Gamma & ::= & \bullet \mid x :_{\mathsf{m}} \mathsf{T} \mid \Gamma_1,\ \Gamma_2
\end{array}
$$

Figure 1.6: Grammar of linear $\lambda$-calculus in modal presentation ($\lambda_{Lm}$)

$$
\begin{array}{rcl}
\mathsf{n} \cdot \bullet & \triangleq & \bullet \\
\mathsf{n} \cdot (x :_{\mathsf{m}} \mathsf{T},\ \Gamma) & \triangleq & (x :_{\mathsf{n}\cdot\mathsf{m}} \mathsf{T}),\ \mathsf{n}\cdot\Gamma
\end{array}
$$

$$
\begin{array}{rcll}
\bullet + \Gamma & \triangleq & \Gamma & \\
(x :_{\mathsf{m}} \mathsf{T},\ \Gamma_1) + \Gamma_2 & \triangleq & x :_{\mathsf{m}} \mathsf{T},\ (\Gamma_1 + \Gamma_2) & \text{if } x \notin \Gamma_2 \\
(x :_{\mathsf{m}} \mathsf{T},\ \Gamma_1) + (x :_{\mathsf{m}'} \mathsf{T},\ \Gamma_2) & \triangleq & x :_{\mathsf{m}+\mathsf{m}'} \mathsf{T},\ (\Gamma_1 + \Gamma_2) &
\end{array}
$$

We will now present the concrete *modal* formulation of the intuitionistic $\lambda$-calculus, which we denote $\lambda_{Lm}$. In fact, we do not use a full semiring here, since we do not require a zero element to encode linearity as we did in $\lambda_{L1}$ and $\lambda_{L2}$. Instead, we work with a *semiring without zero* to represent modes—henceforth still referred to as a *semiring*—which still retains all the algebraic benefits described in ⟦Ghica and Smith, 2014⟧. Our particular semiring consists of only two elements: $1$, representing variables that must be managed linearly, and $\omega$, representing unrestricted variables. We give the following operation tables:

| $+$ | $1$ | $\omega$ |
|---|---|---|
| $1$ | $\omega$ | $\omega$ |
| $\omega$ | $\omega$ | $\omega$ |

| $\cdot$ | $1$ | $\omega$ |
|---|---|---|
| $1$ | $1$ | $\omega$ |
| $\omega$ | $\omega$ | $\omega$ |

Now that we have a completely defined the semiring for modes, we can move to the grammar and typing rules of $\lambda_{Lm}$, given in Figures 1.6 and 1.7. The main change compared to $\lambda_{L2}$ is that the constructor for $!_{\mathsf{m}}\mathsf{T}$ is now $\mathsf{Mod}_{\mathsf{m}}\ t$ (instead of $\mathsf{Many}\ t$ for $!\mathsf{T}$).

In $\lambda_{Lm}$ we are back to a single identity rule $\lambda_{Lm}$–TY/ID, that asks for $x$ to be in the typing context with a mode $\mathsf{m}$ that must be *compatible with a single linear use* (we note $1 \lessapprox \mathsf{m}$). In our semiring, with only two elements, we have both $1 \lessapprox 1$ and $1 \lessapprox \omega$, so $x$ can actually have any mode (either linear or unrestricted, so encompassing both $\lambda_{L2}$–TY/ID$_{\text{LIN}}$ and $\lambda_{L2}$–TY/ID$_{\text{UR}}$), but in more complex modal systems, not all modes have to be compatible with the unit of the semiring[5]. Rules $\lambda_{Lm}$–TY/ID and $\lambda_{Lm}$–TY/1I also allow to discard any context composed only of unrestricted bindings, denoted by $\omega\cdot\Gamma$ (equivalent to notation $!\Gamma$ in $\lambda_{L1}$), so they are performing implicit weakening as in $\lambda_{L2}$.

Every rule of $\lambda_{Lm}$ that mentions two subterms uses the newly defined $+$ operator on typing contexts in the conclusion of the rule. If a same variable $x$ is required in both $\Gamma_1$ and $\Gamma_2$ (either with mode $1$ or $\omega$), then $\Gamma_1 + \Gamma_2$ will contain binding $x :_{\omega} \mathsf{T}$. Said differently, the superterm will automatically deduce whether $x$ needs to be linear or unrestricted based on how (many) subterms use $x$, thanks to the $+$ operator. It's

[5] See Section 2.5.2 for a case of a semiring where not all modes are compatible with the unit.

$\boxed{\Gamma \vdash t : \mathtt{T}}$ *(Typing judgment for terms)*

$$\frac{1 \lessdot \mathtt{m}}{\omega\cdot\Gamma,\ x :_{\mathtt{m}} \mathtt{T} \vdash x : \mathtt{T}}\ \text{ID} \qquad \frac{\Gamma,\ x :_{\mathtt{m}} \mathtt{T} \vdash u : \mathtt{U}}{\Gamma \vdash \lambda x\ {}_{\mathtt{m}}\!\mapsto u : \mathtt{T}\ {}_{\mathtt{m}}\!\multimap \mathtt{U}}\ {\multimap}\text{I} \qquad \frac{}{\omega\cdot\Gamma \vdash () : \mathtt{1}}\ \mathtt{1}\text{I}$$

$$\frac{\Gamma \vdash t_1 : \mathtt{T}_1}{\Gamma \vdash \mathtt{Inl}\, t_1 : \mathtt{T}_1 \oplus \mathtt{T}_2}\ \oplus\text{I}_1 \qquad \frac{\Gamma \vdash t_2 : \mathtt{T}_2}{\Gamma \vdash \mathtt{Inr}\, t_2 : \mathtt{T}_1 \oplus \mathtt{T}_2}\ \oplus\text{I}_2 \qquad \frac{\Gamma_1 \vdash t_1 : \mathtt{T}_1 \quad \Gamma_2 \vdash t_2 : \mathtt{T}_2}{\Gamma_1 + \Gamma_2 \vdash (t_1\,,\, t_2) : \mathtt{T}_1 \otimes \mathtt{T}_2}\ \otimes\text{I}$$

$$\frac{\Gamma \vdash t : \mathtt{T}}{\mathtt{n}\cdot\Gamma \vdash \mathtt{Mod}_{\mathtt{n}}\, t : !_{\mathtt{n}}\mathtt{T}}\ !\text{I} \qquad \frac{\Gamma_1 \vdash t : \mathtt{T} \quad \Gamma_2 \vdash t' : \mathtt{T}\ {}_{\mathtt{m}}\!\multimap \mathtt{U}}{\mathtt{m}\cdot\Gamma_1 + \Gamma_2 \vdash t'\ t : \mathtt{U}}\ {\multimap}\text{E} \qquad \frac{\Gamma_1 \vdash t : \mathtt{1} \qquad \Gamma_2 \vdash u : \mathtt{U}}{\Gamma_1 + \Gamma_2 \vdash t \,;\, u : \mathtt{U}}\ \mathtt{1}\text{E}$$

$$\frac{\Gamma_1 \vdash t : \mathtt{T}_1 \oplus \mathtt{T}_2 \quad \Gamma_2,\ x_1 :_{1} \mathtt{T}_1 \vdash u_1 : \mathtt{U} \quad \Gamma_2,\ x_2 :_{1} \mathtt{T}_2 \vdash u_2 : \mathtt{U}}{\Gamma_1 + \Gamma_2 \vdash \mathtt{case}\ t\ \mathtt{of}\ \{\mathtt{Inl}\, x_1 \mapsto u_1\,,\ \mathtt{Inr}\, x_2 \mapsto u_2\} : \mathtt{U}}\ \oplus\text{E} \qquad \frac{\Gamma_1 \vdash t : \mathtt{T}_1 \otimes \mathtt{T}_2 \quad \Gamma_2,\ x_1 :_{1} \mathtt{T}_1,\ x_2 :_{1} \mathtt{T}_2 \vdash u : \mathtt{U}}{\Gamma_1 + \Gamma_2 \vdash \mathtt{case}\ t\ \mathtt{of}\ (x_1\,,\, x_2) \mapsto u : \mathtt{U}}\ \otimes\text{E}$$

$$\frac{\Gamma_1 \vdash t : !_{\mathtt{n}}\mathtt{T} \quad \Gamma_2,\ x :_{\mathtt{n}} \mathtt{T} \vdash u : \mathtt{U}}{\Gamma_1 + \Gamma_2 \vdash \mathtt{case}\ t\ \mathtt{of}\ \mathtt{Mod}_{\mathtt{n}}\, x \mapsto u : \mathtt{U}}\ !\text{E}$$

Figure 1.7: Typing rules for linear λ-calculus in modal presentation ($\lambda_{Lm}$–TY)

a vastly different approach than the linear context split for subterms in $\lambda_{L1,2}$ and unrestricted context duplication for subterms in $\lambda_{L2}$. We would argue that, thanks to the algebraic properties of the mode semiring, it makes the presentation of the system smoother (at the theoretical level at least; implementing type checking for such a linear type system can be trickier).

Much as in $\lambda_{L2}$, the exponential modality $!_{\mathtt{m}}$ is eliminated by a $\mathtt{case}\ t\ \mathtt{of}\ \mathtt{Mod}_{\mathtt{m}}\, x \mapsto u$ expression, that binds the scrutinee to a new variable $x$ with mode $\mathtt{m}$. The original boxed value can be recreated if needed with $\mathtt{Mod}_{\mathtt{m}}\, x$; indeed, as in $\lambda_{L2}$, elimination rule for $!_{\mathtt{m}}$ is *not* dereliction.

The last specificity of this system is that the function arrow $\multimap$ now carries a mode $\mathtt{m}$ to which it binds its argument. Previously, in $\lambda_{L1,2}$, the argument of a function was always linear, meaning that the corresponding variable binding in the function body was linear. Here, in $\lambda_{Lm}$, we can bind the argument at any mode $\mathtt{m}$—and the rule $\lambda_{Lm}$–TY/$\multimap$E for function application reflects this: the typing context $\Gamma_1$ for the argument passed to the function is multiplied by $\mathtt{m}$ in the conclusion of the rule. Actually, this does not change the expressivity of the system compared to previous presentations. For instance, the function $\lambda x\ {}_{\mathtt{m}}\!\mapsto u : \mathtt{T}\ {}_{\mathtt{m}}\!\multimap \mathtt{U}$ can be encoded as $\lambda x' \mapsto \mathtt{case}\ x'\ \mathtt{of}\ \mathtt{Mod}_{\mathtt{m}}\, x \mapsto u : !_{\mathtt{m}}\mathtt{T} \multimap \mathtt{U}$ using just the purely linear function arrow (at call site, we simply wrap the argument explicitly in an exponential, i.e., replacing $f\ x$ by $f'\ (\mathtt{Mod}_{\mathtt{m}}\, x)$). But since every variable binding must be assigned a mode in this presentation anyway, it is natural to allow this mode $\mathtt{m}$ to be whatever the programmer chooses—not just defaulting to $1$— thereby reducing the noise of explicit wrapping and unwrapping from modalities.

Let's update our running example. The term syntax does not change, we only see an evolution in typing contexts (as we now have only one, with mode annotations, instead of two):

$$\dfrac{\dfrac{1 \lesssim 1}{y :_{\omega} \mathtt{U},\ x :_{1} \mathtt{!T} \vdash x : \mathtt{!T}}\text{ID} \qquad \dfrac{\dfrac{1 \lesssim 1}{y :_{\omega} \mathtt{U},\ x' :_{1} \mathtt{T} \vdash x' : \mathtt{T}}\text{ID} \quad \dfrac{1 \lesssim 1}{y :_{\omega} \mathtt{U},\ x' :_{1} \mathtt{T} \vdash x' : \mathtt{T}}\text{ID}}{y :_{\omega} \mathtt{U},\ x' :_{\omega} \mathtt{T} \vdash (x' , x') : \mathtt{T} \otimes \mathtt{T}}\otimes\text{I}}{\dfrac{y :_{\omega} \mathtt{U},\ x :_{1} \mathtt{!T} \vdash \mathbf{case}\ x\ \mathbf{of}\ \mathrm{Mod}_{\omega}\ x' \mapsto (x' , x') : \mathtt{T} \otimes \mathtt{T}}{y :_{\omega} \mathtt{U} \vdash \lambda x\ {}_{1}{\mapsto}\ \mathbf{case}\ x\ \mathbf{of}\ \mathrm{Mod}_{\omega}\ x' \mapsto (x' , x') : \mathtt{!T}\ {}_{1}{\multimap}\ \mathtt{T} \otimes \mathtt{T}}\multimap\text{I}}\ !\text{E}$$

In $\lambda_{Lm}$, we are only forced to carry the unrestricted, unused part of the typing context ($y :_{\omega} \mathtt{U}$ in this case) towards at least one of the leaves, not all of them, thanks to the way the $+$ works on contexts. For instance, the following is equally valid:

$$\dfrac{\dfrac{1 \lesssim 1}{y :_{\omega} \mathtt{U},\ x :_{1} \mathtt{!T} \vdash x : \mathtt{!T}}\text{ID} \qquad \dfrac{\dfrac{1 \lesssim 1}{x' :_{1} \mathtt{T} \vdash x' : \mathtt{T}}\text{ID} \quad \dfrac{1 \lesssim 1}{x' :_{1} \mathtt{T} \vdash x' : \mathtt{T}}\text{ID}}{x' :_{\omega} \mathtt{T} \vdash (x' , x') : \mathtt{T} \otimes \mathtt{T}}\otimes\text{I}}{\dfrac{y :_{\omega} \mathtt{U},\ x :_{1} \mathtt{!T} \vdash \mathbf{case}\ x\ \mathbf{of}\ \mathrm{Mod}_{\omega}\ x' \mapsto (x' , x') : \mathtt{T} \otimes \mathtt{T}}{y :_{\omega} \mathtt{U} \vdash \lambda x\ {}_{1}{\mapsto}\ \mathbf{case}\ x\ \mathbf{of}\ \mathrm{Mod}_{\omega}\ x' \mapsto (x' , x') : \mathtt{!T}\ {}_{1}{\multimap}\ \mathtt{T} \otimes \mathtt{T}}\multimap\text{I}}\ !\text{E}$$

We also could have chosen to type $x'$ at mode $\omega$ in the two identical instances of the $\lambda_{Lm}$–TY/ID rule at the top of the tree; it has no impact on the rest of the derivation as both $1 + 1$ and $\omega + \omega$ gives $\omega$ as a result. The fact that we now have a few choices to make when typing a term in $\lambda_{L1}$–TY—which we did not have in $\lambda_{L2}$–TY—is the only non-trivial aspect when attempting to show that the two systems are equivalent.

Finally, we can give an alternative and more concise version of this function, taking advantage of the *mode on function arrow* feature that we just discussed above. This new version has type $\mathtt{T}\ {}_{\omega}{\multimap}\ \mathtt{T} \otimes \mathtt{T}$ instead of $\mathtt{!T}\ {}_{1}{\multimap}\ \mathtt{T} \otimes \mathtt{T}$:

$$\dfrac{\dfrac{\dfrac{1 \lesssim 1}{y :_{\omega} \mathtt{U},\ x' :_{1} \mathtt{T} \vdash x' : \mathtt{T}}\text{ID} \quad \dfrac{1 \lesssim 1}{y :_{\omega} \mathtt{U},\ x' :_{1} \mathtt{T} \vdash x' : \mathtt{T}}\text{ID}}{y :_{\omega} \mathtt{U},\ x' :_{\omega} \mathtt{T} \vdash (x' , x') : \mathtt{T} \otimes \mathtt{T}}\otimes\text{I}}{y :_{\omega} \mathtt{U} \vdash \lambda x'\ {}_{\omega}{\mapsto}\ (x' , x') : \mathtt{T}\ {}_{\omega}{\multimap}\ \mathtt{T} \otimes \mathtt{T}}\multimap\text{I}$$

## 1.5 Deep modes: projecting modes through fields of data structures

In original linear logic from Girard (see ILL and presentation $\lambda_{L1}$ above), there is only a one-way morphism between $\mathtt{!T} \otimes \mathtt{!U}$ and $\mathtt{!(T \otimes U)}$; we cannot go from $\mathtt{!(T \otimes U)}$ to $\mathtt{!T} \otimes \mathtt{!U}$. In other terms, an unrestricted pair $\mathtt{!(T \otimes U)}$ does not allow for unrestricted use of its components. The pair can be duplicated or discarded at will, but to be used, it needs to be derelicted first, to become $\mathtt{T} \otimes \mathtt{U}$, that no longer allow to duplicate or discard any of $\mathtt{T}$ or $\mathtt{U}$. As a result, $\mathtt{T}$ and $\mathtt{U}$ will have to be used exactly the same number of times, even though they are part of an unrestricted pair!

Situation is no different in $\lambda_{L2}$ (resp. $\lambda_{Lm}$): although the pair of type $\mathtt{!(T \otimes U)}$ can be bound in an unrestricted binding $x : \mathtt{T} \otimes \mathtt{U} \in \mho$ (resp. $x :_{\omega} \mathtt{T} \otimes \mathtt{U}$ in $\lambda_{Lm}$) and be used as this without need for dereliction, when it will be pattern-matched on (through the rule $\lambda_{L2,m}$–TY/$\otimes$E), implicit dereliction will still happen, and linear-only bindings will be made for its components: $x_1 : \mathtt{T},\ x_2 : \mathtt{U}$ in the linear typing context (resp. $x_1 :_1 \mathtt{T},\ x_2 :_1 \mathtt{U}$ in $\lambda_{Lm}$).

Let's say we want to write the **fst** function, that extracts the first element of a pair. In a non-linear language, we would probably write it this way:

$$\mathtt{fst_{nl}}\ :\ \mathtt{T}\otimes\mathtt{U}\to\mathtt{T}$$
$$\mathtt{fst_{nl}}\ \triangleq\ \mathtt{case}\ \lambda x \mapsto x\ \mathtt{of}\ (x_1\,,\,x_2)\mapsto x_1$$

But in $\lambda_{Lm}$, or even $\lambda_{L1,2}$, we are not allowed to do this. A naive idea would be to give the function the signature $\mathtt{T}\otimes\mathtt{U}\ {}_{\omega}\!\multimap \mathtt{T}$, but that still does not help: as we just said, even with an unrestricted pair, we have to consume both elements of the pair the same number of times. The only valid solution is to indicate, in the signature, that the second element of the pair—that we discard—will not be used linearly, by wrapping it in a $!_{\omega}$ modality:

$$\mathtt{fst_{m}}\ :\ \mathtt{T}\otimes(!_{\omega}\mathtt{U})\ {}_{1}\!\multimap \mathtt{T}$$
$$\mathtt{fst_{m}}\ \triangleq\ \mathtt{case}\ \lambda x\ {}_{1}\!\mapsto x\ \mathtt{of}\ (x_1\,,\,ux_2)\mapsto$$
$$\mathtt{case}\ ux_2\ \mathtt{of}\ \mathtt{Mod}_{\omega}\ x_2\ \mapsto x_1$$

In a real programming context, where linearity is not always the main concern, it gets quite tiresome to have to insert $!_{\omega}$ modalities everywhere in the types, and $\mathtt{case}\ t\ \mathtt{of}\ \mathtt{Mod_m}\ x\ \mapsto \ldots$ everywhere in the terms. Ideally we would like programs valid in the usual λ-calculus to have a valid interpretation in our linear λ-calculus with almost no change required, especially to the syntax.

We can do that with *deep modes*. Deep modes is a common feature in practical modal languages, that consists in allowing modes or their corresponding graded modalities to commute with other data types such as sums and products. It has, for instance, been implemented in Linear Haskell [[Bernardy et al., 2018]], Granule [[Orchard, Liepelt, and Eades III, 2019a]], and the more recent Modal OCaml [[Lorenzen, White, et al., 2024]]. It has also been discussed in [[Jack Hughes, Daniel Marshall, et al., 2021; Jack Hughes, Vollmer, and Orchard, 2021]].

With deep modes, we obtain the following equivalences:

$$\begin{aligned} !_{\mathtt{m}}(\mathtt{T}\otimes\mathtt{U}) &\simeq (!_{\mathtt{m}}\mathtt{T})\otimes(!_{\mathtt{m}}\mathtt{U}) \\ !_{\mathtt{m}}(\mathtt{T}\oplus\mathtt{U}) &\simeq (!_{\mathtt{m}}\mathtt{T})\oplus(!_{\mathtt{m}}\mathtt{U}) \\ !_{\mathtt{m}}(!_{\mathtt{n}}\mathtt{T}) &\simeq !_{(\mathtt{m}\cdot\mathtt{n})}\mathtt{T} \end{aligned}$$

In our modal linear λ-calculus extended with deep modes (further denoted $\lambda_{Ldm}$), functions of type $\mathtt{T}\ {}_{\omega}\!\multimap \mathtt{U}$ can represent all the ones having type $\mathtt{T}\to\mathtt{U}$ in STLC. If we conveniently use compatible syntax for terms and types in both systems, then there is almost no change required to go from a function valid in STLC to one valid in $\lambda_{Ldm}$; we just need to add a few modes annotations (both in terms and types). This contrasts with a faithful interpretation of Girard's linear logic, like we just showed before with $\lambda_{Lm}$, where we need to introduce (and then destruct) modalities explicitly whenever a non-linear control flow occurs (as in $\mathtt{fst_m}$).

Revisiting this **fst** function, with deep modes in $\lambda_{Ldm}$, we are able to write:

$$\mathtt{fst_{dm}}\ :\ \mathtt{T}\otimes\mathtt{U}\ {}_{\omega}\!\multimap \mathtt{T}$$
$$\mathtt{fst_{dm}}\ \triangleq\ \mathtt{case}_{\omega}\ \lambda x\ {}_{\omega}\!\mapsto x\ \mathtt{of}\ (x_1\,,\,x_2)\mapsto x_1$$

This is precisely the same implementation as $\mathtt{fst_{nl}}$, with just extra mode annotations!

$\boxed{\Gamma \vdash t : \mathsf{T}}$ *(Typing judgment for terms)*

$$\frac{\Gamma_1 \vdash t : \mathsf{T}_1 \oplus \mathsf{T}_2 \qquad \Gamma_2,\ x_1 :_{\mathsf{m}} \mathsf{T}_1 \vdash u_1 : \mathsf{U} \qquad \Gamma_2,\ x_2 :_{\mathsf{m}} \mathsf{T}_2 \vdash u_2 : \mathsf{U}}{\mathsf{m} \cdot \Gamma_1 + \Gamma_2 \vdash \mathbf{case}_{\mathsf{m}}\ t\ \mathbf{of}\ \{\mathsf{Inl}\, x_1 \mapsto u_1\ ,\ \mathsf{Inr}\, x_2 \mapsto u_2\} : \mathsf{U}}\ \oplus\mathrm{E}$$

$$\frac{\Gamma_1 \vdash t : \mathsf{T}_1 \otimes \mathsf{T}_2 \qquad \Gamma_2,\ x_1 :_{\mathsf{m}} \mathsf{T}_1,\ x_2 :_{\mathsf{m}} \mathsf{T}_2 \vdash u : \mathsf{U}}{\mathsf{m} \cdot \Gamma_1 + \Gamma_2 \vdash \mathbf{case}_{\mathsf{m}}\ t\ \mathbf{of}\ (x_1\ ,\ x_2) \mapsto u : \mathsf{U}}\ \otimes\mathrm{E} \qquad \frac{\Gamma_1 \vdash t : !_{\mathsf{n}} \mathsf{T} \qquad \Gamma_2,\ x :_{\mathsf{m} \cdot \mathsf{n}} \mathsf{T} \vdash u : \mathsf{U}}{\mathsf{m} \cdot \Gamma_1 + \Gamma_2 \vdash \mathbf{case}_{\mathsf{m}}\ t\ \mathbf{of}\ \mathsf{Mod}_{\mathsf{n}}\, x \mapsto u : \mathsf{U}}\ !\mathrm{E}$$

Figure 1.8: Typing rules for deep modes in linear λ-calculus in modal presentation ($\lambda_{Ldm}$–TY)

**Formal changes in the grammar and typing rules for deep modes** The only change needed on the grammar between $\lambda_{Lm}$ and $\lambda_{Ldm}$ is that the **case** constructs in $\lambda_{Ldm}$ take a mode $\mathsf{m}$ to which they consume the scrutinee , which is propagated to the variable bindings for the field(s) of the scrutinee in the body of the **case**. The new typing rules for **case** are presented in Figure 1.8, all the other rules of linear λ-calculus with deep modes, $\lambda_{Ldm}$, are identical to those of Figure 1.7.

For these new **case** rules, we observe that the typing context $\Gamma_1$ in which the scrutinee types is scaled by $\mathsf{m}$ in the conclusion of these rules. This is very similar to the application rule $\lambda_{Lm}$–TY/⊸E : it makes sure that the resources required to type $t$ are consumed $\mathsf{m}$ times if we want to use $t$ at mode $\mathsf{m}$. Given that $x_2 :_1 \mathsf{T}_2 \vdash ((), x_2) : \mathbf{1} \otimes \mathsf{T}_2$ (the pair uses $x_2$ exactly once), if we want to extract the pair components with mode $\omega$ to drop $x_2$ (as in $\mathbf{fst}_{\mathbf{dm}}$), then $\mathbf{case}_{\omega}\ ((), x_2)\ \mathbf{of}\ (x_1, x_2) \mapsto x_2$ will require context $\omega \cdot (x_2 :_1 \mathsf{T}_2)$ i.e. $x_2 :_{\omega} \mathsf{T}_2$. In other terms, we cannot use parts of a structure having dependencies on linear variables in an unrestricted way (as that would break linearity).

The linear λ-calculus with deep modes, $\lambda_{Ldm}$, will be the basis for our core contribution that follows in next chapter: the destination calculus $\lambda_d$.

## 1.6 Operational semantics of linear λ-calculus (with deep modes)

Many semantics presentations exist for the λ-calculus. In Figure 1.9 we present a small-step reduction system for $\lambda_{Ldm}$, in *reduction semantics* style inspired from [Felleisen, 1987] and subsequent [Biernacka and Danvy, 2007; Danvy and Nielsen, 2004], that is, a semantics in which the evaluation context $E$ is manipulated explicitly and syntactically as a stack. We define an evaluation context $E$, that is, a stack, as either the empty one $[\,]$, or a composition $E' \circ e$ where $e$—at the top of the stack—represents the innermost *evaluation context component* (i.e. a term missing a subterm to be complete).

We represent a running program by a pair $E[t]$ of an evaluation context $E$, and a term $t$ under focus. We call such a pair $E[t]$ a *command*, borrowing the terminology from Curien and Herbelin [2000].

Small-step evaluation rules are of three kinds:

- focusing rules (F) that split the current term under focus in two parts: an evaluation context component that is pushed on the stack $E$ for later use, and a subterm that is put under focus;
- unfocusing rules (U) that recreate a larger term once the term under focus is a value, by popping the most recent evaluation component from the stack and merging it with the value;

$\boxed{E\,[\,t\,] \longrightarrow E'\,[\,t'\,]}$ *(Small-step evaluation)*

$$E\,[\,\mathsf{Inl}\;t\,] \longrightarrow (E \circ \mathsf{Inl}\;[\,])\,[\,t\,] \qquad \star \quad \oplus\mathrm{I}_1\mathrm{F}$$

$$(E \circ \mathsf{Inl}\;[\,])\,[\,v\,] \longrightarrow E\,[\,\mathsf{Inl}\;v\,] \qquad \oplus\mathrm{I}_1\mathrm{U}$$

$$E\,[\,\mathsf{Inr}\;t\,] \longrightarrow (E \circ \mathsf{Inr}\;[\,])\,[\,t\,] \qquad \star \quad \oplus\mathrm{I}_2\mathrm{F}$$

$$(E \circ \mathsf{Inr}\;[\,])\,[\,v\,] \longrightarrow E\,[\,\mathsf{Inr}\;v\,] \qquad \oplus\mathrm{I}_2\mathrm{U}$$

$$E\,[\,(t_1\,,\,t_2)\,] \longrightarrow (E \circ ([\,]\,,\,t_2))\,[\,t_1\,] \qquad \star \quad \otimes\mathrm{IF}_1$$

$$(E \circ ([\,]\,,\,t_2))\,[\,v_1\,] \longrightarrow E\,[\,(v_1\,,\,t_2)\,] \qquad \otimes\mathrm{IU}_1$$

$$E\,[\,(v_1\,,\,t_2)\,] \longrightarrow (E \circ (v_1\,,\,[\,]))\,[\,t_2\,] \qquad \star \quad \otimes\mathrm{IF}_2$$

$$(E \circ (v_1\,,\,[\,]))\,[\,v_2\,] \longrightarrow E\,[\,(v_1\,,\,v_2)\,] \qquad \otimes\mathrm{IU}_2$$

$$E\,[\,\mathsf{Mod}_n\;t\,] \longrightarrow (E \circ \mathsf{Mod}_n\;[\,])\,[\,t\,] \qquad \star \quad !\mathrm{IF}$$

$$(E \circ \mathsf{Mod}_n\;[\,])\,[\,v\,] \longrightarrow E\,[\,\mathsf{Mod}_n\;v\,] \qquad !\mathrm{IU}$$

$$E\,[\,t'\;t\,] \longrightarrow (E \circ t'\;[\,])\,[\,t\,] \qquad \star \quad \multimap\mathrm{EF}_1$$

$$(E \circ t'\;[\,])\,[\,v\,] \longrightarrow E\,[\,t'\;v\,] \qquad \multimap\mathrm{EU}_1$$

$$E\,[\,t'\;v\,] \longrightarrow (E \circ [\,]\;v)\,[\,t'\,] \qquad \star \quad \multimap\mathrm{EF}_2$$

$$(E \circ [\,]\;v)\,[\,v'\,] \longrightarrow E\,[\,v'\;v\,] \qquad \multimap\mathrm{EU}_2$$

$$E\,[\,(\lambda x\;{}_{m}\!\mapsto u)\;v\,] \longrightarrow E\,[\,u[x := v]\,] \qquad \multimap\mathrm{EC}$$

$$E\,[\,t \fatsemi u\,] \longrightarrow (E \circ [\,] \fatsemi u)\,[\,t\,] \qquad \star \quad 1\mathrm{EF}$$

$$(E \circ [\,] \fatsemi u)\,[\,v\,] \longrightarrow E\,[\,v \fatsemi u\,] \qquad 1\mathrm{EU}$$

$$E\,[\,() \fatsemi u\,] \longrightarrow E\,[\,u\,] \qquad 1\mathrm{EC}$$

$$E\,[\,\mathsf{case}_m\;t\;\mathsf{of}\;\{\mathsf{Inl}\,x_1 \mapsto u_1\,,\;\mathsf{Inr}\,x_2 \mapsto u_2\}\,] \longrightarrow (E \circ \mathsf{case}_m\;[\,]\;\mathsf{of}\;\{\mathsf{Inl}\,x_1 \mapsto u_1\,,\;\mathsf{Inr}\,x_2 \mapsto u_2\})\,[\,t\,] \qquad \star \quad \oplus\mathrm{EF}$$

$$(E \circ \mathsf{case}_m\;[\,]\;\mathsf{of}\;\{\mathsf{Inl}\,x_1 \mapsto u_1\,,\;\mathsf{Inr}\,x_2 \mapsto u_2\})\,[\,v\,] \longrightarrow E\,[\,\mathsf{case}_m\;v\;\mathsf{of}\;\{\mathsf{Inl}\,x_1 \mapsto u_1\,,\;\mathsf{Inr}\,x_2 \mapsto u_2\}\,] \qquad \oplus\mathrm{EU}$$

$$E\,[\,\mathsf{case}_m\;(\mathsf{Inl}\,v_1)\;\mathsf{of}\;\{\mathsf{Inl}\,x_1 \mapsto u_1\,,\;\mathsf{Inr}\,x_2 \mapsto u_2\}\,] \longrightarrow E\,[\,u_1[x_1 := v_1]\,] \qquad \oplus\mathrm{EC}_1$$

$$E\,[\,\mathsf{case}_m\;(\mathsf{Inr}\,v_2)\;\mathsf{of}\;\{\mathsf{Inl}\,x_1 \mapsto u_1\,,\;\mathsf{Inr}\,x_2 \mapsto u_2\}\,] \longrightarrow E\,[\,u_2[x_2 := v_2]\,] \qquad \oplus\mathrm{EC}_2$$

$$E\,[\,\mathsf{case}_m\;t\;\mathsf{of}\;(x_1\,,\,x_2) \mapsto u\,] \longrightarrow (E \circ \mathsf{case}_m\;[\,]\;\mathsf{of}\;(x_1\,,\,x_2) \mapsto u)\,[\,t\,] \qquad \star \quad \otimes\mathrm{EF}$$

$$(E \circ \mathsf{case}_m\;[\,]\;\mathsf{of}\;(x_1\,,\,x_2) \mapsto u)\,[\,v\,] \longrightarrow E\,[\,\mathsf{case}_m\;v\;\mathsf{of}\;(x_1\,,\,x_2) \mapsto u\,] \qquad \otimes\mathrm{EU}$$

$$E\,[\,\mathsf{case}_m\;(v_1\,,\,v_2)\;\mathsf{of}\;(x_1\,,\,x_2) \mapsto u\,] \longrightarrow E\,[\,u[x_1 := v_1][x_2 := v_2]\,] \qquad \otimes\mathrm{EC}$$

$$E\,[\,\mathsf{case}_m\;t\;\mathsf{of}\;\mathsf{Mod}_n\,x \mapsto u\,] \longrightarrow (E \circ \mathsf{case}_m\;[\,]\;\mathsf{of}\;\mathsf{Mod}_n\,x \mapsto u)\,[\,t\,] \qquad \star \quad !\mathrm{EF}$$

$$(E \circ \mathsf{case}_m\;[\,]\;\mathsf{of}\;\mathsf{Mod}_n\,x \mapsto u)\,[\,v\,] \longrightarrow E\,[\,\mathsf{case}_m\;v\;\mathsf{of}\;\mathsf{Mod}_n\,x \mapsto u\,] \qquad !\mathrm{EU}$$

$$E\,[\,\mathsf{case}_m\;\mathsf{Mod}_n\,v\;\mathsf{of}\;\mathsf{Mod}_n\,x \mapsto u\,] \longrightarrow E\,[\,u[x := v]\,] \qquad !\mathrm{EC}$$

$\star$ : only allowed if the term that would become the new focus is not already a value

Figure 1.9: Small-step semantics for $\lambda_{Ldm}$ ($\lambda_{Ldm}$-SEM)

- contraction rules (C) that do not operate on the stack $E$ but just transform the term under focus when it's a redex (following a terminology introduced in [Biernacka and Danvy, 2007; Danvy and Nielsen, 2004]).

To have a fully deterministic and straightforward reduction system, focusing rules can only trigger when the subterm that would be focused is not already a value (denoted by $\star$ in Figure 1.9).

Data constructors only have focusing and unfocusing rules, as they do not describe computations that can be reduced. In that regard, $\mathsf{Mod}_{m}\ t$ is treated as an unary data constructor. Once a data constructor is focused, it is evaluated fully to a value form.

In most cases, unfocusing and contraction rules could be merged into a single step. For example, a contraction could be triggered as soon as we have $\left(E \circ \mathbf{case}_{m}\ [\,]\ \mathbf{of}\ (x_1\, ,\, x_2) \mapsto u\right)\left[(v_1\, ,\, v_2)\right]$, without the need to recreate the term $\mathbf{case}_{m}\ (v_1\, ,\, v_2)\ \mathbf{of}\ (x_1\, ,\, x_2) \mapsto u$. However, we prefer the presentation in three distinct steps, that despite being more verbose, clearly shows that contraction rules do not modify the evaluation context.

The rest of the system is very standard. In particular, contraction rules of $\lambda_{Ldm}$—that capture the essence of the calculus—are very similar to those of lambda-calculi with sum and product types (and strict evaluation strategy).

For instance, let's see the reduction for function application. We reuse the same example of our function with signature $!\mathsf{T} \multimap (\mathsf{T} \otimes \mathsf{T})$, here applied to the argument $\mathsf{Mod}_{\omega}\ (\mathsf{Inl}\ ())$:

$$
\begin{array}{llll}
 & [\,]\left[(\lambda x \ {}_{1}\!\mapsto \mathbf{case}\ x\ \mathbf{of}\ \mathsf{Mod}_{\omega}\ x' \mapsto (x'\, ,\, x'))\ (\mathsf{Mod}_{\omega}\ (\mathsf{Inl}\ ()))\right] & & \\
\longrightarrow & [\,]\left[(\mathbf{case}\ \mathsf{Mod}_{\omega}\ (\mathsf{Inl}\ ())\ \mathbf{of}\ \mathsf{Mod}_{\omega}\ x' \mapsto (x'\, ,\, x'))\right] & \multimap\text{EC} & \text{with substitution } x := \mathsf{Mod}_{\omega}\ (\mathsf{Inl}\ ()) \\
\longrightarrow & [\,]\left[(\mathsf{Inl}\ ()\, ,\, \mathsf{Inl}\ ())\right] & !\text{EC} & \text{with substitution } x' := \mathsf{Inl}\ ()
\end{array}
$$

Because we have a function in value form already, with simple body, applied to a value, we do not need to focus into a subterm and unfocus back later; we see only see a succession of contraction steps.

A command of the form $[\,]\left[v\right]$ with $v$ a value is the stopping point of the reduction for all well-typed programs. If we have a non-value term $t$ in $[\,]\left[t\right]$ then a focusing or contraction step should trigger; and if we have a non-empty evaluation context $E$ in $E\left[v\right]$ then an unfocusing step should trigger. Only ill-shaped term or value can cause the reduction to be stuck, which is a symptom of a wrongly-typed program.

## 1.7 Linear types in Haskell

As part of this PhD work, we intend to implement prototypes in an industrial functional programming language as we explore the destination-passing world. Haskell is a suitable target for that, thanks to its support for linear types. In what follows, we give the reader a quick overview of Linear Haskell and how the concepts discussed above map into the Haskell world.

Linear Haskell [Bernardy et al., 2018] is a language extension for the Glasgow Haskell Compiler (GHC) that introduces the linear function arrow, `t ⊸ u` and modifies the type checker of GHC so that it can enforce linearity requirements. A linear function of type `t ⊸ u` guarantees that the argument of the function will be consumed exactly once when the result of the function is consumed exactly once. The regular function arrow `t -> u` is still available when Linear Haskell is enabled; but does not guarantee how many times its argument will be consumed when its result is consumed once. Actually Linear Haskell is based on a modal formalism, like $\lambda_{Ldm}$, so there is also a multiplicity-polymorphic arrow, `t %m -> u`, much like $\mathsf{T}\ {}_{m}\!\multimap \mathsf{U}$; so `t ⊸ u` and `t -> u` are sugar for `t %1 -> u` and `t %ω -> u` respectively.

A value is said to be *consumed once* (or *consumed linearly*) when it is pattern-matched on and its sub-components are consumed once; or when it is passed as an argument to a linear function whose result is consumed once. A function is said to be *consumed once* when it is applied to an argument and when the result is consumed exactly once. We say that a variable `x` is *used linearly* in an expression `u` when consuming `u` once implies consuming `x` exactly once. Conversely, a value that is not used linearly is said to be *used non-linearly* or *used in an unrestricted fashion*.

**Unrestricted Values** Linear Haskell introduces a wrapper named `Ur` which is used to indicate that a value in a linear context does not have to be used linearly. `Ur t` is the Haskell equivalent for `!T` in $\lambda_{L1,2}$ or $!_\omega$`T` in $\lambda_{Ldm}$, and there is an equivalence between `Ur t ⊸ u` and `t → u`. As in $\lambda_{L2}$ and $\lambda_{Ldm}$, we can pattern-match on a term of type `Ur t` with `case term of Ur x ⊸ term'` to obtain an unrestricted variable binding `x` of type `t`.

As usual with linear type systems, only values already wrapped in `Ur` or coming from the left of a non-linear arrow can be put in another `Ur` without breaking linearity. This echoes rules $\lambda_{L1}$–TY/!P or $\lambda_{L2,m}$–TY/!I where a term wrapped in `Many` or `Mod`$_\omega$ must only depend on an unrestricted context. The only exceptions to this rule are terms of types that implement the `Movable` typeclass[6] such as `Int` or `()`. `Movable` provides the function `move :: t ⊸ Ur t`, that is, unconditional promotion to `Ur` for types implementing this typeclass.

**Operators** Some Haskell operators and notations are often used in the rest of this article:

`(⨾) :: () ⊸ u ⊸ u` is used to chain a linear operation returning `()` with one returning a value of type `u` without breaking linearity. It's the Haskell equivalent of $\lambda_{Ldm}$–TY/1E. It is not part of the core Haskell syntax, but can be defined simply as `case term1 of () -> term2`.

`Class ⇒ ...` is notation for typeclass constraints (resolved implicitly by the compiler).

`@t` in `f @t ...` is a type application; it allows to specify an explicit type for polymorphic functions.

In code excerpts, types are in blue, while terms, variables, and data constructors (sometimes having the same name as the type they belong to, like `Ur` (type) / `Ur` (data constructor)) are in black, to stay as consistent as possible with previous chapters. Unfortunately, while type variables `T`, `U` are in uppercase in our formalism, we have no choice than to use lowercase letters in Haskell `t`, `u`.

## 1.8 Uniqueness and Linear Scopes

In the introduction, we said that we will use linear type systems to control the number of times a resource—in particular a destination—is used. A more usual approach to resource management in functional programming is to encapsulate resources within monads. While monads can ensure controlled access, they are contaminating: once a computation touches a monadic value, the surrounding code is typically forced to adopt the same monadic structure, even if it does not manipulate the resource directly. This tends to blur the boundaries between pure and effectful code, making programs harder to read, compose, and refactor. Linear type systems, by contrast, promise precise control over resource usage, without the pervasive overhead, as programs can stay in "direct style".

But linear type systems, at first, may seem insufficient to do that job fully. In a linear type system, we can only promise what we do to a value we receive. We cannot say much about what has been done to it previously. And thus, we cannot ensure it has been used exactly once *overall*.

[6]The `Movable` typeclass is not part of the Linear Haskell language extension, but instead defined in the `linear-base` library.

For instance, given a function of type `t ⊸ u`, if we apply it to an argument of type `t`, we know we will not duplicate this value of type `t` (as long as we use the result of type `u` exactly once). But we cannot say whether the `t` has been duplicated or not beforehand: that's a major issue.

Stating facts about *the past* of a value is the land of *uniqueness typing* or unique type systems. In a unique type system, the main property is not how many times a function uses a value to produce its output, but rather, whether a value has been aliased or not. Uniqueness (or non-uniqueness) is a property on values, while linearity is a property on functions[7]. Having uniqueness or non-uniqueness property on values allows for easier control of resources that should always stay unique during their lifetime. The most common example is efficient implementation of functional arrays. If an array is handled uniquely and is not shared, then we can implement array updates as in-place memory mutations ⟦Danielle Marshall, Vollmer, and Orchard, 2022⟧. By being able to state uniqueness of values, statically through the type system, we can allow predictable compiler optimizations.

However, uniqueness typing does not have a theoretical ground as rich as linear types. Unlike linearity, uniqueness properties have first been studied informally, and only later studied on the logical side, with ⟦Harrington, 2006⟧. There are also two downsides of uniqueness typing for what we want to do later. First, it's not easy with a unique type system to force the use of a resource (and we definitely require a way to force a destination to be used and not just be dropped). With uniqueness typing, we can only state or prove that it is not aliased. Secondly, and more importantly perhaps, higher-order functions (capturing lambda abstractions) present many edge-cases in unique type systems, as detailed in Section 5 of ⟦Harrington, 2006⟧ or mentioned briefly in ⟦Danielle Marshall, Vollmer, and Orchard, 2022⟧ (that's one reason why they prefer a linear system as a basis with a uniqueness topping instead of a uniqueness system with a linearity topping).

Actually, Danielle Marshall, Vollmer, and Orchard ⟦2022⟧ argue that modern programming languages would benefit from integrating both linear and uniqueness typing, as linearity and uniqueness appear to be half-dual, half-complementary notions that together are enough for a very precise control of resource usage in a language. More recently, in that sense, Lorenzen, White, et al. ⟦2024⟧ included both *affinity* (the resource must be used at most once) and uniqueness in their modal system, and studied their interactions with one another.

In Chapter 3 however, our main target language will be Haskell, which does not have uniqueness typing, only linear types, as detailed in Section 1.7. Fortunately, we can emulate uniqueness control in a language equipped with linear types, using *linear scopes* or the *scope function trick*, as one might call it.

**Linear scopes** As we said before, in Linear Haskell, as in most linear type systems, functions can be made to use their arguments linearly or not. In other terms, linearity is a contract that links the input and output of a function. But we cannot say anything a priori about what happened to the value passed as a parameter before the function call, or what happens to the function result after. Linearity control can only be enforced on *variables*[8].

Let say we want to design a file API without monads (that's actually one interesting use of linear/unique types).

[7] At least in the presentations above that follows from Girard's Linear Logic. There exists approaches of Linear Type systems that are more focused on the programming side, for which linearity is a property of values and their types, not functions.

[8] Top-level bindings/definitions are not treated as normal variables, but rather as global constants. Consequently, they are not subject to linearity constraints, which allows for the problematic example we present next.

Assuming the following simplified API, one could write this small program:

```
openFile :: String ⊸ File
closeFile :: File ⊸ ()

file :: File  -- no way to force this value to be used exactly once
file = openFile "test.txt"

nonLinearUseButValid :: () =
  closeFile file ⨾ closeFile file  -- valid even if file is consumed twice
```

With call-by-value or call-by-need semantics, the side-effects of `openFile` will be produced only once, while the side effects of `closeFile` will be produced twice, resulting in an error.

The solution to that is to forbid the consumer from creating and directly accessing a value of the resource type for which we want to enforce linearity/uniqueness.

Instead, we force the consumer of a resource to pass a continuation representing what they want to do on the resource, so that we can check through its signature that it is indeed a linear continuation:

```
withFile :: String ⊸ (File ⊸ t) ⊸ t
closeFile :: File ⊸ ()
nonLinearUseAndEffectivelyRejected :: () =
  withFile "test.txt" (\file → closeFile file ⨾ closeFile file) -- not linear
```

The `File` type is in positive position in the signature of `withFile`, so the function `withFile` should somehow know how to produce a `File`, but this is opaque for the user. What matters is that a file can only be accessed by providing a linear continuation to `withFile`.

Still, this is not enough; because `\file → file` is indeed a linear continuation, one could use `withFile "test.txt" (\file → file)` to leak a `File`, and then use it in a non-linear fashion in the outside world. Hence we must forbid `File` from appearing anywhere in the return type of the continuation.

To do that, we ask the return type to be wrapped in the unrestricted modality `Ur` (the equivalent in the Linear Haskell realm to the ! modality): because the value of type `File` comes from the left of a linear arrow, it cannot be promoted to an `Ur` without breaking linearity, so it cannot leave the scope and has to be used exactly once in this scope. That's why we speak of *linear scopes*. Here's the updated example:

```
withFile' :: String ⊸ (File ⊸ Ur t) ⊸ Ur t
closeFile :: File ⊸ ()
exampleOk :: Ur () = withFile' "test.txt" (\file → closeFile file ⨾ Ur ())
exampleFail :: Ur File =
  withFile' "test.txt" (\file → Ur file) -- leaking file, effectively rejected
```

Still, this solution is not perfect. We effectively prevent any linear resource from leaving the scope, even resources that have nothing to do with the file API and have been created in outer scopes! This issue is talked in length by Spiwack [2023b].

It manifests itself in the following example, which gets rejected:

```
readLine :: File ⊸ (Ur String, File)
writeLine File ⊸ String ⊸ File

exampleFail' :: Ur () =
  withFile' "test.txt" (\file1 →
    ...
    withFile' "test2.txt" (\file2 →
      let (Ur line, file1') = readLine file1
          file2' = writeLine file2 line
       in closeFile file2' ⨾ Ur file1')
    ... -- other operations with file1' here
  )
```

The problem here is that we cannot return `file1'` from the inner scope, because it's a linear resource that cannot be promoted to `Ur`. However, it would be perfectly safe to do so; it's just that we do not have precise enough tools at type level to convince the compiler of this fact. In that case, the most obvious workaround is to extend the scope of `withFile' "test2.txt" (file2 → ...)` as long as we need to operate on `file1'`. Very unfortunate and unpractical indeed to have to prolongate the inner scope because we still need to operate on resources from the outer one! (c.f. *sticky end of linear scopes* in ⟦Spiwack, 2023b⟧)

One easy improvement we can do is to have a single linear scope, providing a linear token type, and let this token type be used to instantiate any resource for which we want to enforce linearity/uniqueness. That way, we can just have one big linear scope, and write most of the code in direct style, still without monads!

```
type Token
dup :: Token ⊸ (Token, Token)
drop :: Token ⊸ ()
withToken :: (Token ⊸ Ur t) ⊸ Ur t

-- A first linear API
openFile' :: String ⊸ Token ⊸ File
closeFile :: File ⊸ ()

-- A second linear API
openDB :: String ⊸ String ⊸ Token ⊸ DB
closeDB :: DB ⊸ ()

exampleOk1 :: Ur () =
  withToken (\tok → let file = openFile' "test.txt" tok in closeFile file ⨾ Ur ())
exampleOk2 :: Ur () =
  withToken (\tok → let (tok1, tok2) = dup tok
                          file = openFile' "test.txt" tok1
                          db = openDB "localhost" 3306 tok2
                       in closeFile file ⨾ closeDB db ⨾ Ur ())
exampleRejected :: Ur File =
  withToken (\tok →  Ur (openFile' "test.txt" tok)) -- leaking file which linearly depends on tok
```

What might be surprising here is that we are allowed to duplicate or drop a token (with `dup` or `drop`), which we announced as a linear resource!

In fact, the goal of the `Token` type is to tie every resource to a linearity requirement. The important part is that the `Token` type should not have a way to be promoted to `Ur`, otherwise tokens and other linear resources could leak and live outside of the linear scope of `withToken`.

The linear `dup` function creates two tokens out of one, but still tie the resulting tokens to the input one through a linear arrow. Allowing to duplicate tokens does not allow the consumer to duplicate the resources we want to control (files, database handles, etc.). So we are safe on that front.

With this approach, dealing with multiple linear resource types is much more uniform. We can use any remaining token to instantiate one or several new linear resources. As we only have one big linear scope, with every linear resource in it, there is no issue with nested scopes that need to be extended. When a program is sliced into many sub-functions, these functions do not even have to deal or know about the linear scope; they just have to pass around and receive tokens, an can otherwise be written in direct style.

Although it's out of scope for this document, Spiwack et al. ⟦2022⟧ goes further on the idea of linear tokens and scopes, with *Linear constraints*, which are roughly a way to pass tokens around in an implicit fashion, to make programming with linear types even more convenient[9].

In next chapters, we will reuse the *linear scope with tokens* strategy to enforce linearity and uniqueness on linear resources, to allow safe effects in pure, direct-style APIs.

## 1.9 Conclusion

In this chapter we introduced the linear $\lambda$-calculus, from its roots in linear logic, to more practical and computation-centric presentations. In particular, we laid theoretical foundations for a modal linear $\lambda$-calculus, which will be reused in the rest of this document.

Furthermore, we explored Linear Haskell, as a concrete target for experimenting with linear types. In particular, we illustrated how linear types can help with managing unique resources, that must be used but also must not be duplicated, even though it requires a bit of care to avoid the inherent limitations of linear types (in that, they cannot guarantee per se what happened to a value in the past).

We are now in comfortable ground and have all the tool needed to explore both theoretical and practical aspect of functional destination passing.

[9] At the time of writing this document, Linear Constraints are seriously considered to be added into GHC, the main Haskell compiler ⟦Spiwack, 2023a⟧.

# Chapter 2

# A formal functional language based on first-class destinations: $\lambda_d$

In the previous chapter we laid out important building blocks for the functional language that we develop here, namely $\lambda_d$. Let's see why destination passing is valuable in a pure functional programming context, and how $\lambda_d$ constitutes a good foundational layer for this.

The results presented in this chapter have also been published at the OOPSLA 2025 conference ⟦Bagrel and Spiwack, 2025a⟧.

## 2.1 Destination passing in a functional setting

In destination-passing style, a function does not return a value: it takes as an argument a location—named *destination*—where the output of the function ought to be written. In this chapter, we will denote destinations by $\lfloor \mathsf{T} \rfloor$ where $\mathsf{T}$ is the type of what can be stored in the destination. A function of type $\mathsf{T} \to \mathsf{U}$ would have signature $\mathsf{T} \to \lfloor \mathsf{U} \rfloor \to \mathbf{1}$ when transformed into destination-passing style. Instead of returning a $\mathsf{U}$, it consumes a destination of type $\lfloor \mathsf{U} \rfloor$ and returns just *unit* (type $\mathbf{1}$). This style is common in systems programming, where destinations $\lfloor \mathsf{U} \rfloor$ are more commonly known as *out parameters* (in C, $\lfloor \mathsf{U} \rfloor$ would typically be a pointer of type $\mathsf{U}*$).

The reason why systems programs rely on destinations so much is that using destinations can save calls to the memory allocator, which are costly, in a context where every gain of performance is highly valuable. If a function returns a $\mathsf{U}$, it has to allocate the space for a $\mathsf{U}$. But with destinations, the caller is responsible for finding space for a $\mathsf{U}$. The caller may simply ask for the space to the memory allocator, in which case we have saved nothing; but it can also reuse the space of an existing $\mathsf{U}$ which it does not need anymore, or space in an array, or even space in a region of memory that the allocator does not have access to, like a memory-mapped file. In fact, as we will see in Section 2.5, destination-passing style is always more general than its direct style counterpart (though it might be slightly more verbose); there are no drawbacks to use it as we can always recover regular direct style functions in a systematic way from the destination-passing ones.

So far we mostly considered destination passing in imperative contexts, such as systems programming, but we argue that destination passing also presents many benefits for functional programming languages, even pure ones. Where destinations truly shine in functional programming is that they let us borrow imperative-like programming techniques to get a more expressive language, with more control over processing, and with careful attention, we can still preserve the nice properties of functional languages such as purity and (apparent) immutability.

Thus, the goal here will be to modify a standard functional programming language just enough to be able to build immutable structures by destination passing without endangering purity and memory safety. Destinations in this particular setting become *write-once-only* references into a structure that contains holes and that cannot be read yet. Quite more restrictive than C pointers (that can be used in read and write fashion without restrictions), but still powerful enough to bring new programming techniques into the functional world.

More precisely, there are two key elements that contribute to the extra expressiveness of destination passing in functional contexts:

- structures can be built in any order. Not only from the leaves to the root, like in ordinary functional programming, but also from the root to the leaves, or any combination thereof. This can be done in ordinary functional programming using function composition in a form of continuation-passing; however destinations act as a very direct optimization of this scheme. This line of work was pioneered by Minamide ⟦1998⟧;

- when destinations are first-class values, they can be passed and stored like ordinary values. The consequence is not only that the order in which a structure is built is arbitrary, but also that this order can be determined dynamically during the runtime of the program. Here is the main novelty of our approach compared to earlier work.

In this chapter, we introduce $\lambda_d$, a pure functional calculus that is based on the very concept of destinations at its core. We intend $\lambda_d$ to serve as a foundational, theoretical calculus to reason about safe destinations in a functional setting; thus, we will only cover very briefly the implementation concerns in this chapter, as $\lambda_d$ is not really meant to be implemented as a real programming language. Actual implementation of destination passing in an existing functional language will be the focus of the next chapters.

$\lambda_d$ subsumes all the functional systems with destinations from the literature, from ⟦Minamide, 1998⟧ to ⟦Lorenzen, Leijen, Swierstra, and Lindley, 2024⟧. As such we expect that these systems or their extensions can be justified simply by giving them a translation into $\lambda_d$, in order to get all the safety results and metatheory of $\lambda_d$ for free. Indeed, we proved type safety theorems for $\lambda_d$ with the Coq proof assistant, as described in Section 2.7.

## 2.2 Working with destinations in $\lambda_d$

Let's introduce and get familiar with $\lambda_d$, a modal, linear, simply typed $\lambda$-calculus with destinations. We borrow most of the syntax of the previous chapter, especially from $\lambda_{Ldm}$. We still use linear logic's $\mathsf{T} \oplus \mathsf{U}$ and $\mathsf{T} \otimes \mathsf{U}$ for sums and products, and linear function arrow $\multimap$, since $\lambda_d$ is linearly typed. The main difference with previous calculi being that $\lambda_d$ does not have first-class data constructors like $\mathsf{Inl}\ t$ or $(t_1\ , t_2)$, as they are replaced with the more general destination-filling operators that we will discover in the next paragraphs.

### 2.2.1 Building up a vocabulary

In its simplest form, destination passing, much like continuation passing, is using a location, received as an argument, to write the output value instead of returning it. For instance, the following is a destination-passing version of the identity function:

$$\texttt{dId} : \mathsf{T} \multimap \lfloor \mathsf{T} \rfloor \multimap \mathsf{1}$$
$$\texttt{dId}\; x\; d \triangleq d \blacktriangleleft x$$

We can think of a destination as a reference to an uninitialized memory location, and $d \blacktriangleleft x$ (read "fill $d$ with $x$") as writing $x$ to the memory location pointed to by $d$.

Performing $d \blacktriangleleft x$ is the simplest way to use a destination: with that we fill it with a value directly. But we can also fill a destination piecewise, by specifying just the outermost constructor that we want to fill it with:

$$\texttt{fillWithInl} : \lfloor \mathsf{T} \oplus \mathsf{U} \rfloor \multimap \lfloor \mathsf{T} \rfloor$$
$$\texttt{fillWithInl}\; d \triangleq d \triangleleft \mathtt{Inl}$$

In this example, we are filling a destination for type $\mathsf{T} \oplus \mathsf{U}$ by setting the outermost constructor to left variant Inl. We think of $d \triangleleft \mathtt{Inl}$ (read "fill $d$ with Inl") as allocating memory to store a block of the form Inl $\square$, and write the address of that block to the location that $d$ points to. Because we have just created an Inl constructor with no argument yet, we return a new destination of type $\lfloor \mathsf{T} \rfloor$ pointing to the uninitialized argument of Inl. Uninitialized memory, when part of a structure or value, like $\square$ in Inl $\square$, is called a *hole*.

Notice that with **fillWithInl** we are constructing the structure from the outermost constructor inward: we have written a value of the form Inl $\square$ into a hole, but we have yet to describe what goes in the new hole $\square$. Such data constructors with uninitialized arguments are called *hollow constructors*[10]. This is opposite to how functional programming usually works, where values are built from the innermost constructors outward: first we build a value $v$ and only then can we use Inl to make Inl $v$. Being able to build structures from the outermost constructor inward is a key ingredient to the extra expressiveness that we promised earlier.

Yet, everything we have shown so far could, in principle, have been expressed using continuations. So it's worth asking: how exactly are destinations different from continuations? Part of the answer lies in our intention to effectively implement destinations as pointers to uninitialized memory (see Section 2.9 but also Chapters 3 and 4). But where destinations really differ from continuations is that we can easily split a destination for a data structure into several destinations for each of its parts. Then, the user is required to fill *all* the resulting destinations. One example of such splitting is the **fillWithPair** function, which writes a hollow pair constructor into a destination:

$$\texttt{fillWithPair} : \lfloor \mathsf{T} \otimes \mathsf{U} \rfloor \multimap \lfloor \mathsf{T} \rfloor \otimes \lfloor \mathsf{U} \rfloor$$
$$\texttt{fillWithPair}\; d \triangleq d \triangleleft (,)$$

After using **fillWithPair**, both the first field *and* the second field of the pair must be filled, using the returned destinations of type $\lfloor \mathsf{T} \rfloor$ and $\lfloor \mathsf{U} \rfloor$ respectively.

[10] The full triangle $\blacktriangleleft$ is used to fill a destination with a fully-formed value, while the *hollow* triangle $\triangleleft$ is used to fill a destination with a *hollow constructor*.

In contrast, a continuation-passing style equivalent of **fillWithPair** would need signature $(T \otimes U \multimap S) \multimap (T \multimap S) \otimes (U \multimap S)$. But **fillWithPairCPS** : $(T \otimes U \multimap S) \multimap (T \multimap S) \otimes (U \multimap S)$ cannot even be defined, as we would need to collect the results of the two created continuations before we can consume the input continuation of type $(T \otimes U \multimap S)$ (thus requiring a form of synchronization and state-passing).

Similarly to continuation-passing transformation though, we can already note that there is a sort of duality between destination-filling operators and the corresponding data constructors. Usual Inl constructor has signature $T \multimap T \oplus U$, while destination-filling ◁Inl has signature $\lfloor T \oplus U \rfloor \multimap \lfloor T \rfloor$. Similarly, pair constructor (,) has signature $T \multimap U \multimap T \otimes U$, while destination-filling ◁(,) has signature $\lfloor T \otimes U \rfloor \multimap \lfloor T \rfloor \otimes \lfloor U \rfloor$. Assuming some flexibility with currying, we see that the types of arguments and results switch sides around the arrow, and get wrapped/unwrapped from $\lfloor \cdot \rfloor$ when we go from the constructor to the destination-filling operator and vice-versa. This observation will generalize to all destination-filling operators and the corresponding constructors.

**Structures with holes** It is crucial to note that while a destination is used to build a structure, the destination refers only to a specific part of the structure that has not been defined yet; not the structure as a whole. Consequently, the (root) type of the structure being built will often be different from the type of the destination at hand. A destination of type $\lfloor T \rfloor$ only indicates that some bigger structure has at least a hole of type $T$ somewhere in it. The type of the structure itself never appears in the signature of destination-filling functions (for instance, using **fillWithPair** only indicates that the structure being operated on has a hole of type $T \otimes U$ that is being written to).

Thus, we still need a type to tie the structure under construction—left implicit by destination-filling primitives—with the destinations representing its holes. To represent this, $\lambda_d$ introduces a type $S \ltimes \lfloor T \rfloor$ for a structure of type $S$ missing a value of type $T$ to be complete. There can be several holes in $S$—resulting in several destinations on the right hand side—and as long as there remain holes in $S$, it cannot be read. For instance, $S \ltimes (\lfloor T \rfloor \otimes \lfloor U \rfloor)$ represents an $S$ that misses both a $T$ and a $U$ to be complete (thus to be readable). The right hand side of $\ltimes$ is not restricted to pair and destination types only; it can be of arbitrarily complex type.

The general form $S \ltimes T$ is read "$S$ ampar $T$". The name "ampar" stands for "asymmetric memory par". The "par" ⅋ operator originally comes from (classical) linear logic, and is an associative and commutative operator that subsumes linear implication and is more expressive: $T \multimap S$ is equivalent to $T^\perp ⅋ S$ or $S ⅋ T^\perp$. Similarly, $T_1 \multimap T_2 \multimap S$ is equivalent to $T1^\perp ⅋ T2^\perp ⅋ S$. Function input's types—which are in negative position—are wrapped in the dualizing operator $\cdot^\perp$ when a function is put into "par" form. Here we take a similar approach: Minamide [[1998]] first observed that structures with holes are akin to linear functions $T \multimap S$, where $T$ represents the missing part; so we decide to represent them in a more expressive fashion under the form $S \ltimes \lfloor T \rfloor$ (echoing $S ⅋ T^\perp$).

In this presentation, the *missing part of type* $T$ appears as a first-class type $\lfloor T \rfloor$, a.k.a the *destination* type. Reusing terminology from Danielle Marshall and Orchard [[2022]], we can say that $\lfloor T \rfloor$ is a *lax inverse* of $T$ for $\otimes$, because given $d : \lfloor T \rfloor$ and $v : T$, we have $d \blacktriangleleft v : 1$[11].

The asymmetric nature of our memory par is an unfortunate byproduct of wanting simple sequential semantics. Indeed, it seems that having a symmetric memory par would require concurrent semantics, as the structure being constructed and the destinations pointing to it could be consumed alike, meaning that we could have to pause and wait for a hole to be filled[12].

[11] Conversely, $\lfloor T \rfloor$ would be a *co-lax inverse* of $T$ for $\ltimes$, as $\lambda un \mapsto un \fatsemi \mathbf{new}_\ltimes$ has type $1 \multimap T \ltimes \lfloor T \rfloor$

[12] [[DeYoung, Pfenning, and Pruiksma, 2020]] features such a system with destinations and a concurrent execution model

Destinations, albeit being first-class, always exist within the context of a structure with holes. A destination is both a witness of a hole present in the structure, and a handle to write to it. Crucially, destinations are otherwise ordinary values that lives in the right side of the corresponding ampar. To access the destinations of an ampar, $\lambda_d$ provides a $\mathtt{upd}_\ltimes$ construction, which lets us apply a function to the right-hand side of the ampar. It is in the body of $\mathtt{upd}_\ltimes : \mathsf{S} \ltimes \mathsf{T} \multimap (\mathsf{T} \multimap \mathsf{U}) \multimap \mathsf{S} \ltimes \mathsf{U}$ that functions operating on destinations can be called to update the structure:

$$
\begin{aligned}
&\mathtt{fillWithInl}' \;:\; \mathsf{S} \ltimes \lfloor \mathsf{T} \oplus \mathsf{U} \rfloor \multimap \mathsf{S} \ltimes \lfloor \mathsf{T} \rfloor \\
&\mathtt{fillWithInl}'\ x \;\triangleq\; \mathtt{upd}_\ltimes\ x\ \mathtt{with}\ d \mapsto d \triangleleft \mathtt{Inl} \\
&\mathtt{fillWithPair}' \;:\; \mathsf{S} \ltimes \lfloor \mathsf{T} \otimes \mathsf{U} \rfloor \multimap \mathsf{S} \ltimes (\lfloor \mathsf{T} \rfloor \otimes \lfloor \mathsf{U} \rfloor) \\
&\mathtt{fillWithPair}'\ x \;\triangleq\; \mathtt{upd}_\ltimes\ x\ \mathtt{with}\ d \mapsto \mathtt{fillWithPair}\ d
\end{aligned}
$$

To tie this up, we need a way to introduce and to eliminate structures with holes. Structures with holes are introduced with $\mathtt{new}_\ltimes$ which creates a value of type $\mathsf{T} \ltimes \lfloor \mathsf{T} \rfloor$. $\mathtt{new}_\ltimes$ is somewhat similar to the linear identity function: it is a hole (of type $\mathsf{T}$) that needs a value of type $\mathsf{T}$ to be a complete value of type $\mathsf{T}$. Memory-wise, it is an uninitialized block large enough to host a value of type $\mathsf{T}$, and a destination pointing to it. Conversely, structures with holes are eliminated with[13] $\mathtt{from}'_\ltimes : \mathsf{S} \ltimes \mathsf{1} \multimap \mathsf{S}$: if all the destinations have been consumed and only unit remains on the right side, then $\mathsf{S}$ no longer has holes and thus is just a normal data structure.

Equipped with these new operators, we can finally show how to derive traditional constructors from piecemeal filling. Indeed, as we said earlier, $\lambda_d$ does not have primitive constructor forms; constructors in $\lambda_d$ are syntactic sugar. We show here the definition of $\mathtt{Inl}$ and $(,)$, but the other constructors are derived similarly. Operator $\fatsemi : \mathsf{1} \multimap \mathsf{U} \multimap \mathsf{U}$, present in second example, is used to chain operations returning unit type $\mathsf{1}$. For easier reading, we also provide type annotation when $\mathtt{new}_\ltimes$ is used:

$$
\begin{aligned}
&\mathtt{Inl} \;:\; \mathsf{T} \multimap \mathsf{T} \oplus \mathsf{U} \\
&\mathtt{Inl}\ x \;\triangleq\; \mathtt{from}'_\ltimes\,(\mathtt{upd}_\ltimes\ (\mathtt{new}_\ltimes : (\mathsf{T} \oplus \mathsf{U}) \ltimes \lfloor \mathsf{T} \oplus \mathsf{U} \rfloor)\ \mathtt{with}\ d \mapsto d \triangleleft \mathtt{Inl} \blacktriangleleft x) \\
&(,) \;:\; \mathsf{T} \multimap \mathsf{U} \multimap \mathsf{T} \otimes \mathsf{U} \\
&(x\,,\,y) \;\triangleq\; \mathtt{from}'_\ltimes\,(\mathtt{upd}_\ltimes\ (\mathtt{new}_\ltimes : (\mathsf{T} \otimes \mathsf{U}) \ltimes \lfloor \mathsf{T} \otimes \mathsf{U} \rfloor)\ \mathtt{with}\ d \mapsto \mathtt{case}\ (d \triangleleft (,))\ \mathtt{of}\ (d_1\,,\,d_2) \mapsto d_1 \blacktriangleleft x \fatsemi d_2 \blacktriangleleft y)
\end{aligned}
$$

**Memory safety and purity** We must reassure the reader here. Of course, using destinations in an unrestricted fashion is not memory safe. We need a linear discipline on destinations for them to be safe. Otherwise, we can encounter two sorts of issues:

- if destinations are not written at least once, as in:

  $$
  \begin{aligned}
  &\mathtt{forget} \;:\; \mathsf{T} \\
  &\mathtt{forget} \;\triangleq\; \mathtt{from}'_\ltimes\,(\mathtt{upd}_\ltimes\ (\mathtt{new}_\ltimes : \mathsf{T} \ltimes \lfloor \mathsf{T} \rfloor)\ \mathtt{with}\ d \mapsto ())
  \end{aligned}
  $$

  then reading the result of $\mathtt{forget}$ would lead to reading a location pointed to by a destination that we never used, in other words, reading uninitialized memory. This must be prevented at all cost;

[13] As the name suggest, there is a more general elimination $\mathtt{from}_\ltimes$. It will be discussed in Section 2.5.

- if destinations are written several times, as in:

$$\begin{aligned}
&\mathtt{ambiguous1} \;:\; \mathtt{Bool}\\
&\mathtt{ambiguous1} \triangleq \mathtt{from}'_{\ltimes}(\mathtt{upd}_{\ltimes}\,(\mathtt{new}_{\ltimes} : \mathtt{Bool} \ltimes \lfloor \mathtt{Bool} \rfloor)\;\mathtt{with}\; d \mapsto d \blacktriangleleft \mathtt{true} \,\fatsemi\, d \blacktriangleleft \mathtt{false})\\
&\mathtt{ambiguous2} \;:\; \mathtt{Bool}\\
&\mathtt{ambiguous2} \triangleq \mathtt{from}'_{\ltimes}(\mathtt{upd}_{\ltimes}\,(\mathtt{new}_{\ltimes} : \mathtt{Bool} \ltimes \lfloor \mathtt{Bool} \rfloor)\;\mathtt{with}\; d \mapsto \mathtt{let}\; x := (d \blacktriangleleft \mathtt{false})\;\mathtt{in}\; d \blacktriangleleft \mathtt{true} \,\fatsemi\, x)
\end{aligned}$$

  then, we have `ambiguous1` that returns `false` and `ambiguous2` that returns `true` due to evaluation order, so we break *let-expansion* that is supposed to be valid in a pure language (including $\lambda_d$).

Actually, we will see in Section 2.3 that a linear discipline is not even enough to ensure fully safe use of destinations in $\lambda_d$. But before dealing with this, let's get more familiar with $\lambda_d$ through more complex examples.

### 2.2.2 Tail-recursive map

Let's see how destinations can be used to build usual data structures. For these examples, we suppose that $\lambda_d$ has equirecursive types and a fixed-point operator (though this is not included in the formal system of Section 2.5).

**Linked lists** We define lists as a fixpoint, as usual: $\mathtt{List}\ \mathtt{T} \triangleq 1 \oplus (\mathtt{T} \otimes (\mathtt{List}\ \mathtt{T}))$. For convenience, we also define filling operators $\triangleleft[]$ and $\triangleleft(::)$, as macros that use the primitive destination-filling operators for sum, product and unit types:

$$\begin{aligned}
&\triangleleft[] \;:\; \lfloor \mathtt{List}\ \mathtt{T} \rfloor \multimap 1\\
&d \triangleleft [] \triangleq d \triangleleft \mathrm{Inl} \triangleleft ()
\end{aligned}
\qquad\qquad
\begin{aligned}
&\triangleleft(::) \;:\; \lfloor \mathtt{List}\ \mathtt{T} \rfloor \multimap \lfloor \mathtt{T} \rfloor \otimes \lfloor \mathtt{List}\ \mathtt{T} \rfloor\\
&d \triangleleft (::) \triangleq d \triangleleft \mathrm{Inr} \triangleleft (,)
\end{aligned}$$

Just like we did in Section 2.2.1 we can recover traditional constructors systematically from destination-filling operators, using $\mathtt{new}_{\ltimes}$, $\mathtt{upd}_{\ltimes}$ and $\mathtt{from}'_{\ltimes}$:

$$\begin{aligned}
&(::) \;:\; \mathtt{T} \otimes (\mathtt{List}\ \mathtt{T}) \multimap \mathtt{List}\ \mathtt{T}\\
&x :: xs \triangleq \mathtt{from}'_{\ltimes}(\mathtt{upd}_{\ltimes}\,(\mathtt{new}_{\ltimes} : (\mathtt{List}\ \mathtt{T}) \ltimes \lfloor \mathtt{List}\ \mathtt{T} \rfloor)\;\mathtt{with}\; d \mapsto\\
&\qquad\qquad \mathtt{case}\;(d \triangleleft (::))\;\mathtt{of}\;(dx\,,dxs) \mapsto dx \blacktriangleleft x \,\fatsemi\, dxs \blacktriangleleft xs)
\end{aligned}$$

**A tail-recursive map function** List are ubiquitous in functional programming, so the fact that the most natural way to write a `map` function on lists is not tail recursive (hence consumes unbounded stack space) is quite unpleasant. Map can be made tail-recursive in two passes: first build the result list in reverse order, then reverse it. But thanks to destinations, we are able to avoid this two-pass process altogether, as they let us extend the tail of the result list directly. We give the complete implementation in Listing 2.1.

The tail-recursive function is $\mathtt{map}'$, it has type $(\mathtt{T} \multimap \mathtt{U})\ {}_{\omega}\!\!\multimap \mathtt{List}\ \mathtt{T} \multimap \lfloor \mathtt{List}\ \mathtt{U} \rfloor \multimap 1$[14]. That is, instead of returning a resulting list, it takes a destination as an input and fills it with the result. At each recursive call, $\mathtt{map}'$ creates a new hollow cons cell to fill the destination. A destination pointing to the tail of the

[14]In fact, we will develop a more complex system of modes starting from Section 2.5 to ensure scope control of destinations, and thus the proper signature will be $(\mathtt{T} \multimap \mathtt{U})\ {}_{\omega\infty}\!\!\multimap \mathtt{List}\ \mathtt{T}\ {}_{1\uparrow}\!\!\multimap \lfloor \mathtt{List}\ \mathtt{U} \rfloor \multimap 1$.

$$\begin{array}{l}
\texttt{List T} \triangleq 1 \oplus (\texttt{T} \otimes (\texttt{List T})) \\
\texttt{map}' : (\texttt{T} \multimap \texttt{U})\ {}_{\omega}\!\multimap \texttt{List T} \multimap \lfloor \texttt{List U} \rfloor \multimap 1 \\
\texttt{map}'\ f\ l\ dl \triangleq \\
\quad \texttt{case}\ l\ \texttt{of}\ \{ \\
\quad\quad [] \mapsto dl \triangleleft [] , \\
\quad\quad x :: xs \mapsto \texttt{case}\ (dl \triangleleft (::))\ \texttt{of} \\
\quad\quad\quad (dx , dxs) \mapsto dx \blacktriangleleft f\ x \ ⨟\ \texttt{map}'\ f\ xs\ dxs \} \\
\texttt{map} : (\texttt{T} \multimap \texttt{U})\ {}_{\omega}\!\multimap \texttt{List T} \multimap \texttt{List U} \\
\texttt{map}\ f\ l \triangleq \texttt{from}'_{\ltimes}\ (\texttt{upd}_{\ltimes}\ (\texttt{new}_{\ltimes} : (\texttt{List U}) \ltimes \lfloor \texttt{List U} \rfloor)\ \texttt{with}\ dl \mapsto \texttt{map}'\ f\ l\ dl)
\end{array}$$

Listing 2.1: Tail-recursive **map** function on lists in $\lambda_d$

new cons cell is also created, on which **map′** is called (tail) recursively. This is really the same algorithm that you could write to implement map on a mutable list in an imperative language. Nevertheless $\lambda_d$ is a pure language with only immutable types[15].

To obtain the regular **map** function, all is left to do is to call $\texttt{new}_{\ltimes}$ to create an initial destination, and $\texttt{from}'_{\ltimes}$ when all destinations have been filled to extract the completed list; much like when we make constructors out of filling operators, like (::) above.

### 2.2.3 Functional queues, with destinations

Implementations for a tail-recursive map are present in most of the previous works, from ⟦Minamide, 1998⟧, to recent work ⟦Bour, Clément, and Scherer, 2021; Leijen and Lorenzen, 2025⟧. Tail-recursive map does not need the full power of $\lambda_d$'s first-class destinations: it just needs a notion of structure with a (single) hole. We will slowly works towards an example which uses first-class destinations at their full potential.

**Difference lists** Just like in any programming language, the iterated concatenation of linked lists $((xs_1 +\!\!+ xs_2) +\!\!+ \ldots) +\!\!+ xs_n$ is quadratic in $\lambda_d$. The usual solution to improve this complexity is to use *difference lists*. The name *difference lists* covers many related implementations for the concept of a "linked list missing a queue". The idea is that a difference list carries the same elements as a list would, but can be easily extended by the back in constant time as we retain a way to set a value for its queue later. In pure functional languages, a difference list is usually represented as a function instead ⟦John Hughes, 1986⟧, as we usually do not have write pointers. A singleton difference list is then $\lambda ys \mapsto x :: ys$, and concatenation of difference lists is function composition. A difference list is turned into a list by setting its queue to be the empty list, or in the functional case, by applying it to the empty list. The consequence is that no matter how many compositions we have, each cons cell will be allocated a single time, making the iterated concatenation linear indeed.

The problem is that in the functional implementation, each concatenation still allocates a closure. If we are building a difference list from singletons and composition, there's roughly one composition per cons cell, so iterated composition effectively performs two traversals of the list. In $\lambda_d$, we actually have write pointers, in the form of *destinations*, so we can do better by representing a difference list as a list with a hole, much like in an imperative setting. A singleton difference list becomes $x :: \square$, and

[15] We consider data structures in $\lambda_d$ to be immutable thanks to the fact that structure with holes cannot be scrutinized while they have not been fully completed, so destination-related mutations are completely opaque and unobservable.

```
DList T ≜ (List T) ⋉ ⌊List T⌋
append : DList T ⊸ T ⊸ DList T
ys append y ≜
    upd⋉ ys with dys ↦ case (dys ◁ (::)) of
        (dy , dys′) ↦ dy ◀ y ⨾ dys′
concat : DList T ⊸ DList T ⊸ DList T
ys concat ys′ ≜ upd⋉ ys with d ↦ d ◁∘ ys′
toList : DList T ⊸ List T
toList ys ≜ from′⋉ (upd⋉ ys with d ↦ d ◁ [])
```

```
Queue T ≜ (List T)⊗(DList T)
singleton : T ⊸ Queue T
singleton x ≜ (Inr (x :: []) , (new⋉ : DList T))
enqueue : Queue T ⊸ T ⊸ Queue T
q enqueue y ≜
    case q of (xs , ys) ↦ (xs , ys append y)
dequeue : Queue T ⊸ 1⊕(T⊗(Queue T))
dequeue q ≜
    case q of {
        ((x :: xs) , ys) ↦ Inr (x , (xs , ys)) ,
        ([] , ys) ↦ case (toList ys) of {
            [] ↦ Inl () ,
            x :: xs ↦ Inr (x , (xs , (new⋉ : DList T)))}}
```

Listing 2.2: Difference list and queue implementation in $\lambda_d$

concatenation is filling the hole with another difference list, using composition operator $\triangleleft\circ$. The detailed implementation is given on the left of Listing 2.2. This encoding for difference lists makes no superfluous traversal: concatenation is just an $O(1)$ in-place update.

**Efficient queue using difference lists** In an immutable functional language, a queue can be implemented as a pair of lists (*front* , *back*) ⟦Hood and Melville, 1981⟧. The list *back* stores new elements in reverse order ($O(1)$ prepend). We pop elements from *front*, except when it is empty, in which case we set the queue to (reverse *back* , []), and pop from the new front.

For their simple implementation, Hood-Melville queues are surprisingly efficient: the cost of the reverse operation is $O(1)$ amortized for a single-threaded use of the queue. Still, it would be better to get rid of this full traversal of the back list. Taking a step back, this *back* list that has to be reversed before it is accessed is really merely a representation of a list that can be extended from the back. And we already know an efficient implementation for this: difference lists.

So we can give an improved version of the simple functional queue using destinations. This implementation is presented on the right-hand side of Listing 2.2. Note that contrary to an imperative programming language, we cannot implement the queue as a single difference list: as mentioned earlier, our type system prevents us from reading the front elements of difference lists. Just like for the simple functional queue, we need a pair of one list that we can read from, and one that we can extend. Nevertheless this implementation of queues is both pure, as guaranteed by the $\lambda_d$ type system, and nearly as efficient as what an imperative programming language would afford.

## 2.3 Scope escape of destinations

In Section 2.2, we made an implicit assumption: establishing a linear discipline on destinations ensures that all destinations will eventually find their way to the left of a fill operator ◀ or ◁, so that the associated holes get written to. This turns out to be slightly incomplete.

To see why, let's consider the type $\lfloor\lfloor \mathtt{T} \rfloor\rfloor$: the type of a destination pointing to a hole where a destination is expected. Think of it as an equivalent of the pointer type $\mathtt{T} **$ in the C language. Destinations are indeed ordinary values, so they can be stored in data structures, and before they get stored, holes stand in their place in the structure. For instance, if we have $d : \lfloor \mathtt{T} \rfloor$ and $dd : \lfloor\lfloor \mathtt{T} \rfloor\rfloor$, we can form $dd \blacktriangleleft d$: $d$ will be stored in the structure pointed to by $dd$.

Should we count $d$ as linearly used here? The alternatives do not seem promising:

- If we count this as a non-linear use of $d$, then $dd \blacktriangleleft d$ is rejected since destinations (like $d$ here) can only be used linearly. This choice is fairly limiting, as it would prevent us from storing destinations in structures with holes, as we will do, crucially, in Section 2.4.
- If we do not count this use of $d$ at all, we can write $dd \blacktriangleleft d \mathbin{⨟} d \blacktriangleleft v$ so that $d$ is both stored for later use *and* filled immediately (resulting in the corresponding hole being potentially written to twice), which is unsound, as discussed at the end of Section 2.2.1.

So linear use it is. But this creates a problem: there's no way, within our linear type system, to distinguish between "a destination that has been used on the left of a triangle so its corresponding hole has been filled" and "a destination that has been stored and its hole still exists at the moment". This oversight may allow us to read uninitialized memory!

Let's compare two examples. We assume a simple store semantics for now where structures with holes stay in the store until they are completed. A reduction step goes from a pair $\mathcal{S}|t$ of a store and a term to a new pair $\mathcal{S}'|t'$, and store are composed of named bindings of the form $\{l := v\}$ with $v$ a value that may contain holes. We will also need the $\mathtt{alloc} : (\lfloor \mathtt{T} \rfloor \multimap 1) \multimap \mathtt{T}$ operator. The semantics of $\mathtt{alloc}$ is: allocate a structure with a single root hole in the store, call the supplied function with the destination to the root hole as an argument; when the function has consumed all destinations (so only unit remains), pop the structure from the store to obtain a complete $\mathtt{T}$. $\mathtt{alloc}\ f$ corresponds, in type and behavior, to $\mathtt{from}'_{\ltimes}\,(\mathtt{upd}_{\ltimes}\ \mathtt{new}_{\ltimes}\ \mathtt{with}\ d \mapsto f\ d)$.

In the following snippets, structures with holes are bound to names $l$ and $ld$ in the store; holes are given names too and denoted by $\boxed{h}$ and $\boxed{hd}$, and concrete destinations are denoted by $\to h$ and $\to hd$.

When the building scope of $l : \mathtt{Bool}$ is parent to the one of $ld : \lfloor \mathtt{Bool} \rfloor$, everything works well because $ld$, that contains destination pointing to $\boxed{h}$, has to be consumed before $l$ can be read. In other terms, storing an older destination into a fresher one is not a problem:

$$
\begin{array}{rrl}
 & \{\} & |\ \mathtt{alloc}\ (\lambda d \mapsto (\mathtt{alloc}\ (\lambda dd \mapsto dd \blacktriangleleft d) : \lfloor \mathtt{Bool} \rfloor) \blacktriangleleft \mathtt{true}) \\
\longrightarrow & \{l := \boxed{h}\} & |\ (\mathtt{alloc}\ (\lambda dd \mapsto dd \blacktriangleleft \to h) : \lfloor \mathtt{Bool} \rfloor) \blacktriangleleft \mathtt{true} \mathbin{⨟} \mathtt{deref}\ l \\
\longrightarrow & \{l := \boxed{h}, ld := \boxed{hd}\} & |\ (\to hd \blacktriangleleft \to h \mathbin{⨟} \mathtt{deref}\ ld) \blacktriangleleft \mathtt{true} \mathbin{⨟} \mathtt{deref}\ l \\
\longrightarrow & \{l := \boxed{h}, ld := \to h\} & |\ \mathtt{deref}\ ld \blacktriangleleft \mathtt{true} \mathbin{⨟} \mathtt{deref}\ l \\
\longrightarrow & \{l := \boxed{h}\} & |\ \to h \blacktriangleleft \mathtt{true} \mathbin{⨟} \mathtt{deref}\ l \\
\longrightarrow & \{l := \mathtt{true}\} & |\ \mathtt{deref}\ l \\
\longrightarrow & \{\} & |\ \mathtt{true}
\end{array}
$$

However, when $ld$'s scope is parent to $l$'s, i.e. when a newer destination is stored into an older one, then we can write a linearly typed yet unsound program:

$$
\begin{array}{rrl}
 & \{\} & |\ \mathtt{alloc}\ (\lambda dd \mapsto \mathtt{case}\ (\mathtt{alloc}\ (\lambda d \mapsto dd \blacktriangleleft d) : \mathtt{Bool})\ \mathtt{of}\ \{\mathtt{true} \mapsto ()\,,\ \mathtt{false} \mapsto ()\}) \\
\longrightarrow & \{ld := \boxed{hd}\} & |\ \mathtt{case}\ (\mathtt{alloc}\ (\lambda d \mapsto \to hd \blacktriangleleft d) : \mathtt{Bool})\ \mathtt{of}\ \{\mathtt{true} \mapsto ()\,,\ \mathtt{false} \mapsto ()\} \mathbin{⨟} \mathtt{deref}\ ld \\
\longrightarrow & \{ld := \boxed{hd}, l := \boxed{h}\} & |\ \mathtt{case}\ (\to hd \blacktriangleleft \to h \mathbin{⨟} \mathtt{deref}\ l)\ \mathtt{of}\ \{\mathtt{true} \mapsto ()\,,\ \mathtt{false} \mapsto ()\} \mathbin{⨟} \mathtt{deref}\ ld \\
\longrightarrow & \{ld := \to h, l := \boxed{h}\} & |\ \mathtt{case}\ (\mathtt{deref}\ l)\ \mathtt{of}\ \{\mathtt{true} \mapsto ()\,,\ \mathtt{false} \mapsto ()\} \mathbin{⨟} \mathtt{deref}\ ld \\
\longrightarrow & \{ld := \to h\} & |\ \mathtt{case}\ \boxed{h}\ \mathtt{of}\ \{\mathtt{true} \mapsto ()\,,\ \mathtt{false} \mapsto ()\} \mathbin{⨟} \mathtt{deref}\ ld
\end{array}
$$

Here the expression $dd \blacktriangleleft d$ results in $d$ escaping its scope for the parent one, so $l$ is just uninitialized memory (the hole $\boxed{h}$) when we dereference it. This example must be rejected by our type system.

A neighboring issue is when sibling destinations interact with one another:

$$
\begin{array}{lrl}
 & \{\} & \mid\ \mathtt{alloc}\ (\lambda d \mapsto \mathtt{case}\ (d \triangleleft (,))\ \mathtt{of}\ (d' : \lfloor \mathsf{Bool} \rfloor\, ,\, dd' : \lfloor\lfloor \mathsf{Bool} \rfloor\rfloor) \mapsto dd' \blacktriangleleft d') : \mathsf{Bool} \otimes \lfloor \mathsf{Bool} \rfloor \\
\longrightarrow & \{l := \boxed{h}\} & \mid\ \mathtt{case}\ (\rightarrow\!h \triangleleft (,))\ \mathtt{of}\ (d' : \lfloor \mathsf{Bool} \rfloor\, ,\, dd' : \lfloor\lfloor \mathsf{Bool} \rfloor\rfloor) \mapsto dd' \blacktriangleleft d' \ ⨾\ \mathtt{deref}\ l \\
\longrightarrow & \{l := (\boxed{h'}\, ,\, \boxed{hd'})\} & \mid\ \mathtt{case}\ (\rightarrow\!h'\, ,\, \rightarrow\!hd')\ \mathtt{of}\ (d' : \lfloor \mathsf{Bool} \rfloor\, ,\, dd' : \lfloor\lfloor \mathsf{Bool} \rfloor\rfloor) \mapsto dd' \blacktriangleleft d' \ ⨾\ \mathtt{deref}\ l \\
\longrightarrow & \{l := (\boxed{h'}\, ,\, \boxed{hd'})\} & \mid\ \rightarrow\!hd' \blacktriangleleft \rightarrow\!h' \ ⨾\ \mathtt{deref}\ l \\
\longrightarrow & \{l := (\boxed{h'}\, ,\, \rightarrow\!h')\} & \mid\ \mathtt{deref}\ l \\
\longrightarrow & \{\} & \mid\ (\boxed{h'}\, ,\, \rightarrow\!h')
\end{array}
$$

Here also, we have a hole $\boxed{h'}$ escaping from the store, which means that if we pattern-match on the result of the code block above, we will also end up reading uninitialized memory!

Again, the problem is that, using purely a linear type system, we can only reject the two last examples if we also reject the first, sound one. In this case, the type $\lfloor\lfloor \mathsf{T} \rfloor\rfloor$ would become practically useless: such destinations could never be filled. This is not the direction we want to take: we really want to be able to store destinations in data structures with holes. So we want $t$ in $d \blacktriangleleft t$ to be allowed to be linear. Without further restrictions, it would not be sound, so to address this, we need an extra control system to prevent scope escape; it cannot be just linearity. More specifically, we must ensure that if a destination $d$ is stored inside another destination $dd$, then $d$ originates from a scope that is strictly older—that is, one that was opened strictly before the scope of $dd$. This condition rules out the two problematic examples we have just discussed.

For $\lambda_d$, we decided to use a system of *ages* to represent which scope a resource originates from. Ages are described in detail in Section 2.5. In the rest of this document, we often refer to *scope escape* to indistinctly designate the two issues presented above: either the case $dd \blacktriangleleft d$ where $d$ comes from a strictly younger scope, or the case where $d$ comes from the same scope as $dd$, as both are often avoided using the same techniques or tricks.

## 2.4 Breadth-first tree traversal

Before turning to the formal details, let us look at a full-fledged example, which uses the whole expressive power of $\lambda_d$. We focus here on breadth-first tree relabeling: “ *Given a tree, create a new one of the same shape, but with the values at the nodes replaced by the numbers* $1 \ldots |T|$ *in breadth-first order.* ”

This is not a very natural problem for functional programming, as breadth-first traversal implies that the order in which the structure must be built (left-to-right, top-to-bottom) is not the same as the structural order of a functional tree—building the leaves first and going up to the root. So it usually requires fancy functional workarounds ⟦Gibbons, 1993; Gibbons et al., 2023; Okasaki, 2000⟧.

We may feel drawn to solve this exercise in an efficient imperative-like fashion, where a queue drives the processing order. That's the standard algorithm taught at university, where the next node to process is dequeued from the front of the queue, and its children nodes are enqueued at the back of the queue for later processing, achieving breadth-first traversal. For that, Minamide's system ⟦Minamide, 1998⟧ where structures with holes are represented as linear functions cannot help. We really need destinations as first-class values to borrow from this imperative power.

```
go : (S ω∞⊸ T1 ⊸ (!ω∞ S)⊗T2) ω∞⊸ S ω∞⊸ Queue (Tree T1⊗⌊Tree T2⌋) ⊸ 1
go f st q ≜rec case (dequeue q) of {
                  Inl () ↦ () ,
                  Inr ((tree , dtree) , q′) ↦ case tree of {
                      Nil ↦ dtree ◁ Nil ⨾ go f st q′ ,
                      Node x tl tr ↦ case (dtree ◁ Node) of
                          (dy , (dtl , dtr)) ↦ case (f st x) of
                              (Modω∞ st′ , y) ↦
                                  dy ◀ y ⨾
                                  go f st′ (q′ enqueue (tl , dtl) enqueue (tr , dtr))}}
mapAccumBfs : (S ω∞⊸ T1 ⊸ (!ω∞ S)⊗T2) ω∞⊸ S ω∞⊸ Tree T1 1∞⊸ Tree T2
mapAccumBfs f st tree ≜ from′⋉ (upd⋉ (new⋉ : (Tree T2) ⋉ ⌊Tree T2⌋) with dtree ↦
                                   go f st (singleton (tree , dtree)))
relabelDps : Tree 1 1∞⊸ Tree Nat
relabelDps tree ≜ mapAccumBfs (λst ω∞↦ λun ↦ un ⨾ (Modω∞ (succ st) , st)) 1 tree
```

Listing 2.3: Breadth-first tree traversal using destination passing in $\lambda_d$

Listing 2.3 presents how we can implement this breadth-first tree traversal in $\lambda_d$, thanks to first-class destinations. We assume the data type `Tree T` has been defined as `Tree T` ≜ `1⊕(T⊗((Tree T)⊗(Tree T)))`; and we refer to the constructors of `Tree T` as `Nil` and `Node`, defined as syntactic sugar as we did for the other constructors before. We also assume some encoding of the type `Nat` of natural numbers. Remember that `Queue T` is the efficient queue type from Section 2.2.3.

The core idea of our algorithm is that we hold a queue of pairs, storing each an input subtree and (a destination to) its corresponding output subtree. When the element (*tree* , *dtree*) at the front of the queue has been processed, the children nodes of *tree* and children destinations of *dtree* are enqueued to be processed later, much as the original imperative algorithm.

We implement the actual breadth-first relabeling **relabelDps** as an instance of a more general breadth-first traversal function **mapAccumBfs**, which applies any state-passing style transformation of labels in breadth-first order. In **mapAccumBfs**, we create a new destination *dtree* into which we will write the result of the traversal, then call **go**. The **go** function is in destination-passing style, but what's remarkable is that **go** takes an unbounded number of destinations as arguments, since there are as many destinations as items in the queue. This is where we use the fact that destinations are ordinary values.

This freshly gained expressivity has a cost though: we need a type system that combines linearity control *and* age control to make the system sound, otherwise we run into the issues of scope escape described in the previous section. We will combine both linearity and the aforementioned ages system in the same *mode semiring*[16]. You already know the linearity annotations $1$ and $\omega$ from Chapter 1; here we also introduce the new age annotation $\infty$, that indicates that the associated argument cannot carry destinations. Arguments with no modes displayed on them, or function arrows with no modes, default to the unit of the semiring; in particular they are linear, and can capture destinations.

[16]We said in Section 1.4 that the modal approach with semirings would make the type system easier to extend, here is the proof.

*Core grammar of terms:*

$$
\begin{aligned}
t, u \; ::= \; & x \;\mid\; t'\,t \;\mid\; t \mathbin{⨾} t' \\
\mid\; & \mathbf{case}_{m}\ t\ \mathbf{of}\ \{\mathsf{Inl}\,x_1 \mapsto u_1\,,\ \mathsf{Inr}\,x_2 \mapsto u_2\} \;\mid\; \mathbf{case}_{m}\ t\ \mathbf{of}\ (x_1\,,\,x_2) \mapsto u \;\mid\; \mathbf{case}_{m}\ t\ \mathbf{of}\ \mathsf{Mod}_{n}\ x \mapsto u \\
\mid\; & \mathbf{upd}_{\ltimes}\ t\ \mathbf{with}\ x \mapsto t' \;\mid\; \mathbf{to}_{\ltimes}\ t \;\mid\; \mathbf{from}_{\ltimes}\ t \;\mid\; \mathbf{new}_{\ltimes} \\
\mid\; & t \triangleleft () \;\mid\; t \triangleleft \mathsf{Inl} \;\mid\; t \triangleleft \mathsf{Inr} \;\mid\; t \triangleleft (,) \;\mid\; t \triangleleft \mathsf{Mod}_{m} \;\mid\; t \triangleleft (\boldsymbol{\lambda} x \mapsto_{m} u) \;\mid\; t \triangleleft\!\circ\, t' \;\mid\; t \blacktriangleleft t'
\end{aligned}
$$

*Grammar of types, modes and contexts:*

$$
\begin{aligned}
\mathsf{T}, \mathsf{U}, \mathsf{S} \; ::= \; & \lfloor_{n}\mathsf{T}\rfloor \quad \textit{(destination)} \\
\mid\; & \mathsf{S} \ltimes \mathsf{T} \quad \textit{(ampar)} \\
\mid\; & 1 \;\mid\; \mathsf{T}_1 \oplus \mathsf{T}_2 \;\mid\; \mathsf{T}_1 \otimes \mathsf{T}_2 \;\mid\; !_{m}\mathsf{T} \;\mid\; \mathsf{T}\ {}_{m}\!\!\multimap \mathsf{U} \\
\Gamma \; ::= \; & \bullet \;\mid\; x :_{m} \mathsf{T} \;\mid\; \Gamma_1,\ \Gamma_2
\end{aligned}
$$

$$
\begin{aligned}
m, n \; & ::= \; \mathsf{pa} \quad \textit{(pair of multiplicity and age)} \\
\mathsf{p} \; & ::= \; 1 \;\mid\; \omega \\
\mathsf{a} \; & ::= \; \uparrow^{k} \;\mid\; \infty \\
\nu \triangleq \uparrow^{0} & \qquad \uparrow \triangleq \uparrow^{1}
\end{aligned}
$$

*Ordering on modes:*

$$\mathsf{pa} \lesssim \mathsf{p'a'} \iff \mathsf{p} \lesssim_{\mathsf{p}} \mathsf{p'} \;\wedge\; \mathsf{a} \lesssim_{\mathsf{a}} \mathsf{a'}$$

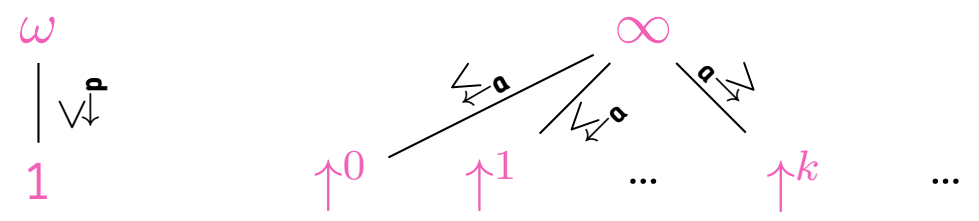

*Operations on modes:*

| + | 1 | ω |
|---|---|---|
| 1 | ω | ω |
| ω | ω | ω |

| · | 1 | ω |
|---|---|---|
| 1 | 1 | ω |
| ω | ω | ω |

$$(\mathsf{pa})\cdot(\mathsf{p'a'}) \triangleq (\mathsf{p}\cdot\mathsf{p'})(\mathsf{a}\cdot\mathsf{a'})$$

| + | $\uparrow^{k}$ | $\infty$ |
|---|---|---|
| $\uparrow^{j}$ | if $k = j$ then $\uparrow^{k}$ else $\infty$ | $\infty$ |
| $\infty$ | $\infty$ | $\infty$ |

| · | $\uparrow^{k}$ | $\infty$ |
|---|---|---|
| $\uparrow^{j}$ | $\uparrow^{k+j}$ | $\infty$ |
| $\infty$ | $\infty$ | $\infty$ |

$$(\mathsf{pa}) + (\mathsf{p'a'}) \triangleq (\mathsf{p} + \mathsf{p'})(\mathsf{a} + \mathsf{a'})$$

*Operations on typing contexts:*

$$
\begin{aligned}
n \cdot \bullet \; & \triangleq \; \bullet \\
n \cdot (x :_{m} \mathsf{T},\ \Gamma) \; & \triangleq \; (x :_{n\cdot m} \mathsf{T}),\ n \cdot \Gamma
\end{aligned}
$$

$$
\begin{aligned}
\bullet + \Gamma \; & \triangleq \; \Gamma \\
(x :_{m} \mathsf{T},\ \Gamma_1) + \Gamma_2 \; & \triangleq \; x :_{m} \mathsf{T},\ (\Gamma_1 + \Gamma_2) \qquad \text{if } x \notin \Gamma_2 \\
(x :_{m} \mathsf{T},\ \Gamma_1) + (x :_{m'} \mathsf{T},\ \Gamma_2) \; & \triangleq \; x :_{m+m'} \mathsf{T},\ (\Gamma_1 + \Gamma_2)
\end{aligned}
$$

Figure 2.1: Terms, types and modes of $\lambda_d$

## 2.5 Type system

$\lambda_d$ is a simply-typed $\lambda$-calculus with unit ($1$), product ($\otimes$) and sum ($\oplus$) types, like $\lambda_{Ldm}$ described in Chapter 1. It also features a mode-polymorphic function arrow ${}_{m}\!\!\multimap$. Its most distinctive features however are the destination $\lfloor_{m}\mathsf{T}\rfloor$ and ampar $\mathsf{S} \ltimes \mathsf{T}$ types which we have introduced in Sections 2.2 to 2.4. The mode annotation $m$ on the destination type $\lfloor_{m}\mathsf{T}\rfloor$ indicates the mode a value must have to be written to the corresponding hole; so far it has been hidden in the examples because we only considered destinations that could accept values with the unit mode.

To ensure that destinations are used soundly, we need both to enforce the linearity of destinations but also to prevent destinations from escaping their scope, as discussed in Section 2.3. To that effect, $\lambda_d$ tracks the *age* of destinations, that is how many nested scopes have been open between the current expression and the scope from which a destination originates. We will see in Section 2.5.2 that scopes are introduced by the $\mathbf{upd}_{\ltimes}\ t\ \mathbf{with}\ x \mapsto t'$ construct. For instance, if we have a term $\mathbf{upd}_{\ltimes}\ t_1\ \mathbf{with}\ x_1 \mapsto \mathbf{upd}_{\ltimes}\ t_2\ \mathbf{with}\ x_2 \mapsto \mathbf{upd}_{\ltimes}\ t_3\ \mathbf{with}\ x_3 \mapsto x_1$, then we say that the innermost occurrence of $x_1$ has age $\uparrow^{2}$ because two nested $\mathbf{upd}_{\ltimes}$ separate the definition and use site of $x_1$.

```
Inl t ≜ from'⋉ (upd⋉ new⋉ with d ↦
                  d ◁ Inl ◀ t)
Inr t ≜ from'⋉ (upd⋉ new⋉ with d ↦
                  d ◁ Inr ◀ t)
Mod_m t ≜ from'⋉ (upd⋉ new⋉ with d ↦
                  d ◁ Mod_m ◀ t)
λx m↦ u ≜ from'⋉ (upd⋉ new⋉ with d ↦
                  d ◁ (λx m↦ u))

from'⋉ t ≜
  case (from⋉ (upd⋉ t with un ↦ un ⨟ Mod_1∞ ())) of
    (st , ex) ↦ case ex of
      Mod_1∞ un ↦ un ⨟ st
() ≜ from'⋉ (upd⋉ new⋉ with d ↦ d ◁ ())
(t1 , t2) ≜ from'⋉ (upd⋉ new⋉ with d ↦ case (d ◁ (,)) of
              (d1 , d2) ↦ d1 ◀ t1 ⨟ d2 ◀ t2)
```

Listing 2.4: Desugaring rules for terms in $\lambda_d$

For $\lambda_d$, we take the same modal approach as for $\lambda_{Ldm}$, but we enrich our mode semiring to have an *age* axis. Thanks to the algebraic nature of the modal approach, for most typing rules, we will be able to reuse those of $\lambda_{Ldm}$ without any modification as just the elements of the semiring change, not the properties nor the structure of the semiring.

The syntax of $\lambda_d$ terms is presented in Figure 2.1, and the syntactic sugar that we used in Sections 2.2 to 2.4 is presented in Listing 2.4.

### 2.5.1 Modes and the age semiring

Our mode semiring (more precisely, a commutative additive semigroup $+$ and a multiplicative monoid $(\cdot, 1)$, with the usual distributivity law $n \cdot (m_1 + m_2) = n \cdot m_1 + n \cdot m_2$, and equipped with a partial order $\lesssim$, such that $+$ and $\cdot$ are order-preserving) is, as promised, the product of a multiplicity semiring, to track linearity, and an age semiring, to prevent scope escape. The resulting compound structure is also a semiring.

The multiplicity semiring has elements $1$ (linear) and $\omega$ (unrestricted), it's the same semiring as in ⟦Atkey, 2018⟧ or ⟦Bernardy et al., 2018⟧, and the same as described in detail in Chapter 1. The full description of the multiplicity semiring is given again in Figure 2.1.

Ages are more interesting. We write ages as $\uparrow^k$ (with $k$ a natural number), for "defined $k$ scopes ago". We also have an age $\infty$ for variables that do not originate from a $\texttt{upd}_{\ltimes}\ t\ \texttt{with}\ x \mapsto t'$ construct i.e. that are not destinations, and can be freely used in and returned by any scope. The main role of age $\infty$ is thus to act as a guarantee that a value does not contain destinations. Finally, we will write $\nu \triangleq \uparrow^0$ ("now") for the age of destinations that originate from the current scope; and $\uparrow \triangleq \uparrow^1$.

The operations or order on ages are not the usual ones on natural numbers though. Indeed, it's crucial that $\lambda_d$ tracks the precise age of variables, and usually, ordering on modes usually implies a form of subtyping (where a variable having mode $n$ can be used in place where a variable with mode $m$ is expected if $m \lesssim n$). Here we do not want variables from two scopes ago to be used as if they were from one scope ago. The ordering reflects this with finite ages being arranged in a flat order (so $\uparrow^1$ and $\uparrow^2$ are not comparable), with $\infty$ being bigger than all of them. Multiplication of ages reflects nesting of scope, as such, (finite) ages are multiplied by adding their numerical exponents $\uparrow^k \cdot \uparrow^j = \uparrow^{k+j}$. In the typing rules, the most common form of scope nesting is opening a scope, which is represented by multiplying the age of existing bindings by $\uparrow$ (that is, adding 1 to the ages seen as a natural numbers). When a same variable is shared between two subterms, $+$ takes the least upper bound for the age order above, in other terms, the variable must be at the same age in both subterms, or have age $\infty$ otherwise, which let it assume whichever age it needs in each subterm.

The unit of the new mode semiring is the pair of the units from the multiplicity and age semirings, ie. $1\nu$. We will usually omit the mode annotations when the mode is that unit.

Finally, as in $\lambda_{Lm}$ and $\lambda_{Ldm}$ from Chapter 1, operations on modes are lifted to typing contexts, still following the insights from [Ghica and Smith, 2014] (see Figure 2.1).

Anticipating on next section, let's see how *plus* works on context, modes, and especially ages. Let's type the expression $d \blacktriangleleft x \fatsemi x$. Most of the typing here behaves the same as $\lambda_{Ldm}$–TY, the only particularity is that the operator $\blacktriangleleft$ scales the typing context of its right operand by $\uparrow$ (which corresponds to asking it to be from the previous scope, to prevent scope escape of destinations). We have the following typing tree:

$$\dfrac{\dfrac{\dfrac{1\nu \lesssim 1\nu}{d :_{1\nu} \lfloor T \rfloor \vdash d : \lfloor T \rfloor}\,\textsc{id}_V \qquad \dfrac{1\nu \lesssim 1\nu}{x :_{1\nu} T \vdash x : T}\,\textsc{id}_V}{d :_{1\nu} \lfloor T \rfloor,\ x :_{1\uparrow} T \vdash d \blacktriangleleft x : 1}\,\lfloor\ \rfloor E_L \qquad \dfrac{1\nu \lesssim 1\nu}{x :_{1\nu} T \vdash x : T}\,\textsc{id}_V}{d :_{1\nu} \lfloor T \rfloor,\ x :_{\omega\infty} T \vdash d \blacktriangleleft x \fatsemi x : T}\,1E$$

At the root of the tree, with operator $\fatsemi$, the typing context of the two initial branches are summed, so we have $(d :_{1\nu} \lfloor T \rfloor,\ x :_{1\uparrow} T) + (x :_{1\nu} T) = d :_{1\nu} \lfloor T \rfloor,\ x :_{(1+1)(\nu+\uparrow)} T = d :_{1\nu} \lfloor T \rfloor,\ x :_{\omega\infty} T$. In other terms, because we need $x$ to be of age $\uparrow$ in the first branch, but of age $\nu$ in the second branch—as, we recall, $\nu = \uparrow^0$ and $\uparrow = \uparrow^1$ are *not* comparable, so we cannot use the rule $\lambda_d$–TY/$\textsc{id}_V$ on a variable with mode $1\uparrow$— then $x$ must be of age $\infty$[17].

Alternatively, we can have:

$$\dfrac{\dfrac{\dfrac{1\nu \lesssim 1\nu}{d :_{1\nu} \lfloor T \rfloor \vdash d : \lfloor T \rfloor}\,\textsc{id}_V \qquad \dfrac{1\nu \lesssim \omega\infty}{x :_{\omega\infty} T \vdash x : T}\,\textsc{id}_V}{d :_{1\nu} \lfloor T \rfloor,\ x :_{\omega\infty} T \vdash d \blacktriangleleft x : 1}\,\lfloor\ \rfloor E_L \qquad \dfrac{1\nu \lesssim \omega\infty}{x :_{\omega\infty} T \vdash x : T}\,\textsc{id}_V}{d :_{1\nu} \lfloor T \rfloor,\ x :_{\omega\infty} T \vdash d \blacktriangleleft x \fatsemi x : T}\,1E$$

where $x$ is given mode $\omega\infty$ from the leaves to the root. We can make use of a variable with mode $\omega\infty$ because we have $1\nu \lesssim \omega\infty$, which matches the premise of rule $\lambda_d$–TY/$\textsc{id}_V$.

### 2.5.2 Typing rules

Core typing rules of $\lambda_d$ are given in Figure 2.2. Rules for elimination of $\multimap$, $1$, $\oplus$, $\otimes$ and $!_m$ are the same as in $\lambda_{Ldm}$ of Chapter 1, so we will not cover them again here.

One notable difference with $\lambda_{Ldm}$ though, is, as announced, that introduction rules for data structures, a.k.a. data constructors for types $1$, $\oplus$, $\otimes$ and $!_m$, are derived from elimination rules of destinations, and thus are not part of the core language. They are presented on the distinct Figure 2.3. The introduction form for functions $\lambda_d$–TY$_s$/$\multimap$I is also derived from elimination form $\lambda_d$–TY/$\lfloor\multimap\rfloor$E for the corresponding destinations of function (this way, all types except ampar and destinations themselves have their introduction forms derived from a destination-elimination one).

[17] Age requirements are only there for scope-sensitive resources, that is, just destinations and the structures they are stored in. Age $\infty$ is meant to designate non scope-sensitive resources, and consequently, is also the result of incompatible age requirements (as only non scope-sensitive resources can fit incompatible age requirements).

$\boxed{\Gamma \vdash t : \mathsf{T}}$ *(Typing judgment for terms)*

$$\frac{1\nu \lessdot \mathsf{m}}{\omega\nu \cdot \Gamma,\ x :_{\mathsf{m}} \mathsf{T} \vdash x : \mathsf{T}}\ \textsc{IdV} \qquad \frac{\Gamma_1 \vdash t : \mathsf{T} \quad \Gamma_2 \vdash t' : \mathsf{T}\ {}_{\mathsf{m}}\!\multimap \mathsf{U}}{\mathsf{m} \cdot \Gamma_1 + \Gamma_2 \vdash t'\ t : \mathsf{U}}\ {\multimap}\mathrm{E} \qquad \frac{\Gamma_1 \vdash t : \mathsf{1} \qquad \Gamma_2 \vdash u : \mathsf{U}}{\Gamma_1 + \Gamma_2 \vdash t \fatsemi u : \mathsf{U}}\ \mathsf{1}\mathrm{E}$$

$$\frac{\Gamma_1 \vdash t : \mathsf{T}_1 \oplus \mathsf{T}_2 \quad \Gamma_2,\ x_1 :_{\mathsf{m}} \mathsf{T}_1 \vdash u_1 : \mathsf{U} \quad \Gamma_2,\ x_2 :_{\mathsf{m}} \mathsf{T}_2 \vdash u_2 : \mathsf{U}}{\mathsf{m} \cdot \Gamma_1 + \Gamma_2 \vdash \mathtt{case}_{\mathsf{m}}\ t\ \mathtt{of}\ \{\mathtt{Inl}\, x_1 \mapsto u_1\ ,\ \mathtt{Inr}\, x_2 \mapsto u_2\} : \mathsf{U}}\ {\oplus}\mathrm{E}$$

$$\frac{\Gamma_1 \vdash t : \mathsf{T}_1 \otimes \mathsf{T}_2 \quad \Gamma_2,\ x_1 :_{\mathsf{m}} \mathsf{T}_1,\ x_2 :_{\mathsf{m}} \mathsf{T}_2 \vdash u : \mathsf{U}}{\mathsf{m} \cdot \Gamma_1 + \Gamma_2 \vdash \mathtt{case}_{\mathsf{m}}\ t\ \mathtt{of}\ (x_1\ ,\ x_2) \mapsto u : \mathsf{U}}\ {\otimes}\mathrm{E} \qquad \frac{\Gamma_1 \vdash t : {!}_{\mathsf{n}}\mathsf{T} \quad \Gamma_2,\ x :_{\mathsf{m}\cdot\mathsf{n}} \mathsf{T} \vdash u : \mathsf{U}}{\mathsf{m} \cdot \Gamma_1 + \Gamma_2 \vdash \mathtt{case}_{\mathsf{m}}\ t\ \mathtt{of}\ \mathtt{Mod}_{\mathsf{n}}\ x \mapsto u : \mathsf{U}}\ {!}\mathrm{E}$$

$$\frac{\Gamma_1 \vdash t : \mathsf{U} \ltimes \mathsf{T} \quad 1{\uparrow} \cdot \Gamma_2,\ x :_{1\nu} \mathsf{T} \vdash t' : \mathsf{T}'}{\Gamma_1 + \Gamma_2 \vdash \mathtt{upd}_{\ltimes}\ t\ \mathtt{with}\ x \mapsto t' : \mathsf{U} \ltimes \mathsf{T}'}\ {\ltimes}\textsc{Upd} \qquad \frac{\Gamma \vdash u : \mathsf{U}}{\Gamma \vdash \mathtt{to}_{\ltimes}\ u : \mathsf{U} \ltimes \mathsf{1}}\ {\ltimes}\textsc{To}$$

$$\frac{\Gamma \vdash t : \mathsf{U} \ltimes ({!}_{1\infty}\mathsf{T})}{\Gamma \vdash \mathtt{from}_{\ltimes}\ t : \mathsf{U} \otimes ({!}_{1\infty}\mathsf{T})}\ {\ltimes}\textsc{From} \qquad \frac{}{\omega\nu \cdot \Gamma \vdash \mathtt{new}_{\ltimes} : \mathsf{T} \ltimes \lfloor_{1\nu}\mathsf{T}\rfloor}\ {\ltimes}\textsc{New} \qquad \frac{\Gamma \vdash t : \lfloor_{\mathsf{n}}\mathsf{1}\rfloor}{\Gamma \vdash t \triangleleft () : \mathsf{1}}\ \lfloor\mathsf{1}\rfloor\mathrm{E}$$

$$\frac{\Gamma \vdash t : \lfloor_{\mathsf{n}}\mathsf{T}_1 \oplus \mathsf{T}_2\rfloor}{\Gamma \vdash t \triangleleft \mathtt{Inl} : \lfloor_{\mathsf{n}}\mathsf{T}_1\rfloor}\ \lfloor\oplus\rfloor\mathrm{E}_1 \qquad \frac{\Gamma \vdash t : \lfloor_{\mathsf{n}}\mathsf{T}_1 \oplus \mathsf{T}_2\rfloor}{\Gamma \vdash t \triangleleft \mathtt{Inr} : \lfloor_{\mathsf{n}}\mathsf{T}_2\rfloor}\ \lfloor\oplus\rfloor\mathrm{E}_2 \qquad \frac{\Gamma \vdash t : \lfloor_{\mathsf{n}}\mathsf{T}_1 \otimes \mathsf{T}_2\rfloor}{\Gamma \vdash t \triangleleft (,) : \lfloor_{\mathsf{n}}\mathsf{T}_1\rfloor \otimes \lfloor_{\mathsf{n}}\mathsf{T}_2\rfloor}\ \lfloor\otimes\rfloor\mathrm{E}$$

$$\frac{\Gamma \vdash t : \lfloor_{\mathsf{n}}{!}_{\mathsf{n}'}\mathsf{T}\rfloor}{\Gamma \vdash t \triangleleft \mathtt{Mod}_{\mathsf{n}'} : \lfloor_{\mathsf{n}'\cdot\mathsf{n}}\mathsf{T}\rfloor}\ \lfloor{!}\rfloor\mathrm{E} \qquad \frac{\Gamma_1 \vdash t : \lfloor_{\mathsf{n}}\mathsf{T}\ {}_{\mathsf{m}}\!\multimap \mathsf{U}\rfloor \quad \Gamma_2,\ x :_{\mathsf{m}} \mathsf{T} \vdash u : \mathsf{U}}{\Gamma_1 + (1{\uparrow}\cdot\mathsf{n}) \cdot \Gamma_2 \vdash t \triangleleft (\boldsymbol{\lambda} x\ {}_{\mathsf{m}}\!\mapsto u) : \mathsf{1}}\ \lfloor{\multimap}\rfloor\mathrm{E}$$

$$\frac{\Gamma_1 \vdash t : \lfloor_{1\nu}\mathsf{U}\rfloor \quad \Gamma_2 \vdash t' : \mathsf{U} \ltimes \mathsf{T}}{\Gamma_1 + 1{\uparrow} \cdot \Gamma_2 \vdash t \mathbin{\triangleleft\!\circ} t' : \mathsf{T}}\ \lfloor\,\rfloor\mathrm{E_C} \qquad \frac{\Gamma_1 \vdash t : \lfloor_{\mathsf{n}}\mathsf{T}\rfloor \quad \Gamma_2 \vdash t' : \mathsf{T}}{\Gamma_1 + (1{\uparrow}\cdot\mathsf{n}) \cdot \Gamma_2 \vdash t \blacktriangleleft t' : \mathsf{1}}\ \lfloor\,\rfloor\mathrm{E_L}$$

Figure 2.2: Typing rules of $\lambda_d$ ($\lambda_d$–TY)

$\boxed{\Gamma \vdash t : \mathsf{T}}$ *(Derived typing judgment for syntactic sugar forms)*

$$\frac{\Gamma \vdash t : \mathsf{T} \ltimes \mathsf{1}}{\Gamma \vdash \mathtt{from}'_{\ltimes}\, t : \mathsf{T}}\ \ltimes\textsc{from}' \qquad \frac{}{\omega\nu \cdot \Gamma \vdash () : \mathsf{1}}\ \mathsf{1}\mathrm{I} \qquad \frac{\Gamma_2,\ x :_{\mathsf{m}} \mathsf{T} \vdash u : \mathsf{U}}{\Gamma_2 \vdash \lambda x\ {}_{\mathsf{m}}\!\mapsto u : \mathsf{T}\ {}_{\mathsf{m}}\!\multimap \mathsf{U}}\ \multimap\mathrm{I}$$

$$\frac{\Gamma_2 \vdash t : \mathsf{T}_1}{\Gamma_2 \vdash \mathtt{Inl}\, t : \mathsf{T}_1 \oplus \mathsf{T}_2}\ \oplus\mathrm{I}_1 \qquad \frac{\Gamma_2 \vdash t : \mathsf{T}_2}{\Gamma_2 \vdash \mathtt{Inr}\, t : \mathsf{T}_1 \oplus \mathsf{T}_2}\ \oplus\mathrm{I}_2 \qquad \frac{\Gamma_2 \vdash t : \mathsf{T}}{\mathsf{m} \cdot \Gamma_2 \vdash \mathtt{Mod}_{\mathsf{m}}\, t : !_{\mathsf{m}} \mathsf{T}}\ !\mathrm{I}$$

$$\frac{\Gamma_{21} \vdash t_1 : \mathsf{T}_1 \qquad \Gamma_{22} \vdash t_2 : \mathsf{T}_2}{\Gamma_{21} + \Gamma_{22} \vdash (t_1\,,\, t_2) : \mathsf{T}_1 \otimes \mathsf{T}_2}\ \otimes\mathrm{I}$$

Figure 2.3: Derived typing rules for syntactic sugar ($\lambda_d$–TY$_s$)

Rules $\lambda_d$–TY/IDV, $\lambda_d$–TY/$\ltimes$NEW and $\lambda_d$–TY$_s$/1I, that are the potential leaves of the typing tree, are where weakening for unrestricted variables is allowed to happen. That's why they allow to discard any typing context of the form $\omega\nu \cdot \Gamma$. It's very similar to what happened in $\lambda_{Ldm}$–TY/ID or $\lambda_{Ldm}$–TY/1I, except that $\Gamma$ is scaled by $\omega\nu$ instead of $\omega$ now.

Rule $\lambda_d$–TY/IDV, in addition to weakening, allows for coercion of the mode $\mathsf{m}$ of the variable used, with ordering constraint $1\nu \lessdot \mathsf{m}$ as defined in Figure 2.1. Unlike in $\lambda_{Ldm}$, here not all modes $\mathsf{m}$ are compatible with $1\nu$. Notably, mode coercion still does not allow for a finite age to be changed to another, as $\uparrow^j$ and $\uparrow^k$ are not comparable w.r.t. $\lessdot_{\mathsf{a}}$ when $j \neq k$. So for example we cannot use a variable with mode $1\uparrow$ in most contexts.

For pattern-matching, with rules $\lambda_d$–TY/$\oplus$E, $\lambda_d$–TY/$\otimes$E and $\lambda_d$–TY/!E, we keep the same *deep mode* approach as in $\lambda_{Ldm}$: **case** expressions are parametrized by a mode $\mathsf{m}$ by which the typing context $\Gamma_1$ of the scrutinee is multiplied. The variables which bind the subcomponents of the scrutinee then inherit this mode.

Let's now move on to the real novelties of the typing rules for $\lambda_d$, that is, rules for operators acting on ampars and destinations.

**Rules for scoping** As destinations always exist in the context of a structure with holes, and must stay in that context, we need a formal notion of *scope*. Scopes are created by $\lambda_d$–TY/$\ltimes$UPD, as destinations are only ever accessed through $\mathbf{upd}_{\ltimes}$. More precisely, $\mathbf{upd}_{\ltimes}\ t$ **with** $x \mapsto t'$ creates a new scope which spans over $t'$. In that scope, $x$ has age $\nu$ (now), and the ages of the existing bindings in $\Gamma_2$ are multiplied by $\uparrow$ (i.e. we add 1 to ages seen as a numbers). That is represented by $t'$ typing in $1\uparrow \cdot \Gamma_2,\ x :_{1\nu} \mathsf{T}$ while the parent term $\mathbf{upd}_{\ltimes}\ t$ **with** $x \mapsto t'$ types in unscaled contexts $\Gamma_1 + \Gamma_2$. This difference of age between $x$—introduced by $\mathbf{upd}_{\ltimes}$, containing destinations—and $\Gamma_2$ lets us see what originates from older scopes. Specifically, distinguishing the age of destinations is crucial when typing filling primitives to avoid the pitfalls of Section 2.3.

Figure 2.4 illustrates scopes introduced by $\mathbf{upd}_{\ltimes}$, and how the typing rules for $\mathbf{upd}_{\ltimes}$ and $\blacktriangleleft$ interact. For the first time here we see ampar values $/v_2\ {}_{,}v_1/$, represented as pair-like objects with a structure with holes $v_2$ on the left, and an arbitrary structure $v_1$ containing destinations on the right (ampar values will be formally introduced in Section 2.6.1). With $\mathbf{upd}_{\ltimes}$ we enter a new scope where the destinations are accessible, but the structure with holes remains in the outer scope. As a result, when filling a destination with $\lambda_d$–TY/⌊ ⌋E$_\text{L}$, for instance $d_{11}$ ◀ $x_0$ in Figure 2.4, we type $d_{11}$ in the new scope, while we type $x_0$ in the outer scope, as it's being moved to the structure with holes on the left of the ampar, which lives in the outer scope too. This is the opposite of the scaling that **upd**$_\ltimes$ does: while **upd**$_\ltimes$ creates a new scope for its body, operator ◀, and similarly, ◁∘ and ◁($\boldsymbol{\lambda} x$ $_\text{m}$↦ $u$), transfer their right operand to the outer scope. In other words, the right-hand side of ◀ or ◁ is an enclave for the parent scope.

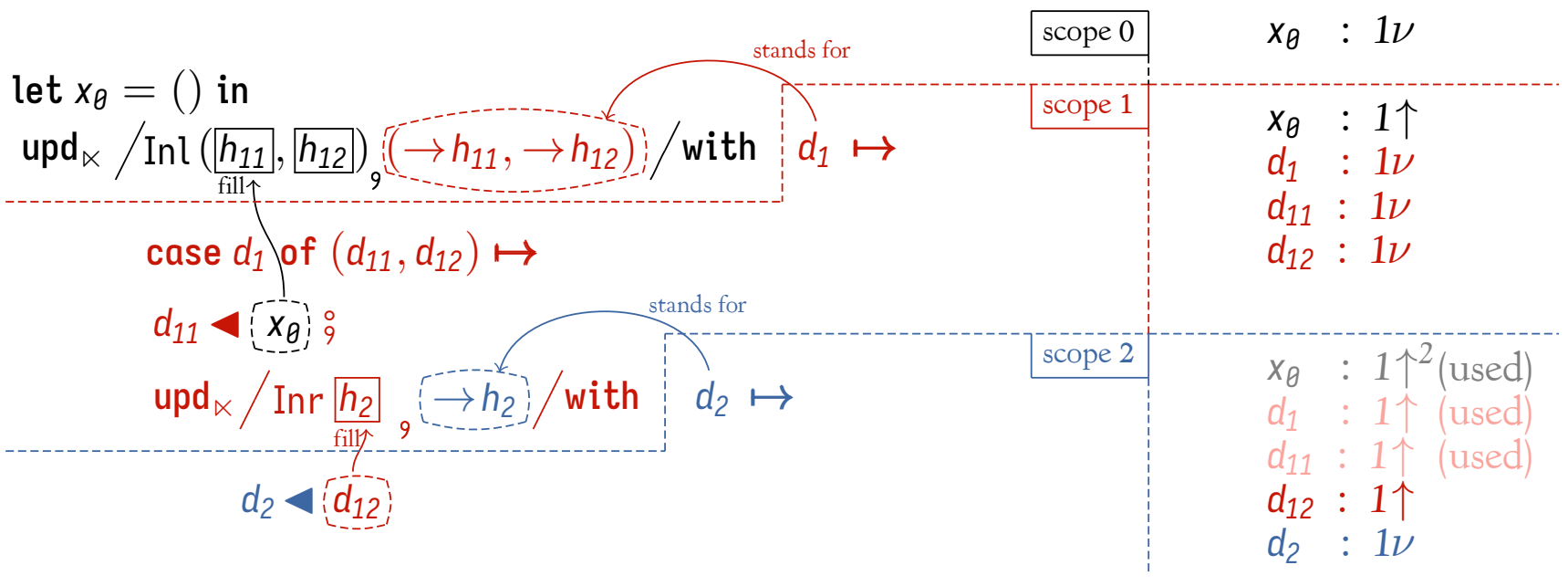

Figure 2.4: Evolution of ages through nested scopes

When using **from**$'_\ltimes$ (rule $\lambda_d$–TY$_\text{s}$/⋉FROM'), the left of an ampar is extracted to the current scope. This is the fundamental reason why the left of an ampar has to "take place" in the current scope. We know the structure is complete and can be extracted because the right-hand side is of type unit (1) and thus, no destination on the right-hand side means no hole can remain on the left. **from**$'_\ltimes$ is implemented in terms of **from**$_\ltimes$ in Listing 2.4 to keep the core calculus tidier (and limit the number of typing rules, evaluation contexts, etc), but it can be implemented much more efficiently in a real-world implementation.

When an ampar is eliminated with the more general **from**$_\ltimes$ in rule $\lambda_d$–TY/⋉FROM however, we extract both sides of the ampar to the current scope, even though the right-hand side is normally in a different scope. This is only safe to do because the right-hand side is required to have type $!_{1\infty}$T, which means it is scope-insensitive: it cannot contain any scope-controlled resource. Furthermore, it ensures that the right-hand side cannot contain destinations, so the structure has to be complete and thus is ready to be read.

In $\lambda_d$–TY/⋉TO, on the other hand, there is no need to bother with scopes: the operator **to**$_\ltimes$ wraps an already completed structure in a trivial ampar whose left side is the structure (that continues to type in the current scope), and right-hand side is unit.

The remaining operators ◁(), ◁Inl, ◁Inr, ◁Mod$_\text{m}$, ◁(,) from rules of the form $\lambda_d$–TY/⌊ ⌋E are the other destination-filling primitives. They write a hollow constructor to the hole pointed by the destination operand, and return the potential new destinations that are created for new holes in the hollow constructor (or unit if there is none).

The operator ◁∘ from rule $\lambda_d$–TY/⌊ ⌋E$_\text{C}$ let us fill a destination with an incomplete structure. This is the main way we can compose structures with holes together, reminiscing of the hcomp operator from [Minamide, 1998]. Unlike ◀, the operator ◁∘ does not return 1 but instead returns the remaining destinations of the incomplete structure that has been used to fill the hole.

Type-wise, one specificity of ◁∘, compared to other destination-filling primitives, is that it only accepts a left-hand-side operand of type ⌊$_{1\nu}$U⌋—that is, a destination whose corresponding hole has the unit mode. This restriction is deliberate. Suppose we have $d$ ◁∘($ampar$ : U ⋉ ⌊$_{1\nu}$T⌋) where $d$ : ⌊$_{\omega\nu}$U⌋, meaning $d$ refers to a hole behind a modality $!_{\omega\nu}$. The expression $d$ ◁∘ $ampar$ would return the destination of type ⌊$_{1\nu}$T⌋ (such a type signature claims that the associated hole is of mode $1\nu$) that constituted the right-hand side of *ampar*. However, this destination actually points to a hole in a structure that has now been moved behind a modality $!_{\omega\nu}$, and must therefore only accept values of mode $\omega\nu$.

There is no simple way to update the mode annotations of such destinations on the fly when using $\triangleleft\!\circ$ on a destination with a modality $m$ different than $1\nu$. The problem becomes even more delicate when foreign destinations—those referencing holes from others ampars—are stored on the right-hand side: their modes, on the other hand, must not be modified at all. To avoid these complexities and ensure soundness, we restrict $\triangleleft\!\circ$ to operated only on destinations whose corresponding holes have mode $1\nu$.

**Putting into practice** As an exercise, now that we have all the typing rules in place, we can work out the typing tree for the synthetic destination-filling operator $\triangleleft(::)$ and the corresponding synthetic constructor $(::)$ that we used since Section 2.2.2. We recall that $\mathsf{List\ T} \triangleq 1\oplus(\mathsf{T}\otimes(\mathsf{List\ T}))$. Let's start with the definition and then the typing tree of $\triangleleft(::)$:

$$\triangleleft(::)\ :\ \lfloor_{m}\mathsf{List\ T}\rfloor \multimap (\lfloor_{m}\mathsf{T}\rfloor\otimes\lfloor_{m}\mathsf{List\ T}\rfloor)$$
$$d \triangleleft (::)\ \triangleq\ d \triangleleft \mathsf{Inr} \triangleleft (,)$$

$A$-TY:

$$\dfrac{\dfrac{\dfrac{\dfrac{\dfrac{1\nu \lesssim 1\nu}{d :_{1\nu} \lfloor_{m}\mathsf{List\ T}\rfloor \vdash d : \lfloor_{m}\mathsf{List\ T}\rfloor}\ \textsc{idV}}{d :_{1\nu} \lfloor_{m}\mathsf{List\ T}\rfloor \vdash d : \lfloor_{m}1\oplus(\mathsf{T}\otimes(\mathsf{List\ T}))\rfloor}\ \text{unfold } \mathsf{List\ T} \text{ definition on the right}}{d :_{1\nu} \lfloor_{m}\mathsf{List\ T}\rfloor \vdash d \triangleleft \mathsf{Inr} : \lfloor_{m}\mathsf{T}\otimes(\mathsf{List\ T})\rfloor}\ \lfloor\oplus\rfloor E_2}{d :_{1\nu} \lfloor_{m}\mathsf{List\ T}\rfloor \vdash d \triangleleft \mathsf{Inr} \triangleleft (,) : \lfloor_{m}\mathsf{T}\rfloor\otimes\lfloor_{m}\mathsf{List\ T}\rfloor}\ \lfloor\otimes\rfloor E}{\bullet \vdash \lambda d \mapsto d \triangleleft \mathsf{Inr} \triangleleft (,) : \lfloor_{m}\mathsf{List\ T}\rfloor \multimap \lfloor_{m}\mathsf{T}\rfloor\otimes\lfloor_{m}\mathsf{List\ T}\rfloor}\ \lambda_d\text{-TY}_s/\!\multimap\! I$$

Now, let's move on to $(::)$ synthetic constructor:

$$(::)\ :\ \mathsf{T} \multimap \mathsf{List\ T} \multimap \mathsf{List\ T}$$
$$x :: xs\ \triangleq\ \mathsf{from}'_{\ltimes}(\mathsf{upd}_{\ltimes}\ \mathsf{new}_{\ltimes}\ \mathsf{with}\ d \mapsto \mathsf{case}\ (d \triangleleft (::))\ \mathsf{of}\ (dx\,,\,dxs) \mapsto dx \blacktriangleleft x\ ⨾\ dxs \blacktriangleleft xs)$$

Given that the typing tree will grow very large, we will first type the subexpression
$B \triangleq \mathsf{case}\ (d \triangleleft (::))\ \mathsf{of}\ (dx\,,\,dxs) \mapsto dx \blacktriangleleft x\ ⨾\ dxs \blacktriangleleft xs$, then type the whole expression corresponding to operator $(::)$.

$B$-TY:

$$\dfrac{\dfrac{\dfrac{1\nu \lesssim 1\nu}{d :_{1\nu} \lfloor\mathsf{List\ T}\rfloor \vdash d : \lfloor\mathsf{List\ T}\rfloor}\ \textsc{idV}}{d :_{1\nu} \lfloor\mathsf{List\ T}\rfloor \vdash d \triangleleft (::) : \lfloor\mathsf{T}\rfloor\otimes\lfloor\mathsf{List\ T}\rfloor}\ A\text{-TY} \quad \dfrac{\dfrac{\dfrac{1\nu \lesssim 1\nu}{dx :_{1\nu} \lfloor\mathsf{T}\rfloor \vdash dx : \lfloor\mathsf{T}\rfloor}\ \textsc{idV} \quad \dfrac{1\nu \lesssim 1\nu}{x :_{1\nu} \mathsf{T} \vdash x : \mathsf{T}}\ \textsc{idV}}{dx :_{1\nu} \lfloor\mathsf{T}\rfloor,\ 1\!\uparrow\!\cdot x :_{1\nu} \mathsf{T} \vdash dx \blacktriangleleft x : 1}\ \lfloor\rfloor E_L \quad \dfrac{\dfrac{1\nu \lesssim 1\nu}{dxs :_{1\nu} \lfloor\mathsf{List\ T}\rfloor \vdash dxs : \lfloor\mathsf{List\ T}\rfloor}\ \textsc{idV} \quad \dfrac{1\nu \lesssim 1\nu}{xs :_{1\nu} \mathsf{List\ T} \vdash xs : \mathsf{List\ T}}\ \textsc{idV}}{dxs :_{1\nu} \lfloor\mathsf{List\ T}\rfloor,\ 1\!\uparrow\!\cdot xs :_{1\nu} \mathsf{List\ T} \vdash dxs \blacktriangleleft xs : 1}\ \lfloor\rfloor E_L}{dx :_{1\nu} \lfloor\mathsf{T}\rfloor,\ dxs :_{1\nu} \lfloor\mathsf{List\ T}\rfloor,\ 1\!\uparrow\!\cdot(x :_{1\nu} \mathsf{T},\ xs :_{1\nu} \mathsf{List\ T}) \vdash dx \blacktriangleleft x\ ⨾\ dxs \blacktriangleleft xs : 1}\ 1E}{d :_{1\nu} \lfloor\mathsf{List\ T}\rfloor,\ 1\!\uparrow\!\cdot(x :_{1\nu} \mathsf{T},\ xs :_{1\nu} \mathsf{List\ T}) \vdash \mathsf{case}\ (d \triangleleft (::))\ \mathsf{of}\ (dx\,,\,dxs) \mapsto dx \blacktriangleleft x\ ⨾\ dxs \blacktriangleleft xs : 1}\ \otimes E$$

$C$-TY:

$$\dfrac{\dfrac{\dfrac{\dfrac{\dfrac{}{\bullet \vdash \mathsf{new}_{\ltimes} : \mathsf{List\ T} \ltimes \lfloor\mathsf{List\ T}\rfloor}\ \ltimes\textsc{new} \quad \dfrac{}{d :_{1\nu} \lfloor\mathsf{List\ T}\rfloor,\ 1\!\uparrow\!\cdot(x :_{1\nu} \mathsf{T},\ xs :_{1\nu} \mathsf{List\ T}) \vdash B : 1}\ B\text{-TY}}{x :_{1\nu} \mathsf{T},\ xs :_{1\nu} \mathsf{List\ T} \vdash \mathsf{upd}_{\ltimes}\ \mathsf{new}_{\ltimes}\ \mathsf{with}\ d \mapsto B : \mathsf{List\ T} \ltimes 1}\ \ltimes\textsc{upd}}{x :_{1\nu} \mathsf{T},\ xs :_{1\nu} \mathsf{List\ T} \vdash \mathsf{from}'_{\ltimes}(\mathsf{upd}_{\ltimes}\ \mathsf{new}_{\ltimes}\ \mathsf{with}\ d \mapsto B) : \mathsf{List\ T}}\ \ltimes\textsc{from}'}{x :_{1\nu} \mathsf{T} \vdash \lambda xs \mapsto \mathsf{from}'_{\ltimes}(\mathsf{upd}_{\ltimes}\ \mathsf{new}_{\ltimes}\ \mathsf{with}\ d \mapsto B) : \mathsf{List\ T} \multimap \mathsf{List\ T}}\ \lambda_d\text{-TY}_s/\!\multimap\! I}{\bullet \vdash \lambda x \mapsto \lambda xs \mapsto \mathsf{from}'_{\ltimes}(\mathsf{upd}_{\ltimes}\ \mathsf{new}_{\ltimes}\ \mathsf{with}\ d \mapsto B) : \mathsf{T} \multimap \mathsf{List\ T} \multimap \mathsf{List\ T}}\ \lambda_d\text{-TY}_s/\!\multimap\! I$$

If we look at the typing tree $C$-TY from the root to the leaves, we see that when going under $\mathtt{upd}_\ltimes$, variable bindings for $x$ and $xs$ are multiplied by $1\uparrow$ (their age is increased by one). But this is not an issue; in fact, it's required by the typing rule $\lambda_d$–TY/$\lfloor\ \rfloor E_L$ that the right operand is of age $\uparrow$ (as always, to prevent scope escape), as we see in typing tree $B$-TY.

Also, as in a regular functional programming language, modes and more specifically ages do *not* show up in the signature of synthetic constructor $(::)$; only in its internal typing derivations, because the use of destinations to define $(::)$ is, from this point of view, just implementation details hidden to the user.

## 2.6 Operational semantics

We give the operational semantics of $\lambda_d$ in the *reduction semantics* style that we used in Chapter 1, that's to say with explicit syntactic manipulation of evaluation contexts. However this time we will be more thorough. In particular, we will introduce the grammar of evaluation contexts formally, and we will also define typing rules for evaluation contexts and commands, so as to prove that typing is preserved throughout the reduction. As before, commands $E[t]$ are used to represent running programs; they're described in detail in Section 2.6.2.

But first, we need a class of runtime values (we will often just say *values*), and their corresponding typing rules, as $\lambda_d$ currently lacks any way to represent destinations or holes, or really any kind of value (for instance $\mathsf{Inl}\,()$ has been, so far, just syntactic sugar for a term $\mathtt{from}'_\ltimes\,(\mathtt{upd}_\ltimes\ \mathtt{new}_\ltimes\ \mathtt{with}\ d \mapsto \ldots)$). It's a peculiarity of $\lambda_d$ that values (in particular, data constructors) are just used during reduction; usually they are also allowed in the term syntax of functional languages as in $\lambda_{Ldm}$.

### 2.6.1 Runtime values and new typing context forms

The syntax of runtime values is given in Figure 2.5. It features constructors for all of our basic types, as well as a primitive function form $\lambda_v x\ {}_{m}\!\mapsto u$ (which has more restrictions than the syntactic-sugar form $\lambda x\ {}_{m}\!\mapsto u$; for instance, it cannot contain free variables). The more interesting values are holes $\boxed{h}$, destinations $\rightarrow h$, and ampars ${}_H\langle v_2\ {}_{,}\,v_1\rangle$, which were not present in $\lambda_{Ldm}$, and that we will describe in the rest of the section. In order for the operational semantics to use substitution, which requires substituting variables with values, we also extend the syntax of terms to include values, and we add the corresponding typing rule $\lambda_d$–TY/FROMVAL (at the top of Figure 2.6).

Destinations and holes are two faces of the same coin, as seen in Section 2.2.1, and we must ensure that throughout the reduction, a destination always points to a hole, and a hole is always the target of exactly one destination. Thus, the new idea of our system is to feature *hole bindings* $\boxed{h} :_{n} \mathsf{T}$ and *destination bindings* $\rightarrow h :_{m} \lfloor_{n} \mathsf{T}\rfloor$ in typing contexts in addition to the usual variable bindings $x :_{m} \mathsf{T}$. In both cases, we call $h$ a *hole name*. We complement this by new typing context forms: we now have $\Gamma$, $\Theta$ and $\Delta$. $\Gamma$ is the form of typing context used when typing terms; we allow it to contain variable bindings (as before) but also destination bindings. On the other hand, the new context form $\Theta$ can contain both destination bindings and hole bindings, but *not a destination binding and a hole binding for the same hole name*. $\Theta$ cannot contain variable bindings either. Finally, a typing context $\Delta$ can only contain destination bindings.

We extend our previous context operations $+$ and $\cdot$ to act on the new binding forms, as described in Figure 2.5. Context addition is still very partial; for instance, $(\boxed{h} :_{n} \mathsf{T}) + (\rightarrow h :_{m} \lfloor_{n'} \mathsf{T}\rfloor)$ is not defined, as $h$ is present on both sides but with different binding forms.

*Grammar extended with values:*

$$t, u \ ::= \ \ldots \ \mid \ v$$

$$\begin{array}{rcll} v & ::= & \boxed{h} & \textit{(hole)} \\ & \mid & \rightarrow h & \textit{(destination)} \\ & \mid & {}_{H}/ v_2 \, {}_{,} v_1 / & \textit{(ampar value)} \\ & \mid & () \ \mid \ \mathtt{Inl}\, v \ \mid \ \mathtt{Inr}\, v \ \mid \ \mathtt{Mod}_{m}\, v \ \mid \ (v_1 , v_2) \ \mid \ \lambda_v x \, {}_{m}\!\mapsto u & \end{array}$$

*Extended grammar of typing contexts:*

$$\begin{array}{rclllll} \Delta & ::= & \bullet & \mid \ \rightarrow h :_{m} \lfloor_{n} T \rfloor & & \mid & \Delta_1, \ \Delta_2 \\ \Gamma & ::= & \bullet & \mid \ \rightarrow h :_{m} \lfloor_{n} T \rfloor & \mid \ x :_{m} T & \mid & \Gamma_1, \ \Gamma_2 \\ \Theta & ::= & \bullet & \mid \ \rightarrow h :_{m} \lfloor_{n} T \rfloor & \mid \ \boxed{h} :_{n} T & \mid & \Theta_1, \ \Theta_2 \end{array}$$

*Operations on new typing context forms:*

$$\begin{array}{rcll} n' \cdot (\boxed{h} :_{n} T, \ \Theta) & \triangleq & (\boxed{h} :_{n' \cdot n} T), \ n' \cdot \Theta & \\ n' \cdot (\rightarrow h :_{m} \lfloor_{n} T \rfloor, \ \Gamma) & \triangleq & (\rightarrow h :_{n' \cdot m} \lfloor_{n} T \rfloor), \ n' \cdot \Gamma & \dagger \\ & & & \\ (\boxed{h} :_{n} T, \ \Theta_1) + \Theta_2 & \triangleq & \boxed{h} :_{n} T, \ (\Theta_1 + \Theta_2) & \text{if } h \notin \Theta_2 \\ (\boxed{h} :_{n} T, \ \Theta_1) + (\boxed{h} :_{n'} T, \ \Theta_2) & \triangleq & \boxed{h} :_{n + n'} T, \ (\Theta_1 + \Theta_2) & \\ (\rightarrow h :_{m} \lfloor_{n} T \rfloor, \ \Gamma_1) + \Gamma_2 & \triangleq & \rightarrow h :_{m} \lfloor_{n} T \rfloor, \ (\Gamma_1 + \Gamma_2) & \text{if } h \notin \Gamma_2 \ \dagger \\ (\rightarrow h :_{m} \lfloor_{n} T \rfloor, \ \Gamma_1) + (\rightarrow h :_{m'} \lfloor_{n} T \rfloor, \ \Gamma_2) & \triangleq & \rightarrow h :_{m + m'} \lfloor_{n} T \rfloor, \ (\Gamma_1 + \Gamma_2) & \dagger \\ & & & \\ \rightarrow^{-1}(\bullet) & \triangleq & \bullet & \\ \rightarrow^{-1}(\rightarrow h :_{1\nu} \lfloor_{n} T \rfloor, \ \Delta) & \triangleq & (\boxed{h} :_{n} T), \ \rightarrow^{-1}(\Delta) & \end{array}$$

$\dagger$ : *same rule is also true for $\Theta$ or $\Delta$ replacing $\Gamma$*

Figure 2.5: Runtime values, new typing context forms, and operators

*Typing values as terms:*

$$\frac{\Delta \vdash_{\mathsf{v}} v : \mathsf{T}}{\omega\nu \cdot \Gamma,\ \Delta \vdash v : \mathsf{T}}\ \lambda_d\text{–TY/FROMVAL}$$

---

$\boxed{\Theta \vdash_{\mathsf{v}} v : \mathsf{T}}$ *(Typing judgment for values)*

$$\frac{}{\boxed{h} :_{1\nu} \mathsf{T} \vdash_{\mathsf{v}} \boxed{h} : \mathsf{T}}\ \text{IDH} \qquad \frac{1\nu \lessdot \mathsf{m}}{\rightarrow h :_{\mathsf{m}} \lfloor_{\mathsf{n}} \mathsf{T} \rfloor \vdash_{\mathsf{v}} \rightarrow h : \lfloor_{\mathsf{n}} \mathsf{T} \rfloor}\ \text{IDD} \qquad \frac{}{\cdot \vdash_{\mathsf{v}} () : \mathsf{1}}\ \mathsf{1}\text{I}$$

$$\frac{\Delta,\ x :_{\mathsf{m}} \mathsf{T} \vdash u : \mathsf{U}}{\Delta \vdash_{\mathsf{v}} \lambda_{\mathsf{v}} x\ {}_{\mathsf{m}}\!\mapsto u : \mathsf{T}\ {}_{\mathsf{m}}\!\multimap \mathsf{U}}\ \multimap\text{I} \qquad \frac{\Theta \vdash_{\mathsf{v}} v_1 : \mathsf{T}_1}{\Theta \vdash_{\mathsf{v}} \mathsf{Inl}\, v_1 : \mathsf{T}_1 \oplus \mathsf{T}_2}\ \oplus\text{I}_1 \qquad \frac{\Theta \vdash_{\mathsf{v}} v_2 : \mathsf{T}_2}{\Theta \vdash_{\mathsf{v}} \mathsf{Inr}\, v_2 : \mathsf{T}_1 \oplus \mathsf{T}_2}\ \oplus\text{I}_2$$

$$\frac{\Theta_1 \vdash_{\mathsf{v}} v_1 : \mathsf{T}_1 \quad \Theta_2 \vdash_{\mathsf{v}} v_2 : \mathsf{T}_2}{\Theta_1 + \Theta_2 \vdash_{\mathsf{v}} (v_1\,,\, v_2) : \mathsf{T}_1 \otimes \mathsf{T}_2}\ \otimes\text{I} \qquad \frac{\Theta \vdash_{\mathsf{v}} v' : \mathsf{T}}{\mathsf{n} \cdot \Theta \vdash_{\mathsf{v}} \mathsf{Mod}_{\mathsf{n}}\, v' : !_{\mathsf{n}} \mathsf{T}}\ !\text{I} \qquad \frac{1\uparrow \cdot \Delta_1,\ \Delta_3 \vdash_{\mathsf{v}} v_1 : \mathsf{T} \quad \Delta_2,\ \rightarrow^{-1}\Delta_3 \vdash_{\mathsf{v}} v_2 : \mathsf{U}}{\Delta_1,\ \Delta_2 \vdash_{\mathsf{v}} {}_{\mathsf{hnames}(\Delta_3)}\!/v_2\ {}_{\text{,}}\, v_1/ : \mathsf{U} \ltimes \mathsf{T}}\ \ltimes\text{I}$$

Figure 2.6: Typing rules for runtime values ($\lambda_d$–TY$_\text{V}$)

One of the main goals of $\lambda_d$ is to ensure that a hole value is never read. We do not distinguish values with holes from fully-defined values at the syntactic level: instead types prevent holes from being read. The type system maintains this invariant by simply not allowing any hole bindings in the context when typing terms (indeed, typing rules $\lambda_d$–TY from Figure 2.2 still hold, and for instance, typing contexts must be of shape Γ). In fact, the only place where holes are introduced, is in typing rules for values, in Figure 2.6, and more precisely, in the left-hand side $v_2$ of an ampar ${}_H/v_2\ {}_{,}v_1/$ in rule $\lambda_d$–TY$_\text{V}$/$\ltimes$I.

Specifically, holes come from the operator $\rightarrow^{-1}$, which represents the hole bindings matching the specified set of destination bindings. It's a partial, pointwise operation on typing contexts Δ, as defined in Figure 2.5. Note that $\rightarrow^{-1}\Delta$ is undefined if any destination binding in Δ has a mode other than $1\nu$ (so that this operation is bijective).

Furthermore, in $\lambda_d$–TY$_\text{V}$/$\ltimes$I, the holes $\rightarrow^{-1}\Delta_3$ and the corresponding destinations $\Delta_3$ of the ampar are bound together and consequently removed from the ampar's typing context: this is how we ensure that, indeed, there's one destination per hole and one hole per destination. We annotate, with subscript $H$, the set of hole names that an ampar ${}_H/v_2\ {}_{,}v_1/$ bounds. That's why, in rule $\lambda_d$–TY$_\text{V}$/$\ltimes$I, the resulting ampar carries the set of hole names $\mathsf{hnames}(\Delta_3)$ (we use the operator $\mathsf{hnames}(\bullet)$ to denote the holes names present in a given typing context).

That being said, an ampar can also store destinations that are not its own, from other scopes; they are represented by $1\uparrow \cdot \Delta_1$ and $\Delta_2$ in the respective typing contexts of $v_1$ and $v_2$ in$\lambda_d$–TY$_\text{V}$/$\ltimes$I.

Rule $\lambda_d$–TY$_V$/IDH indicates that a hole must have mode $1\nu$ in the typing context to be well-typed; in particular mode coercion is not allowed here, and neither is weakening. Only when a hole is behind an exponential, that mode can change to some arbitrary mode $\mathsf{n}$. The mode of a hole constrains which values can be written to it, e.g. in $\boxed{h} :_{\mathsf{n}} \mathsf{T} \vdash_{\mathsf{v}} \mathtt{Mod}_{\mathsf{n}}\, \boxed{h} : !_{\mathsf{n}}\mathsf{T}$, only a value with mode $\mathsf{n}$ (more precisely, a value typed in a context of the form $\mathsf{n} \cdot \Theta$) can be written to $\boxed{h}$.

Surprisingly, in $\lambda_d$–TY$_V$/IDD, we see that a destination can be typed with any mode $\mathsf{m}$ coercible to $1\nu$. We did this to mimic the rule $\lambda_d$–TY/IDV and make the general modal substitution lemma expressible for $\lambda_d$[18]. We formally proved however that throughout the reduction of a well-typed closed program, $\mathsf{m}$ will never be of multiplicity $\omega$ or age $\infty$—a destination is always linear and of finite age—so mode coercion is never actually used; and we used this at our advantage to make the formal proof of the substitution lemma much easier. The other mode $\mathsf{n}$, appearing in $\lambda_d$–TY$_V$/IDD, is not the mode of the destination binding; instead it is part of the type $\lfloor_{\mathsf{n}}\mathsf{T}\rfloor$ and corresponds to the mode of values that we can write to the corresponding $\boxed{h}$; so for it no coercion can take place.

**Other salient points** While values are typed in contexts $\Theta$ allowing both destination and hole bindings, when using a value as a term in $\lambda_d$–TY/FROMVAL, it's only allowed to have free destinations, but no free holes (as we require a typing context of form $\Delta$ for the value).

Notice, also, that values (used as terms, or not) cannot have free variables, since contexts $\Theta$ only contain hole and destination bindings, but no variable binding. That values are closed is a standard feature of denotational semantics or abstract machine semantics. This is true even for function values ($\lambda_d$–TY$_V$/$\multimap$I), which, also is prevented from containing free holes. Since a function's body is unevaluated, it's unclear what it'd mean for a function to contain holes; at the very least it'd complicate our system a lot, and we are unaware of any benefit supporting free holes in functions could bring.

One might wonder how we can represent a curried function $\boldsymbol{\lambda} x \mapsto \boldsymbol{\lambda} y \mapsto x\ \mathtt{concat}\ y$ at the value level, as the inner abstraction captures the free variable $x$. The answer is that such a function, at value level, is encoded as $\boldsymbol{\lambda}_{\mathsf{v}} x \mapsto \mathtt{from}'_{\ltimes}(\mathtt{upd}_{\ltimes}\ \mathtt{new}_{\ltimes}\ \mathtt{with}\ d \mapsto d \triangleleft (\boldsymbol{\lambda} y \mapsto x\ \mathtt{concat}\ y))$, where the inner closure is not yet in value form. As $d \triangleleft (\boldsymbol{\lambda} y \mapsto x\ \mathtt{concat}\ y)$ is part of term syntax, it's allowed to have free variable $x$.

### 2.6.2 Evaluation contexts and commands

A running program is represented by a *command* $E[t]$, that is, a pair of an evaluation context $E$, and an (extended) term $t$ under focus, as in Section 1.6.

The grammar of evaluation contexts is given in Figure 2.7. As for $\lambda_{Ldm}$, an evaluation context $E$ is the composition of an arbitrary number of focusing components $e$. It might seem surprising that we do not need a notion of store or state in our semantics to represent the mutation induced by filling destinations. In fact, destination filling only requires a very tame notion of state—so tame that we can simply represent writing to a hole by a substitution *on the evaluation context*, instead of using more heavy store semantics.

It's important to note, however, that unlike in $\lambda_{Ldm}$, command $E[t]$ of $\lambda_d$ will not always have a corresponding term. The reason is that the focusing components are all directly derived from the term syntax, except for the *open ampar* focusing component ${}^{\mathrm{op}}_{H}/v_2\ {}_{,}[\,]/$ that does not have a corresponding term construct. This focusing component indicates that an ampar is currently being processed by $\mathtt{upd}_{\ltimes}$, with its left-hand side $v_2$ (the structure being built) being attached to the open ampar focusing component,

[18] Generally, in modal systems, if $x :_{\mathsf{m}} \mathsf{T},\ \Gamma \vdash u : \mathsf{U}$ and $\Delta \vdash v : \mathsf{T}$ then $\mathsf{m} \cdot \Delta,\ \Gamma \vdash u[x := v] : \mathsf{U}$ (see in particular [[Abel and Bernardy, 2020]]). In $\lambda_d$ we have $x :_{\omega\infty} \lfloor_{\mathsf{n}}\mathsf{T}\rfloor \vdash () : 1$ and $\rightarrow h :_{1\nu} \lfloor_{\mathsf{n}}\mathsf{T}\rfloor \vdash \rightarrow h : \lfloor_{\mathsf{n}}\mathsf{T}\rfloor$ so $\omega\infty \cdot (\rightarrow h :_{1\nu} \lfloor_{\mathsf{n}}\mathsf{T}\rfloor) \vdash ()[x := \rightarrow h] : 1$ should be valid.

$$
\begin{array}{rcl}
e & ::= & t'\,[\,] \mid [\,]\,v \mid [\,]\,\mathbin{\fatsemi}\,u \\
& \mid & \mathtt{case}_{\mathtt{m}}\,[\,]\,\mathtt{of}\,\{\mathtt{Inl}\,x_1 \mapsto u_1\,,\ \mathtt{Inr}\,x_2 \mapsto u_2\} \mid \mathtt{case}_{\mathtt{m}}\,[\,]\,\mathtt{of}\,(x_1\,,x_2) \mapsto u \mid \mathtt{case}_{\mathtt{m}}\,[\,]\,\mathtt{of}\,\mathtt{Mod}_{\mathtt{n}}\,x \mapsto u \\
& \mid & \mathtt{upd}_{\ltimes}\,[\,]\,\mathtt{with}\,x \mapsto t' \mid \mathtt{to}_{\ltimes}\,[\,] \mid \mathtt{from}_{\ltimes}\,[\,] \mid [\,] \lhd\!\circ\, t' \mid v \lhd\!\circ\,[\,] \mid [\,] \blacktriangleleft t' \mid v \blacktriangleleft [\,] \\
& \mid & [\,] \lhd () \mid [\,] \lhd \mathtt{Inl} \mid [\,] \lhd \mathtt{Inr} \mid [\,] \lhd (,) \mid [\,] \lhd \mathtt{Mod}_{\mathtt{m}} \mid [\,] \lhd (\lambda x \,{}_{\mathtt{m}}\!\mapsto u) \\
& \mid & {}^{\mathrm{op}}_{H}/v_2\,{}_{\fatsemi}[\,]/ \qquad \textit{(open ampar focusing component)} \\
E & ::= & [\,] \mid E \circ e
\end{array}
$$

Figure 2.7: Grammar of evaluation contexts

while its right-hand side (containing destinations) is either in subsequent focusing components, or in the term under focus. Ampars being open during the evaluation of $\mathtt{upd}_{\ltimes}$'s body and closed back afterwards is counterpart to the notion of scopes in typing rules.

**Typing of evaluation contexts** Evaluation contexts are typed in a context $\Delta$ that can only contain destination bindings. The typing context $\Delta$ of an evaluation context $E$ is exactly the typing context that the term $t$ has to use to form a valid $E[t]$. In other words, while $\Gamma \vdash t : \mathtt{T}$ *requires* the bindings of $\Gamma$, judgment $\Delta \dashv E : \mathtt{T} \rightarrowtail \mathtt{U_0}$ *provides* the bindings of $\Delta$. Typing rules for evaluation contexts, and the sole typing rule for commands, are given in Figure 2.8.

An evaluation context has a context type $\mathtt{T} \rightarrowtail \mathtt{U_0}$. The meaning of $E : \mathtt{T} \rightarrowtail \mathtt{U_0}$ is that given $t : \mathtt{T}$, $E[t]$ returns a value of type $\mathtt{U_0}$. Composing an evaluation context $E : \mathtt{T} \rightarrowtail \mathtt{U_0}$ with a new focusing component never affects the type $\mathtt{U_0}$ of the future command; only the type $\mathtt{T}$ of the focus is altered.

All typing rules for evaluation contexts can be derived systematically from the ones for the corresponding term (except for the rule $\lambda_d$–TY$_{\mathrm{E}}$/$\ltimes$OP, as there is no direct term equivalent to the *open ampar* focusing component). Let's take the rule $\lambda_d$–TY$_{\mathrm{E}}$/$\otimes$E as an example:

$\lambda_d$–TY$_{\mathrm{E}}$/$\otimes$E

$$
\frac{\mathtt{m}\cdot\Delta_1,\ \Delta_2 \dashv E : \mathtt{U} \rightarrowtail \mathtt{U_0} \qquad \Delta_2,\ x_1 :_{\mathtt{m}} \mathtt{T_1},\ x_2 :_{\mathtt{m}} \mathtt{T_2} \vdash u : \mathtt{U}}{\Delta_1 \dashv E \circ \mathtt{case}_{\mathtt{m}}\,[\,]\,\mathtt{of}\,(x_1\,,x_2) \mapsto u : (\mathtt{T_1} \otimes \mathtt{T_2}) \rightarrowtail \mathtt{U_0}}
$$

$\lambda_d$–TY/$\otimes$E

$$
\frac{\Gamma_1 \vdash t : \mathtt{T_1} \otimes \mathtt{T_2} \qquad \Gamma_2,\ x_1 :_{\mathtt{m}} \mathtt{T_1},\ x_2 :_{\mathtt{m}} \mathtt{T_2} \vdash u : \mathtt{U}}{\mathtt{m}\cdot\Gamma_1 + \Gamma_2 \vdash \mathtt{case}_{\mathtt{m}}\,t\,\mathtt{of}\,(x_1\,,x_2) \mapsto u : \mathtt{U}}
$$

- the typing context $\mathtt{m}\cdot\Delta_1,\ \Delta_2$ in the premise for $E$ in $\lambda_d$–TY$_{\mathrm{E}}$/$\otimes$E corresponds to $\mathtt{m}\cdot\Gamma_1 + \Gamma_2$ in the conclusion of $\lambda_d$–TY/$\otimes$E;
- the typing context $\Delta_2,\ x_1 :_{\mathtt{m}} \mathtt{T_1},\ x_2 :_{\mathtt{m}} \mathtt{T_2}$ in the premise for term $u$ in $\lambda_d$–TY$_{\mathrm{E}}$/$\otimes$E corresponds to the typing context $\Gamma_2,\ x_1 :_{\mathtt{m}} \mathtt{T_1},\ x_2 :_{\mathtt{m}} \mathtt{T_2}$ for the same term in $\lambda_d$–TY/$\otimes$E;
- the typing context $\Delta_1$ in the conclusion for $E \circ \mathtt{case}_{\mathtt{m}}\,[\,]\,\mathtt{of}\,(x_1\,,x_2) \mapsto u$ in $\lambda_d$–TY$_{\mathrm{E}}$/$\otimes$E corresponds to the typing context $\Gamma_1$ in the premise for $t$ in $\lambda_d$–TY/$\otimes$E (the term $t$ is located where the focus $[\,]$ is in $\lambda_d$–TY$_{\mathrm{E}}$/$\otimes$E).

We think of the typing rule for an evaluation context as a rotation of the typing rule for the associated term, where the typing contexts of one subterm and the conclusion are swapped, and the typing contexts of the other potential subterms are kept unchanged (with the difference that typing contexts for evaluation contexts are of shape $\Delta$ instead of $\Gamma$).

$\boxed{\Delta \dashv E : \mathsf{T} \rightarrowtail \mathsf{U_0}}$ *(Typing judgment for evaluation contexts)*

$$\dfrac{}{\cdot \dashv [\,] : \mathsf{U_0} \rightarrowtail \mathsf{U_0}}\ \textsc{id}$$

$$\dfrac{\mathsf{m} \cdot \Delta_1,\ \Delta_2 \dashv E : \mathsf{U} \rightarrowtail \mathsf{U_0} \qquad \Delta_2 \vdash t' : \mathsf{T}\ {}_{\mathsf{m}}\!\multimap \mathsf{U}}{\Delta_1 \dashv E \circ t'\,[\,] : \mathsf{T} \rightarrowtail \mathsf{U_0}}\ {\multimap}\mathrm{E_1}$$

$$\dfrac{\mathsf{m} \cdot \Delta_1,\ \Delta_2 \dashv E : \mathsf{U} \rightarrowtail \mathsf{U_0} \qquad \Delta_1 \vdash v : \mathsf{T}}{\Delta_2 \dashv E \circ [\,]\,v : (\mathsf{T}\ {}_{\mathsf{m}}\!\multimap \mathsf{U}) \rightarrowtail \mathsf{U_0}}\ {\multimap}\mathrm{E_2}$$

$$\dfrac{\Delta_1,\ \Delta_2 \dashv E : \mathsf{U} \rightarrowtail \mathsf{U_0} \qquad \Delta_2 \vdash u : \mathsf{U}}{\Delta_1 \dashv E \circ [\,] \fatsemi u : \mathsf{1} \rightarrowtail \mathsf{U_0}}\ \mathsf{1}\mathrm{E}$$

$$\dfrac{\mathsf{m} \cdot \Delta_1,\ \Delta_2 \dashv E : \mathsf{U} \rightarrowtail \mathsf{U_0} \qquad \Delta_2,\ x_1 :_{\mathsf{m}} \mathsf{T_1} \vdash u_1 : \mathsf{U} \qquad \Delta_2,\ x_2 :_{\mathsf{m}} \mathsf{T_2} \vdash u_2 : \mathsf{U}}{\Delta_1 \dashv E \circ \mathbf{case}_{\mathsf{m}}\ [\,]\ \mathbf{of}\ \{\mathsf{Inl}\,x_1 \mapsto u_1\,,\ \mathsf{Inr}\,x_2 \mapsto u_2\} : (\mathsf{T_1} \oplus \mathsf{T_2}) \rightarrowtail \mathsf{U_0}}\ \oplus\mathrm{E}$$

$$\dfrac{\mathsf{m} \cdot \Delta_1,\ \Delta_2 \dashv E : \mathsf{U} \rightarrowtail \mathsf{U_0} \qquad \Delta_2,\ x_1 :_{\mathsf{m}} \mathsf{T_1},\ x_2 :_{\mathsf{m}} \mathsf{T_2} \vdash u : \mathsf{U}}{\Delta_1 \dashv E \circ \mathbf{case}_{\mathsf{m}}\ [\,]\ \mathbf{of}\ (x_1\,,\, x_2) \mapsto u : (\mathsf{T_1} \otimes \mathsf{T_2}) \rightarrowtail \mathsf{U_0}}\ \otimes\mathrm{E}$$

$$\dfrac{\mathsf{m} \cdot \Delta_1,\ \Delta_2 \dashv E : \mathsf{U} \rightarrowtail \mathsf{U_0} \qquad \Delta_2,\ x :_{\mathsf{m} \cdot \mathsf{m}'} \mathsf{T} \vdash u : \mathsf{U}}{\Delta_1 \dashv E \circ \mathbf{case}_{\mathsf{m}}\ [\,]\ \mathbf{of}\ \mathsf{Mod}_{\mathsf{m}'}\,x \mapsto u : {!_{\mathsf{m}'}}\mathsf{T} \rightarrowtail \mathsf{U_0}}\ !\mathrm{E}$$

$$\dfrac{\Delta_1,\ \Delta_2 \dashv E : \mathsf{U} \ltimes \mathsf{T}' \rightarrowtail \mathsf{U_0} \qquad \mathsf{1}{\uparrow} \cdot \Delta_2,\ x :_{\mathsf{1}\nu} \mathsf{T} \vdash t' : \mathsf{T}'}{\Delta_1 \dashv E \circ \mathbf{upd}_{\ltimes}\ [\,]\ \mathbf{with}\ x \mapsto t' : (\mathsf{U} \ltimes \mathsf{T}) \rightarrowtail \mathsf{U_0}}\ \ltimes\textsc{upd}$$

$$\dfrac{\Delta \dashv E : (\mathsf{U} \ltimes \mathsf{1}) \rightarrowtail \mathsf{U_0}}{\Delta \dashv E \circ \mathbf{to}_{\ltimes}\ [\,] : \mathsf{U} \rightarrowtail \mathsf{U_0}}\ \ltimes\textsc{to}$$

$$\dfrac{\Delta \dashv E : (\mathsf{U} \otimes ({!_{\mathsf{1}\infty}}\mathsf{T})) \rightarrowtail \mathsf{U_0}}{\Delta \dashv E \circ \mathbf{from}_{\ltimes}\ [\,] : (\mathsf{U} \ltimes ({!_{\mathsf{1}\infty}}\mathsf{T})) \rightarrowtail \mathsf{U_0}}\ \ltimes\textsc{from}$$

$$\dfrac{\Delta \dashv E : \mathsf{1} \rightarrowtail \mathsf{U_0}}{\Delta \dashv E \circ [\,] \triangleleft () : \lfloor_{\mathsf{n}} \mathsf{1} \rfloor \rightarrowtail \mathsf{U_0}}\ \lfloor \mathsf{1} \rfloor\mathrm{E}$$

$$\dfrac{\Delta \dashv E : \lfloor_{\mathsf{n}} \mathsf{T_1} \rfloor \rightarrowtail \mathsf{U_0}}{\Delta \dashv E \circ [\,] \triangleleft \mathsf{Inl} : \lfloor_{\mathsf{n}} \mathsf{T_1} \oplus \mathsf{T_2} \rfloor \rightarrowtail \mathsf{U_0}}\ \lfloor \oplus \rfloor\mathrm{E_1}$$

$$\dfrac{\Delta \dashv E : \lfloor_{\mathsf{n}} \mathsf{T_2} \rfloor \rightarrowtail \mathsf{U_0}}{\Delta \dashv E \circ [\,] \triangleleft \mathsf{Inr} : \lfloor_{\mathsf{n}} \mathsf{T_1} \oplus \mathsf{T_2} \rfloor \rightarrowtail \mathsf{U_0}}\ \lfloor \oplus \rfloor\mathrm{E_2}$$

$$\dfrac{\Delta \dashv E : (\lfloor_{\mathsf{n}} \mathsf{T_1} \rfloor \otimes \lfloor_{\mathsf{n}} \mathsf{T_2} \rfloor) \rightarrowtail \mathsf{U_0}}{\Delta \dashv E \circ [\,] \triangleleft (,) : \lfloor_{\mathsf{n}} \mathsf{T_1} \otimes \mathsf{T_2} \rfloor \rightarrowtail \mathsf{U_0}}\ \lfloor \otimes \rfloor\mathrm{E}$$

$$\dfrac{\Delta \dashv E : \lfloor_{\mathsf{m} \cdot \mathsf{n}} \mathsf{T} \rfloor \rightarrowtail \mathsf{U_0}}{\Delta \dashv E \circ [\,] \triangleleft \mathsf{Mod}_{\mathsf{m}} : \lfloor_{\mathsf{n}} {!_{\mathsf{m}}}\mathsf{T} \rfloor \rightarrowtail \mathsf{U_0}}\ \lfloor ! \rfloor\mathrm{E}$$

$$\dfrac{\Delta_1,\ (\mathsf{1}{\uparrow}{\cdot}\mathsf{n}) \cdot \Delta_2 \dashv E : \mathsf{1} \rightarrowtail \mathsf{U_0} \qquad \Delta_2,\ x :_{\mathsf{m}} \mathsf{T} \vdash u : \mathsf{U}}{\Delta_1 \dashv E \circ [\,] \triangleleft (\lambda x\ {}_{\mathsf{m}}\!\mapsto u) : \lfloor_{\mathsf{n}} \mathsf{T}\ {}_{\mathsf{m}}\!\multimap \mathsf{U} \rfloor \rightarrowtail \mathsf{U_0}}\ \lfloor \multimap \rfloor\mathrm{E}$$

$$\dfrac{\Delta_1,\ \mathsf{1}{\uparrow} \cdot \Delta_2 \dashv E : \mathsf{T} \rightarrowtail \mathsf{U_0} \qquad \Delta_2 \vdash t' : \mathsf{U} \ltimes \mathsf{T}}{\Delta_1 \dashv E \circ [\,] \triangleleft\!\circ\, t' : \lfloor_{\mathsf{1}\nu} \mathsf{U} \rfloor \rightarrowtail \mathsf{U_0}}\ \lfloor\,\rfloor\mathrm{E_{C1}}$$

$$\dfrac{\Delta_1,\ \mathsf{1}{\uparrow} \cdot \Delta_2 \dashv E : \mathsf{T} \rightarrowtail \mathsf{U_0} \qquad \Delta_1 \vdash v : \lfloor_{\mathsf{1}\nu} \mathsf{U} \rfloor}{\Delta_2 \dashv E \circ v \triangleleft\!\circ [\,] : \mathsf{U} \ltimes \mathsf{T} \rightarrowtail \mathsf{U_0}}\ \lfloor\,\rfloor\mathrm{E_{C2}}$$

$$\dfrac{\Delta_1,\ (\mathsf{1}{\uparrow}{\cdot}\mathsf{n}) \cdot \Delta_2 \dashv E : \mathsf{1} \rightarrowtail \mathsf{U_0} \qquad \Delta_2 \vdash t' : \mathsf{T}}{\Delta_1 \dashv E \circ [\,] \blacktriangleleft t' : \lfloor_{\mathsf{n}} \mathsf{T} \rfloor \rightarrowtail \mathsf{U_0}}\ \lfloor\,\rfloor\mathrm{E_{L1}}$$

$$\dfrac{\Delta_1,\ (\mathsf{1}{\uparrow}{\cdot}\mathsf{n}) \cdot \Delta_2 \dashv E : \mathsf{1} \rightarrowtail \mathsf{U_0} \qquad \Delta_1 \vdash v : \lfloor_{\mathsf{n}} \mathsf{T} \rfloor}{\Delta_2 \dashv E \circ v \blacktriangleleft [\,] : \mathsf{T} \rightarrowtail \mathsf{U_0}}\ \lfloor\,\rfloor\mathrm{E_{L2}}$$

$$\dfrac{\mathsf{hnames}(E)\ \#\#\ \mathsf{hnames}(\Delta_3) \qquad \Delta_1,\ \Delta_2 \dashv E : (\mathsf{U} \ltimes \mathsf{T}') \rightarrowtail \mathsf{U_0} \qquad \Delta_2,\ {\rightarrow^{-1}}\Delta_3 \vdash_{\mathsf{v}} v_2 : \mathsf{U}}{\mathsf{1}{\uparrow} \cdot \Delta_1,\ \Delta_3 \dashv E \circ {}^{\mathrm{op}}_{\mathsf{hnames}(\Delta_3)}\langle v_2\,, [\,] \rangle : \mathsf{T}' \rightarrowtail \mathsf{U_0}}\ \ltimes\textsc{op}$$

---

$\boxed{\vdash E[\,t\,] : \mathsf{T}}$ *(Typing judgment for commands)*

$$\dfrac{\Delta \dashv E : \mathsf{T} \rightarrowtail \mathsf{U_0} \qquad \Delta \vdash t : \mathsf{T}}{\vdash E[\,t\,] : \mathsf{U_0}}\ \lambda_d\text{–}\textsc{ty}_{\textsc{cmd}}$$

Figure 2.8: Typing rules of evaluation contexts ($\lambda_d$–$\textsc{ty}_\textsc{e}$) and commands ($\lambda_d$–$\textsc{ty}_{\textsc{cmd}}$)

### 2.6.3 Reduction semantics

Now that every piece is in place, let's focus on the reduction rules of $\lambda_d$, displayed in Figures 2.9 and 2.10. As in Section 1.6 for $\lambda_{Ldm}$, focus (F suffix), unfocus (U suffix) and contraction (C suffix) rules of $\lambda_d$ are triggered in a purely deterministic fashion. Once a subterm is a value, it cannot be focused on again.

Figure 2.9 presents all the rules that do not incur any substitution on the evaluation context.

Reduction rules for function application and pattern-matching are the same as in $\lambda_{Ldm}$. Reduction rules for $\texttt{to}_\ltimes$ are pretty straightforward: once we have a value at hand, we embed it in a trivial ampar with just unit on the right. Rules for $\texttt{from}_\ltimes$ are fairly symmetric: once we have an ampar with a value of the shape $\texttt{Mod}_{1\infty}\ v_1$ on the right, we can extract both the left and right side out of the ampar shell, into a normal pair.

Finally, $\texttt{new}_\ltimes$ has only a single rule that transforms it into the "identity" ampar object with just one hole on the left, and the corresponding destination on the right.

Rules of Figure 2.10 are all related to $\texttt{upd}_\ltimes$ and destination-filling operators, whose contraction rules modify the evaluation context deeply, instead of just pushing or popping a focusing composant. For that, we need to introduce the special substitution $E(\!|h :=_H v|\!)$ that is used to update structures under construction, that are attached to open ampar focusing components in the stack. Such a substitution is triggered when a destination $\rightarrow h$ is filled in the term under focus, typically in destination-filling primitives reductions, and results in the value $v$ being written to hole $\boxed{h}$. The value $v$ may contain holes itself (e.g. when the hollow constructor $\texttt{Inl}\ \boxed{h'+1}$ is being written to the hole $\boxed{h}$ in $\lambda_d$–SEM/$\lfloor\oplus\rfloor$E$_1$C), hence the set $H$ tracks the potential hole names introduced by value $v$, and is used to update the hole name set of the corresponding (open) ampar. Proper definition of $E(\!|h :=_H v|\!)$ is given at the bottom of Figure 2.9.

$\lambda_d$–SEM/$\lfloor 1\rfloor$EC and $\lambda_d$–SEM/$\lfloor\multimap\rfloor$EC of Figure 2.10 do not create any new hole; they only write a value to an existing one. On the other hand, rules $\lambda_d$–SEM/$\lfloor\oplus\rfloor$E$_1$C, $\lambda_d$–SEM/$\lfloor\oplus\rfloor$E$_2$C, $\lambda_d$–SEM/$\lfloor !\rfloor$EC and $\lambda_d$–SEM/$\lfloor\otimes\rfloor$EC all write a hollow constructor to the hole $\boxed{h}$ that contains new holes. Thus, we need to generate fresh names for these new holes, and also return a destination for each new hole with a matching name.

The substitution $E(\!|h :=_H v|\!)$ should only be performed if $h$ is a globally unique name; otherwise we break the promise of a write-once memory model. To this effect, we allow name shadowing while an ampar is in value form (which is why $\texttt{new}_\ltimes$ is allowed to reduce to the same $_{\{1\}}/\boxed{1}\ _{,}\rightarrow 1/$ every time), but as soon as an ampar is open (when it becomes an open ampar focusing component), it should have globally unique hole names. This restriction is enforced in rule $\lambda_d$–TYE/$\ltimes$OP by the premise $\texttt{hnames}(E)\ \#\#\ \texttt{hnames}(\Delta_3)$, requiring hole name sets from $E$ and $\Delta_3$ to be disjoint when an open ampar focusing component is created during reduction of $\texttt{upd}_\ltimes$. Likewise, any hollow constructor written to a hole should have globally unique hole names. We assume that hole names are natural numbers for simplicity's sake.

To obtain globally fresh names, in the premises of the corresponding rules, we first set $h' = \texttt{max}(\texttt{hnames}(E) \cup \{h\})+1$ or similar definitions for $h''$ and $h'''$ (see at the bottom of Figure 2.10) to find a new unused name[19]. Then we use either the *shifted set* $H\pm h'$ or the *conditional shift operator* $h[H\pm h']$ as defined at the bottom of Figure 2.9 to replace all names or just a specific one with fresh unused name(s). We extend *conditional shift* $\bullet[H\pm h']$ to arbitrary values, terms, and typing contexts in the obvious way (keeping in mind that $_{H'}/v_2\ _{,}v_1/$ binds the names in $H'$).

[19]Sometimes we do not take the smallest unused name possible; it's mostly the result of using whatever made the formal proofs easier. For instance, in the formal proofs with Coq, renaming is implemented in terms of more general permutations, which make it more easily interoperable with typing contexts, represented as functions, but adds a few extra requirements to make the lemma hold.

$\boxed{E\left[\,t\,\right] \longrightarrow E'\left[\,t'\,\right]}$ *(Small-step evaluation of commands)*

| Rule | | Name |
|---|---|---|
| $E\left[\,t'\ t\,\right] \longrightarrow \left(E \circ t'\ [\,]\right)\left[\,t\,\right]$ | $\star$ | $\multimap$EF$_1$ |
| $\left(E \circ t'\ [\,]\right)\left[\,v\,\right] \longrightarrow E\left[\,t'\ v\,\right]$ | | $\multimap$EU$_1$ |
| $E\left[\,t'\ v\,\right] \longrightarrow \left(E \circ [\,]\ v\right)\left[\,t'\,\right]$ | $\star$ | $\multimap$EF$_2$ |
| $\left(E \circ [\,]\ v\right)\left[\,v'\,\right] \longrightarrow E\left[\,v'\ v\,\right]$ | | $\multimap$EU$_2$ |
| $E\left[\,(\boldsymbol{\lambda}_{\mathsf{v}} x\ {}_{\mathrm{m}}\!\mapsto u)\ v\,\right] \longrightarrow E\left[\,u[x := v]\,\right]$ | | $\multimap$EC |
| $E\left[\,t\ ⨾\ u\,\right] \longrightarrow \left(E \circ [\,]\ ⨾\ u\right)\left[\,t\,\right]$ | $\star$ | 1EF |
| $\left(E \circ [\,]\ ⨾\ u\right)\left[\,v\,\right] \longrightarrow E\left[\,v\ ⨾\ u\,\right]$ | | 1EU |
| $E\left[\,()\ ⨾\ u\,\right] \longrightarrow E\left[\,u\,\right]$ | | 1EC |
| $E\left[\,\mathtt{case}_{\mathrm{m}}\ t\ \mathtt{of}\ \{\mathtt{Inl}\,x_1 \mapsto u_1\,,\ \mathtt{Inr}\,x_2 \mapsto u_2\}\,\right] \longrightarrow \left(E \circ \mathtt{case}_{\mathrm{m}}\ [\,]\ \mathtt{of}\ \{\mathtt{Inl}\,x_1 \mapsto u_1\,,\ \mathtt{Inr}\,x_2 \mapsto u_2\}\right)\left[\,t\,\right]$ | $\star$ | $\oplus$EF |
| $\left(E \circ \mathtt{case}_{\mathrm{m}}\ [\,]\ \mathtt{of}\ \{\mathtt{Inl}\,x_1 \mapsto u_1\,,\ \mathtt{Inr}\,x_2 \mapsto u_2\}\right)\left[\,v\,\right] \longrightarrow E\left[\,\mathtt{case}_{\mathrm{m}}\ v\ \mathtt{of}\ \{\mathtt{Inl}\,x_1 \mapsto u_1\,,\ \mathtt{Inr}\,x_2 \mapsto u_2\}\,\right]$ | | $\oplus$EU |
| $E\left[\,\mathtt{case}_{\mathrm{m}}\ (\mathtt{Inl}\,v_1)\ \mathtt{of}\ \{\mathtt{Inl}\,x_1 \mapsto u_1\,,\ \mathtt{Inr}\,x_2 \mapsto u_2\}\,\right] \longrightarrow E\left[\,u_1[x_1 := v_1]\,\right]$ | | $\oplus$EC$_1$ |
| $E\left[\,\mathtt{case}_{\mathrm{m}}\ (\mathtt{Inr}\,v_2)\ \mathtt{of}\ \{\mathtt{Inl}\,x_1 \mapsto u_1\,,\ \mathtt{Inr}\,x_2 \mapsto u_2\}\,\right] \longrightarrow E\left[\,u_2[x_2 := v_2]\,\right]$ | | $\oplus$EC$_2$ |
| $E\left[\,\mathtt{case}_{\mathrm{m}}\ t\ \mathtt{of}\ (x_1\,,\,x_2) \mapsto u\,\right] \longrightarrow \left(E \circ \mathtt{case}_{\mathrm{m}}\ [\,]\ \mathtt{of}\ (x_1\,,\,x_2) \mapsto u\right)\left[\,t\,\right]$ | $\star$ | $\otimes$EF |
| $\left(E \circ \mathtt{case}_{\mathrm{m}}\ [\,]\ \mathtt{of}\ (x_1\,,\,x_2) \mapsto u\right)\left[\,v\,\right] \longrightarrow E\left[\,\mathtt{case}_{\mathrm{m}}\ v\ \mathtt{of}\ (x_1\,,\,x_2) \mapsto u\,\right]$ | | $\otimes$EU |
| $E\left[\,\mathtt{case}_{\mathrm{m}}\ (v_1\,,\,v_2)\ \mathtt{of}\ (x_1\,,\,x_2) \mapsto u\,\right] \longrightarrow E\left[\,u[x_1 := v_1][x_2 := v_2]\,\right]$ | | $\otimes$EC |
| $E\left[\,\mathtt{case}_{\mathrm{m}}\ t\ \mathtt{of}\ \mathtt{Mod}_{\mathrm{n}}\ x \mapsto u\,\right] \longrightarrow \left(E \circ \mathtt{case}_{\mathrm{m}}\ [\,]\ \mathtt{of}\ \mathtt{Mod}_{\mathrm{n}}\ x \mapsto u\right)\left[\,t\,\right]$ | $\star$ | !EF |
| $\left(E \circ \mathtt{case}_{\mathrm{m}}\ [\,]\ \mathtt{of}\ \mathtt{Mod}_{\mathrm{n}}\ x \mapsto u\right)\left[\,v\,\right] \longrightarrow E\left[\,\mathtt{case}_{\mathrm{m}}\ v\ \mathtt{of}\ \mathtt{Mod}_{\mathrm{n}}\ x \mapsto u\,\right]$ | | !EU |
| $E\left[\,\mathtt{case}_{\mathrm{m}}\ \mathtt{Mod}_{\mathrm{n}}\ v'\ \mathtt{of}\ \mathtt{Mod}_{\mathrm{n}}\ x \mapsto u\,\right] \longrightarrow E\left[\,u[x := v']\,\right]$ | | !EC |
| $E\left[\,\mathtt{to}_{\ltimes}\ u\,\right] \longrightarrow \left(E \circ \mathtt{to}_{\ltimes}\ [\,]\right)\left[\,u\,\right]$ | $\star$ | $\ltimes$toF |
| $\left(E \circ \mathtt{to}_{\ltimes}\ [\,]\right)\left[\,v_2\,\right] \longrightarrow E\left[\,\mathtt{to}_{\ltimes}\ v_2\,\right]$ | | $\ltimes$toU |
| $E\left[\,\mathtt{to}_{\ltimes}\ v_2\,\right] \longrightarrow E\left[\,{}_{\{\}}/v_2\ {}_{,}()/\,\right]$ | | $\ltimes$toC |
| $E\left[\,\mathtt{from}_{\ltimes}\ t\,\right] \longrightarrow \left(E \circ \mathtt{from}_{\ltimes}\ [\,]\right)\left[\,t\,\right]$ | $\star$ | $\ltimes$fromF |
| $\left(E \circ \mathtt{from}_{\ltimes}\ [\,]\right)\left[\,v\,\right] \longrightarrow E\left[\,\mathtt{from}_{\ltimes}\ v\,\right]$ | | $\ltimes$fromU |
| $E\left[\,\mathtt{from}_{\ltimes}\ {}_{\{\}}/v_2\ {}_{,}\mathtt{Mod}_{1\infty}\ v_1/\,\right] \longrightarrow E\left[\,(v_2\,,\ \mathtt{Mod}_{1\infty}\ v_1)\,\right]$ | | $\ltimes$fromC |
| $E\left[\,\mathtt{new}_{\ltimes}\,\right] \longrightarrow E\left[\,{}_{\{1\}}/\boxed{1}\ {}_{,}{\rightarrow}1/\,\right]$ | | $\ltimes$newC |

$\star$ : only allowed if the term that would become the new focus is not already a value

---

*Name set shift and conditional name shift:*

$$H{\pm}h' \triangleq \{h{+}h' \mid h \in H\}$$

$$h[H{\pm}h'] \triangleq \begin{cases} h{+}h' & \text{if } h \in H \\ h & \text{otherwise} \end{cases}$$

---

*Special substitution for open ampars:*

$$\left(E \circ {}^{\mathrm{op}}_{\{h\}\sqcup H}/v_2\ {}_{,}[\,]/\right)\llparenthesis h :=_{H'} v' \rrparenthesis = E \circ {}^{\mathrm{op}}_{H\sqcup H'}/v_2 \llparenthesis h :=_{H'} v' \rrparenthesis\ {}_{,}[\,]/$$

$$\left(E \circ e\right)\llparenthesis h :=_{H'} v' \rrparenthesis = E\llparenthesis h :=_{H'} v' \rrparenthesis \circ e \qquad \text{if } h \notin e$$

Figure 2.9: Small-step semantics for $\lambda_d$ ($\lambda_d$–sem), part 1

| | | |
|---|---|---|
| $E\left[\mathtt{upd}_{\ltimes}\ t\ \mathtt{with}\ x \mapsto t'\right] \longrightarrow \left(E \circ \mathtt{upd}_{\ltimes}\ [\,]\ \mathtt{with}\ x \mapsto t'\right)\left[t\right]$ | $\star$ | $\ltimes$UPDF |
| $\left(E \circ \mathtt{upd}_{\ltimes}\ [\,]\ \mathtt{with}\ x \mapsto t'\right)\left[v\right] \longrightarrow E\left[\mathtt{upd}_{\ltimes}\ v\ \mathtt{with}\ x \mapsto t'\right]$ | | $\ltimes$UPDU |
| $E\left[\mathtt{upd}_{\ltimes}\ {}_{H}/v_2\ {}_{9}v_1/\ \mathtt{with}\ x \mapsto t'\right] \longrightarrow \left(E \circ {}^{\mathrm{op}}_{H \pm h'''}/v_2[H{\pm}h''']\ {}_{9}[\,]/\right)\left[t'[x := v_1[H{\pm}h''']]\right]$ | $\star$ | $\ltimes$OP |
| $\left(E \circ {}^{\mathrm{op}}_{H}/v_2\ {}_{9}[\,]/\right)\left[v_1\right] \longrightarrow E\left[{}_{H}/v_2\ {}_{9}v_1/\right]$ | | $\ltimes$CL |
| $E\left[t \lhd ()\right] \longrightarrow \left(E \circ [\,] \lhd ()\right)\left[t\right]$ | $\star$ | $\lfloor 1 \rfloor$EF |
| $\left(E \circ [\,] \lhd ()\right)\left[v\right] \longrightarrow E\left[v \lhd ()\right]$ | | $\lfloor 1 \rfloor$EU |
| $E\left[\rightarrow h \lhd ()\right] \longrightarrow E(\!\vert h :=_{\{\}}\ ()\vert\!)\left[()\right]$ | | $\lfloor 1 \rfloor$EC |
| $E\left[t \lhd \mathtt{Inl}\right] \longrightarrow \left(E \circ [\,] \lhd \mathtt{Inl}\right)\left[t\right]$ | $\star$ | $\lfloor \oplus \rfloor$E$_{\mathrm{I}}$F |
| $\left(E \circ [\,] \lhd \mathtt{Inl}\right)\left[v\right] \longrightarrow E\left[v \lhd \mathtt{Inl}\right]$ | | $\lfloor \oplus \rfloor$E$_{\mathrm{I}}$U |
| $E\left[\rightarrow h \lhd \mathtt{Inl}\right] \longrightarrow E(\!\vert h :=_{\{h'+1\}}\ \mathtt{Inl}\ \boxed{h'+1}\vert\!)\left[\rightarrow h'+1\right]$ | $\star$ | $\lfloor \oplus \rfloor$E$_{\mathrm{I}}$C |
| $E\left[t \lhd \mathtt{Inr}\right] \longrightarrow \left(E \circ [\,] \lhd \mathtt{Inr}\right)\left[t\right]$ | $\star$ | $\lfloor \oplus \rfloor$E$_{2}$F |
| $\left(E \circ [\,] \lhd \mathtt{Inr}\right)\left[v\right] \longrightarrow E\left[v \lhd \mathtt{Inr}\right]$ | | $\lfloor \oplus \rfloor$E$_{2}$U |
| $E\left[\rightarrow h \lhd \mathtt{Inr}\right] \longrightarrow E(\!\vert h :=_{\{h'+1\}}\ \mathtt{Inr}\ \boxed{h'+1}\vert\!)\left[\rightarrow h'+1\right]$ | $\star$ | $\lfloor \oplus \rfloor$E$_{2}$C |
| $E\left[t \lhd \mathtt{Mod}_{\mathrm{m}}\right] \longrightarrow \left(E \circ [\,] \lhd \mathtt{Mod}_{\mathrm{m}}\right)\left[t\right]$ | $\star$ | $\lfloor ! \rfloor$EF |
| $\left(E \circ [\,] \lhd \mathtt{Mod}_{\mathrm{m}}\right)\left[v\right] \longrightarrow E\left[v \lhd \mathtt{Mod}_{\mathrm{m}}\right]$ | | $\lfloor ! \rfloor$EU |
| $E\left[\rightarrow h \lhd \mathtt{Mod}_{\mathrm{m}}\right] \longrightarrow E(\!\vert h :=_{\{h'+1\}}\ \mathtt{Mod}_{\mathrm{m}}\ \boxed{h'+1}\vert\!)\left[\rightarrow h'+1\right]$ | $\star$ | $\lfloor ! \rfloor$EC |
| $E\left[t \lhd (,)\right] \longrightarrow \left(E \circ [\,] \lhd (,)\right)\left[t\right]$ | $\star$ | $\lfloor \otimes \rfloor$EF |
| $\left(E \circ [\,] \lhd (,)\right)\left[v\right] \longrightarrow E\left[v \lhd (,)\right]$ | | $\lfloor \otimes \rfloor$EU |
| $E\left[\rightarrow h \lhd (,)\right] \longrightarrow E(\!\vert h :=_{\{h'+1,h'+2\}}\ (\boxed{h'+1}, \boxed{h'+2})\vert\!)\left[(\rightarrow h'+1, \rightarrow h'+2)\right]$ | $\star$ | $\lfloor \otimes \rfloor$EC |
| $E\left[t \lhd (\boldsymbol{\lambda} x\ {}_{\mathrm{m}}\!\mapsto u)\right] \longrightarrow \left(E \circ [\,] \lhd (\boldsymbol{\lambda} x\ {}_{\mathrm{m}}\!\mapsto u)\right)\left[t\right]$ | $\star$ | $\lfloor \multimap \rfloor$EF |
| $\left(E \circ [\,] \lhd (\boldsymbol{\lambda} x\ {}_{\mathrm{m}}\!\mapsto u)\right)\left[v\right] \longrightarrow E\left[v \lhd (\boldsymbol{\lambda} x\ {}_{\mathrm{m}}\!\mapsto u)\right]$ | | $\lfloor \multimap \rfloor$EU |
| $E\left[\rightarrow h \lhd (\boldsymbol{\lambda} x\ {}_{\mathrm{m}}\!\mapsto u)\right] \longrightarrow E(\!\vert h :=_{\{\}}\ \boldsymbol{\lambda}_{\mathbf{v}} x\ {}_{\mathrm{m}}\!\mapsto u\vert\!)\left[()\right]$ | | $\lfloor \multimap \rfloor$EC |
| $E\left[t \lhd\!\circ\ t'\right] \longrightarrow \left(E \circ [\,] \lhd\!\circ\ t'\right)\left[t\right]$ | $\star$ | $\lfloor\ \rfloor$E$_{\mathrm{C}}$F$_{\mathrm{I}}$ |
| $\left(E \circ [\,] \lhd\!\circ\ t'\right)\left[v\right] \longrightarrow E\left[v \lhd\!\circ\ t'\right]$ | | $\lfloor\ \rfloor$E$_{\mathrm{C}}$U$_{\mathrm{I}}$ |
| $E\left[v \lhd\!\circ\ t'\right] \longrightarrow \left(E \circ v \lhd\!\circ [\,]\right)\left[t'\right]$ | $\star$ | $\lfloor\ \rfloor$E$_{\mathrm{C}}$F$_{2}$ |
| $\left(E \circ v \lhd\!\circ [\,]\right)\left[v'\right] \longrightarrow E\left[v \lhd\!\circ\ v'\right]$ | | $\lfloor\ \rfloor$E$_{\mathrm{C}}$U$_{2}$ |
| $E\left[\rightarrow h \lhd\!\circ_{H}\ /v_2\ {}_{9}v_1/\right] \longrightarrow E(\!\vert h :=_{(H \pm h'')}\ v_2[H{\pm}h'']\vert\!)\left[v_1[H{\pm}h'']\right]$ | $\star$ | $\lfloor\ \rfloor$E$_{\mathrm{C}}$C |
| $E\left[t \blacktriangleleft t'\right] \longrightarrow \left(E \circ [\,] \blacktriangleleft t'\right)\left[t\right]$ | $\star$ | $\lfloor\ \rfloor$E$_{\mathrm{L}}$F$_{\mathrm{I}}$ |
| $\left(E \circ [\,] \blacktriangleleft t'\right)\left[v\right] \longrightarrow E\left[v \blacktriangleleft t'\right]$ | | $\lfloor\ \rfloor$E$_{\mathrm{L}}$U$_{\mathrm{I}}$ |
| $E\left[v \blacktriangleleft t'\right] \longrightarrow \left(E \circ v \blacktriangleleft [\,]\right)\left[t'\right]$ | $\star$ | $\lfloor\ \rfloor$E$_{\mathrm{L}}$F$_{2}$ |
| $\left(E \circ v \blacktriangleleft [\,]\right)\left[v'\right] \longrightarrow E\left[v \blacktriangleleft v'\right]$ | | $\lfloor\ \rfloor$E$_{\mathrm{L}}$U$_{2}$ |
| $E\left[\rightarrow h \blacktriangleleft v\right] \longrightarrow E(\!\vert h :=_{\{\}}\ v\vert\!)\left[()\right]$ | | $\lfloor\ \rfloor$E$_{\mathrm{L}}$C |

$$\text{where} \quad \left\{ \begin{array}{lcl} h' & = & \mathtt{max}(\mathtt{hnames}(E) \cup \{h\})+1 \\ h'' & = & \mathtt{max}(H \cup (\mathtt{hnames}(E) \cup \{h\}))+1 \\ h''' & = & \mathtt{max}(H \cup \mathtt{hnames}(E))+1 \end{array} \right.$$

$\star$ : only allowed if the term that would become the new focus is not already a value

Figure 2.10: Small-step semantics for $\lambda_d$ ($\lambda_d$–SEM), part 2

Rules $\lambda_d$–SEM/$\ltimes$OP and $\lambda_d$–SEM/$\ltimes$CL dictate how and when an ampar (a value) is converted to an open ampar (a focusing component) and vice-versa, and they make use of the shifting strategy we have just introduced. With $\lambda_d$–SEM/$\ltimes$OP, the hole names bound by the ampar gets renamed to fresh ones, and the left-hand side gets attached to the focusing component ${}^{\text{op}}_{H\pm h''}/v_2[H\pm h''] \,,[\,]/$ while the right-hand side (containing destinations) is substituted in the body of the $\mathtt{upd}_{\ltimes}$ statement (which becomes the new term under focus). The rule $\lambda_d$–SEM/$\ltimes$CL triggers when the body of a $\mathtt{upd}_{\ltimes}$ statement has reduced to a value. In that case, we can close the ampar, by popping the focusing component from the stack $E$ and merging back with $v_2$ to form an ampar value again.

In rule $\lambda_d$–SEM/$\lfloor\,\rfloor$E$_\text{C}$C, we write the left-hand side $v_2$ of an ampar value ${}_{H}/v_2 \,, v_1/$ to a hole $\boxed{h}$ that is part of a structure with holes somewhere inside $E$. This results in the composition of two structures with holes. Because we dissociate $v_2$ and $v_1$ that were previously bound together by the ampar connective ($v_2$ is merged with another structure, while $v_1$ becomes the new focus), their hole names are no longer bound, so we need to make them globally unique, as we do when an ampar is opened with $\mathtt{upd}_{\ltimes}$. This renaming is carried out by the conditional shift $v_2[H\pm h'']$ and $v_1[H\pm h'']$.

**Putting into practice** Let's reuse the example from Section 2.5.2 with the synthetic $(::)$ *cons* constructor, and see the reduction of expression $() :: \mathtt{Inl}\,()$. Here, $\mathtt{Inl}\,()$ is the value form corresponding to *nil* (given our encoding of lists), and we also assume that $()$ in the expression is the value form of Figure 2.5, not the syntactic constructor defined in Listing 2.4. Finally, we consider $\mathtt{from}'_{\ltimes}$ to be a primitive (for brevity) with its contraction rule $\lambda_d$–SEM/$\ltimes$FROM'C : $E\left[\,\mathtt{from}'_{\ltimes}({}_{\{\}}/v_2 \,, ()/)\,\right] \;\longrightarrow\; E\left[\,v_2\,\right]$ and trivial focus and unfocus rules.

$$
\begin{array}{lll}
 & [\,]\big[() :: \mathtt{Inl}\,()\big] & \\
\longrightarrow & [\,]\big[\mathtt{from}'_{\ltimes}(\mathtt{upd}_{\ltimes}\,\mathtt{new}_{\ltimes}\,\mathtt{with}\,d \mapsto \mathtt{case}\,(d \triangleleft \mathtt{Inr} \triangleleft (,))\,\mathtt{of}\,(dx\,,dxs) \mapsto dx \blacktriangleleft () \,⨟\, dxs \blacktriangleleft \mathtt{Inl}\,())\big] & \text{expanding } (::) \text{ and } \triangleleft(::) \text{ definitions} \\
\longrightarrow & ([\,] \circ \mathtt{from}'_{\ltimes}[\,])\big[\mathtt{upd}_{\ltimes}\,\mathtt{new}_{\ltimes}\,\mathtt{with}\,d \mapsto \mathtt{case}\,(d \triangleleft \mathtt{Inr} \triangleleft (,))\,\mathtt{of}\,(dx\,,dxs) \mapsto dx \blacktriangleleft () \,⨟\, dxs \blacktriangleleft \mathtt{Inl}\,()\big] & \ltimes\text{FROM'F} \\
\longrightarrow & ([\,] \circ \mathtt{from}'_{\ltimes}[\,] \circ \mathtt{upd}_{\ltimes}[\,]\,\mathtt{with}\,d \mapsto \mathtt{case}\,(d \triangleleft \mathtt{Inr} \triangleleft (,))\,\mathtt{of}\,(dx\,,dxs) \mapsto dx \blacktriangleleft () \,⨟\, dxs \blacktriangleleft \mathtt{Inl}\,())\big[\mathtt{new}_{\ltimes}\big] & \ltimes\text{UPDF} \\
\longrightarrow & ([\,] \circ \mathtt{from}'_{\ltimes}[\,] \circ \mathtt{upd}_{\ltimes}[\,]\,\mathtt{with}\,d \mapsto \mathtt{case}\,(d \triangleleft \mathtt{Inr} \triangleleft (,))\,\mathtt{of}\,(dx\,,dxs) \mapsto dx \blacktriangleleft () \,⨟\, dxs \blacktriangleleft \mathtt{Inl}\,())\big[{}_{\{1\}}/\boxed{1}\,,\rightarrow\!1/\big] & \ltimes\text{NEWC} \\
\longrightarrow & ([\,] \circ \mathtt{from}'_{\ltimes}[\,])\big[\mathtt{upd}_{\ltimes}\,{}_{\{1\}}/\boxed{1}\,,\rightarrow\!1/\,\mathtt{with}\,d \mapsto \mathtt{case}\,(d \triangleleft \mathtt{Inr} \triangleleft (,))\,\mathtt{of}\,(dx\,,dxs) \mapsto dx \blacktriangleleft () \,⨟\, dxs \blacktriangleleft \mathtt{Inl}\,()\big] & \ltimes\text{UPDU} \\
\longrightarrow & ([\,] \circ \mathtt{from}'_{\ltimes}[\,] \circ {}^{\text{op}}_{\{2\}}/\boxed{2}\,,[\,]/)\big[\mathtt{case}\,(\rightarrow\!2 \triangleleft \mathtt{Inr} \triangleleft (,))\,\mathtt{of}\,(dx\,,dxs) \mapsto dx \blacktriangleleft () \,⨟\, dxs \blacktriangleleft \mathtt{Inl}\,()\big] & \ltimes\text{OP} \\
\longrightarrow & ([\,] \circ \mathtt{from}'_{\ltimes}[\,] \circ {}^{\text{op}}_{\{2\}}/\boxed{2}\,,[\,]/ \circ \mathtt{case}_{1\nu}[\,]\,\mathtt{of}\,(dx\,,dxs) \mapsto dx \blacktriangleleft () \,⨟\, dxs \blacktriangleleft \mathtt{Inl}\,())\big[\rightarrow\!2 \triangleleft \mathtt{Inr} \triangleleft (,)\big] & \otimes\text{EF} \\
\longrightarrow & ([\,] \circ \mathtt{from}'_{\ltimes}[\,] \circ {}^{\text{op}}_{\{2\}}/\boxed{2}\,,[\,]/ \circ \mathtt{case}_{1\nu}[\,]\,\mathtt{of}\,(dx\,,dxs) \mapsto dx \blacktriangleleft () \,⨟\, dxs \blacktriangleleft \mathtt{Inl}\,() \circ [\,] \triangleleft (,))\big[\rightarrow\!2 \triangleleft \mathtt{Inr}\big] & \lfloor\otimes\rfloor\text{EF} \\
\longrightarrow & ([\,] \circ \mathtt{from}'_{\ltimes}[\,] \circ {}^{\text{op}}_{\{2\}}/\mathtt{Inr}\,\boxed{4}\,,[\,]/ \circ \mathtt{case}_{1\nu}[\,]\,\mathtt{of}\,(dx\,,dxs) \mapsto dx \blacktriangleleft () \,⨟\, dxs \blacktriangleleft \mathtt{Inl}\,() \circ [\,] \triangleleft (,))\big[\rightarrow\!4\big] & \lfloor\oplus\rfloor\text{E}_2\text{C with sub. } E(\!|2 :=_{\{4\}} \mathtt{Inr}\,\boxed{4}|\!) \\
\longrightarrow & ([\,] \circ \mathtt{from}'_{\ltimes}[\,] \circ {}^{\text{op}}_{\{4\}}/\mathtt{Inr}\,\boxed{4}\,,[\,]/ \circ \mathtt{case}_{1\nu}[\,]\,\mathtt{of}\,(dx\,,dxs) \mapsto dx \blacktriangleleft () \,⨟\, dxs \blacktriangleleft \mathtt{Inl}\,())\big[\rightarrow\!4 \triangleleft (,)\big] & \lfloor\otimes\rfloor\text{EU} \\
\longrightarrow & ([\,] \circ \mathtt{from}'_{\ltimes}[\,] \circ {}^{\text{op}}_{\{6,7\}}/\mathtt{Inr}\,(\boxed{6}\,,\boxed{7})\,,[\,]/ \circ \mathtt{case}_{1\nu}[\,]\,\mathtt{of}\,(dx\,,dxs) \mapsto dx \blacktriangleleft () \,⨟\, dxs \blacktriangleleft \mathtt{Inl}\,())\big[(\rightarrow\!6\,,\rightarrow\!7)\big] & \lfloor\otimes\rfloor\text{EC with sub. } E(\!|4 :=_{\{6,7\}} (\boxed{6}\,,\boxed{7})|\!) \\
\longrightarrow & ([\,] \circ \mathtt{from}'_{\ltimes}[\,] \circ {}^{\text{op}}_{\{6,7\}}/\mathtt{Inr}\,(\boxed{6}\,,\boxed{7})\,,[\,]/)\big[\mathtt{case}\,(\rightarrow\!6\,,\rightarrow\!7)\,\mathtt{of}\,(dx\,,dxs) \mapsto dx \blacktriangleleft () \,⨟\, dxs \blacktriangleleft \mathtt{Inl}\,()\big] & \otimes\text{EU} \\
\longrightarrow & ([\,] \circ \mathtt{from}'_{\ltimes}[\,] \circ {}^{\text{op}}_{\{6,7\}}/\mathtt{Inr}\,(\boxed{6}\,,\boxed{7})\,,[\,]/)\big[\rightarrow\!6 \blacktriangleleft () \,⨟\, \rightarrow\!7 \blacktriangleleft \mathtt{Inl}\,()\big] & \otimes\text{EC} \\
\longrightarrow & ([\,] \circ \mathtt{from}'_{\ltimes}[\,] \circ {}^{\text{op}}_{\{6,7\}}/\mathtt{Inr}\,(\boxed{6}\,,\boxed{7})\,,[\,]/ \circ [\,] \,⨟\, \rightarrow\!7 \blacktriangleleft \mathtt{Inl}\,())\big[\rightarrow\!6 \blacktriangleleft ()\big] & 1\text{EF} \\
\longrightarrow & ([\,] \circ \mathtt{from}'_{\ltimes}[\,] \circ {}^{\text{op}}_{\{7\}}/\mathtt{Inr}\,((),\boxed{7})\,,[\,]/ \circ [\,] \,⨟\, \rightarrow\!7 \blacktriangleleft \mathtt{Inl}\,())\big[()\big] & \lfloor\,\rfloor\text{E}_1\text{C with sub. } E(\!|6 :=_{\{\}} ()|\!) \\
\longrightarrow & ([\,] \circ \mathtt{from}'_{\ltimes}[\,] \circ {}^{\text{op}}_{\{7\}}/\mathtt{Inr}\,((),\boxed{7})\,,[\,]/)\big[() \,⨟\, \rightarrow\!7 \blacktriangleleft \mathtt{Inl}\,()\big] & 1\text{EU} \\
\longrightarrow & ([\,] \circ \mathtt{from}'_{\ltimes}[\,] \circ {}^{\text{op}}_{\{7\}}/\mathtt{Inr}\,((),\boxed{7})\,,[\,]/)\big[\rightarrow\!7 \blacktriangleleft \mathtt{Inl}\,()\big] & 1\text{EC} \\
\longrightarrow & ([\,] \circ \mathtt{from}'_{\ltimes}[\,] \circ {}^{\text{op}}_{\{\}}/\mathtt{Inr}\,((),\mathtt{Inl}\,())\,,[\,]/)\big[()\big] & \lfloor\,\rfloor\text{E}_1\text{C with sub. } E(\!|7 :=_{\{\}} \mathtt{Inl}\,()|\!) \\
\longrightarrow & ([\,] \circ \mathtt{from}'_{\ltimes}[\,])\big[{}_{\{\}}/\mathtt{Inr}\,((),\mathtt{Inl}\,())\,,()/\big] & \ltimes\text{CL} \\
\longrightarrow & [\,]\big[\mathtt{from}'_{\ltimes}({}_{\{\}}/\mathtt{Inr}\,((),\mathtt{Inl}\,())\,,()/)\big] & \ltimes\text{FROM'U} \\
\longrightarrow & [\,]\big[\mathtt{Inr}\,((),\mathtt{Inl}\,())\big] & \ltimes\text{FROM'C}
\end{array}
$$

At the end of the reduction, we get the empty evaluation context with value $\texttt{Inr}\,(()\,,\,\texttt{Inl}\,())$ on focus, which is the value form for the singleton list containing just unit, as expected.

We also see how, throughout the reduction, the contractions of destination-filling operations under focus trigger global substitutions $E\,(\!|h :=_H v|\!)$ on the evaluation context $E$. As said before, these substitutions are responsible for mutating the structure under construction, which is (always) attached on *open ampar* focusing component, somewhere in the evaluation context $E$.

### 2.6.4 Type safety

With the semantics now defined, we can state the usual type safety theorems:

**Theorem 1** (Type preservation). *If* $\vdash E[t] : \mathsf{T}$ *and* $E[t] \longrightarrow E'[t']$ *then* $\vdash E'[t'] : \mathsf{T}$.

**Theorem 2** (Progress). *If* $\vdash E[t] : \mathsf{T}$ *and* $\forall v, E[t] \neq [\,][v]$ *then* $\exists E', t'.\ E[t] \longrightarrow E'[t']$.

As seen in the example above, a command of the form $[\,][v]$ cannot be reduced further, as it only contains a fully determined value, and no pending computation. This it is the stopping point of the reduction, and any well-typed command eventually reaches this form.

Given the way we typed holes and destinations in Section 2.6.1, we have the guarantee that a closed program does not contain any free holes or destinations, and that any holes and destinations remaining in $[\,][v]$ at the end of the reduction must be properly balanced and wrapped within an ampar.

## 2.7 Formal proof of type safety

We've formalized $\lambda_d$ and proved type preservation and progress theorems with the Coq proof assistant in ⟦Bagrel and Spiwack, 2025b⟧.

Turning to a proof assistant was a pragmatic choice: typing context handling in $\lambda_d$ can be quite tedious, and it was hard, without computer assistance, to make sure that we had not made mistakes in our proofs. The version of $\lambda_d$ that we have proved is written in Ott ⟦Sewell et al., 2007⟧, and the same Ott file is used as a source for the typing rules displayed in this document, making sure that we have proved the same system as we are presenting; though some simplifications are applied by a script during compilation for visual clarity.

Most of the proof was done by myself with little prior experience with Coq. However the development sped up dramatically thanks to the help of my industrial advisor, who was able to use his long prior experience with Coq to introduce the good core lemmas upon which we could easily base many others.

In total the proof is about 7000 lines long, and contains nearly 500 lemmas. Many of the cases of the type preservation and progress lemmas are similar. To handle such repetitive cases, the use of a large-language-model based autocompletion system has proven quite effective, at the detriment of elegance of the proofs. For instance, we do not have any abstract formalization of semirings: it was more expedient to brute-force the properties we needed by hand.

There are nonetheless a few points of interest in our Coq development, that we present in the next paragraphs.

**Representation of typing contexts** We represent contexts as finite-domain functions, rather than syntactic lists, as this works much better when defining sums of contexts. While there are a handful of finite function libraries in the ecosystem, we needed *finite dependent functions* (because the type of binders depends on whether we are binding a variable name or a hole name). This did not exist, but for our limited purposes, it turned out to be feasible to roll our own (about 1000 lines of proofs). The underlying data type is actual functions: this was simpler to develop, but, as we will see soon, it makes equality more complex than with a bespoke data type.

The definitions of the finite function and context types are provided below:

```
Definition Support {A B} (l : list A) (f : forall x:A, option (B x)) : Prop
  := forall (x:A) (y:B x), f x = Some y -> List.In x l.

Record T A B := {
  underlying :> forall x:A, option (B x);
  supported : exists l : list A, Support l underlying;
}.
Inductive name : Type :=
  | name_Var (x:var)
  | name_DH (h:hname).

Definition binding_type_of (n : name) : Type :=
match n with
  | name_Var _ => binding_var
  | name_DH _ => binding_dh
end.

Inductive binding_var : Type :=
  | binding_Var (m:mode) (T:type).

Inductive binding_dh : Type :=
  | binding_Dest (m:mode) (T:type) (n:mode)
  | binding_Hole (T:type) (n:mode).

Definition ctx : Type := T name binding_type_of.
```

The Support predicate indicates that a list is an over-approximation of the effective domain of our partial function. We then use the classical *constructive indefinite description* axiom to retrieve a suitable support list for a given context when needed:

```
(* Axiom *)
ClassicalEpsilon.constructive_indefinite_description
  : forall (A : Type) (P : A -> Prop), (exists x : A, P x) -> {x : A | P x}

Definition a_support {A B} (f : T A B) : list A
  := constructive_indefinite_description _ f.(supported).
```

We also introduce dom ctx as a filter over a_support ctx to retain only the elements in the effective domain of our partial function ctx.

We also need to reason about equality of typing contexts in proofs. The definition of our partial function type would naturally lend itself well to setoid equality, but unfortunately this is not supported by the inference rules we generate through Ott.

As a workaround, we assume the classical *functional extensionality* and *proof irrelevance* axioms, allowing us to state that two contexts are equal as long as they assign the same binding to all variable and hole names:

```
(* Axioms *)
ProofIrrelevance.proof_irrelevance : forall (P : Prop) (p1 p2 : P), p1 = p2
FunctionalExtensionality.functional_extensionality_dep
  : forall (A : Type) (B : A -> Type) (f g : forall x : A, B x),
    (forall x : A, f x = g x) -> f = g

Lemma ext_eq' : forall {A B} (f g : T A B), f.(underlying) = g.(underlying) -> f = g.
Proof. intros A B [f f_supp] [g g_supp] h_ext. cbn in *. subst g. f_equal.
  apply proof_irrelevance.
Qed.

Lemma ext_eq : forall {A B} (f g : T A B), (forall x, f x = g x) -> f = g.
Proof. intros * h_ext. apply ext_eq'.
  apply functional_extensionality_dep.
  assumption.
Qed.
```

In effect, we recover a form of extensional equality that behaves morally like setoid equality, but integrates smoothly with the infrastructure we generate through Ott.

**Partial addition of contexts** As seen in Sections 2.5.2 and 2.6.1, addition of contexts is highly partial. For example, two bindings with the same name can only be combined if they also have the same type, and we cannot combine a hole binding with a destination binding for the same hole name.

Rather than encoding addition as a binary function returning an optional context—which would integrate poorly with the rules generated by Ott—we model it as a total function to contexts. Instead, to capture failure cases, we introduce a special error mode in our mode semiring. A binding carrying this error mode signals that the whole context is considered invalid:

```
Definition mode : Type := option (mul * age).
Notation "'💀'" := None.
Notation "'¹ν'" := (Some (pair Lin (Fin 0))).
Notation "'¹↑'" := (Some (pair Lin (Fin 1))).
Notation "'¹∞'" := (Some (pair Lin Inf)).
...

Definition mode_plus (m1 m2: mode) : mode :=
match m1, m2 with
  | None, _ ⇒ None
  | _, None ⇒ None
  | Some (p1, a1), Some (p2, a2) ⇒ Some (mul_plus p1 p2, age_plus a1 a2)
end.

Inductive IsValid : mode -> Prop
  := IsValidProof : forall (pa : mul * age), IsValid (Some pa).
```

```

Definition ValidOnly (G: ctx) : Prop
  := forall (n : name) (binding: binding_type_of n), G n = Some binding -> IsValid (mode_of binding).
...

Lemma ValidOnly_union_forward
  : forall (G1 G2 : ctx), ValidOnly G1 -> ValidOnly G2 -> G1 # G2 -> ValidOnly (G1 + G2).
Lemma ValidOnly_union_backward
  : forall (G1 G2 : ctx), ValidOnly (G1 + G2) -> ValidOnly G1 /\ ValidOnly G2.
...
```

The cost of this approach is that most typing rules must now include explicit well-formedness conditions—such as `IsValid` and `ValidOnly`—to ensure that the input contexts do not contain any faulty bindings. Additional lemmas are required to propagate these validity conditions through context operations, using properties such as disjointness to guarantee the absence of errors in the resulting contexts.

## 2.8 Translation of prior work into $\lambda_d$ terms and types

We announced in Section 2.1 that $\lambda_d$ subsumes existing systems. Let's see how.

**A Functional Representation of Data Structures with a Hole** Minamide's system [Minamide, 1998] introduces two primitives, **happ** and **hcomp**, as well as the hole abstraction form, $\hat{\lambda} x \mapsto t$, representing structures with (a) hole, where $x$ appears exactly once in $t$.

The paper indicates that hole abstractions cannot have lambda abstraction, application, or case expression operating on the variable of the abstraction, but otherwise allow these constructs to appear in the body of the hole abstraction if they are operating on other variables or values.

In a well-typed, closed program, we can evaluate the body of the hole abstraction (under the $\hat{\lambda}$) as far as possible, until it is almost a value except for the single occurence of $x$ (we know $x$ cannot cause this reduction to be stuck, as it cannot appear in the evaluation path of an application or case expression). Fully evaluated hole abstractions then directly translate to our ampar value form: we interpret $\hat{\lambda} x \mapsto t$ as ${}_{\{h\}}\langle t[x := \boxed{h}] \,{}_{9}\, {\to}h \rangle$. In other terms, we take the body of the hole abstraction as the left-hand side of the ampar, with the single occurrence of the variable replaced by a hole $\boxed{h}$, and just use the corresponding destination ${\to}h$ as the right-hand side. The type of hole abstractions, $(\mathsf{T}, \mathsf{S})\mathbf{hfun}$, is translated as $\mathsf{S} \ltimes \lfloor \mathsf{T} \rfloor$.

In the general case, where the hole abstraction may not be fully evaluated, we interpret it by the following sequence of operations:

- spawning a new ampar with $\mathbf{new}_{\ltimes}$;
- opening it with $\mathbf{upd}_{\ltimes}$;
- building the data constructor in which the hole variable appear by a chain of fill expressions $d \triangleleft \ldots$ or $d \blacktriangleleft \ldots$ (following the same recipe that we used to recover normal data constructors or compound constructors from destination-filling operators; see (::) at the beginning of Section 2.2.2);
- returning, as the result of the body of $\mathbf{upd}_{\ltimes}$, the destination corresponding to the field of the data constructor in which we want the hole to be;
- other constructs, like case expressions or applications related to other variables or values are transposed untouched into the body of $\mathbf{upd}_{\ltimes}$.

It's a bit tedious to give a formal description of the translation, but the process is nonetheless systematic, and becomes quite clear on an example, with Minamide's term on the left, and $\lambda_d$ one on the right:

$$\begin{array}{l|l}
\texttt{ha} : (\texttt{T}, \texttt{List T})\texttt{hfun} & \texttt{ha}_{\lambda_d} : (\texttt{List T}) \ltimes \lfloor \texttt{T} \rfloor \\
\texttt{ha} \triangleq \hat{\lambda} x \mapsto \texttt{0} :: x :: (f\ 2) & \texttt{ha}_{\lambda_d} \triangleq \texttt{case upd}_\ltimes\ \texttt{new}_\ltimes\ \texttt{with}\ d \mapsto \\
 & \qquad (d \triangleleft (::))\ \texttt{of}\ (d_0 , dt) \mapsto \\
 & \qquad\quad \texttt{case}\ (dt \triangleleft (::))\ \texttt{of}\ (d_1 , dt') \mapsto \\
 & \qquad\qquad d_0 \blacktriangleleft \texttt{0}\ ⨟\ dt' \blacktriangleleft (f\ 2)\ ⨟\ d_1
\end{array}$$

Once $(f\ 2)$ has reduced to a value $v$, the $\lambda_d$'s version will be evaluated to the promised ampar form ${}_H/ \texttt{0} :: \boxed{h} :: v\ {}_{9} \to h/$.

Because destinations are not part of [Minamide, 1998], any variable in a program from Minamide's system can be given age $\infty$ (so mode $1\infty$ or $\omega\infty$ depending on whether it is linear or not), except of course the variables bound by hole abstractions.

The translation of **happ** and **hcomp** is even more direct:

$$\begin{array}{l}
\texttt{happ} : (\texttt{T}, \texttt{S})\texttt{hfun} \multimap \texttt{T} \multimap \texttt{S} \\
\texttt{happ}_{\lambda_d} : \texttt{S} \ltimes \lfloor \texttt{T} \rfloor \multimap \texttt{T} \multimap \texttt{S} \\
\texttt{happ}_{\lambda_d}\ ampar\ x \triangleq \texttt{from}'_\ltimes (\texttt{upd}_\ltimes\ ampar\ \texttt{with}\ d \mapsto d \blacktriangleleft x)
\end{array}$$

$$\begin{array}{l}
\texttt{hcomp} : (\texttt{T}, \texttt{S})\texttt{hfun} \multimap (\texttt{U}, \texttt{T})\texttt{hfun} \multimap (\texttt{U}, \texttt{S})\texttt{hfun} \\
\texttt{hcomp}_{\lambda_d} : \texttt{S} \ltimes \lfloor \texttt{T} \rfloor \multimap \texttt{T} \ltimes \lfloor \texttt{U} \rfloor \multimap \texttt{S} \ltimes \lfloor \texttt{U} \rfloor \\
\texttt{hcomp}_{\lambda_d}\ ampar_1\ ampar_2 \triangleq \texttt{upd}_\ltimes\ ampar_1\ \texttt{with}\ d \mapsto d \triangleleft\!\circ\ ampar_2
\end{array}$$

With this translation we can completly express Minamide's system in $\lambda_d$ with no loss of flexiblity.

**The Functional Essence of Imperative Binary Search Trees** The portion of the work from Leijen and Lorenzen [2025] and Lorenzen, Leijen, Swierstra, and Lindley [2024] about *first-class constructor contexts* is fairly similar, in both expressiveness and presentation, to Minamide's *hole abstractions*. A hole abstraction $\hat{\lambda} x \mapsto t$ is written **ctx** $t[x := _]$, whose body is the same as $t$ but with an underscore _ (or a square □ in some versions of their work) in place of the named hole variable $x$. Similarly, **happ** is replaced by the ++. operator (notice the period), and **hcomp** by the ++ operator. Consequently, the translation for terms from [Leijen and Lorenzen, 2025; Lorenzen, Leijen, Swierstra, and Lindley, 2024] system to ours is the same as for Minamide's system.

For types, [Leijen and Lorenzen, 2025; Lorenzen, Leijen, Swierstra, and Lindley, 2024] only define $\texttt{ctx}<\texttt{T}>$, where the hole and the resulting structure (once the hole will be filled) are of the same type $\texttt{T}$. This translates to $\texttt{T} \ltimes \lfloor \texttt{T} \rfloor$ in $\lambda_d$.

## 2.9 Implementation of $\lambda_d$ using in-place memory mutations

The formal language presented in Sections 2.5 and 2.6 is not meant to be implemented as-is. Practical implementation of most $\lambda_d$'s ideas will be the focus of Chapters 3 and 4.

First, $\lambda_d$ does not have recursion, this would have obscured the formal presentation of the system. However, adding a standard form of recursion does not create any complication.

Secondly, ampars are not managed linearly in $\lambda_d$; only destinations are. That is to say that an ampar can be wrapped in an exponential, e.g. $\mathtt{Mod}_{\omega\nu}\ {}_{\{h\}}\langle 0 :: \boxed{h}\ ,\!\rightarrow\! h\rangle$ (representing a difference list $0{::}\square$ that can be used non-linearly), and then used twice, each time in a different way:

$$\begin{array}{l} \mathtt{case}\ \mathtt{Mod}_{\omega\nu}\ {}_{\{h\}}\langle 0 :: \boxed{h}\ ,\!\rightarrow\! h\rangle\ \mathtt{of}\ \mathtt{Mod}_{\omega\nu}\ x \mapsto \\ \quad \mathtt{let}\ x_1 := x\ \mathtt{append}\ 1\ \mathtt{in} \\ \quad \mathtt{let}\ x_2 := x\ \mathtt{append}\ 2\ \mathtt{in} \qquad\qquad \longrightarrow^* \quad 0 :: 1 :: 0 :: 2 :: [\,] \\ \qquad \mathtt{toList}\ (x_1\ \mathtt{concat}\ x_2) \end{array}$$

It may seem counter-intuitive at first, but this program is valid and safe in $\lambda_d$. Thanks to the renaming discipline we detailed in Section 2.6.3, every time an ampar is operated over with $\mathtt{upd}_\ltimes$, its hole names are renamed to fresh ones. One way we can implement this in practice is to allocate a fresh copy of $x$ every time we call $\mathtt{upd}_\ltimes$ on it (recall that **append** is implemented in terms of $\mathtt{upd}_\ltimes$), in a *copy-on-write* fashion. This way filling destinations is still implemented as mutation.

However, this is a long way from the efficient implementation promised in Section 2.2. Copy-on-write can be optimized using fully-in-place functional programming ⟦Lorenzen, Leijen, and Swierstra, 2023⟧, where, thanks to reference counting, we do not need to perform a copy when the difference list is not aliased. This idea is further explored, for structure with holes specifically, in Chapter 7 of their more recent work ⟦Leijen and Lorenzen, 2025⟧. But that will not be the direction we will follow in the following development, as we do not want to deal explicitly with reference counting.

An alternative is to refine the linear type system further in order to guarantee that ampars are unique and avoid copy-on-write altogether. We held back from doing that in the formalization of $\lambda_d$ as, again, it obfuscates the presentation of the system without adding much in return.

To make ampars linear, we follow the recipe we developed in Section 1.8, highly inspired from ⟦Spiwack, 2023b; Spiwack et al., 2022⟧, and introduce a new built-in type **Token**, together with the primitive **dup**, **drop**, and **withToken**. We also switch $\mathtt{new}_\ltimes$ for $\mathtt{new}_{\ltimes\mathtt{IP}}$:

$$\begin{array}{l} \mathtt{dup} : \mathtt{Token} \multimap \mathtt{Token} \otimes \mathtt{Token} \\ \mathtt{drop} : \mathtt{Token} \multimap \mathbf{1} \\ \mathtt{withToken} : (\mathtt{Token}\ {}_{1\infty}\!\multimap\ !_{\omega\infty}\mathsf{T}) \multimap\ !_{\omega\infty}\mathsf{T} \\ \mathtt{new}_{\ltimes\mathtt{IP}} : \mathtt{Token} \multimap \mathsf{T} \ltimes \lfloor \mathsf{T} \rfloor \end{array}$$

Ampars produced by $\mathtt{new}_{\ltimes\mathtt{IP}}$ have a linear dependency on a **Token**. If the same ampar, originally created with $\mathtt{new}_{\ltimes\mathtt{IP}}\ tok$, were to be used twice in a block $t$, then $t$ would require a typing context $\{tok :_{\omega\infty} \mathtt{Token}\}$, thus the block would be rejected by **withToken**. In the other hand, duplicating $tok$ into $tok_1$ and $tok_2$ first, and then using each new token to create a different ampar would be linearly valid.

Now that ampars are managed linearly, we can change the allocation and renaming mechanisms:

- the hole name for a new ampar is chosen fresh right from the start (this corresponds to a new heap allocation);
- adding a new hollow constructor still requires freshness for its hole names (this corresponds to a new heap allocation too);
- Using $\mathtt{upd}_\ltimes$ over an ampar and filling destinations or composing two ampars using ◁∘ no longer requires any renaming: we have the guarantee that all the names involved are globally fresh, and can only be used once, so it actually corresponds to an in-place memory update.

In Chapters 3 and 4, dedicated to the implementation of DPS in a functional setting, we will treat ampars as linear resources, in the same style as $\lambda_d$ extended with **Tokens** and $\mathtt{new}_{\ltimes\mathtt{IP}}$.

**From binary products and sums to more efficient memory forms** In $\lambda_d$ we only have *binary* product types and sum types. However, it's very straightforward to extend the language and implement destination-based building for n-ary sums of n-ary products, with constructors for each variant having multiple fields directly, instead of each field needing an extra indirection as in the binary sum of products $1 \oplus (S \otimes (T \otimes U))$. In fact, we will do that in next chapter. However, we still require field's values to be represented by pointers. It's crucial for the semantics of the $\triangleleft\circ$ operator.

## 2.10 Conclusion

$\lambda_d$ is a purely functional calculus which treats destinations as first class values that can be passed around and used in very flexible ways. It supports data structures with multiple holes, and allows both composition of data structures with holes and storing destinations in data structures that, themselves, have holes.

With $\lambda_d$, we can now apply some techniques and algorithms traditionally confined to imperative languages within a purely functional setting, all while preserving safety guarantees. Indeed, thanks to a linear type system augmented with a system of ages, the mutations introduced by destination use are opaque and controlled.

However, we do not anticipate that a system of ages like the one of $\lambda_d$ will actually be used in a programming language: it's unlikely that destinations are so central to the design of a programming language that it's worth baking them so deeply in the type system. Perhaps a compiler that makes heavy use of destinations in its optimizer could use $\lambda_d$ as a typed intermediate representation. But, more realistically, our expectation is that $\lambda_d$ can be used as a theoretical framework to analyze destination-passing systems: if a destination passing system can be expressed in terms of $\lambda_d$ primitives, then it is sound.

In the chapters ahead, we aim to determine which (hopefully moderate) restrictions we have to impose on $\lambda_d$'s flexiblity to port most of its fundamental concepts safely into a real-world functional programming language, namely, Haskell—whose type system is not as rich as $\lambda_d$'s one, designed on purpose for destination control.

# Chapter 3

# A first implementation of destination passing in Haskell

In the previous chapter, we saw how we can build a $\lambda$-calculus from the ground up with first-class support for destination passing, and that still achieves memory safety. It's now time to put that into practice in an broadly used programming language.

This chapter revisits and refines the results presented in [Bagrel, 2024a], which was published at JFLA 2024 prior to the completion of the theoretical work presented in Chapter 2.

## 3.1 Implementation target: Linear Haskell

Designing an industrial-grade programming language *just* to add support for destinations would be a gigantic task, with much risk to fail. Instead, we base our work on an existing functional programming language, for instance, Haskell[20], and we aim to implement as many ideas from $\lambda_d$ (see Chapter 2) as possible in this language. Haskell is equipped with support for linear types through its main compiler, *GHC*, since version 9.0.1 [Bernardy et al., 2018]. We kindly redirect the reader to Section 1.7 for a primer on Linear Haskell and how to use it to control resource use. Moreover, Haskell is a *pure* functional language, which means our destination-passing API will have to be pure, with (naturally impure) memory writes safely encapsulated within it, as explained in the introduction, so that no mutation or incomplete structure can be observed by the user.

The main challenge is that Haskell, albeit incorporating linear types, does not have any concept of scope or age control. So we will not be able to avoid the fundamental issues presented in Section 2.3 without imposing limitations on the flexibility of the system.

[20]There are not many other industrial-grade functional programming languages, and most of them do not have support for linear types, which are the first piece required to make a system like $\lambda_d$ safe. Thanks to recent work from Lorenzen, White, et al. [2024], OCaml will soon have support for *affinity*, but not for *linearity* (despite the aforementioned work using the term linearity). Affinity ensures that a resource can be used at most once, but does not guarantee it will be used exactly once; it may be discarded without use. This would be problematic for us, since we rely on the guarantee that destinations cannot be dropped: it ensures that when an `ampar` no longer contains destinations on the right, all holes in the corresponding structure have been filled.

## 3.2 Safe destination passing with just linear types

As we have seen in Section 2.3, linear types are not enough, alone, to make $\lambda_d$ safe. We definitely need ages and scope control. In this chapter, we will take a radical decision: we restrict destinations to only be used to build non-linear (a.k.a. unrestricted) data structures, so that the issue of scope escape disappears completely. This also rules out the potential interaction between sibling destinations also described in Section 2.3. In exchange, we will not be able to store destinations inside data structures built using ampars. We further use the name *DPS Haskell* to refer to this prototype destination-passing API we develop in Haskell.

In practice, it means that a destination can only be filled with an unrestricted value (which forbids filling a destination with another destination, as destinations are linear resources). We will use the Haskell type `data UDest t` to represent destinations for unrestricted values, and type `data UAmpar s t` to represent ampars where the `s` side is unrestricted.

Because there is no difference when creating the spine[21] of a data structures that is linear or unrestricted, our fill functions that add a new hollow constructor to an existing structure will be exactly the same as in $\lambda_d$. Only the signature of fill operators acting on the leaves of data structures will change in the implementation compared to the formal definition. The mapping of operators and types between $\lambda_d$ and DPS Haskell is given in Figure 3.1

In $\lambda_d$, destinations always have finite ages. As a result, we use age $\infty$ as a marker that something does not contain destinations. This way, we can still enforce scope control independently of linearity. This is particularly visible in rule $\lambda_d$–TY/$\ltimes$FROM (Figure 2.2), where a linear resource can be extracted from the right of an ampar as long as it has age $\infty$.

In DPS Haskell however, we do not have ages. So we use the fact that destinations are always linear too, and we employ the $\omega$ multiplicity as a marker that something does not contain destinations. Of course, this does not come for free; by doing so we conflate the absence of destination with non-linearity. Controlling scopes that way means we must be more conservative to preserve safety, which restrains the possibilities of a few operators.

Now, let's see how programs look like in DPS Haskell. We will start by reimplementing examples from the previous chapter, to see how they take form in a practical setting.

## 3.3 A glimpse of DPS Haskell programming

The following subsections present three typical cases in which DPS programming brings expressiveness or performance benefits over a more traditional functional implementation. We will start by revisiting some examples from Section 2.2.

### 3.3.1 Efficient difference lists

Linked lists are a staple of functional programming, but they are not efficient for concatenation, as we have seen in Section 2.2.3, especially when the concatenation calls are nested to the left. For more efficient concatenation, we can define destination-based difference lists, which are effectively lists with a hole at the end.

[21] The spine refers to the main chain of connected data constructors that form the overall shape of a data structure, connecting its "leaves".

*Destination-filling operators:*

| $\lambda_d$ | DPS Haskell | | |
|---|---|---|---|
| $d \triangleleft ()$ | `fill @'() d` | or | `d &fill @'()` |
| $d \triangleleft$ `Inl` | `fill @'Inl d` | or | `d &fill @'Inl` |
| $d \triangleleft$ `Inr` | `fill @'Inr d` | or | `d &fill @'Inr` |
| $d \triangleleft (,)$ | `fill @'(,) d` | or | `d &fill @'(,)` |
| $d \triangleleft$ `Mod`$_{\omega\infty}$ | `fill @'Ur d` | or | `d &fill @'Ur` |
| $d \triangleleft (\lambda x \;{}_{m}\!\!\mapsto term)$ | `fillLeaf (\x → term) d` | or | `d &fillLeaf (\x → term)` |
| $d \triangleleft\!\circ\; term$ | `fillComp term d` | or | `d &fillComp term` |
| $d \blacktriangleleft term$ | `fillLeaf term d` | or | `d &fillLeaf term` |

(one can see `&fill @'` as the equivalent of $\triangleleft$ in $\lambda_d$, `&fillLeaf` as $\blacktriangleleft$, and `&fillComp` as $\triangleleft\!\circ$)

*Types:*

| $\lambda_d$ | DPS Haskell |
|---|---|
| $\lfloor_{\omega\infty} \mathsf{T} \rfloor$ | `UDest t` |
| $!_{\omega\infty} \mathsf{S} \ltimes \mathsf{T}$ | `UAmpar s t` |

Figure 3.1: Mappings between $\lambda_d$ and DPS Haskell

This example will be the opportunity to show how DPS Haskell operates on memory. So far, with $\lambda_d$, we worked on a theoretical memory model with global substitutions (on the whole evaluation context) and no representation of allocations. Here, with DPS Haskell, we assume that we have a proper heap. We pay no attention right now to garbage collection / deallocation; instead we will deal with this topic in Section 3.5.1.

Following the mapping tables of Section 3.2, we know the DPS Haskell representation of difference lists is `type DList t = UAmpar [t] (UDest [t])`. As announced previously, because our difference lists are built using `UAmpar`s, they cannot host linear resources. That's the major limitation of this first approach of DPS programming for Haskell: we cannot use efficient destination-based data structures to store linear data. For instance, the breadth-first tree traversal example developed in Chapter 2 uses such destination-based difference lists to store destinations, and thus, will not type as is in DPS Haskell without modifications.

As with ampars in $\lambda_d$, the left side of the `UAmpar` carries the structures being built, while the right side carries the destinations of the structure: the `UDest [t]` must be filled (with an unrestricted `[t]`) to get a readable (unrestricted) `[t]`.

The implementation of destination-backed difference lists is presented in Listing 3.1 (we recall that the $\lambda_d$'s version is presented in Listing 2.2). Note that Haskell uses `::` for typing, and `(:)` for the list *cons* constructor, which is the opposite of what these symbols mean in $\lambda_d$.

- `newDList` is just an alias for `newUAmpar @[t]`, to linearly exchange a token for a new `UAmpar [t] (UDest [t])`—that is, exactly `DList t`. Recall that UAmpars are treated as linear resources now, and are created in exchange of a linear token, following the principles described in Sections 1.8 and 2.9. There is no data in this new UAmpar yet, it's just a hole and a destination pointing to it, as we see in Figure 3.2;

```haskell
-- we recall here the definition of list type in Haskell
data [t] = []          -- nil constructor
         | (:) t [t]  -- cons constructor

-- we reuse the same linear token API as in Section 1.8
withToken :: (Token ⊸ Ur t) ⊸ Ur t
-- we recall the signature of `updWith` in Haskell
updWith :: ∀ s t u. UAmpar s t ⊸ (t ⊸ u) ⊸ UAmpar s u

type DList t = UAmpar [t] (UDest [t])

newDList :: Token ⊸ DList t
newDList = newUAmpar @[t]

append :: DList t ⊸ t → DList t
dlist `append` x =
  dlist `updWith` \d → case (d &fill @'(:)) of
    (dh, dt) → (dh &fillLeaf x) ⨾ dt

concat :: DList t ⊸ DList t ⊸ DList t
dlist1 `concat` dlist2 = dlist1 `updWith` \dt1 → (dt1 &fillComp dlist2)

toList :: DList t ⊸ Ur [t]
toList dlist = fromUAmpar' (dlist `updWith` \dt → dt &fill @'[])
```

Listing 3.1: Implementation of difference lists in DPS Haskell

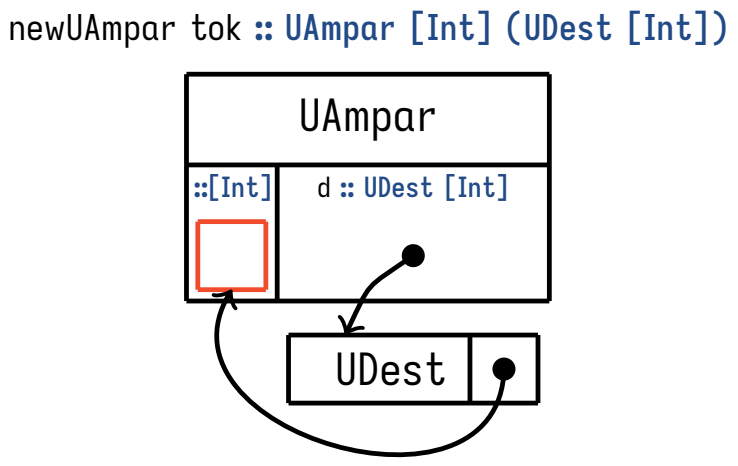

Figure 3.2: Memory representation of `newUAmpar tok` in DPS Haskell

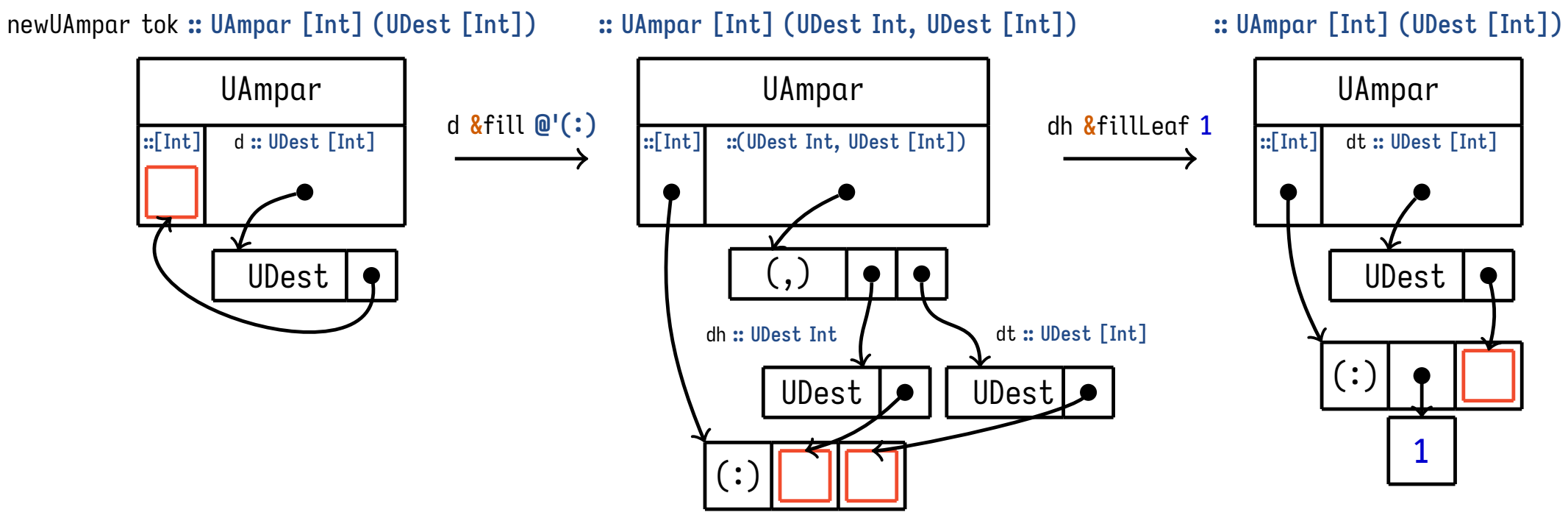

Figure 3.3: Memory behavior of append newUAmpar 1 in DPS Haskell

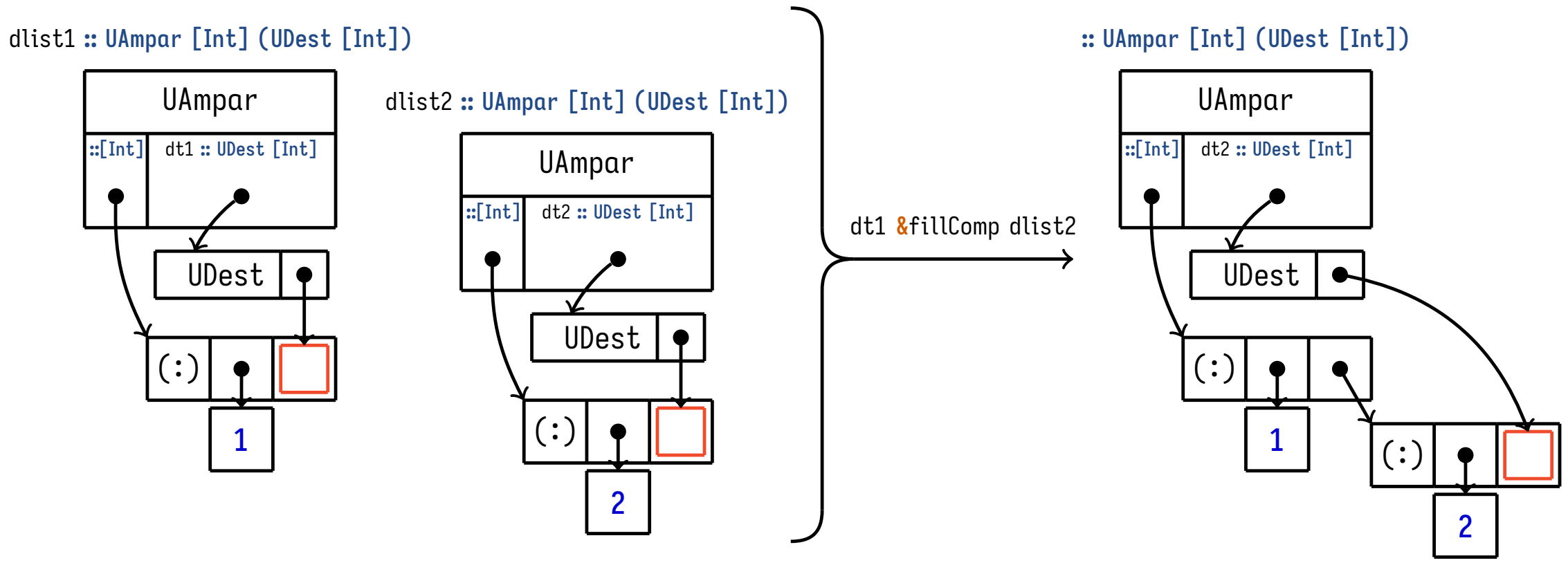

Figure 3.4: Memory behavior of concat dlist1 dlist2 in DPS Haskell (based on fillComp)

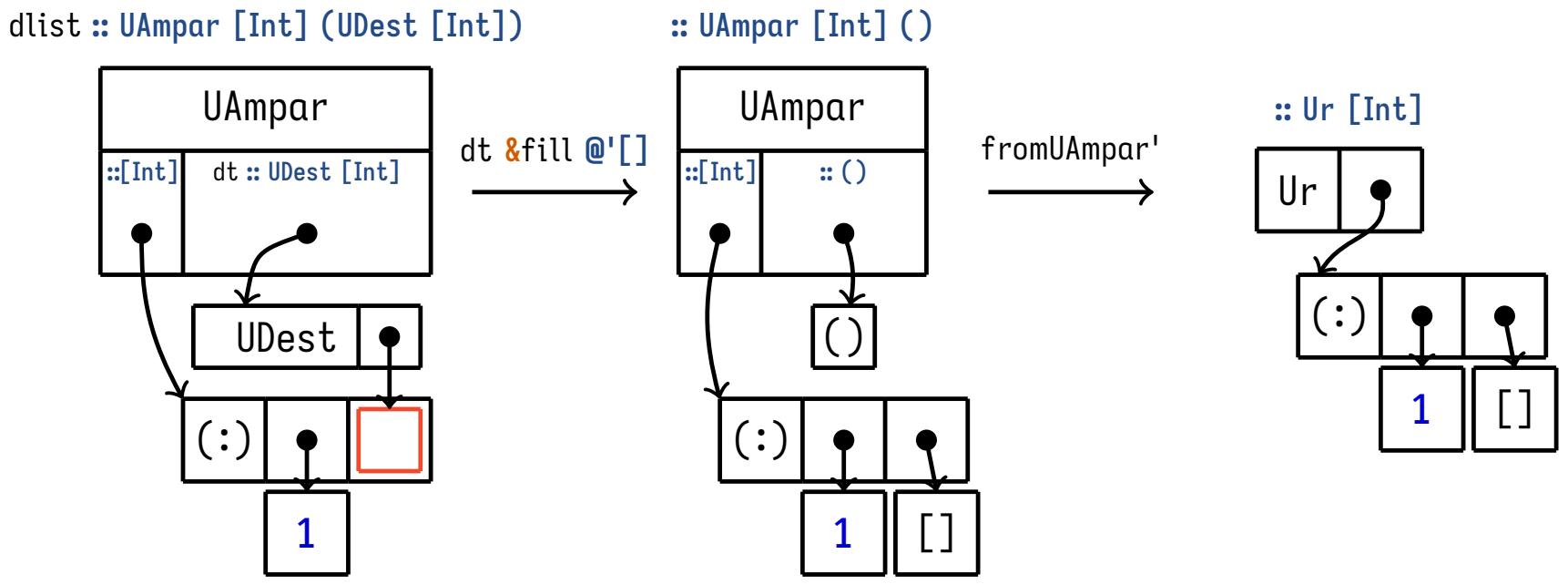

Figure 3.5: Memory behavior of toList dlist in DPS Haskell

- `append` (Figure 3.3) adds an element at the tail position of a difference list. For this, it first uses `fill @'(:)` to fill the hole at the end of the list represented by `d :: UDest [t]` with a hollow `(:)` constructor that has two new holes, pointed to by `dh :: UDest t` and `dt :: UDest [t]`. Then, `fillLeaf` fills the hole represented by `dh` with the value of type `t` to append. The hole at the end of the resulting difference list is the one pointed by `dt :: UDest [t]` which has not been filled yet, and stays on the right side of the resulting UAmpar.
- `concat` (Figure 3.4) concatenates two difference lists, `dlist1` and `dlist2`. It uses `fillComp` to fill the destination `dt1` of the first difference list with the root of the second difference list `dlist2`. The resulting `UAmpar` object hence has the same root as the first list, holds the elements of both lists, and inherits the hole of the second list. Memory-wise, `concat` just writes the address of the second list into the hole of the first one.
- `toList` (Figure 3.5) completes the UAmpar structure by plugging a new *nil* `[]` constructor into its hole with `fill @'[]`, and then removes the `UAmpar` wrapper as the structure is now complete, using `fromUAmpar'`. The completed list is wrapped in `Ur`, as it is allowed to be used non-linearly—it's always the case with UAmpars—and so that it can also escape the linear scope created by `withToken` if needed.

Linearity of UAmpars, enforced by the linear token technique, is essential[22] to allow safe, in-place, memory updates when destination-filling operations occur, as shown in Figures 3.2 to 3.5. Otherwise, we could first complete a difference list with `let Ur l = toList dlist`, then add a new cons cell to `dlist` with `append dlist x` (actually reusing the destination inside `dlist` for the second time). This would create a hole inside `l`, although it is of type `[t]` so we would be able to pattern-match on it, and might get a segfault!

The performance of this implementation (compared to a functional encoding of difference lists) is discussed in Section 3.6.1.

### 3.3.2 Breadth-first tree traversal

Let's move on now to breadth-first tree traversal. We want to traverse a binary tree and update its node values so that they are numbered in a breadth-first order.

In Section 2.4, we took full advantage of $\lambda_d$'s flexibility, and implemented the breadth-first traversal using a queue to drive the processing order, as we would do in an imperative programming language. This queue contained both input subtrees and destinations to output subtrees, in breadth-first order. Furthermore, the queue was implemented efficiently using a pair of a queue and a ampar-based difference list.

In DPS Haskell though, destinations cannot be filled with linear elements; so, an ampar-based difference list would not be able to store the destinations we need to store (the one representing the output subtrees), and consequently, neither would an ampar-based efficient queue.

So we have no choice than to revert to the regular, non-ampar-based data structures of Haskell to build a suitable Hood-Melville ⟦Hood and Melville, 1981⟧ queue. We see here that destination passing in Haskell cannot fully replace constructors unlike in $\lambda_d$; both data constructors and destination-filling primitives need to coexist[23]. The implementation of Hood-Melville queues is presented in Listing 3.2.

[22]In $\lambda_d$ we could choose to use a smart renaming technique for hole names (morally equivalent to copy-on-write) so that ampars do not have to be managed linearly. Here, as we aim for an efficient implementation, the idea of copying ampars and updating the target of destinations on-the-go does not seem promising.

[23]Destination-based data structure building is more general than constructor-based structure building as long as we can fill destinations with linear resources, as in $\lambda_d$. When we add the restriction that destination can only be filled with unrestricted elements, as we do in DPS Haskell, then it is no longer more general than constructor-based structure building, and we need to conserve both building ways in the language to stay at the same level of expressiveness.

```haskell
data HMQueue t = HMQueue [t] [t]

newHMQueue :: HMQueue t
newHMQueue = HMQueue [] []

singleton :: t ⊸ HMQueue t
singleton x = HMQueue [x] []

toList :: HMQueue t ⊸ [t]
toList (HMQueue front backRev) = front ++ reverse backRev

enqueue :: HMQueue t ⊸ t ⊸ HMQueue t
enqueue (HMQueue front backRev) x = HMQueue front (x : backRev)

dequeue :: HMQueue t ⊸ Maybe (t, HMQueue t)
dequeue (HMQueue front backRev) = case front of
  [] → case reverse backRev of
    [] → Nothing
    (x : front') → Just (x, HMQueue front' [])
  (x : front') → Just (x, HMQueue front' backRev)
```

Listing 3.2: Implementation of Hood-Melville queue in Haskell

As before, for the breadth-first traversal, we will keep a queue of pairs of a tree to be relabeled and of the destination where the relabeled result is expected, and process each of them when their turn comes. The only thing that changes is the implementation for the queue. The DPS Haskell implementation of breadth-first tree traversal is provided in Listing 3.3.

Except from the choice of queue implementation, and the fact that ampars must be created from `Tokens`, this implementation is very much the same as in Section 2.4. In the signature of `go` function, the linear arrow enforces the fact that every destination ever put in the queue is eventually consumed at some point, which guarantees that the output tree is complete after the function has run.

As we will see in Section 3.6, this imperative-style implementation of breadth-first traversal in DPS Haskell, despite not using efficient queue representations such as difference lists, still shows substantial performance gains compared to the functional implementation from Gibbons et al. ⟦2023⟧, which we must admit, while pretty elegant, was not designed with performance as a primary focus.

## 3.4 DPS Haskell API and design concerns

Listing 3.4 presents the pure API of DPS Haskell. This API is sufficient to implement all the examples of Section 3.3. This section explains its various parts in detail and how it compares to the calculus of Chapter 2.

```
data Tree t = Nil | Node t (Tree t) (Tree t)

relabelDps :: Token ⊸ Tree t → Tree Int
relabelDps tree = fst (mapAccumBfs (\st _ → (st + 1, st)) 1 tree)

mapAccumBfs :: ∀ s t u. Token ⊸ (s → t → (s, u)) → s → Tree t → Ur (Tree u, s)
mapAccumBfs tok f s0 tree =
  fromUAmpar (newUAmpar @(Tree u) tok `updWith` \dtree → go s0 (singleton (Ur tree, dtree)))
  where
    go :: s → Queue (Ur (Tree t), UDest (Tree u)) ⊸ Ur s
    go st q = case dequeue q of
      Nothing → Ur st
      Just ((utree, dtree), q') → case utree of
        Ur Nil → dtree &fill @'Nil ⨟ go st q'
        Ur (Node x tl tr) → case (dtree &fill @'Node) of
          (dy, dtl, dtr) →
            let q'' = q' `enqueue` (Ur tl, dtl) `enqueue` (Ur tr, dtr)
                (st', y) = f st x
             in dy &fillLeaf y ⨟ go st' q''
```

Listing 3.3: Implementation of breadth-first tree traversal in DPS Haskell

```
data Token
dup :: Token ⊸ (Token, Token)
drop :: Token ⊸ ()
withToken :: ∀ t. (Token ⊸ Ur t) ⊸ Ur t

data UAmpar s t
newUAmpar :: ∀ s. Token ⊸ UAmpar s (UDest s)
tokenBesides :: ∀ s t. UAmpar s t ⊸ (UAmpar s t, Token)
toUAmpar :: ∀ s. Token ⊸ s → UAmpar s ()
fromUAmpar :: ∀ s t. UAmpar s (Ur t) ⊸ Ur (s, t)
fromUAmpar' :: ∀ s. UAmpar s () ⊸ Ur s
updWith :: ∀ s t u. UAmpar s t ⊸ (t ⊸ u) ⊸ UAmpar s u

data UDest t
type family UDestsOf lCtor t  -- returns dests associated to fields of constructor
fill :: ∀ lCtor t. UDest t ⊸ UDestsOf lCtor t
fillComp :: ∀ s t. UAmpar s t ⊸ UDest s ⊸ t
fillLeaf :: ∀ t. t → UDest t ⊸ ()
```

Listing 3.4: DPS Haskell API, initial version

### 3.4.1 The UAmpar type

As with Ampars from the previous chapter, UAmpar structures, that serve as a wrapper for structures with holes, can be freely passed around and stored, but need to be completed before any reading i.e. pattern-matching can be made on them. As a result, `UAmpar s t` is defined as an opaque data type, with no public constructor.

In `UAmpar s t`, `s` stands for the type of the structure being built, and `t` is the type of what needs to be linearly consumed before the structure can be read. Eventually, when complete, the structure will be wrapped in `Ur`; this illustrates the fact that it has been made only from unrestricted elements.

The `newUAmpar` operator is the main way for a user of the API to create a new UAmpar. Its signature is equivalent to the one of $\texttt{new}_{\ltimes\texttt{IP}}$ from Section 2.9.

Also, sometimes, when programming in DPS style, we will not have a `Token` at hand, but only an existing `UAmpar`. In this case, we can piggyback on the linearity of that existing UAmpar (as it linearly depends, directly or indirectly, on a token) to get a new linear `Token` so that we can spawn a new UAmpar, or any other linear resource really. This is the goal of the `tokenBesides` function.

The operator `tokenBesides` is particularly useful when implementing efficient queues in DPS Haskell (as in Section 2.2.3):

```
type DList t = UAmpar [t] (UDest [t])
data EffQueue t = EffQueue [t] (DList t)

newEffQueue :: Token ⊸ EffQueue t
newEffQueue tok = EffQueue [] (newUAmpar @[t] tok)

dequeue :: EffQueue t ⊸ Maybe (t, EffQueue t)
dequeue (EffQueue front back) = case front of
  [] → case (toList back) of
    Ur [] → Nothing
    Ur (x : xs) → Just (x, (EffQueue xs (newUAmpar @[t] tok)))
  (x : xs) → Just (x, (EffQueue xs back))
```

In the second branch of the inner-most `case` in `dequeue`, when we have transformed the back difference list (from which we used to write) into the new front list (from which we will now read), we have to create a new back difference list, that is, a new ampar. However, we do not have a token at hand. It would be quite unpractical for a function like `dequeue` to ask for a linear token to be passed; it's better if tokens are only requested when spawning a new data structure, not when operating on one. That's why `tokenBesides` comes handy: we can instead reuse the linearity requirement of the existing UAmpar:

```
tokenBesides :: ∀ s t. UAmpar s t ⊸ (UAmpar s t, Token)

dequeue :: EffQueue t ⊸ Maybe (t, EffQueue t)
dequeue (EffQueue front back) = case front of
  [] → case tokenBesides back of (back, tok) → case (toList back) of
    Ur [] → drop tok ⨾ Nothing
    Ur (x : xs) → Just (x, (EffQueue xs (newDList tok)))
  (x : xs) → Just (x, (EffQueue xs back))
```

This new version is not perfect either though. We have to create a new token before even knowing if we will need it. Indeed, we can only know if a new token is really needed when we have read the content of the old UAmpar, and we can only read it when it is no longer a UAmpar… so at that point we cannot spawn a new token any longer. So the solution is to always spawn a new token before consuming the old UAmpar, taking the risk that if the queue is really empty, we will have to discard the fresh token immediately (with `drop`) before returning `Nothing`. Implicit token passing, as proposed by Linear Constraints ⟦Spiwack, 2023a; Spiwack et al., 2022⟧, seems to really be the way forward to get rid of this kind of issues altogether.

**Getting out of an UAmpar** Same as in $\lambda_d$, the structure encapsulated within a UAmpar can be released in two ways: with `fromUAmpar'`, the value on the `t` side must be unit (`()`), and just the complete `s` is returned, wrapped in `Ur`. With `fromUAmpar`, the type on the `t` side must be of the form `Ur t'`, and then a pair `Ur (s, t')` is returned. We recall that in $\lambda_d$, we had $\texttt{from}'_{\ltimes} : \mathsf{S} \ltimes \mathsf{1} \multimap \mathsf{S}$ and $\texttt{from}_{\ltimes} : \mathsf{S} \ltimes (!_{1\infty}\mathsf{T}) \multimap (\mathsf{S} \otimes !_{1\infty}\mathsf{T})$, where the left side, when extracted from the ampar, is not wrapped in any modality.

Here in DPS Haskell, it is actually safe to wrap the structure that has been built in `Ur` because, as we have said previously, its leaves either come from non-linear sources (as `fillLeaf :: t → UDest t ⊸ ()` consumes its first argument non-linearly) or are made of 0-ary constructors added with `fill`, both of which can be used in an unrestricted fashion safely; and its spine, made of hollow constructors, can be duplicated at will.

### 3.4.2 Destination-filling functions

The last part of the API is the one in charge of actually building the structures in a top-down fashion. To fill a hole represented by `UDest t`, three functions are available:

`fillLeaf :: ∀ t. t → UDest t ⊸ ()` uses a value of type `t` to fill the hole represented by the destination, as ◀ from $\lambda_d$. However, if the destination is consumed linearly, as in $\lambda_d$, the value to fill the hole is not (as indicated by the first non-linear arrow). This is key to the fact that UAmpars only host unrestricted data. Memory-wise, when `fillLeaf` is used, the address of the object of type `t` is written into the memory cell pointed to by the destination of type `UDest t` (see Figure 3.8).

`fillComp :: ∀ s t. UAmpar s t ⊸ UDest s ⊸ t` is used to plug two `UAmpar` objects together. The parent `UAmpar` is not represented in the signature of the function, as with the similar operator ◁∘ from $\lambda_d$. Instead, only the hole of the parent UAmpar that will receive the address of the child UAmpar is represented in the signature of the function by `UDest s`; while `UAmpar s t` in the signature refers to the child UAmpar. A call to `fillComp` always takes place in the scope of `` `updWith` `` over the parent one (as this is the only way to obtain the destination that constitutes the left-hand-side operand of `fillComp`):

```
parent :: UAmpar BigStruct (UDest SmallStruct, UDest OtherStruct)
child :: UAmpar SmallStruct (UDest Int)
comp = parent `updWith` \(ds, do) → (ds &fillComp child, do)
     :: UAmpar BigStruct (UDest Int, UDest OtherStruct)
```

The resulting structure `comp` is morally a `BigStruct` like `parent`, that inherited the hole from the child structure (`UDest Int`) and still has its other hole (`UDest OtherStruct`) waiting to be filled. An example of memory behavior of `fillComp` in action can be seen in Figure 3.4.

Finally, `fill :: ∀ lCtor t. UDest t ⊸ UDestsOf lCtor t` is the generic function to fill a destination with hollow constructors. It takes a data constructor as a type parameter (`lCtor`) and allocates a corresponding hollow constructor—that is, a heap object that has the same header as the specified constructor but unspecified fields—and returns one new destination for each field of this constructor. The address of the allocated hollow constructor is written in the destination that is provided to `fill`. An example of the memory behavior of `fill @'(:) :: UDest [t] ⊸ (UDest t, UDest [t])` is given in Figure 3.6 and the one for `fill @'[] :: UDest [t] ⊸ ()` is given in Figure 3.7.

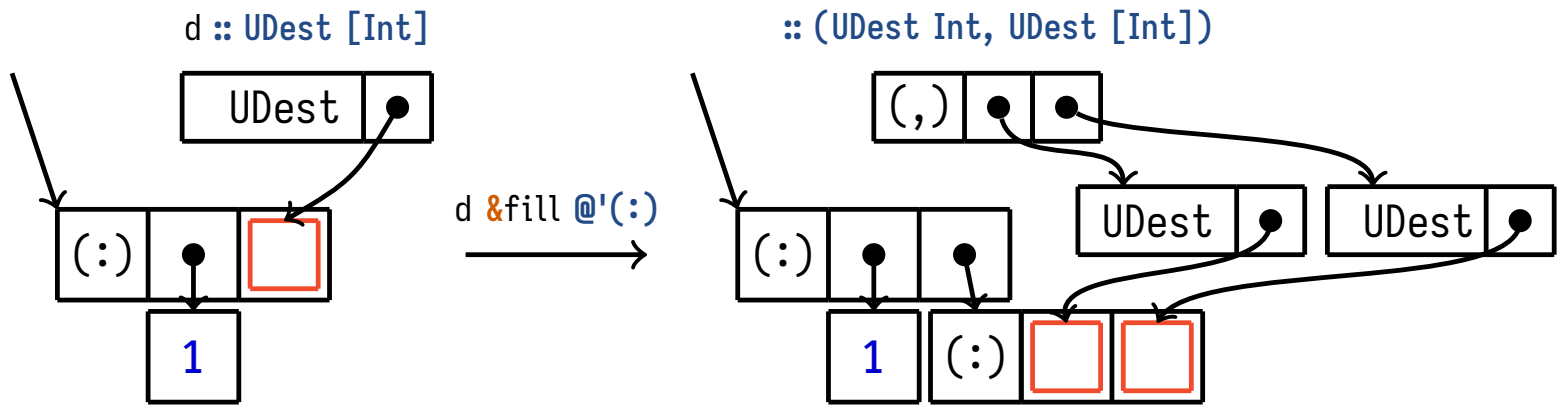

Figure 3.6: Memory behavior of `fill @'(:) :: UDest [t] ⊸ (UDest t, UDest [t])`

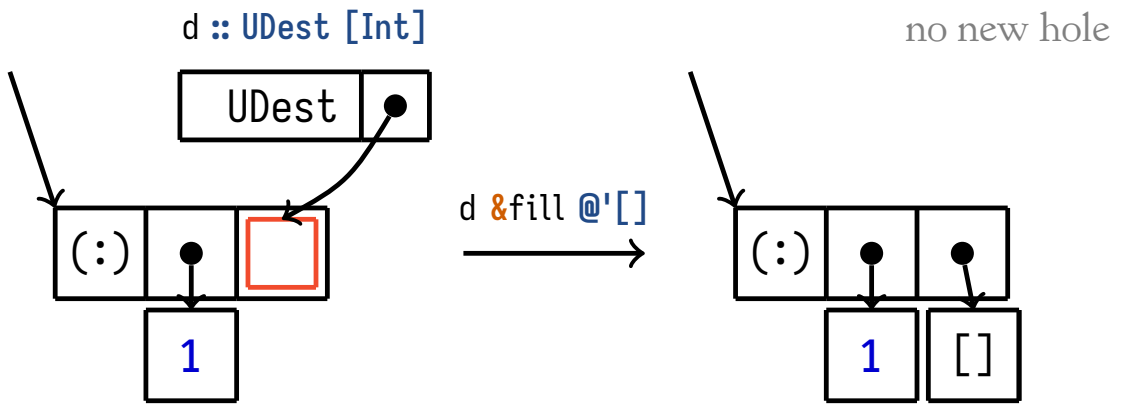

Figure 3.7: Memory behavior of `fill @'[] :: UDest [t] ⊸ ()`

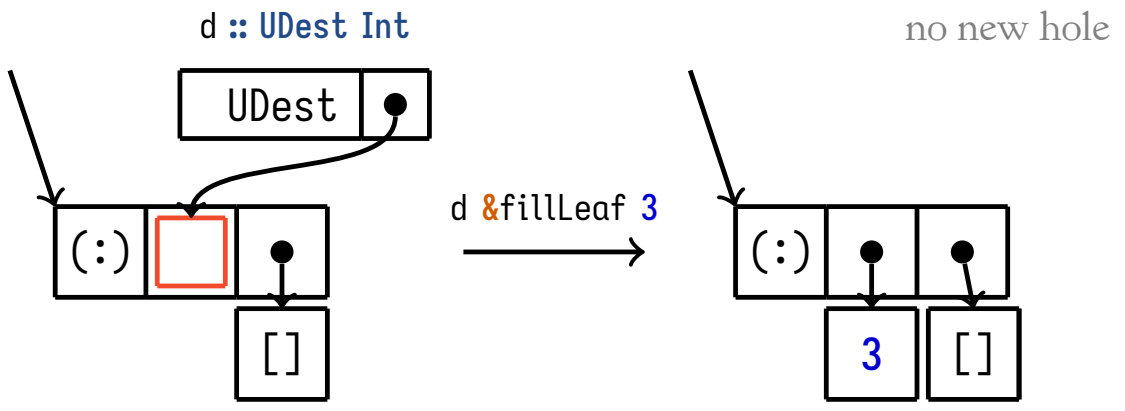

Figure 3.8: Memory behavior of `fillLeaf :: t → UDest [t] ⊸ ()`

The output of the `fill` function is composed of one destination of matching type for each of the fields of the specified constructor; in Haskell this is represented by the type `UDestsOf lCtor t`. `UDestsOf` is a type family (i.e. a function from types to types) whose role is to map a constructor (lifted as a type) to the type of destinations for its fields. For example, `UDestsOf '[] [t] = ()` and `UDestsOf '(:) [t] = (UDest t, UDest [t])`. More generally, the `UDestsOf` typeclass reflects the duality between the types of fields of a constructor and the ones of destinations for a hollow constructor, as first evoked in Section 2.2.1.

## 3.5 Compact regions: a safe space for implementing DPS Haskell

Having incomplete structures in the memory inherently introduces a lot of tension with both the garbage collector and the compiler. Indeed, the garbage collector of GHC assumes that every heap object it traverses is well-formed, whereas UAmpar structures are absolutely ill-formed: they contain uninitialized pointers, which the GC should absolutely not follow. Also, the compiler and GC can make some optimizations because they assume that every object is immutable, while DPS programming breaks that guarantee by mutating constructors after they have been allocated (albeit only one update can happen).

Consequently, we looked for an alternative memory management scheme, mostly independent from the garbage collector, that would let us implement the DPS Haskell API without having to change the garbage collector internals or the memory representation used by GHC.

### 3.5.1 Compact regions

*Compact regions* from Yang et al. [2015] are special memory *arenas* for Haskell. A compact region represents a memory area in the Haskell heap that is almost fully independent from the GC and the rest of the garbage-collected heap. For the GC, each compact region is seen as a single heap object with a single lifetime. The GC can efficiently check whether there is at least one pointer in the garbage-collected heap that points into the region, and while this is the case, the region is kept alive. When this condition is no longer matched, the whole region is discarded. The result is that the GC will not traverse any node from the region: it is treated as one opaque block (even though it is actually implemented as a chain of blocks of the same size, this does not change the principle). Also, compact regions are immobile in memory; the GC will not move them, so a destination to a hole in the compact region can just be implemented as a raw pointer (type `Addr#` in Haskell): `data UDest r t = UDest Addr#`, as we have the guarantee that the UAmpar containing the pointed hole will not move.

By using compact regions to implement DPS programming, we completely elude the concerns of tension between the garbage collector and UAmpar structures that we just mentioned above. Conversely, our destination-passing API also provides a new and powerful way to use compact regions.

In exchange for this implementation decision, we get two extra restrictions. First, every structure in a region must be in a fully-evaluated form. This is contrasting with the usual Haskell evaluation scheme, where everything is lazy by default. Consequently, as regions are strict, any heap object that is copied to a region is first forced into normal form (i.e. it is being fully evaluated). This might not always be a win, sometimes laziness is preferable for better performance.

Secondly, data in a region cannot contain pointers to the garbage-collected heap, or pointers to other regions: it must be self-contained. This forces us to slightly modify the API, to add a phantom type parameter `r` that tags each object with the identifier of the region it belongs to so that our API remains memory-safe, without runtime checks. There are two related consequences: first, when a value from the garbage-collected heap is used as an argument to `fillLeaf`, it has to be fully evaluated and copied into the region instead of making just a pointer update; and secondly, `fillComp` can only plug together two `UAmpar`s that come from the same region.

A typeclass `Region r` is also needed to carry around the details about a region that are required for the implementation. This typeclass has a single method `reflect`, not available to the user, that returns the `RegionInfo` structure associated to identifier `r`.

The `inRegion` function is the new addition to the modified API presented in Listing 3.5. It receives an expression of arbitrary type in which `r` must be a free type variable. Internally, when used, it spawns a new compact region and a rigid type variable $\underline{r}$ (behaving like a fresh, concrete, opaque type) and uses the `reflection` library to provide an instance of `Region` $\underline{r}$ on-the-fly that links $\underline{r}$ and the `RegionInfo` for the new region (through the `reflect` function), and then substitute `r` with $\underline{r}$ when evaluating the body of `inRegion`, so that any call to `reflect @r` becomes `reflect @`$\underline{r}$ and returns the right `RegionInfo`. Actually, `inRegion` is a form of scope function (although this time, it is not linear).

```
data Token
dup :: Token ⊸ (Token, Token)
drop :: Token ⊸ ()
withToken :: ∀ t. (Token ⊸ Ur t) ⊸ Ur t

type Region r :: Constraint
inRegion :: ∀ t. (∀ r. Region r ⇒ t) ⊸ t

data UAmpar r s t
newUAmpar :: ∀ r s. Region r ⇒ Token ⊸ UAmpar s (UDest r s)
tokenBesides :: ∀ r s t. Region r ⇒ UAmpar r s t ⊸ (UAmpar r s t, Token)
toUAmpar :: ∀ r s. Region r ⇒ Token ⊸ s → UAmpar r s ()
fromUAmpar  :: ∀ r s t. Region r ⇒ UAmpar r s (Ur t) ⊸ Ur (s, t)
fromUAmpar' :: ∀ r s. Region r ⇒ UAmpar r s () ⊸ Ur s
updWith :: ∀ r s t u. Region r ⇒ UAmpar r s t ⊸ (t ⊸ u) ⊸ UAmpar r s u

data UDest r t
type family UDestsOf lCtor r t  -- returns dests associated to fields of constructor
fill :: ∀ lCtor r t. Region r ⇒ UDest r t ⊸ UDestsOf lCtor r t
fillComp :: ∀ r s t. Region r ⇒ UAmpar r s t ⊸ UDest r s ⊸ t
fillLeaf :: ∀ r t. Region r ⇒ t → UDest r t ⊸ ()
```

Listing 3.5: DPS Haskell API for compact regions

### 3.5.2 Practical DPS programming in compact regions

We will now study an example, related to implementation concerns, for which DPS programming really shines (and for which compact regions are a really nice fit).

In client-server applications, the following pattern is very frequent: the server receives a request from a client with a serialized payload, the server then deserializes the payload, runs some code, and responds to the request. Most often, the deserialized payload is kept alive for the entirety of the request handling. In a garbage collected language, there's a real cost to this: the garbage collector (GC) will traverse the deserialized payload again and again, although we know that all its internal pointers are alive for the duration of the request.

Instead, we would rather consider the deserialized payload as a single heap object, which does not need to be traversed, and can be freed as a block. Compact regions are, in fact, a very suitable tool for this job, as the GC never follows pointers into a compact region and consider each region as a having the same lifetime for all its contents.

If we use compact regions as-is, with the API provided with GHC, we would first deserialize the payload normally, in the GC heap, then copy it into a compact region and then only keep a reference to the copy. This way, internal pointers of the region copy will never be followed by the GC, and that copy will be collected as a whole later on, whereas the original one in the GC heap will be collected immediately.

However, we are still allocating two copies of the deserialized payload (once in the GC heap, once in the region during the copy). This is wasteful, it would be much better to allocate directly in the region. Fortunately, with our implementation of DPS Haskell with compact regions, we now have a much better way to allocate and build structures directly into these compact regions!

```haskell
1  parseSList ∷ ByteString → Int → [SExpr] → Either Error SExpr
2  parseSList bs i acc = case bs !? i of
3    Nothing → Left (UnexpectedEOFSList i)
4    Just x → if
5      | x == ')' → Right (SList i (reverse acc))
6      | isSpace x → parseSList bs (i + 1) acc
7      | otherwise → case parseSExpr bs i of
8          Left err → Left err
9          Right child → parseSList bs (endPos child + 1) (child : acc)
```

Listing 3.6: Implementation of the S-expression parser in Haskell (no destinations)

```haskell
parseSListDPS ∷ ByteString → Int → UDest [SExpr] ⊸ Either Error Int
parseSListDPS bs i d = case bs !? i of
  Nothing → d &fill @'[] ⨾ Left (UnexpectedEOFSList i)
  Just x → if
    | x == ')' → d &fill @'[] ⨾ Right i
    | isSpace x → parseSListDPS bs (i + 1) d
    | otherwise →
        case d &fill @'(:) of
          (dh, dt) → case parseSExprDps bs i dh of
            Left err → dt &fill @'[] ⨾ Left err
            Right endPos → parseSListDPS bs (endPos + 1) dt
```

Listing 3.7: Implementation of the S-expression parser in DPS Haskell

Let's see how using destinations and compact regions for a parser of S-expressions (representing our request payload) can lead to greater performance. S-expressions are parenthesized lists whose elements are separated by spaces. These elements can be of several types: int, string, symbol (a textual token with no quotes around it), or a list of other S-expressions.

Parsing an S-expression can be done naively with mutually recursive functions:

- `parseSExpr` scans the next character, and either dispatches to `parseSList` if it encounters an opening parenthesis, or to `parseSString` if it encounters an opening quote, or eventually parses the string into a number or symbol;

- `parseSList` calls `parseSExpr` to parse the next token, and then calls itself again until reaching a closing parenthesis, accumulating the parsed elements along the way.

Only the implementation of `parseSList` will be presented here as it is enough for our purpose, but the full implementation of both the naive and destination-based versions of the whole parser can be found in `src/Compact/Pure/SExpr.hs` of ⟦Bagrel, 2023b⟧.

The implementation presented in Listing 3.6 is quite standard: the accumulator `acc` collects the nodes that are returned by `parseSExpr` in the reverse order (because it's the natural building order for a linked list without destinations). When the end of the list is reached (line 5), the accumulator is reversed, wrapped in the `SList` constructor, and returned. We also store, in the parsed result, the index `i` corresponding to the end position of the structure in the input string.

We will see that destinations can bring very significative performance gains with only very little stylistic changes in the code. Accumulators of tail-recursive functions just have to be changed into destinations. Instead of writing elements into a list that will be reversed at the end as we did before, the program in the destination style will directly write the elements into their final location and in the right order!

Code for `parseSListDPS` is presented in Listing 3.7. Let's see what changed compared to the naive implementation:

- even for error cases, we are forced to consume the destination that we receive as an argument (to stay linear), hence we write some sensible default data to it (see line 3);
- the `SExpr` value resulting from `parseSExprDps` is not collected by `parseSListDPS` but instead written directly into its final location by `parseSExprDps` through the passing and filling of destination `dh` (see line 9);
- adding an element of type `SExpr` to the accumulator `[SExpr]` is replaced with writing a new cons cell with `fill @'(:)` into the hole represented by the `UDest [SExpr]`, writing an element to the new *head* destination, and then doing a recursive call with the new *tail* destination passed as an argument (which has type `UDest [SExpr]` again);
- instead of reversing and returning the accumulator at the end of the processing, it is enough to complete the list by writing a nil element to the tail destination (with `fill @'[]`, see line 5), as the list has been built in a top-down approach;
- ultimately, DPS functions just return the offset of the next character to read instead of returning a parsed value (as the parsed value is written directly into the destination received as a parameter).

Thanks to that new implementation which is barely longer (in terms of lines of code) than the naive one, the program runs almost twice as fast, mostly because garbage-collection time goes to almost zero. The detailed benchmark is available in Section 3.6.

We see here that compact regions and destination-passing style are quite symbiotic; compact regions makes the DPS Haskell API easy to implement, and destination-passing style makes compact regions more efficient and flexible to use.

### 3.5.3 Memory representation of `UAmpar`s in Compact regions

As we detailed in Sections 2.6.1 and 3.3.1, we want `UAmpar r s t` to contain a value of type `s` and one of type `t`, and let the value of type `s` free when the one of type `t` has been fully consumed (or linearly transformed into `Ur t'`).

We also know that `newUAmpar` returns an `UAmpar r s (UDest s)`: there is nothing more here than an empty memory cell that will host the root of the future structure of type `s`, which the associated destination of type `UDest s` points to, as presented in Figure 3.2. As we said earlier, whatever goes in the destination is exactly what will be retrieved in the `s` side.

The naive and most direct idea is to represent `UAmpar r s t` in memory as a pair-like constructor with one field of type `s` and one of type `t`, with the first field being there to host the root of the structure being built (as in Figure 3.2). However, the root hole, that will later host the structure of type `s`, might contain garbage data, so the GC must be prevented to read it and follow uninitialized pointers (otherwise we risk a segmentation fault). In our compact region implementation, we can prevent this issue by having the root hole live inside the compact region (so that the GC does not look into it). But we also want the `UAmpar` constructor to live in the garbage-collected heap so that it can sometimes be optimized away by the compiler, and always deallocated as soon as possible. These two requirements are, apparently, incompatible, as the root hole corresponds to the first field of the `UAmpar` constructor in this representation.

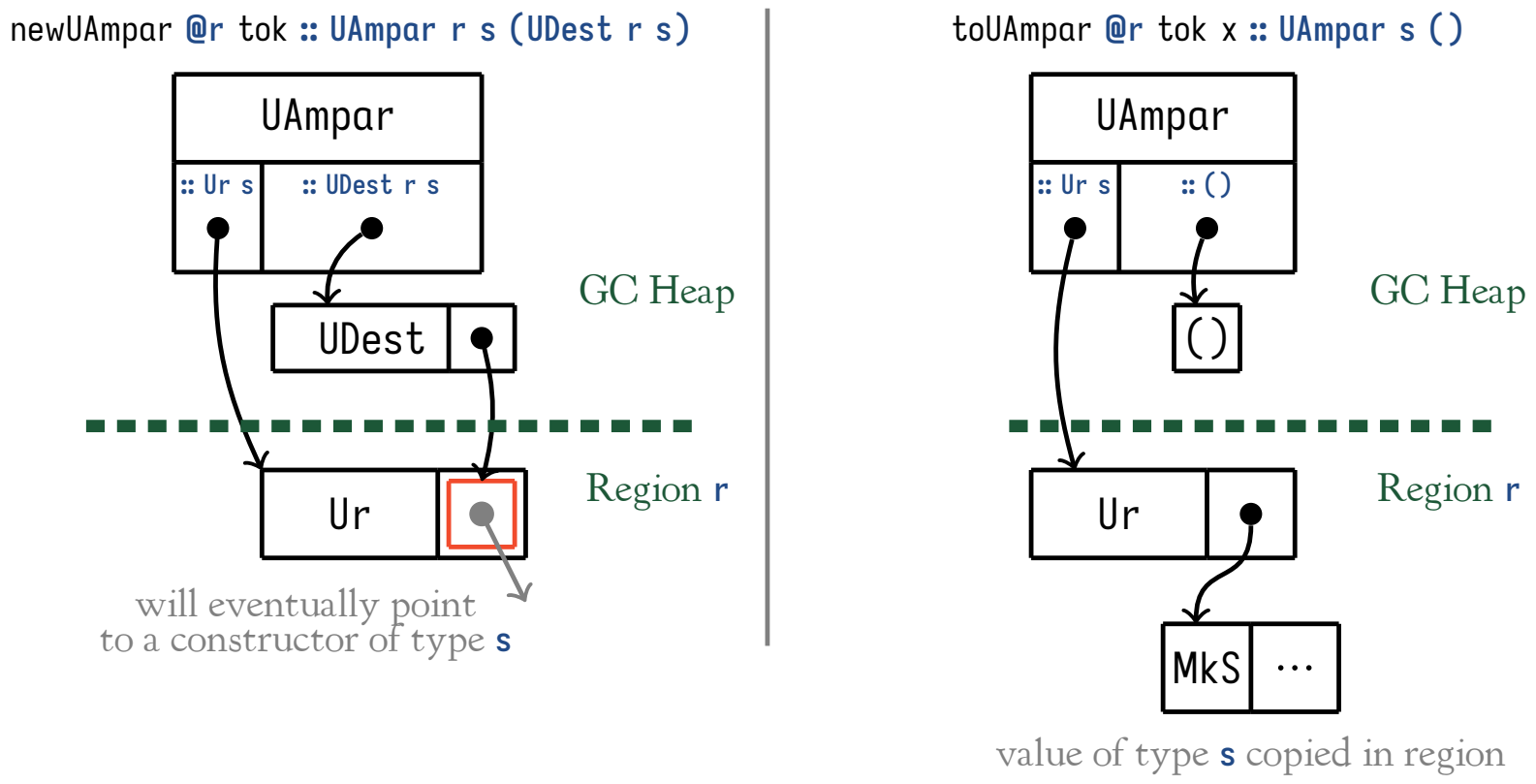

Figure 3.9: Representation of `UAmpar r` using `Ur` (not chosen)

Given that UAmpars hold unrestricted data, one potential workaround is to represent `UAmpar r s t` in memory as a constructor with one field of type `Ur s` and one of type `t` (instead of one of type `s` and one of type `t` previously). It means we need an intermediary `Ur` constructor between the UAmpar wrapper and the structure of type `s` being built, and we can decide to allocate this `Ur` constructor inside the region, with its field of type `s`, also inside the region, serving as the root hole. With this approach, illustrated in Figure 3.9, we have a root hole that will not be read by the GC, while the `UAmpar` wrapper can live in the GC heap without issues and be discarded as soon as possible, as we wanted! We also get a very efficient implementation of `fromUAmpar'`, as the first field of a completed UAmpar is directly the value of type `Ur s` that `fromUAmpar'` should return (it does not help for `fromUAmpar` though). The downside is that every UAmpar will now allocate a few words in the region (to host the `Ur` constructor[24]) that will not be collected by the GC for a long time even if the parent `UAmpar` is collected. In particular, this makes `toUAmpar` quite inefficient memory-wise, as it will have to allocate a `Ur` wrapper in the region that is useless for already complete structures (because already complete structures do not need to have a root hole living in the compact region, as they do not have holes at all).

Another option, that we decided to go for in our real implementation, is to allocate an object inside the compact region to host the root hole of the structure *only* for truly incomplete structures created with `newUAmpar`. For already complete structures that are turned into an `UAmpar` with `toUAmpar`, we skip this allocation; but we preserve the same underlying types for the fields of the UAmpar in both cases (`newUAmpar` and `toUAmpar`). This is made possible by the use of the special indirection object (`stg_IND` in GHC codebase) in the UAmpar returned by `newUAmpar`. A `stg_IND` object is a kind of one-field constructor that is considered to have type `s` although its field also has type `s`.

We illustrate this in Figure 3.10:

- in the pair-like structure returned by `newUAmpar`, the `s` side points to an indirection object, which is allocated in the region, and serves as the root hole for the incomplete structure (because it is in the compact region, the GC will not follow garbage pointers that it might contain at the moment);
- in the pair-like structure returned by `toUAmpar`, the `s` side directly points to the object of type `s` that has been copied to the region, without indirection.

[24]One might ask why `Ur` cannot be a zero-cost `newtype` wrapper. First, if `Ur` were a `newtype`, it would not work for what we are trying to achieve here, as we are relying on the fact that it creates an indirection in memory. Secondly, and more importantly, there are semantics motivations, detailed in ⟦Spiwack, 2024⟧.

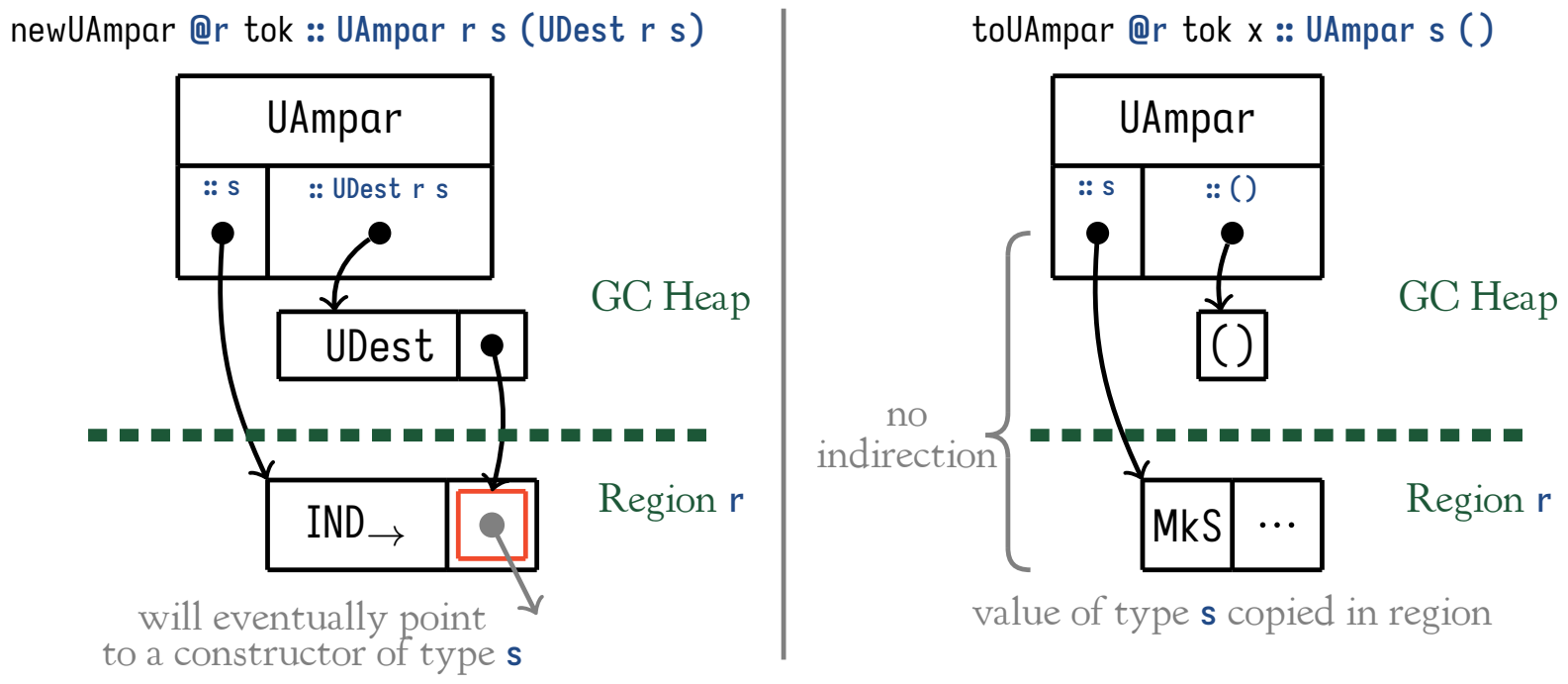

Figure 3.10: Representation of `UAmpar r` using indirections

Compared to the previous solution, this one is more efficient when using `toUAmpar` (as no long-lived garbage is produced), but does no longer give efficiency benefits for `fromUAmpar'` (as we no longer produce an `Ur` object). This is the solution we with with in the original artifact ⟦Bagrel, 2023b⟧.

In ⟦Bagrel, 2023b⟧, we thus have the following private definitions[25] for our API types:

```
data Token = Token  -- same representation as ()
data UAmpar r s t = UAmpar s t  -- same representation as (s, t),
                                -- with sometimes an indirection on the s field
data UDest r t = UDest Addr#  -- boxed Addr#, same representation as Ptr t
```

### 3.5.4 Deriving fill for all constructors with Generics

In $\lambda_d$, it was quite easy to have a dedicated destination-filling function for each corresponding data constructor. Indeed, we only had to account for a finite number of them (one for unit, two for sums, one for pair, and one for exponential; with the one for function being quite optional).

In Haskell however, we want to be able to do destination passing with arbitrary data types, even user-defined ones, so it would not be feasible to implement, manually, one filling function per possible data constructor. So instead, we have to generalize all filling functions into a polymorphic `fill` function that is able to work for any data constructor.

To this extent, we create a typeclass `Fill lCtor t`, and we define `fill` as the only method of this class. Having such a typeclass lets us tie the behaviour of `fill` to the type parameter `lCtor`. The idea is that `lCtor` represents the type-level name of the data constructor we want to use `fill` with. The syntax at call site is of the form `fill @'Ctor`: we lift the constructor `Ctor` into a type-level name with the `'` operator, then make a type application with `@`. We recall that the `fill` function should plug a new hollow constructor of the specified variant into the hole of a structure whose corresponding destination is received as an argument.

Let's now see how we can use the type-level name of the constructor `lCtor` and its base type `t` to obtain all the information we need to write the concrete implementation of the corresponding `fill` function. The first thing we need to know is the shape of the constructor and more precisely the number and types of its fields, as every data constructor for a linked data type can be seen as an $n$-ary product constructor. So we will leverage `GHC.Generics` to find the required information.

[25] In the sense that they are opaque for the consumer of the API.

`GHC.Generics` is a built-in Haskell library that provides compile-time inspection of a type definition through the `Generic` typeclass: list of constructors, their fields, memory representation, etc. And that typeclass can be derived automatically for (almost) any type!

Here's, for example, the `Generic` representation of `Maybe t`:

```
repl> :k! Rep (Maybe a) () -- display the Generic representation of Maybe a
M1 D (MetaData "Maybe" "GHC.Maybe" "base" False) (
  M1 C (MetaCons "Nothing" PrefixI False) U1
  :+: M1 C (MetaCons "Just" PrefixI False) (M1 S [...] (K1 R t)))
```

We see that there are two different constructors (indicated by `M1 C ...` lines): `Nothing` has zero fields (indicated by `U1`) and `Just` has one field of type `t` (indicated by `K1 R t`).

As illustrated with the example above, the generic representation of a type contains most of what we want to know about its constructors. So, with type-level programming techniques[26], in particular with type families, we can extract the parts of this generic representation that are relevant to us (as type-level expressions), and refer to them in the head of our typeclass' instances, to have several specialized implementations of `fill`. For instance, in ⟦Bagrel, 2023b⟧, we have one specialized implementation of `fill` for every possible number of fields in `lCtor`, from 0 to 7[27]. The main reason is that `fill` should return a tuple containing as many new destinations as there are fields in the the chosen constructor, and at the moment, there is no way to abstract over the length of a tuple in Haskell, so we need to hardcode each possible case.

The resolution of the `UDestsOf lCtor t` type family into a concrete tuple type of destinations is made similarly, by type-level expressions based on the generic representation of the base type `t` which `lCtor` belongs to.

At this point we still miss a major ingredient though: the internal Haskell machinery to allocate the proper hollow constructor object in the compact region.

### 3.5.5 Changes to GHC internals and its RTS

We will see here how to allocate a hollow constructor, that is, a hollow heap object for a given constructor, but let's first take a detour to give more context about the internals of the compiler.

Haskell's runtime system (RTS) is written in a mix of C and C-- ⟦Ramsey, Jones, and Lindig, 2005⟧. The RTS has many roles, among which managing threads, organizing garbage collection or managing compact regions. It also defines various primitive operations, named *external primops*, that expose the RTS capabilities as normal functions that can be used in Haskell code. Despite all its responsibilities, the RTS is surprisingly not responsible for the allocation of normal constructors (built in the garbage-collected heap). One reason is that it does not have all the information needed to build a constructor heap object, namely, the info table associated to the constructor.

In Haskell, a heap object is a piece of data in memory, that corresponds to a given instance of a data constructor, or a (partially-applied) function. The info table is what defines both the layout and behavior of a heap object. All heap objects representing the same data constructor (let's say `Just`) have the same info table, which acts as the identity card for this constructor, even when the associated types are different (e.g. `Just x :: Maybe Int` and `Just y :: Maybe Bool` share the same info table). In the compilation process,

[26] see `src/Compact/Pure/Internal.hs:418` in ⟦Bagrel, 2023b⟧

[27] We could go higher, 7 is just an arbitrary limit for the number of fields that is rather common in standard Haskell libraries.

during code generation, all instances of the `Just` constructor will hold the label `Just_con_info`, that is later resolved by the linker into an actual pointer to the shared info table, that all corresponding heap objects for this constructor will carry.

On the other hand, the RTS is a static piece of code that is compiled once when GHC is built, and is then linked into every executable built by GHC. Only when a program is launched does the RTS run (to oversee the program execution), so the RTS has no direct way to access the information that had been emitted during the compilation of the program. In particular, it has no way to inspect the info table labels have long been replaced by actual pointers (which are indistinguishable in memory from any other data). But on the other hand, the RTS is the one which knows how to allocate space inside a compact region for our new hollow constructor.

As a result, we need a way to pass information, or more precisely, the info table pointer, from compile time to the runtime. Consequently, we manage the creation of a new hollow constructor in a compact region through two new primitives:

- one *internal primop* which resolves, at compile time, a data constructor into the (static) value representing its info table pointer.
- one *external primop*, in the RTS, which allocates space inside a compact region for a hollow constructor, and which sets the info table pointer of the hollow constructor to a value received as an argument, that the developer can provide using the internal primop.

All the alterations to GHC that will be showed here are available in full form in ⟦Bagrel, 2023a⟧.

**Internal primop: obtain the info table pointer of a constructor as a static value** Internal primops are macros that are transformed at compile time into C-- code. So, for our new internal primop, in charge of obtaining the info table label (later resolved into a pointer) of a constructor, we must indicate our choice of constructor as a type-level argument so that it can be inspected at compile-time (as, of course, term-level values cannot be inspected easily at compile time). For that, we use the *tick* operator `'` in front of the constructor name, which promotes an object from the term-level namespace (in which data constructors live) to type-level, to be used where a type would normally be expected[28].

Now, we need the internal primop to inspect the type parameter representing our choice of constructor. Usually, in Haskell, we use a parameter of type `Proxy t` (the unit type with a phantom type parameter `t`) as a carrier for a type-level input that we want to pass to a function. Unfortunately, primops are special functions, and due to a quirk of the compiler, they no longer have access to the type of their arguments at the stage in which they are resolved, so designing our primop with a `Proxy lCtor` parameter would not let it read the supplied `lCtor`. However primops can—surprisingly—access their return type. Consequently, we design a custom return type `InfoPtrPlaceholder#` for our primop that carries the output value of the primop (that is to say, an info table pointer); but we also add a phantom type parameter `lCtor` to `InfoPtrPlaceholder#` that has no relation to the underlying value, but just serves as a carrier for our type-level input representing the chosen constructor.

The concrete implementation is summarized in Listing 3.8 (we omit the roughly one hundred lines of boilerplate code scattered across other parts of the GHC codebase that handle primop registration, interface declaration, etc.). This implementation goes a bit further than what we just described above, as `reifyInfoPtr#` can obtain the info table pointer of data constructors (as announced), but also the info table pointer of special compiler objects like the indirection object we used in Section 3.5.3. The primop pattern-matches on the type `resTy` of its return value (which should be of the form `InfoPtrPlaceholder# lCtorOrSym`).

[28]We refer the reader to ⟦Yorgey et al., 2012⟧ for more information about promoting term-level objects to type-level ones in Haskell.

```
case primop of
  [...]
  ReifyInfoPtrOp -> \_ -> -- we do not care about the function argument (# #)
    opIntoRegsTy $ \[res] resTy -> emitAssign (CmmLocal res) $ case resTy of
      -- when the phantom type parameter is a lifted data constructor, extracts it as a DataCon
      TyConApp _infoPtrPlaceholderTyCon [_typeParamKind, TyConApp tyCon _]
        | Just dataCon <- isPromotedDataCon_maybe tyCon ->
          CmmLit (CmmLabel (
            mkConInfoTableLabel (dataConName dataCon) DefinitionSite))
      -- when the phantom type parameter is a Symbol, we extracts the symbol value in 'sym'
      TyConApp _infoPtrPlaceholderTyCon [_typeParamKind, LitTy (StrTyLit sym)] ->
          CmmLit (CmmLabel (
            mkCmmInfoLabel rtsUnitId (fsLit "stg_" `appendFS` sym)))
      _ -> [...] -- error when no pattern matches
```

Listing 3.8: Implementation of `reifyInfoPtr#` in `compiler/GHC/StgToCmm/Prim.hs`

In the case it reads a lifted data constructor (which is the most common case), it resolves the primop call into the label `<ctor>_con_info` which corresponds to the info table pointer of that constructor. In the case `lCtorOrSym` is a symbol (that is, a type-level string literal), it resolves the primop call into the label `stg_<sym>` (e.g. to allocate a `stg_IND` object, as we just mentionned). The returned `InfoPtrPlaceholder# lCtorOrSym` can later be converted back to an `Addr#` using the `unsafeCoerceAddr` function, which despite the name, is not an unsafe coercion when used in this way.

**External primop: allocate a hollow constructor in a region** Our external primop, as part of the RTS, is directly written in C--. Its implementation is presented in Listing 3.9, under the name `stg_compactAddHollowzh`; but in Haskell code we access it through the name `compactAddHollow#`. The primop consists mostly in a glorified call to the `ALLOCATE` macro defined in the `Compact.cmm` file, which tries to do a pointer-bumping allocation in the current block of the compact region if there is enough space, and otherwise adds a new block to the region. Then, it takes the info table pointer of the constructor to allocate, received as its second parameter `W_ info` (as it cannot obtain this information itself), and write it to the first word of the heap object using `SET_HDR`. Then the pointer to the newly allocated hollow constructor is set as the return value.

**Combining both primops to obtain a new hollow constructor** It's time to show the two primops in action. Here is how we could allocate, for example, a hollow `Just` in a compact region:

```
hollowJust :: Maybe a = compactAddHollow#
  region
  (unsafeCoerceAddr (reifyInfoPtr# (# #) :: InfoPtrPlaceholder# 'Just ))
```

It's important to note that the expression `reifyInfoPtr# (# #)` will be replaced at compile-time by a static value representing the info table pointer of `Just`; while the call to `compactAddHollow#` behaves as a normal function call, and will only be executed when the program runs.

```
// compactAddHollow#
//   :: Compact# → Addr# → State# RealWorld → (# State# RealWorld, a #)
stg_compactAddHollowzh(P_ compact, W_ info) {
    W_ pp, ptrs, nptrs, size, tag, hp;
    P_ to, p; p = NULL;  // p is not actually used by ALLOCATE macro
    again: MAYBE_GC(again); STK_CHK_GEN();

    pp = compact + SIZEOF_StgHeader + OFFSET_StgCompactNFData_result;
    ptrs  = TO_W_(%INFO_PTRS(%STD_INFO(info)));
    nptrs  = TO_W_(%INFO_NPTRS(%STD_INFO(info)));
    size = BYTES_TO_WDS(SIZEOF_StgHeader) + ptrs + nptrs;

    ALLOCATE(compact, size, p, to, tag);
    P_[pp] = to;
    SET_HDR(to, info, CCS_SYSTEM);
  #if defined(DEBUG)
    ccall verifyCompact(compact);
  #endif
    return (P_[pp]);
}
```

Listing 3.9: Implementation of `compactAddHollow#` in `rts/Compact.cmm`

**Built-in type family to go from a lifted constructor to the associated symbol** The internal primop `reifyInfoPtr#` that we just introduced uses a constructor lifted at type level as an input. So the `fill` function, which uses `reifyInfoPtr#` and `compactAddHollow#` under the hood, should also use a constructor lifted at type level as a way of selecting the hollow constructor that will be added to a structure with holes.

However, `fill` relies on `GHC.Generics` to obtain various informations on the chosen constructor (see Section 3.5.4), and in the `Generic` representation of a type, constructors are identified by their name (as type-level strings), which are distinct type-level entities than the constructors lifted at type level like we used in `reifyInfoPtr#`.

Consequently, we need a way to translate a constructor lifted at type level into the type-level string representing the name of this constructor[29]. This is the role of the type family `LCtorToSymbol` that we also added to GHC. It just inspects its (type-level) parameter representing a constructor, fetches its associated `DataCon` structure, and returns a type-level string (kind `Symbol`) carrying the constructor name, as presented in Listing 3.10.

**Putting it all together, and further evolutions** All these elements combined provide the low-level backbone needed to implement our `fill` function. They are summarized in Figure 3.11. The implementations presented above were designed to minimize changes to GHC itself, based on the idea that the fewer

[29]We perform the translation in this direction because a type-level string alone does not uniquely identify a constructor, as multiple constructors with the same name but different shapes can exist in a codebase. By using a constructor promoted to type level instead, we leverage GHC import and name resolution logic to avoid any ambiguity.

```haskell
matchFamLCtorToSymbol :: [Type] -> Maybe (CoAxiomRule, [Type], Type)
matchFamLCtorToSymbol [kind, ty]
  | TyConApp tyCon _ <- ty, Just dataCon <- isPromotedDataCon_maybe tyCon =
      let symbolLit = (mkStrLitTy . occNameFS . occName . getName $ dataCon)
       in Just (axLCtorToSymbolDef, [kind, ty], symbolLit)
matchFamLCtorToSymbol tys = Nothing

axLCtorToSymbolDef =
  mkBinAxiom "LCtorToSymbolDef" typeLCtorToSymbolTyCon Just
    (\case { TyConApp tyCon _ -> isPromotedDataCon_maybe tyCon ; _ -> Nothing })
    (\_ dataCon -> Just (mkStrLitTy . occNameFS . occName . getName $ dataCon))
```

Listing 3.10: Implementation of `LCtorToSymbol` in `compiler/GHC/Builtin/Types/Literal.hs`

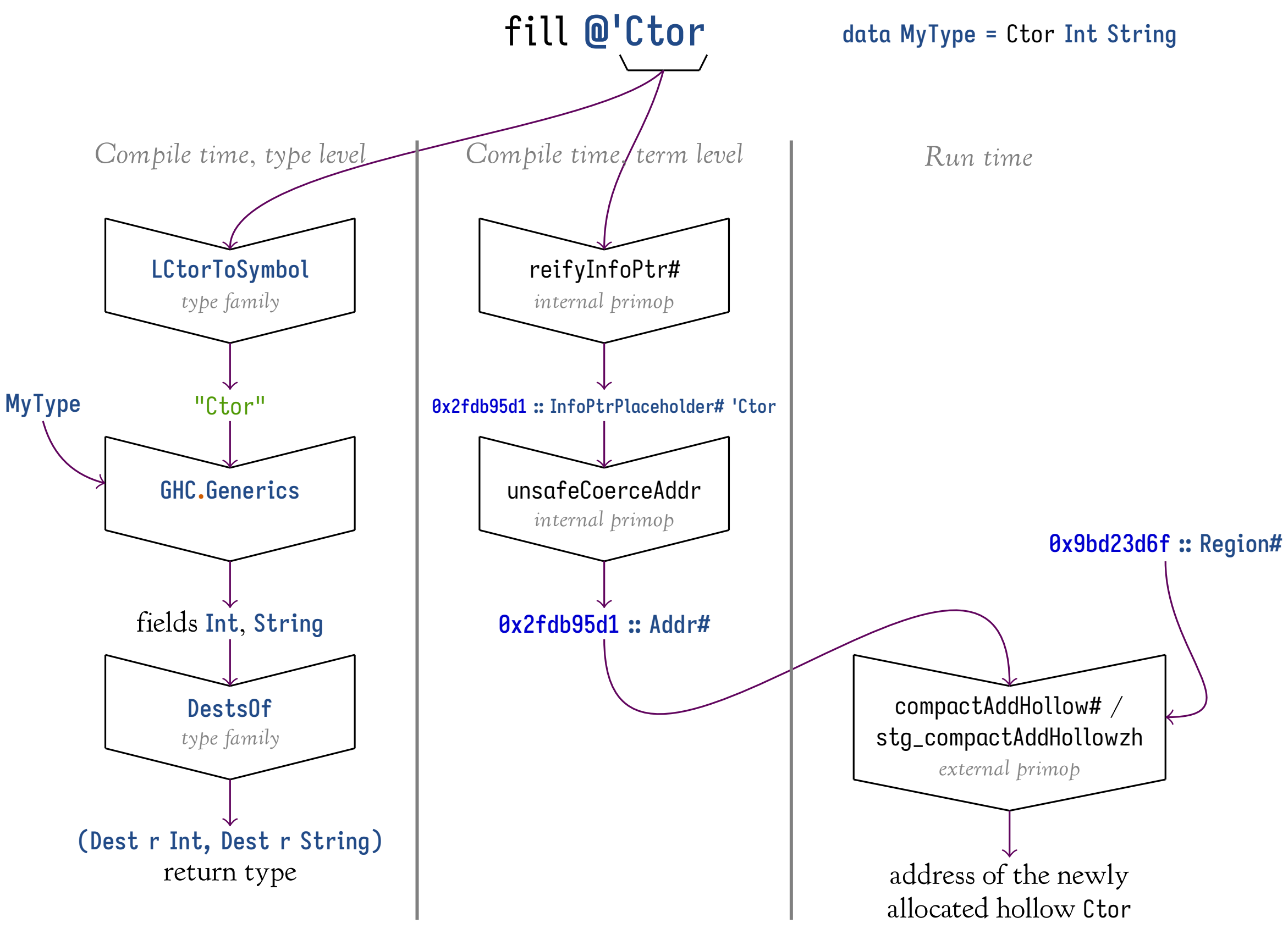

Figure 3.11: Schematic overview of the implementation architecture for `fill`

modifications required, the more realistic it would be for this feature to work well and be accepted into the main version of GHC, especially since destination-passing remains a relatively niche concern for such an industrial-grade compiler.

However, since this initial experimentation [Bagrel, 2023a], a pull request has been opened [Bagrel, 2024b] to gather community input on the topic, and we now appear to be moving toward a slightly more complete set of primitives than the three previously mentioned. In particular, primitive support for `fill` will likely be included in GHC itself (rather than existing solely in library code, as it does currently). At the time of writing, the interface design is still under discussion, and much of the implementation remains to be done (or be adapted from earlier experiments).

## 3.6 Evaluating the performance of DPS Haskell

It's now time to see if our claims of performance are verified with our prototype destination-passing implementation [Bagrel, 2023b].

**Benchmarking methodology** We will compare programs in both usual functional style and DPS style, and see if we can observe differences in terms of speed but also memory efficiency. But first we must be careful in the way we measure time and space usage, especially since Haskell defaults to a lazy evaluation strategy. In particular we need a way to force Haskell to fully evaluate the results of the implementations that we benchmark.

In all DPS implementations, the program output is stored in a compact region (as required). As noted in Section 3.5.1, compact regions enforce strictness; so the structure is automatically fully evaluated. While this may be suboptimal in some real use-cases, it simplifies benchmarking of these implementations at least.

For naive implementations, we must choose a way to force full evaluation. One option is to keep the result in the GC heap and use `Control.DeepSeq.force` to recursively evaluate each thunk. Alternatively, we can copy the result to a compact region using `Data.Compact.compact`, which enforces strictness (through a RTS-defined chunk-forcing procedure that is distinct from `Control.DeepSeq.force`).

Copying into a compact region is only really desirable for a handful of programs, for which the result is held in memory for a long time and used in full (as the unused parts cannot be collected if it is in a compact region). Still, `compact` is sometimes faster than `force`. We therefore benchmark naive implementations primarily with `force`, but also with `compact` (noted with a `.copyCR` suffix) when this appears beneficial—either because storing the result in a compact region is desirable (e.g., in the parser example), or when it yields significantly better performance.

All charts have a relative scale on the Y axis (for both time and memory). The baseline is the time taken and the peak memory used by the method of reference, which is plotted in navy blue. Thus, the absolute scale differs for each input size, with 1.0 corresponding to the reference method's value at that size. For memory charts, all series (*peak*, *allocated*, and *copied*) use the peak memory value of the method of reference as a baseline, so that values across all series at a given size are directly comparable.

**About memory reporting** Accurate memory reporting in Haskell is challenging. When benchmarks are run via `Tasty.Bench`, each is executed repeatedly until time results fall within a target deviation. `Tasty.Bench` also reports three memory metrics: allocated memory, copied volume during garbage collection, and peak memory. Unfortunately, allocated memory does not include allocations in compact regions, making it unsuitable as a baseline. We therefore use peak memory as a baseline, which approximates how RAM hungry a given implementation is.

However, if we launch a full suite of benchmarks with `Tasty.Bench`, then the peak memory value reported can only ever increase across the whole run, and is not reset when benchmarking a different implementation of the suite. We tried to solve this by running the benchmark of each implementation + size separately (instead of launching a full suite), so that each implementation can have a proper peak memory reporting. Yet even then, `Tasty.Bench` always report at least 69 MB of peak memory use, even if the given case uses far less (as we can deduce from the associated allocated and copied values).

Consequently, in addition to the `Tasty` benchmarks, we ran each implementation + size separately, 10 times, with RTS options `+T +s`, to get a more accurate reading of peak memory use. In this benchmarking mode, we observed a minimum of 6 MB of peak memory—accounting for the Haskell runtime and other boilerplate—which is quite better than the 69 MB reported by `Tasty.Bench`.

Still, many programs report exactly 6 MB of peak memory, while allocating between 0 and 3 MB (on average 0.5 MB). This shows that the reported peak memory does not accurately reflect the actual memory used by the implementation itself—it consistently overestimates usage by 3 to 6 MB, typically 5.5 MB. Based on this, we decided to subtract 5.5 MB from all peak memory readings equal to or above 10.5 MB to obtain an adjusted peak memory value. We do not show the data points that have a peak reading below 10.5 MB, as the margin of error, up to 2.5 MB, would represent more than half of the adjusted value.

When we do not have a trusted data point for the peak memory value of the reference method, we instead use an arbitrary origin for that input size (that is computed to preserve the shape of the copied and allocated curves). That way, we can still easily compare the other values with one another.

**Execution environment** We launch the benchmark of each implementation at a given size in isolation, either manually or through `Tasty.Bench`, with RTS options `+T +s`. We do not set any other runtime option; in particuler, we do *not* use multithreading or configure any particular garbage collection option. When launched through `Tasty.Bench`, the tests are repeated until the time results fall within a target deviation of 5 %. When launched manually, we each test is repeated 10 times and we take the average of the results.

All programs are compiled with optimization `-O2`, with a version of GHC 9.12 modified to include the changes initially presented in ⟦Bagrel, 2023a⟧ (which were initially designed for GHC 9.8).

Benchmarks have been conducted on a laptop under Kubuntu 22.04 with Intel i7-1165G7 @ 2.80 GHz CPU, 32 GB RAM, with power supply connected.

### 3.6.1 Concatenating lists and difference lists

We compared three implementations for list concatenation.

The first one, `concatListRight.force`, uses usual linked lists, and has calls to `(++)` nested to the right, giving the most optimal context for list concatenation: it should run in $\mathcal{O}(n)$ time. We only include it as a point of reference, since it is not the most common use-case for concatenation in practice. In real-world programs, list concatenation is often nested to the left, for example when appending lines to a log before eventually flushing it to disk. We chose not to include the version with calls to `(++)` nested to the left, as its time and memory consumption grow so rapidly that it renders the rest of the graph unreadable.

For the two other implementations, the calls to `concat` are nested to the left. The implementation `concatDListFunLeft.force` uses function-backed difference lists, and `concatDListDpsLeft` uses destination-backed ones (from Section 3.3.1). They should still run in $\mathcal{O}(n)$ however, thanks to difference list magic.

We see in Figure 3.12 that the destination-backed difference lists have a slightly reduced memory consumption and lower runtime (about 10 %) compared to the two other implementations, for very large datasets only, in exchange of being quite slower and less memory efficient for all the smaller datasets.

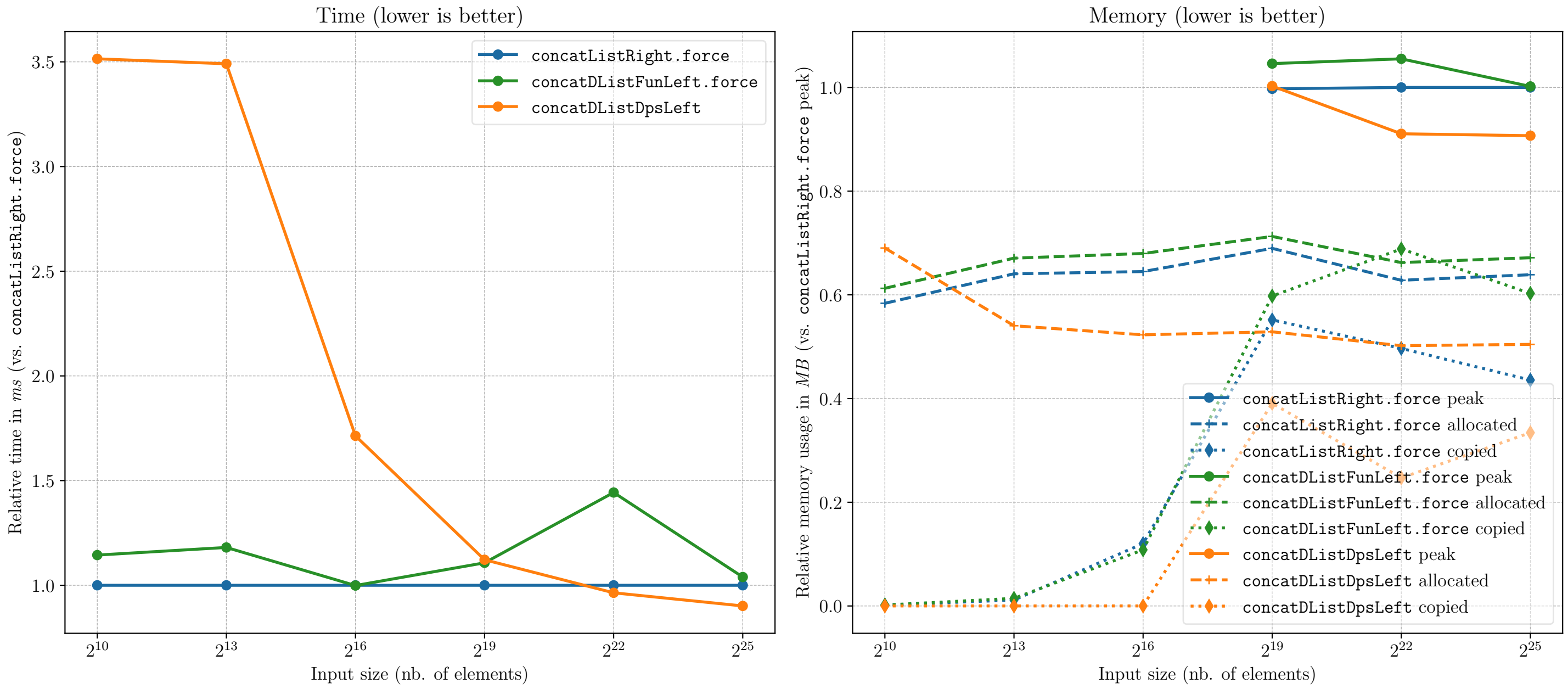

Figure 3.12: Benchmark of iterated concatenations on lists and difference lists

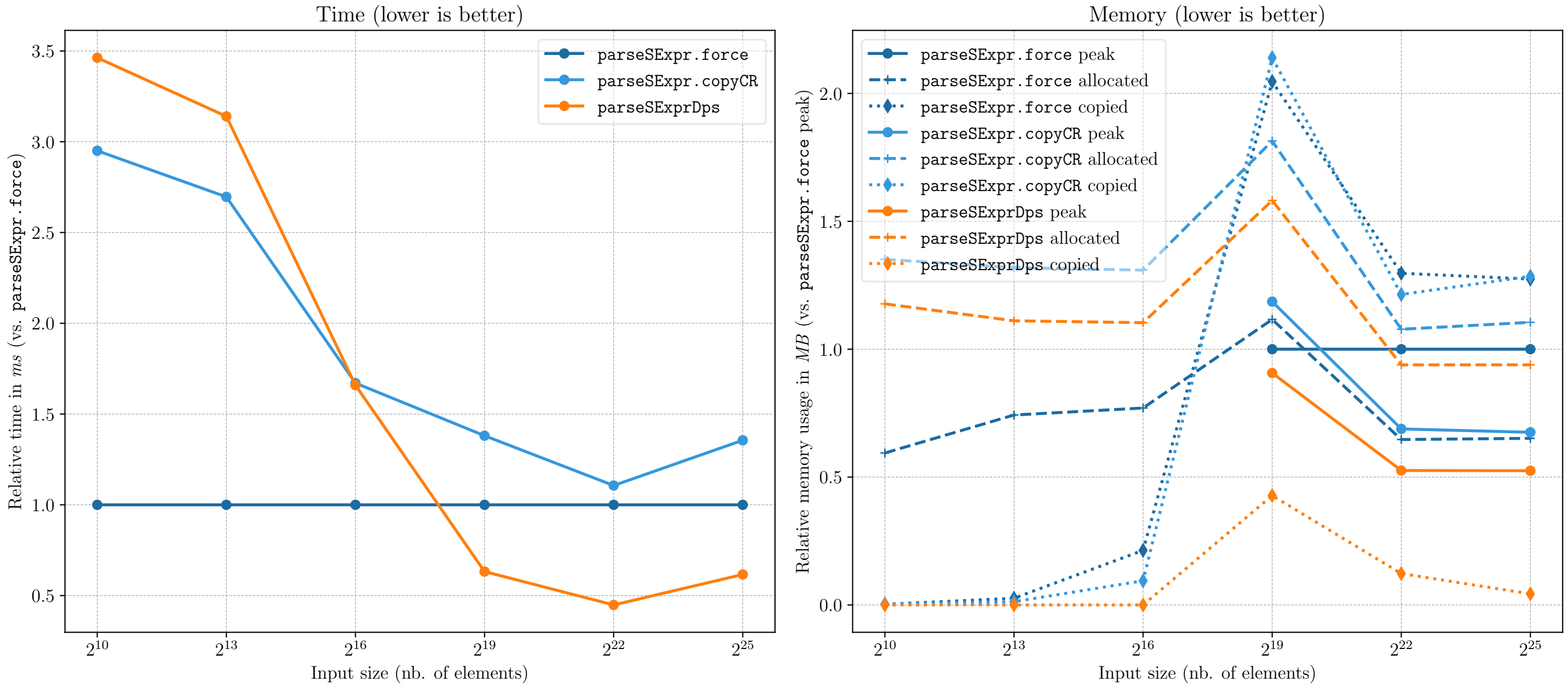

Figure 3.13: Benchmark of S-expression parser

It's a trend that we will see for the rest of the benchmarks: for very large datasets, garbage collection becomes very costly, and the destination-passing implementations shine not really by being faster to create or process data, but rather by avoiding garbage collection on large data trees that are still live. It's much cheaper to collect garbage (e.g. the `Ampar` or `Dest` wrappers in the GC heap) than to preserve non-garbage.

Compared to the benchmarks initially presented in [Bagrel, 2024a], it seems that the performance of destination-based difference lists has improved, through more efficient handling of linear programs by the compiler, albeit the difference with naive implementations still is not substancial enough compared to what we could expect theoretically, especially when other simpler tweaks exist (notably, setting custom GC parameters at program's launch).

### 3.6.2 Breadth-first tree relabeling

For breadth-first tree traversal, we benchmark the destination-passing implementation of Section 3.3.2 against an implementation based on *Phases* applicatives, presented in [Gibbons et al., 2023]—a similarly concise approach to breadth-first tree traversal in a functional setting:

```haskell
import Control.Monad.State.Lazy (runState, state)

data Phases m a where
  Pure :: a -> Phases m a
  Link :: (a -> b -> c) -> m a -> Phases m b -> Phases m c


instance Functor (Phases m) where ...
instance (Applicative m) => Applicative (Phases m) ...

now :: (Applicative m) => m a -> Phases m a
now xs = Link (curry fst) xs (Pure ())

later :: (Applicative m) => Phases m a -> Phases m a
later xs = Link (curry snd) (pure ()) xs

runPhases :: (Applicative m) => Phases m a -> m a
runPhases (Pure x) = pure x
runPhases (Link f xs ys) = pure f <*> xs <*> runPhases ys

traverseBFS :: (Applicative m) => (a -> m b) -> BinTree a -> Phases m (BinTree b)
traverseBFS _ Nil = pure Nil
traverseBFS f (Node x tl tr) =
  pure Node <*> now (f x) <*> later ((traverseBFS f) tl) <*> later ((traverseBFS f) tr)

relabelPh :: BinTree () -> (BinTree Int, Int)
relabelPh tree = runState (runPhases $ traverseBFS (\_ -> state (\s -> (s, s + 1))) tree) 0
```

We see in part Figure 3.14 that the destination-based tree traversal is about five times more efficient, both time-wise and memory-wise, than the implementation from Gibbons et al. [2023]. We must admit though that we did not particularly try to optimize the performance of the *Phases* implementation as much as we did for our DPS framework in general.

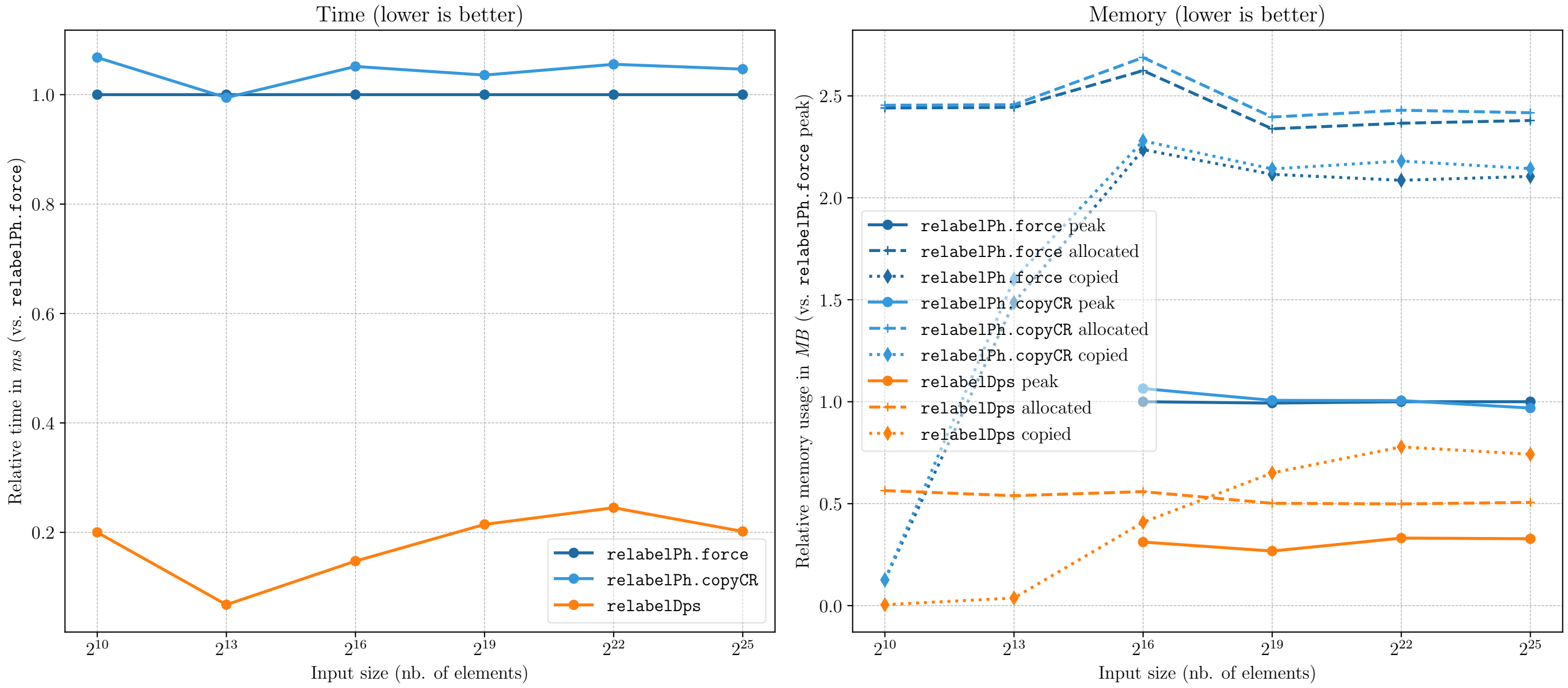

Figure 3.14: Benchmark of breadth-first tree relabeling

This is a good example of a program for which the destination-passing version is almost as natural to write as the usual functional-style one, but performs significantly better even at smaller sizes.

### 3.6.3 Parsing S-expressions

In Figure 3.13, we compare the naive and DPS implementations of the S-expression parser presented in Section 3.5.2. For this particular program, using compact regions is interesting even without destination-passing considerations, as it will reduce the GC load of the application as long as the data is held in memory (which is quite usual with parsed documents). Thus, we benchmark the naive version twice: once in which the results stays in the GC heap and is evaluated with `force` (its name is suffixed with a star), and once to produce a result that we copy in a compact region (`copyCR`).

The DPS implementation starts by being less efficient than the naive versions for small inputs, but gets significantly better than the two other implementations tested as soon as garbage collection kicks in (c.f. copied memory value rising very fast compared to the allocated and peak memory values, around $2^{16}$).

On the dataset of size $2^{19}$ to $2^{25}$, the DPS implementation uses 35-55% less time and 10-45% less memory at its peak than the two naive implementation, despite showing more allocations than `parseSExpr.force`. This indicates that most of the data allocated in the GC heap by the DPS version is wrappers such as `Ampar`s and `Dest`s that just last one generation and thus can be discarded very early by the GC (as intended), without needing to be copied into the next generation, unlike the data nodes allocated by the naive versions that have to be persisted by the GC (by copying them).

It's slightly surprising to see that for this particular example, the `parseSExpr.copyCR` version performs worse than `parseSExpr.force` even for the largest datasets. In a real-world scenario though—where data is processed for some time after having been parsed instead of being discarded immediately—we expect the `copyCR` implementation to get an edge on garbage collection time (and thus, runtime) over the `force` version, but it would probably still be behind the DPS implementation which saves the copy of the result from the GC heap to the compact region.

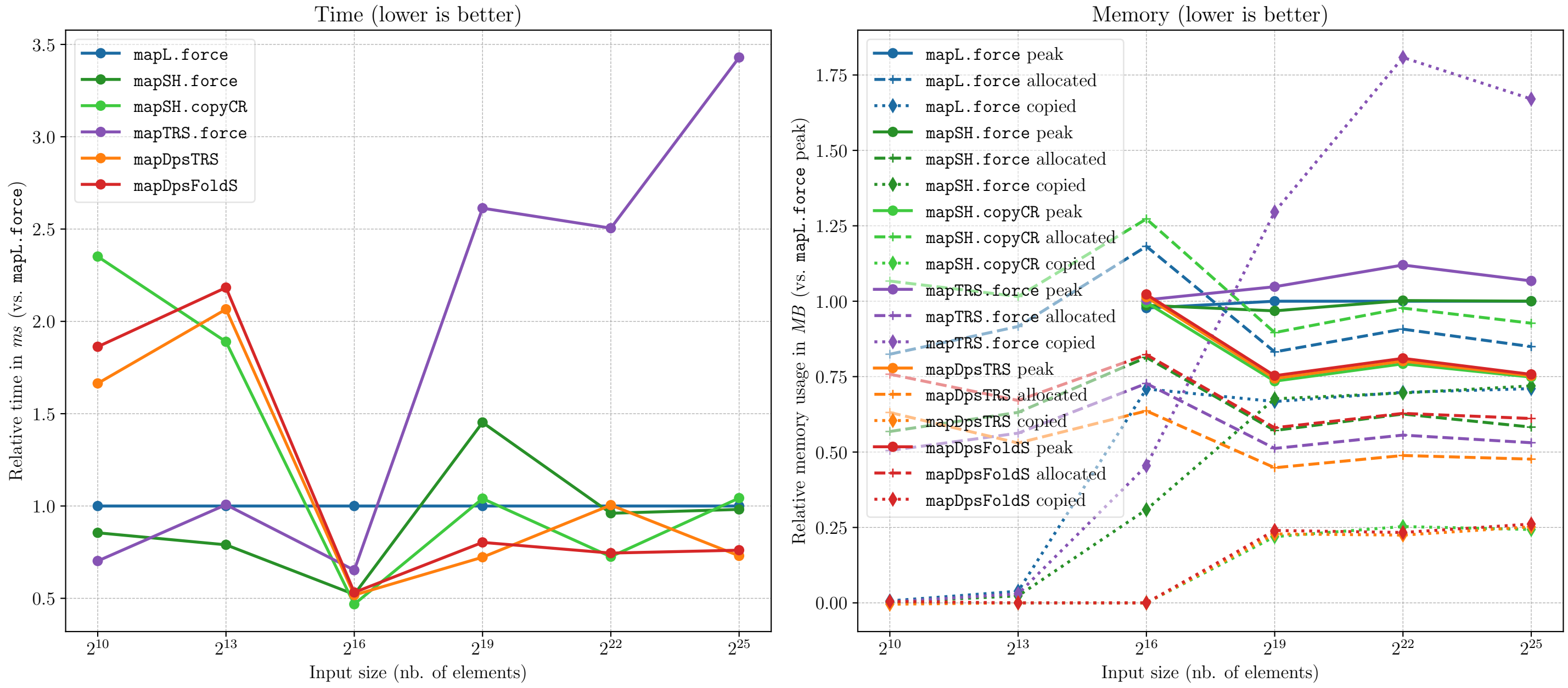

Figure 3.15: Benchmark of map function on list

### 3.6.4 Mapping a function over a list

In a strict functional language such as OCaml, the choice of implementation for the *map* function on lists is crucial as the naive, non tail-recursive one makes the stack grow linearly with the size of the processed list. To alleviate the issue, a strict tail-recursive implementation is possible (`mapTRS`), but it requires an extra $\mathcal{O}(n)$ operation at the end of the processing to reverse the accumulator. In Haskell, the issue is of a lesser importance perhaps, as the lazy-by-default evaluation strategy makes the naive *map* (either `mapL` for full lazy behavior, or `mapSH` for strict head but lazy tail) run efficiently in constant stack space. Still, let see how our DPS versions compare.

With destinations, *map* can be implemented in a tail-recursive fashion too (`mapDpsTRS`), as explained in Section 2.2.2, without requiring the reverse operation that `mapTRS` needs, as the list is built directly in a top-down approach. Interestingly, thanks to destinations, *map* can alternatively be implemented as a left fold (`mapDpsFoldS`), as we also show below! In the following code excerpt, the `!` in front of a variable name denotes a strict binding, so the expression assigned to the variable will be evaluated to weak-head normal form before the body of the `let` is evaluated.

```
append :: UDest [t] ⊸ t → UDest [t]
d `append` x = case d &fill @'(:) of (dh, dt) → dh &fillLeaf x ⨾ dt

mapDpsTRS :: (a → b) → [a] → UDest [b] ⊸ ()
mapDpsTRS _ [] d = d &fill @'[]
mapDpsTRS f (x : xs) d = let !y = f x ; !d' = d `append` y in mapDpsTRS f xs d'

mapDpsFoldS :: (a → b) → [a] → UDest [b] ⊸ ()
mapDpsFoldS f l dl = fill @'[] (foldl' (\d x → let !y = f x in d `append` y) d l)
```

Figure 3.15 presents the results of the different implementations. First, we confirm that the most naive `mapL.force` is performing really well given its simplicity. We also see that the strict functional tail-recursive implementation `mapTRS` is not particularly interesting for a language like Haskell (even in its strict head, lazy tail variant that is not shown in the chart, but got similar results), as it performs worse than all the other implementations both time and memory-wise.

For `mapSH`, the situation is a bit more complex. With the `force` evaluation method, it performs partially better, partially worse than the reference implementation `mapL.force` depending on the input size; but with copy to compact region in guise of deep evaluation, it achieves the smallest possible time for $2^{16}$ and $2^{22}$ sizes, surpassing the destination-based implementations by a few percents, with equally good peak memory use too. We cannot really explain why we get so better results with copy to compact region for this particular implementation though; but it deserved to be included for being the top-performer for a few sizes.

Finally, the destination-based implementations `mapDpsTRS` and `mapDpsFoldS` present both very honorable performance, with 20-45% less time and up to 25% peak memory saving compared to the reference implementation for most datasests starting from $2^{16}$ size. However, as we said before, the performance of DPS Haskell on small lists is still fairly disappointing. Reducing the overhead of destination-passing should probably be a focus of future refinements of the DPS Haskell implementation.

As a side note, Bour, Clément, and Scherer ⟦2021⟧ showed that in OCaml, the equivalent of `mapDpsTRS` is always more performant time-wise (15-50% less time) than the equivalent of `mapTRS`. However, their destination-passing system is used only at the compiler's intermediate language level, which perhaps allow for more aggressive optimizations than what we can achieve at the source level.

## 3.7 Conclusion

Programming with destinations definitely has a place in practical functional programming, as shown by the recent adoption of *Tail Modulo Cons* ⟦Bour, Clément, and Scherer, 2021⟧ in the OCaml compiler. In this chapter, we have shown how we can take most ideas from $\lambda_d$ (Chapter 2) and adapt them to allow safe destination passing in Haskell, without modifying the core principles of this language. We also showed that most of the examples presented in chapter Chapter 2 can be implemented in DPS Haskell with little to no adaptations required, and we also emphasized how our particular implementation of destination passing for Haskell can help to improve the performance of programs in which garbage collection is very costly. To that extent, we rely on the *compact regions* ⟦Yang et al., 2015⟧ memory management scheme for Haskell; but our work also extend the way in which compact regions can be used efficiently.

Finally, we also demonstrated practical gains brought by destination passing for some of the algorithms we presented throughout this manuscript. Still, our DPS Haskell implementation suffers from a few limitations that do not always let destination-based algorithms match their theoretical performance. For example, compact regions are completely secluded from the garbage-collected heap, which requires us to copy data from the latter to the former when filling destination-based data structures, inducing overhead. Also, the relatively high boilerplate cost of creating regions, creating ampars, and creating/eliminating wrappers along the way cannot compete with the well-optimized machinery of the GHC runtime system for lazy data structures on small examples particularly.

In this chapter, we could only use DPS Haskell for structures carrying unrestricted data. This was decided to keep a (relatively) simple API while avoiding the issues of scope escape (Section 2.3). In the next chapter, we will revisit this limitation and find ways to modify DPS Haskell so that we can safely store linear resources in structures with holes.

# Chapter 4

# Extending DPS Haskell to support linear data structures

The main issue with the approach of our DPS Haskell API developed in Chapter 3 is that data structures built through destination filling cannot store linear data. This makes any data structure built using destinations (e.g. for speed benefits/better complexity on operations) less generic than the usual functional data structures that can store both linear and unrestricted data alike.

We recall that this limitation results from having a lot of flexibility in the way destinations can be used, without a system of ages to prevent scope escape in a fine-grained manner. So we had to use *unrestrictedness* as a barrier so that destinations cannot escape their scope.

Previous works however, such as ⟦Minamide, 1998⟧, or ⟦Lorenzen, Leijen, Swierstra, and Lindley, 2024⟧, would allow for efficient queues (see Section 2.2.3) or any structure with holes to store linear data, without issues with scope escape. This happens mainly because they do not have a concept of destinations; in these works we can only fill holes in a structure by interacting with the incomplete structure itself (the ampar). So we ought to find a way for our Haskell implementation to be at least as expressive as these works.

In this chapter, we will start by defining the challenges and risks introduced by destinations for linear data, and then iterate through various design ideas to extend our DPS Haskell API with support for them while maintaining safety. First, we will consider a restricted model where ampars for linear data can only have one hole (i.e., one destination on the right-hand side), *à la* Minamide. Then, we will consider an extension to support any ampar for linear data without restrictions (in particular, allowing several holes), by leveraging linear lenses.

## 4.1 Challenges of destinations for linear data: scope escape, updated

Starting from this point, we will say *linear destination* for an object of type $\lfloor_1 \mathsf{T} \rfloor$ or corresponding `Dest 1 t`, that is, a destination that can host linear data. Similarly, an *unrestricted destination* is an object of type $\lfloor_\omega \mathsf{T} \rfloor$ or `Dest ω t`. Regardless, a destination is always, itself, a resource that must be managed linearly. Finally, for most of this chapter, we omit compact region details, for clarity of exposition.

If we want to allow destinations to be filled with linear data, we have to be extremely careful about potential scope escape (see Section 2.3). Scope escape can only occur when a *linear* destination from an older scope is filled within a newer one. This is because only linear destinations can themselves be filled with other destinations. As discussed in Section 2.3, filling a newer destination with a strictly older one is safe; it is the reverse—filling an older destination with a newer one (or one of the same age)—that causes issues.

A potential remedy would be to disallow linear destinations to be filled using `fillLeaf` or `fillComp` functions (by changing the signature of these functions to work on unrestricted destinations only). That way, linear destinations could only be filled by operating on the ampar they come from, as in ⟦Lorenzen, Leijen, Swierstra, and Lindley, 2024; Minamide, 1998⟧. That way, we are sure we can only fill the freshest linear destinations (as destinations start to age when they are made accessible in a new scope with `updWith`; so destinations still on the right side of a closed ampar are fresher than any other destination in the context). We will use the base name `extend` for this new family of operators that operate directly on an ampar to fill its destinations. Unrestricted destinations, on the other hand, can still be used with the various `fill` functions. We should note that filling a destination with a hollow constructor cannot cause any scope escape issues, so it's safe to use `fill @'Ctor` on both linear and unrestricted destinations.

Let's see what this updated API would look like:

```
data Ampar s t
data Dest m t  -- m is the multiplicity, either 1 or ω
newAmpar :: Token ⊸ Ampar s (Dest 1 s)
updWith :: Ampar s t ⊸ (t ⊸ u) ⊸ Ampar s u

-- for unrestricted destinations
fillLeaf :: t → Dest ω t → ()

-- for linear destinations
extendLeaf :: t ⊸ Ampar s (Dest 1 t) ⊸ Ampar s ()
```

The idea is that `extendLeaf` is a way of opening a new scope, filling the linear destination belonging to that fresh scope, and closing it immediately afterwards, without letting the user interfere. So we could think that in `extendLeaf val (Ampar struct d)`, the linear destination `d` which gets filled by `extendLeaf` would be always strictly younger than `val` which gets stored into it, preventing scope escape (recall the sane example of Section 2.3, where an older destination is stored into a younger one with no issues).

Except...it does not work:

```
newAmpar tok0 `updWith` \(d0 :: Dest 1 (Dest 1 t)) →
  newAmpar tok1 `updWith` \(d1 :: Dest 1 t) →
    (newAmpar tok2 `updWith` \(d2 :: Dest ω ()) → d2 &fillLeaf () ⨾ d0) &extendLeaf d1
```

In this particularly convoluted example—which nonetheless typechecks given the API described above[30]— an old destination `d0` is hidden inside the nest of a fresher one `d2` that we have just consumed. So we have an ampar (``newAmpar tok2 `updWith` \d2 → d2 &fillLeaf () ⨾ d0``) whose right-hand side contains a destination `d0` which is not its own, but comes from a previous scope. However, it is indistinguishable from an ampar whose right-hand side contains only its own destinations, if we just look at the types. So we are allowed to use `extendLeaf` on the cursed ampar, which causes scope escape (as `d1` gets stored in the outermost ampar)!

There seems to be an inevitable friction between having a capturing `updWith` function, and having safe linear destinations, when we do not have a proper age system to ensure there is no cheating with scopes.

[30] It would, however, be rejected by $\lambda_d$, as `d0` would have age $\uparrow^2$ when used as a variable in the innermost scope, which is disallowed by rule $\lambda_d$–TY/ID.

**A non-capturing `updWith` function** If we want to prevent older destinations from hiding in newer ampars, we can ask for `updWith` to *not* capture any linear resource. This way, the only destinations accessible and usable in the scope of `updWith` are the ones coming immediately from the ampar being opened.

This is done by changing the signature from `updWith :: Ampar s t ⊸ (t ⊸ u) ⊸ Ampar s u` to `updWith :: Ampar s t ⊸ (t ⊸ u) → Ampar s u`. Notice the unrestricted function arrow →: it means the linear function supplied by the user must not capture anything linear (but still must use its argument linearly). Adding this drastic restriction has many consequences.

First, obviously, we cannot use any linear resource coming from outer scopes inside the scope of `updWith`. In particular, we cannot use a `Token` coming from the top-level linear scope (we described how convenient it is to just have one top-level linear scope in Section 1.8). Instead, we have to start a new nested linear scope (by calling `withToken`) if we want to create new ampars in the body of `updWith`. Alternatively we can tweak `updWith` to also provide a linear token to the inner scope, which gives `updWithToken :: Ampar s t ⊸ ((t, Token) ⊸ u) → Ampar s u` (the ampar on which `updWithToken` operates is a linear resource on which we can base the chain of linearity that threads through the new token).

As a side note, even with the restriction on `updWith`, we still cannot fill linear destinations using a `fillLeaf` or `fillComp` function, for two main reasons. First, we cannot bring linear resources into scope in the first place, so we rarely have anything useful to fill into a linear destination. Second, allowing `fillLeaf` to operate on linear destinations could enable interactions between sibling destinations originating from the same scope, as described in Section 2.3.

In earlier chapters, the restrictions put in place to prevent proper scope escape also happened to prevent such interaction between sibling destinations; but this is not the case here. Let's see this with an example:

```
fromAmpar' (
  newAmpar @(t, Dest 1 t) tok `updWith` \(d :: Dest 1 (t, Dest 1 t)) →
    case (d &fill @'(,)) of (dt :: Dest 1 t, ddt :: Dest 1 (Dest 1 t)) →
      ddt &fillLeaf dt
)
```

Here, we create a new ampar for a pair of a `t` and a `Dest 1 t`. After filling the root destination `d` with a hollow pair constructor, we get two destinations, `dt` and `ddt`. If we have access to a `fillLeaf` function operating with a linear destination on the left, we can actually put `dt` into `ddt`. Because we have consumed the two destinations and unit is returned, it seems that the ampar is complete, while in fact we have never provided a value of type `t` for the first field of the pair! So the structure is still incomplete while we are allowed to call `fromAmpar'` on it and read it!

This demonstrates that even with the restriction on `updWith` (which prevents capturing linear resources), allowing `fillLeaf` or `fillComp` to operate on linear destinations would still be unsafe.

## 4.2 Completing the family of operators for ampars with a single hole

We've just seen that to use linear destinations safely, we need operators that work on closed ampars directly, instead of destinations. In fact, that is exactly what happens in ⟦Minamide, 1998⟧, or ⟦Lorenzen, Leijen, Swierstra, and Lindley, 2024⟧: they only allow incomplete structures with one hole, and extending the structure happens by manipulating the whole incomplete structure itself, not destinations (as they do not exist in these works).

So as we just did with extendLeaf above, we will be designing new operators for ampars with only one hole, so exactly one destination on their right-hand side.

**The generic extend function** Similarly to the fill @'Ctor function (see Sections 3.4.2 and 3.5.4), we would like an extend @'Ctor function that let us extend an ampar having only one hole with a new arbitrary hollow constructor. However, unlike fill @'Ctor that can return a variable number of destinations (e.g. fill @'(:) returns two new destinations), we must design a function extend @'Ctor that takes several values as an input to fill directly into all except one of the fields of the hollow constructor, so that only one new hole remains, and only one new destination is created[31].

The signature of such a function becomes a bit cumbersome to account for its genericity, as shown below:

```
extend :: ∀ lCtor k t m s fiTys kthFiTy othFiTys.
        (OnlyLinear lCtor t,
        FieldTypes lCtor t ~ fiTys,
        At k fiTys ~ kthFiTy, Remove k fiTys ~ othFiTys) ⇒
        othFiTys %m→ Ampar s (Dest m t) ⊸ Ampar s (Dest m kthFiTy)
```

There's a lot to decipher here. OnlyLinear lCtor t is a constraint that asks for the chosen data constructor to only have linear fields (more about that in a moment). FieldsTypes is a type family that returns the types of the fields for a given constructor (specified lCtor and the type t to which the constructor belongs). For example, FieldsTypes '(:) [Int] resolves to (Int, [Int]). At k is a type family that returns the k$^{\text{th}}$ element (i.e. type) of a tuple type, and similarly, Remove k returns the same tuple type it is given, with the k$^{\text{th}}$ removed.

The symbol ~ is for type equality constraint, and can be used as above as a way to bind type variables to complex type-level expressions in a function or type class signature. Above, we say that fiTys represents all field types of lCtor, kthFiTy is the k$^{\text{th}}$ of these, and othFiTys is all the remaining field types except the k$^{\text{th}}$ one, in a tuple if there are more than one remaining type.

For instance, in extend @'(:) @0 @[Int], kthFiTy is Int, and othFiTys is [Int] (not in tuple form because there is only one remaining type). Conversely, in extend @'(:) @1 @[Int], kthFiTy is [Int], and othFiTys is just Int.

In practice, such a signature means that extend takes an ampar whose right side is a destination of type t, with multiplicity m[32]. It then allocates a new hollow constructor (specified by lCtor), consumes othFiTys at multiplicity m to fill all the fields of the new hollow constructor except the k$^{\text{th}}$ one, and then returns a destination with same multiplicity m for that remaining field of type kthFiTy.

For example, the append :: DList t ⊸ t ⊸ DList t function on difference lists can be expressed directly with extend (see the following code excerpt). So far, DList could only store unrestricted elements, but now, we can use difference lists to carry linear elements! This way, DList (using destinations) is as general as usual lists or the usual pure functional encoding of difference lists.

[31] Alternatively, it's still possible to use updWith to access a linear destination, then use fill @'(:) on it, creating two new linear destinations. But then we cannot use the extendLeaf function teased above to fill these destinations with complete values, because extendLeaf only works on an ampar with just a single destination on the right-hand side, and fillLeaf only works on unrestricted destinations.

[32] That way, we can use extend on an ampar with either a linear or unrestricted destination on the right, even though unrestricted destinations can be manipulated directly in a more flexible fashion by using the functions from the fill family.

```
type DList t = Ampar [t] (Dest 1 [t])

append :: DList t ⊸ t ⊸ DList t
dlist `append` x = dlist &extend @'(:) @1 x
```

We can also use extend to define and use *zipper* trees (in the light of ⟦Huet, 1997; Lorenzen, Leijen, Swierstra, and Lindley, 2024⟧) that can hold linear data:

```
data Tree t = Nil | Node (Tree t) t (Tree t)
type ZipperTree t = Ampar (Tree t) (Dest 1 (Tree t))

nodeL :: ZipperTree t ⊸ t ⊸ Tree t ⊸ ZipperTree t
nodeL parent x rtree = parent &extend @'Node @0 (x, rtree)

nodeR :: ZipperTree t ⊸ Tree t ⊸ t ⊸ ZipperTree t
nodeR parent ltree x = parent &extend @'Node @2 (ltree, x)
```

Note that extend does not support constructors with 0 fields (because extend needs to return a destination, and there are no new destination created when writing a 0-ary constructor into a hole). But 0-ary constructors can be written to the structure like any other complete value anyway, with extendLeaf:

```
extendLeaf :: ∀ s t m. t %m→ Ampar s (Dest m t) ⊸ Ampar s ()
```

Here we have also made extendLeaf multiplicity-polymorphic, compared to the initial version presented in Section 4.1, but its behavior stays the same.

It is a bit pointless in fact to return an Ampar s () to the user from a call to extendLeaf; it's better to directly extract the completed structure of type s:

```
fromAmparWithLeaf :: ∀ s t m. t %m→ Ampar s (Dest m t) ⊸ s
fromAmparWithLeaf = fromAmpar' . extendLeaf

-- we use it to implement the remaining operator for zipper trees
done :: ZipperTree t ⊸ Tree t
done parent = parent &fromAmparWithLeaf Nil
```

**Multiplicity of fields of a constructor and extendUr** Above, in the definition of extend, we considered that the constructor represented by lCtor had only linear fields, as indicated by the constraint OnlyLinear lCtor t. Indeed, the multiplicity m at which we consume the values for the other fields, and at which we return the new destination, are the same as the multiplicity of the parent destination.

In $\lambda_d$, every data constructor has linear fields, except the exponential constructor $\text{Mod}_{m}$ which has a single field of mode m. In Haskell however, constructors have only linear fields by default, but can have unrestricted fields when the corresponding datatype is defined with GADT syntax and that an unrestricted arrow → is following the field's type. An unrestricted field is a way to hold unrestricted data without needing an explicit Ur layer.

Ideally, we should thus refine the extend function further, and be polymorphic over the multiplicity of each field. However, to this date, we cannot get the information about field multiplicities with GHC.Generics—and this is what we use to make DPS Haskell usable with user-defined types. So, for now, it does not seem easy to modify the extend function even more to account for arbitrary multiplicity on fields.

There is also another consequence: it means we have to add a dedicated version of `extend` for the `Ur` constructor:

```
extendUr :: ∀ s t m. Ampar s (Dest m (Ur t)) ⊸ Ampar s (Dest ω t)
```

Given that `m·ω=ω` for any `m`, the `extendUr` function can operate on an ampar with either a linear or unrestricted destination, but always returns an ampar with an unrestricted destination.

Using `extendUr` is the primary way, in this world, to obtain an initial unrestricted destination `Dest ω t`, as `newAmpar` initially spawns an ampar with just a linear destination on the right.

**Composition of incomplete structures with `extendComp`** The last function we need in this setting is `extendComp`, which plugs two ampars together:

```
extendComp :: ∀ s t u. Ampar t u ⊸ Ampar s (Dest 1 t) ⊸ Ampar s u
```

In `extendComp`, the destination in the parent ampar must be linear, as we already mentioned in Section 2.5.2. Otherwise, the following exploit becomes possible:

```
extendCompFaulty :: ∀ s t u m. Ampar t u ⊸ Ampar s (Dest m t) ⊸ Ampar s u

parentAmpar :: Ampar (Ur t) (Dest ω t) = newAmpar tok1 &extendUr
-- resolves to Ampar (Ur h) (Dest →h)

childAmpar :: Ampar Int (Dest 1 Int) = newAmpar tok2
-- resolves to Ampar (h') (Dest →h')

resAmpar :: Ampar (Ur t) (Dest 1 t) = parentAmpar &extendCompFaulty freshAmpar
-- resolves to Ampar (Ur h') (Dest →h')

let Ur (linearRes') = resAmpar &fromAmparWithLeaf linearResourceDoNotDuplicate in
  (linearRes', linearRes')
```

With the code above, we create an ampar `resAmpar` with a linear destination pointing to an unrestricted hole. This happens because the ampar resulting from a composition inherits whatever was on the right of the child ampar (here a linear destination `Dest` → h′ :: `Dest 1 Int`). But the left side of the child ampar has been grafted into the unrestricted hole h of the parent ampar. As a result, h' becomes transitively unrestricted, while its corresponding destination remains linear. This enables the duplication of linear resources, as shown in the last line, so we must prevent it. For the same reason, we cannot define a function `fillComp :: Ampar t u ⊸ Dest ω t ⊸ u` in this setting.[33] In Chapter 3, this issue did not arise because all destinations were unrestricted by design.

What we can do, however, is define alternative versions of `extendComp` and `fillComp` that operate on unrestricted destinations, with a slightly modified signature, that makes them safe this time:

```
extendUComp :: ∀ s t u. Ampar (Ur t) u ⊸ Ampar s (Dest ω t) ⊸ Ampar s u
fillUComp :: ∀ t u. Ampar (Ur t) u ⊸ Dest ω t ⊸ u
```

Here, the child ampar must be an unrestricted structure (as indicated by the leading `Ur`). This ensures that any destination appearing in `u` is already unrestricted. As a result, we eliminate the linearity exploit and recover the same composition behavior we had in Chapter 3.

[33] Recall that we already ruled out defining `fill`-like functions for linear destinations since the first section of this chapter.

## 4.3 Ampars with several holes, storing linear data

Being limited to only one destination on the right-hand side of ampars is not ideal. Indeed, we stay at the same expressivity level as existing work, and we lose part of the flexibility that made $\lambda_d$ innovative in the first place.

Supporting multiple linear destinations on the right-hand side of ampars introduces significant complexity however. In the previous section, with only one destination per ampar, we avoided scope escape by requiring the user to interact with the ampar directly, rather than with the destination itself. This was straightforward, as there was no ambiguity—the single destination was the only possible target of the `extend` operation. By contrast, if we now allow arbitrary types on the right-hand side, but still aim to restrict direct user access to destinations (for the same reasons as above; `updWith` is now non-capturing, and we want to prevent scope escape and malicious interaction between sibling destinations), we must introduce a mechanism for users to indicate which specific destination they intend to operate on. One idea is to ask the user for a linear lens. First, let's define what a lens is.

Lenses are pairs of getters and setters that enable convenient handling of immutable data structures in functional programming languages. A lens of type `Lens t1 t2 u1 u2` has two components (at least conceptually):

- a getter function to access a value of type `u1` somewhere inside a structure of type `t1`, which establishes the *focus* of the lens;
- a function to transform a structure of type `t1` into a structure of type `t2` by replacing the focused value of type `u1` by a value of type `u2`. In particular, when `u1`=`u2`, then `t1`=`t2`, and thus this function acts as an setter for the focused value.

In their simplest expressions, non-linear lenses are represented by the following datatype:

```
data LensNL t1 t2 u1 u2 = LensNL
  { view :: t1 → u1  -- just get u1 out of t1, discard the rest
  , update :: t1 → u2 → t2  -- discard the u1 inside t1 and replace it with a u2 to make a t2
  }
```

Given a data structure `Triple` with 3 fields, let's create a lens to focus on the first field:

```
data Triple s t u = Triple { first :: s, second :: t, third :: u }

fstNL :: LensNL (Triple s t u) (Triple s' t u) s s'
fstNL = LensNL { view = \(Triple x y z) → x
               , update = \(Triple x y z) → x' → Triple x' y z }
```

With a non-linear lens `LensNL` such as above, if we use the `view` component, then we consume a value of type `t1` but discard most of it except the part of type `u1`. If we use the `update` component, it's the opposite; we discard the part of type `u1` but keep all the rest of the original `t1`.

For linear lenses, that we will introduce now, we are not allowed to use the `view` and `update` components separately; otherwise we could drop parts of the original structure of type `t1`, as we have just seen. So instead, we group the two functions `view` and `update` into a single one:

```
newtype Lens t1 t2 u1 u2 = Lens (t1 ⊸ (u1, u2 ⊸ t2))
```

With a linear lens of type `Lens t1 t2 u1 u2`, we can split a structure of type `t1` into a pair of a value of type `u1` (the focus), and a closure of type `u2 ⊸ t2` that carries all the rest of the original `t1`, and is ready to produce a structure of type `t2` if we give it a `u2`. In other terms, this second component of type `u2 ⊸ t2` is really just a functional representation of a `t2` missing a `u2` to be complete. Both of the components of the pair have to be consumed if we want to respect linearity.

It's fairly immediate to update our `Triple` exemple for linear lenses:

```
fst :: Lens (Triple s t u) (Triple s' t u) s s'
fst = Lens (\(Triple x y z) → (x, \x' → Triple x' y z))
--                  captures x ↗          ↖ captures y and z
```

**Lenses to select destinations to operate on** We now introduce a new family of functions, with base name `extendFocused`, that can work on ampars with several holes (thus several destinations), using linear lenses. These linear lenses serve a dual purpose: not only do they let the user select the destination to operate on, but they also give us a way to reinsert the newly created destinations (from the fields of the new hollow constructor) or the byproduct of the filling operation at the exact position previously occupied by the consumed destination, on the right-hand side of the ampar. This is made possible by the bidirectional nature of lenses, acting both as getters and setters.

We begin with the generic `extendFocused @'Ctor` function:

```
extendFocused :: ∀ lCtor u t1 t2 s m.
            (OnlyLinear lCtor u) ⇒
            Lens t1 t2 (Dest m u) (DestsOf lCtor m u) → Ampar s t1 ⊸ Ampar s t2

extendWithNodeOnFst :: Ampar s (Triple (Dest m (Tree t1)) t2 t3)
                  ⊸ Ampar s (Triple (Dest m (Tree t1), Dest m t1, Dest m (Tree t1)) t2 t3)
extendWithNodeOnFst ampar = ampar & extendFocused fst @'Node
```

Here, `extendFocused` takes a lens of type `Lens t1 t2 (Dest m u) (DestsOf lCtor m u)`, i.e. a lens that focuses on a `Dest m u` inside the right-hand side of type `t1` of the ampar, and lets us build a new right-hand side of type `t2` for the resulting ampar, where the destination of type `Dest m u` is replaced by a value of type `DestsOf lCtor m u` (that is, a tuple with the new potential destinations).

The lens is taken as a non-linear argument by `extendFocused` (as indicated by the non-linear arrow →), so that it cannot capture any linear resource. So the lens is forced to be just a purely structural operation that can be applied as many times as we want on various values of type `t1`.

Nonetheless, we could have implemented `extendWithNodeOnFst` in terms of `updWith` and `fill @'Node`[34].

In fact, the added value of linear lenses becomes clear when designing the `extendFocusedLeaf` function, which for the first time (in the Haskell implementation), enables us to fill a linear destination with a value when the said destination is located within an arbitrary structure on the right-hand side of an ampar:

```
extendFocusedLeaf :: ∀ u t1 t2 s m. u %m→ Lens t1 t2 (Dest m u) () → Ampar s t1 ⊸ Ampar s t2

extendWithValOnFst :: t1 ⊸ Ampar s (Triple (Dest 1 t1) t2 t3) ⊸ Ampar s (Triple () t2 t3)
extendWithValOnFst x ampar = ampar &extendFocusedLeaf fst x
```

[34] As said previously, `fill @'Ctor` cannot cause scope escape and thus is safe to use for both linear and unrestricted destinations

No scope escape can happen here. As `updWith` is non-capturing, we are sure that the destinations on the right-hand side of an ampar are the ones pointing to the holes on its left-hand side (eliminating the issue presented at the end of Section 4.1). So like `extendLeaf`, the `extendFocusedLeaf` function can only be used to fill a newer destination with existing (thus, older) values. Likewise, we cannot have sibling destinations interacting with one another. Which means, we have reached the goal stated at the beginning of this chapter: we can build structures having several linear holes in a safe manner in DPS Haskell!

For completeness, we finally define `extendFocusedUr`, `extendFocusedComp` and `extendFocusedUComp`, following the same pattern as for `extendFocusedLeaf`:

```
extendFocusedUr :: ∀ u t1 t2 s m.
                     Lens t1 t2 (Dest m (Ur u)) (Dest ω u) → Ampar s t1 ⊸ Ampar s t2
extendFocusedComp :: ∀ u t1 t2 s u1 u2.
                       Ampar u1 u2 ⊸ Lens t1 t2 (Dest 1 u1) u2 → Ampar s t1 ⊸ Ampar s t2
extendFocusedUComp :: ∀ u t1 t2 s u1 u2.
                        Ampar (Ur u1) u2 ⊸ Lens t1 t2 (Dest ω u1) u2 → Ampar s t1 ⊸ Ampar s t2
```

Both of these functions are very similar to their respective one-hole versions `extendUr` and `extendComp`. In `extendFocusedUr`, we consume a destination of type `Dest m (Ur u)` and replace it with a destination of type `Dest ω u`. In `extendFocusedComp`, we consume a destination of type `Dest 1 u1`, to which we write the left-hand side (of type `u1`) of the child ampar, and we replace this consumed destination with whatever was on the right-hand side (of type `u2`) of the child ampar. The only difference, between the `extend` and the corresponding `extendFocused` functions, is that the filling operation in the latter happens at a focused location on the right-hand side of an ampar instead of being made on the single destination at top level as in the former.

## 4.4 Final API and breadth-first tree traversal, updated

The family of `extendFocused` functions, based on linear lenses, addresses the core challenge of safely filling destinations with linear data. Although the resulting interface is not entirely satisfactory—particularly from an ergonomic standpoint—it still strikes a reasonable compromise between expressiveness and complexity of the interface (given the constraint of working with the existing Haskell type system).

For instance, the DPS system supports two usable levels of nesting for destinations, allowing us to construct and use linear destinations of destinations (type `Dest 1 (Dest m t)`). This is in fact perfectly sufficient to implement non-trivial examples, such as the breadth-first traversal of a tree, but this time using the efficient, destination-based queues in Haskell! Let's revisit these examples one last time. For the final part of this chapter, we also take compact region considerations into account (this mostly consists in adding `r` and `Region r` all over the function's signatures). The code of efficient queues (and difference lists for linear data, that they are made of) is presented in Listing 4.1, and the code for updated breadth-first tree traversal is given in Listing 4.2.

These implementations derive directly from the various iterations and improvements we have made over the different lists, efficient queues and breadth-first traversal examples all along this document.

Notably though, we leverage both the `fill` and the `extend` families of functions in these examples. For instance, we reuse the implementation of difference lists presented in Section 4.2, for which we do not need the flexibility brought by `updWith`, and instead rely on the `extend` functions so that we can store linear data in the difference lists (very similarly to what is presented in [Lorenzen, Leijen, Swierstra, and

```haskell
type DList r t = Ampar r [t] (Dest r 1 [t])

newDList :: Region r ⇒ Token ⊸ DList r t
newDList = newAmpar @[t]

dListToList :: Region r ⇒ DList r t ⊸ [t]
dListToList dlist = fromAmpar' (dlist &extendLeaf [])

append :: Region r ⇒ DList r t ⊸ t ⊸ DList r t
append dlist x = dlist &extend @'(:) @1 x

-----------------------------------------------------------------------------

data EffQueue r t = EffQueue [t] (DList r t)

newEffQueue :: Region r ⇒ Token ⊸ EffQueue r t
newEffQueue tok = EffQueue [] (newDList tok)

singleton :: Region r ⇒ Token ⊸ t ⊸ EffQueue r t
singleton tok x = EffQueue [x] (newDList tok)

queueToList :: Region r ⇒ EffQueue r t ⊸ [t]
queueToList (EffQueue front back) = front ++ dListToList back

enqueue :: Region r ⇒ EffQueue r t ⊸ t ⊸ EffQueue r t
enqueue (EffQueue front back) x = EffQueue front (back `append` x)

dequeue :: Region r ⇒ EffQueue r t ⊸ Maybe (t, EffQueue r t)
dequeue (EffQueue front back) = case front of
  [] → case tokenBesides back of (back, tok) → case (toList back) of
    [] → drop tok ⨾ Nothing
    (x : xs) → Just (x, (EffQueue xs (newDList tok)))
  (x : xs) → Just (x, (EffQueue xs back))
```

Listing 4.1: Efficient queue for linear data in extended DPS Haskell

```
data Tree t = Nil | Node t (Tree t) (Tree t)

relabelDps :: Region r ⇒ Token ⊸ Tree t → Tree Int
relabelDps tree = fst (mapAccumBfs (\st _ → (st + 1, st)) 1 tree)

mapAccumBfs :: ∀ r s t u. Region r ⇒ Token ⊸ (s → t → (s, u)) → s → Tree t → (Tree u, s)
mapAccumBfs tok f s0 tree =
  case fromAmpar (newAmpar @(Ur (Tree u)) tok `updWith` \du → go s0 (singleton (Ur tree, du &fillUr)))
  of (Ur outTree, Ur st) → (outTree, st)
  where
    go :: s → EffQueue r (Ur (Tree t), Dest r ω (Tree u)) ⊸ Ur s
    go st q = case dequeue q of
      Nothing → Ur st
      Just ((utree, dtree), q') → case utree of
        Ur Nil → dtree &fill @'Nil ⨾ go st q'
        Ur (Node x tl tr) → case (dtree &fill @'Node) of
          (dy, dtl, dtr) →
            let q'' = q' `enqueue` (Ur tl, dtl) `enqueue` (Ur tr, dtr)
                (st', y) = f st x
             in dy &fillLeaf y ⨾ go st' q''
```

Listing 4.2: Breadth-first tree traversal using efficient queues in extended DPS Haskell

Lindley, 2024; Minamide, 1998]). On the other hand, the breadth-first tree traversal is still implemented using `updWith` and `fill`-like functions, which are essential for operating on the destinations stored in the queue.[35]

**Full, final API for DPS Haskell with linear destination support** In Listing 4.3, we sum up all the functions we have encountered in this chapter, that are needed to create a Haskell API as flexible as possible, while still being safe, to perform destination passing.

At the time of writing the thesis, we have not implemented this whole API yet, but we envision that there should not be any particular technical issue to do so. Indeed, the behavior of the `extend` operators derives directly from the corresponding `fill` functions, that already have a concrete implementation in [Bagrel, 2023b].

[35]It would not be possible to implement the breadth-first traversal using only `extend` functions with lens support. If it is possible at all, it would require more powerful *optics* (the broader family of abstractions that includes lenses), and the resulting code would likely be much more verbose.

```haskell
data Ampar r s t
newAmpar :: ∀ r s. Region r ⇒ Token ⊸ Ampar s (Dest r 1 s)
tokenBesides :: ∀ r s t. Region r ⇒ Ampar r s t ⊸ (Ampar r s t, Token)
toAmpar :: ∀ r s. Region r ⇒ Token ⊸ s ⊸ Ampar r s () -- now linear in s
fromAmpar  :: ∀ r s t. Region r ⇒ Ampar r s (Ur t) ⊸ (s, Ur t)
fromAmpar' :: ∀ r s. Region r ⇒ Ampar r s () ⊸ s
updWith :: ∀ r s t u. Region r ⇒ Ampar r s t ⊸ (t ⊸ u) → Ampar r s u -- non-capturing now
updWithToken :: ∀ r s t u. Region r ⇒ Ampar r s t ⊸ ((t, Token) ⊸ u) → Ampar r s u

data Dest r m t
type family DestsOf lCtor r m t -- now multiplicity-polymorphic

-- === API for unrestricted destinations only ===
fillLeaf :: ∀ r t. Region r ⇒ t → Dest r ω t ⊸ ()
fillUComp :: ∀ r t u. Region r ⇒ Ampar r (Ur t) u ⊸ Dest r ω t ⊸ u

-- === API for ampars with linear or unrestricted destinations ===
fill :: ∀ lCtor r m t. (Region r, OnlyLinear lCtor t) ⇒ Dest r m t ⊸ DestsOf lCtor r m t
fillUr :: ∀ r m t. Region r ⇒ Dest r m (Ur t) ⊸ Dest r ω t

-- only one destination on the right (simplified but less expressive)
extend :: ∀ lCtor r k t m s fiTys kthFiTy othFiTys.
          (Region r, OnlyLinear lCtor t,
          FieldTypes lCtor t ~ fiTys,
          At k fiTys ~ kthFiTy, Remove k fiTys ~ othFiTys) ⇒
          othFiTys %m→ Ampar r s (Dest r m t) ⊸ Ampar r s (Dest r m kthFiTy)
extendUr :: ∀ r s t m. Region r ⇒ Ampar r s (Dest r m (Ur t)) ⊸ Ampar r s (Dest r ω t)
extendLeaf :: ∀ r s t m. Region r ⇒ t %m→ Ampar r s (Dest r m t) ⊸ Ampar r s ()
extendComp :: ∀ r s t u. Region r ⇒ Ampar r t u ⊸ Ampar r s (Dest r 1 t) ⊸ Ampar r s u
extendUComp :: ∀ r s t u. Region r ⇒ Ampar r (Ur t) u ⊸ Ampar r s (Dest r ω t) ⊸ Ampar r s u

-- with arbitrary type on the right; destination focused with a linear lens
extendFocused :: ∀ lCtor r u t1 t2 s m.
                 (Region r, OnlyLinear lCtor u) ⇒
                 Lens t1 t2 (Dest r m u) (DestsOf lCtor r m u) → Ampar r s t1 ⊸ Ampar r s t2
extendFocusedUr :: ∀ r u t1 t2 s m. (Region r) ⇒
                   Lens t1 t2 (Dest r m (Ur u)) (Dest r ω u) → Ampar r s t1 ⊸ Ampar r s t2
extendFocusedLeaf :: ∀ r u t1 t2 s m. (Region r) ⇒
                     u %m→ Lens t1 t2 (Dest r m u) () → Ampar r s t1 ⊸ Ampar r s t2
extendFocusedComp :: ∀ r u t1 t2 s u1 u2. (Region r) ⇒
                     Ampar r u1 u2 ⊸ Lens t1 t2 (Dest r 1 u1) u2 → Ampar r s t1 ⊸ Ampar r s t2
extendFocusedUComp :: ∀ r u t1 t2 s u1 u2. (Region r) ⇒
                      Ampar r (Ur u1) u2 ⊸ Lens t1 t2 (Dest r ω u1) u2 → Ampar r s t1 ⊸ Ampar r s t2
```

Listing 4.3: Extended DPS Haskell API with support for linear destinations

## 4.5 Conclusion

In this chapter we took time to explore what happens when we relax the restriction we had imposed in Chapter 3, namely, that destinations could only be filled with non-linear data (which, consequently, meant that ampars could only be used to store non-linear data).

In several careful steps, avoiding scope escape and similar dangers at every corner, we refined our DPS Haskell API to lift this limitation. First, by allowing the user to build ampars to store linear data, but only if they have only one hole each, in the light of the foundational works ⟦Lorenzen, Leijen, Swierstra, and Lindley, 2024; Minamide, 1998⟧. This gave birth to the family of `extend` functions, that operate on ampars directly instead of destinations.

In a second step, we proposed yet another set of functions, denoted by `extendFocused`, that let the user fill ampars of arbitrary shapes, with no limit on the number of holes or destinations, with linear data. These functions also operate directly on an ampar, and let the user select the destination they want to focus on, inside the ampar, with a *linear lens*—that is, a convenient object that acts as both a getter and setter for immutable functional data structures.

In the end, we recover a system that is as close to $\lambda_d$ (Chapter 2) as we could safely design within the existing type system of Haskell, with expressivity improvements over the initial DPS Haskell API. Of course, the ergonomics for filling linear destinations are not as smooth as in $\lambda_d$, but our newly gained flexibility let us implement a revised version of breadth-first tree traversal that, this time, used exactly the same algorithm and underlying data structures as the initial version proposed in Section 2.4.

The final API we obtain, after all these refinements, has not been implemented yet in Haskell, though we do not think that it will be particularly challenging, as most of the hard work for low-level operations has already been done in Section 3.5.

# Related Work

In the realm of safe controlled memory updates for functional languages, *functional data structures with holes* and *destination passing* are two very close themes, since *destination passing* provides a flexible and principled way to manipulate such structures. However, despite their conceptual overlap, *functional destination passing* and *functional data structures with holes* remain slightly distinct in their nature.

*Functional destination passing* encompasses techniques where updates to write-once memory cells are exposed through a safe or at least controlled interface. In fact, in previous work, this interface has not often been directly exposed to programmers; instead, in these cases, *destination passing* is employed internally by compilers to optimize functional programs. This includes performance improvements for functional linked data structures (Work 1) and functional arrays (Work 2), which are written in the usual purely functional style but benefit from behind-the-scenes mutations to reduce allocations and improve performance, with no explicit user control. Ultimately, destinations can also be used in the operational semantics of a theoretical lambda-calculus (Work 3).

In contrast, *functional structures with holes* focus on allowing incomplete data structures, a.k.a. structures with holes, to exist as first-class values—that the user can manipulate explicitely—in a purely functional setting, and be completed or refined later. Completing these structures necessarily involves some form of memory update, but many systems (Work 4 and 5) do not include an explicit notion of first-class write pointer (*destination*) to do the mutation. In contrast, this thesis explores how both *functional structures with holes* and *functional destination passing* can be combined effectively, leveraging the strengths of each to achieve more expressive manipulation patterns of functional data, always in a safe manner.

Both of these notions inherently involve mutation and uninitialized data and therefore require robust static guarantees to ensure safety. *Linear* and *uniqueness type systems* are thus commonly employed to enforce strict control over destinations or structures with holes, preventing unsafe updates and premature reads on incomplete structures. However, as demonstrated in this work, *linearity* alone is insufficient to capture the full range of safety and expressiveness we want. Modal type systems (Work 6 and 7) are in that case particularly helpful to let us design a coherent and ergonomic type system, easily, for our specific needs. Complementary approaches such as *permission-based* type systems (Work 8) and *lifetime* management techniques *à la* Rust (Work 9) also play an important role to enforce fine-grained control over resources, which may be leveraged to control our destinations or structures with holes.

## 1 Tail modulo constructor

⟦Bour, Clément, and Scherer, 2021⟧ (and ⟦Allain et al., 2024⟧, expanding the former with a correctness proof) focus on the perennial problem that the map function on linked lists is not tail-recursive, hence consumes stack space. The map function can be made tail recursive, but at the cost of an extra reverse operation on the list. Their observation is that there's a systematic transformation of functions where the only recursive call is under a data constructor application (e.g. *cons* in the map instance) to a destination-passing tail-recursive implementation.

Here, there's no destination in user land, only in the intermediate representation. However, there is a programmatic interface: the programmer annotates a function like

```
let[@tail_mod_cons] rec map =
```

to ask the compiler to perform the translation. The compiler will then throw an error if it cannot, making the optimization behavior entirely predictable. This has been included in the OCaml compiler since version 4.14, and is the one example we know of destinations built in a production-grade functional compiler.

Both our $\lambda_d$ and DPS Haskell API allow to write the result of a tail-modulo-constructor transformation (that is, the destination-passing tail-recursive program) as a well-typed program in userland, which is not possible in OCaml. However, we do not provide a way to do a similar automatic transformation itself.

On the flip-side, tail modulo constructor is too weak to handle our difference lists, as it misses first-class support for structure with holes; they can only exist in the background for the duration of a function being transformed. The tail-modulo-constructor transformation also cannot be applied on our breadth-first tree traversal example (see Sections 2.4 and 3.3.2).

## 2 Destination-passing style for efficient memory management

Shaikhha et al. [2017] present a destination-based intermediate language for a functional array programming language, with destination-specific optimizations, that boasts near-C performance.

This is another significant evidence to date of the benefits of destination-passing style for performance in functional languages, although their work is on array programming, while this PhD manuscript focuses on linked data structures. They can therefore benefit from optimizations that are perhaps less valuable for us, such as allocating one contiguous memory chunk for several arrays.

The other difference between their work and ours is that their language is solely an intermediate language: it would be unsound to program in it manually, as there are no particular safety guarantees. We, on the other hand, are proposing a theoretical system, but also a concrete Haskell API to make it sound for the programmer to use destinations directly.

We see these two aspects as complementing each other: good compiler optimizations are important to alleviate the burden from the programmer and allow high-level abstraction; having the possibility to use destinations in code affords the programmer more control, should they need it.

## 3 Semi-axiomatic sequent calculus

In [DeYoung, Pfenning, and Pruiksma, 2020] constructors return to a destination rather than allocating memory. It is very unlike the other systems described in this chapter in that it's completely founded in the Curry-Howard isomorphism. Specifically it gives an interpretation of a sequent calculus which mixes Gentzen-style deduction rules and Hilbert-style axioms. As a consequence, they feature a *par* connective that is completely symmetric; and, unlike our $\lfloor T \rfloor$ type, their dualization connective is involutive.

The cost of this elegance is that computations may try to pattern-match on a hole, in which case they must wait for the hole to be filled. So the semantics of holes is that of a future or a promise. In turns this requires the semantics of their calculus to be fully concurrent, while $\lambda_d$ only needs a sequential execution model, with a very tame form of state manipulation (linear substitutions on the evaluation context).

# 4 A functional representation of data structures with a hole

The idea of using linear types as a foundation of a functional calculus in which incomplete data structures can exist and be composed as first class values dates back to ⟦Minamide, 1998⟧. Our systems, both theoretically and in the implementations, are strongly inspired by theirs. In ⟦Minamide, 1998⟧, a first-class structure with a hole is called a *hole abstraction*. Hole abstractions are represented by a special kind of linear functions with bespoke restrictions. As with any function, we cannot pattern-match on their output (or pass it to another function) until they have been applied; but they also have the restriction that we cannot pattern-match on their argument—the *hole variable*—as that one can only be used directly as an argument of data constructors, or of other hole abstractions. The type of hole abstractions `(T, S)hfun` (a structure of type `S` missing a `T` to be complete) is thus a weak form of linear function type $T \multimap S$.

In ⟦Minamide, 1998⟧, it's only ever possible to represent structures with a single hole. But this is a rather superficial restriction. The author does not comment on this, but we believe that this restriction only exists for convenience of the exposition: the language is lowered to a language without function abstraction and where composition is performed by combinators and it's easier to write a combinator for single-argument-function composition.

However, we have seen in Section 4.3 that we can actually design combinators for structures with multiple holes, thanks to linear lenses (at the cost of some extra syntactic burden). So having multiple-hole data structures would not have changed their system in any profound way.

The more important difference is that their system is based on a type of linear functions, while our systems in Chapters 2 and 3 exploit the expressivity of the linear logic's "par" type. In classical linear logic, linear implication $T \multimap S$ is reinterpreted as $S \parr T^{\perp}$. We, likewise, reinterpret `(T, S)hfun` as $S \ltimes \lfloor T \rfloor$ or `Ampar s (Dest t)` —a sort of weak "par".

A key consequence is that destinations—as first-class representations of holes—appear naturally in $\lambda_d$ or DPS Haskell, while ⟦Minamide, 1998⟧ does not have them. This means that using ⟦Minamide, 1998⟧, one can implement the examples with difference lists and queues from Sections 2.2.3 and 3.3.1, but could not do our breadth-first traversal example from Sections 2.4 and 3.3.2, since it requires being able to store destinations in a structure.

Nevertheless, we still retain the main restrictions that Minamide ⟦1998⟧ places on hole abstractions. For instance, we cannot pattern-match on `S` in (unapplied) `(T, S)hfun`; so in $\lambda_d$, we cannot act directly on the left-hand side `S` of $S \ltimes T$, only on the right-hand side `T` (the equivalent restrictions are also in place in DPS Haskell). Similarly, hole variables can only be used as arguments of constructors or hole abstractions; it's reflected in $\lambda_d$ by the fact that the only way to act on destinations is via fill operations.

The ability to manipulate and, in particular, to store destinations does come at a cost: $\lambda_d$ requires an additional notion of ages to ensure that destinations are used soundly across scopes. In the absence of such a mechanism, as in DPS Haskell, the use of destinations for linear data must be carefully constrained to preserve soundness. Despite these additional requirements, our systems—both theoretical and practical—are strictly more general than that of Minamide ⟦1998⟧. Their system can be naturally embedded within $\lambda_d$ (see Section 2.8) or expressed in DPS Haskell (see Section 4.2).

## 5 Tail recursion modulo context and The functional essence of imperative binary search Trees

The recent line of work from Leijen and Lorenzen [[2025]] and Lorenzen, Leijen, Swierstra, and Lindley [[2024]] develops in two main axes. The first explores a generalization of Work 1, by considering the *tail-recursion modulo context* transformation, in which the *context* can be instantiated in various forms. For example, when the context is a one-hole data constructor context—equivalent to the hole abstractions from Work 4—they recover the *tail-recursion modulo constructor* transformation. But the framework extends further: they are able to handle monoid contexts, semiring contexts, exponential contexts, and more. Hence, it allows the transformation of a broader class of typically non-tail-recursive functions into efficient tail-recursive equivalents, via a formalized and systematic method, surpassing the scope of the transformation described in Work 1.

The second axis, which is closer to our own, focuses on the semantics and representation of first-class one-hole data constructor contexts, continuing the lineage of Minamide's work (4). In particular, they investigate how such one-hole data constructor contexts (i.e. data structure with one hole) behave under various memory models, particularly in settings where the constructor contexts are not managed linearly! This closely echoes the copy-on-write strategy that also emerged in our work on $\lambda_d$ (see the end of Section 2.6.3 and beginning of Section 2.9), when ampars do not have to be managed linearly. However, their approach goes further: they define formal operational semantics over an explicit memory store, incorporating reference counting. They also propose optimizations to reduce copying overhead, by detecting when a structure with a hole is uniquely referenced and can thus be safely mutated in place—reminiscent of techniques in [[Lorenzen, Leijen, and Swierstra, 2023]].

Furthermore, they implement this *hybrid* strategy—as they call it—in the Koka programming language, and demonstrate performance gains brought by the *tail-recursion modulo context* transformation on various benchmarks. On the *map* example (see Section 3.6.4), their approach achieves performance on par with, or even superior to, our DPS Haskell implementation.

On the other hand, their system remains at the same level of expressiveness as Minamide's: destinations are purely internal and not exposed to the user. As they acknowledge, their model cannot represent structures with multiple holes, and thus, like Minamide's, it cannot support breadth-first tree traversal.

## 6 Oxidizing OCaml

Lorenzen, White, et al. [[2024]] present an extension of the OCaml type system to support modes. Their modes are split along three different "axes", among which affinity and locality are comparable to our multiplicities and ages from Chapter 2. Like our multiplicities, there are two modes for affinity `once` and `many`, though in [[Lorenzen, White, et al., 2024]], `once` supports weakening, whereas $\lambda_d$'s 1 multiplicity is properly linear (proper linearity matters for destination lest we can discard them without filling the corresponding hole, and end up reading uninitialized memory).

Locality tracks scope. There are two locality modes, `local` (does not escape the function) and `global` (can escape the current function). The authors present their locality mode as a drastic simplification of Rust's lifetime system (see Work 9), which nevertheless fits their need.

However, this system is too limited to track the scope of destinations as precisely as $\lambda_d$ needs. The key issue is that if destinations from nested scopes are assigned the same mode (which has to happen when mapping an infinity of ages to just two), we lose the ability to safely distinguish their lifetimes—making it possible to reproduce the counterexamples from Section 2.3. This suggests that implementing $\lambda_d$ in its entirety within modal OCaml is likely infeasible.

That said, if a mechanism could be introduced to prevent discarding resources of mode `once`, then much—if not all—of the destination-passing APIs described in Chapter 3 could still be realized in their version of OCaml, as these APIs rely only on linearity, not full age tracking.

Lorenzen, White, et al. ⟦2024⟧ is not the first attempt to introduce a control system over resources in OCaml. ⟦Scherer et al., 2017⟧ explored integrating an ML-like statically typed language with another featuring linear types and state, in which references and in-place reuse is covered quite explicitly.

## 7 Quantitative program reasoning with graded modal types

Orchard, Liepelt, and Eades III ⟦2019b⟧ present a functional framework with generalized support for *graded modal types*, and the language *Granule* as an instance of the said framework. Graded modal types refer to a type system with graded modalities—that is, modalities indexed with modes, like $!_{\mathrm{m}}$—in which the mode, here called *coeffect*, belongs to a semiring, as in ⟦Abel and Bernardy, 2020; Ghica and Smith, 2014⟧.

Their framework can be instantiated with different semirings, and they showcase a practical instance that supports fine-grained linearity (where usage counts come from $\mathbb{N} \cup \{\omega\}$, rather than the coarser $\{1, \omega\}$), along with a tagging system for public/private data to enable sensitive dataflow analysis. They furthermore include *mode ranges*, and a suitable ordering/subtyping relation on modes (similar to ours), that gives more flexibility in the way modes can be used and combined.

With this, Granule appears to be a promising target for functional destination-passing. While the instantiation they present lacks a notion of ages or explicit scope control, making a full embedding of $\lambda_d$ likely infeasible—it seems on the other hand capable of supporting (at type level at least) much, if not all, of the destination-passing APIs we developed for Haskell in Chapters 3 and 4.

Their general framework, however, could be instantiated with the semiring we developed in Section 2.5.1, and be used as a basis for the type system of $\lambda_d$. But we would still need to extend the set of runtime values to account for structure with holes and destinations, and likewise, introduce a mechanism for state update—such as the linear substitutions on the evaluation context we employed in Section 2.6.3.

## 8 Programming with permissions in Mezzo

Protzenko and Pottier ⟦2013⟧ introduced the Mezzo programming language, which features user-facing mutable data structures (unlike ours, where mutability is hidden from the user). Mezzo's soundness is ensured by a powerful capability system, which governs, in particular, how mutable structures or references can be accessed and shared. In this respect, Mezzo bears a strong resemblance to the Rust programming language.

Of particular relevance to our work is Mezzo's ability to safely "freeze" mutable data structures into immutable ones once construction is complete. This principle is applied, for instance, in Mezzo's standard library for lists, where an efficient *map* function on immutable lists is implemented using destination-passing and mutable lists internally. The capability system guarantees the safe and correct use of destinations in this low-level implementation. An earlier example of destination-passing used as a performance optimization in a mutable setting can also be found in ⟦Larus, 1989⟧.

# 9 Rust lifetimes

The Rust programming language uses a system of lifetimes (see e.g. ⟦Pearce, 2021⟧) that plays a similar role as our system of ages in $\lambda_d$ (Chapter 2).

Rust lifetimes are symbolic. Borrows and moves generate constraints (inequalities of the form $\alpha \leqslant \beta$) on the symbolic lifetimes. For example, that the lifetime of a reference is larger than the lifetime of any structure the reference is stored in. Without enforcing these constraints, Rust would face the same kinds of issues described in Section 2.3: a borrow could outlive the data it refers to, just as destinations in our system could outlive the structures they point to if scope escape were not properly prevented. In Rust, the borrow checker must check that the constraints are solvable, while in $\lambda_d$, ages are set explicitly, with no analysis needed.

Another difference between the two systems is that $\lambda_d$'s ages (and modes in general) are relative. An explicit modality $!_{1\uparrow k}$ must be used when a part has an age different than its parent, and means that the part is $k$ scope older than the parent. On the other hand, Rust's lifetimes are absolute, and the lifetime of a part is tracked independently of the lifetime of its parent.

# Conclusion

## Contributions

At the end of this exotic but exciting journey, we can say with confidence: safe, functional destination passing is not an oxymoron—it exists!

Destination passing is, at its core, the act of passing a write pointer so that a function can write its result directly into a memory cell, rather than returning it. This approach offers performance gains by avoiding intermediate allocations, and expressiveness gains by enabling novel ways of constructing data—the former primarily in imperative contexts, and the latter in functional ones. Yet, at first glance, destination passing appears quite impure: tightly coupled with memory mutation, and seemingly at odds with functional purity.

Our early experiments in Haskell, actually pre-dating the formal development of $\lambda_d$, gave us the insight that linear types can tame the impure nature of destinations. Linearity can ensure that a destination is used exactly once, guaranteeing that once a value has been written to a memory cell, it will not be overwritten. This makes destination passing compatible with immutability, a cornerstone of functional programming.

However, *scope escape* was discovered not long after, and with it, we realized that destinations for linear data would pose a significant challenge for type safety, even with a strict linear discipline being enforced on these destinations. Scope escape occurs when a linear resource is stored in a linear container that outlives the scope in which the resource was introduced. Linear interfaces typically rely on the assumption that each resource type comes with a specific set of destructor-like functions, and that one of these must be called for each resource. If the resource type is opaque, linearity forces us to consume the resource in the scope where it becomes available, and then the only way to do so is by calling one of its designated destructors. But if we can fake the consumption of the resource by storing it linearly outside its scope—performing scope escape—instead of calling a destructor, then we break this expectation. In our case, scope escape happens when a destination `Dest m t`—the linear resource—is filled into a destination `Dest 1 (Dest m t)` from a parent scope—representing the linear container. But scope escape is not an issue specific to destinations: it poses a threat for any kind of resource whose API relies on linearity for safety. Conversely, scope escape can be introduced by any writable form of storage given a linear interface, whether it be destinations, OCaml-style `ref`s, etc.

In early DPS Haskell, we avoided scope escape by disallowing destinations for linear data entirely. Even then, we could still reap quite a few benefits of destination passing: implementing tail-recursive *map*, efficient difference lists, and other patterns involving first-class structures with holes.

Yet this restriction limited expressiveness. As shown in the breadth-first traversal example (Chapter 3), some patterns could not be expressed without destinations for linear data. It became clear that a more complete solution was needed—one that reconciled destinations for linear data with memory safety.

To make progress, we stepped away from implementation concerns and designed a theoretical calculus, $\lambda_d$, so centered around destination passing that conventional functional data constructors are not even necessary. The key achievement was a mechanized proof of soundness for $\lambda_d$. To permit this, our calculus depends on several notable features:

- A linear type system, akin to the one from Linear Haskell, ensuring that destinations are neither duplicated nor discarded—letting us know statically when a structure is fully filled (except if scope escape happens);
- An age system, that allows precise tracking of scopes in which resources are introduced. The age system is the main technical device used to prevent the scope escape issue;
- A modal type system combining linearity and age control into a same mode semiring, inspired by recent work in the literature;
- A novel typing discipline for structures with holes, in which both the holes in a structure and their associated destinations are represented as named bindings in the typing context. When a hole binding and a destination binding share the same name, they cancel each other out. This mechanism allows us to enforce statically that a structure with holes has exactly as many unused destinations as it has holes—no more, no less;
- Operational semantics based on the explicit manipulation of an evaluation context. In particular, linear substitutions on the evaluation context are used to model a lightweight form of state update. Compared to store-based semantics, our approach is simpler, while still faithfully capturing the behavior of destination-induced memory mutations;

Armed with this theory, we could return to our original motivation: practical destination passing in Haskell. Haskell's support for linear types made it a natural fit[36], but its lack of age tracking meant we had to explore new trade-offs. In particular, we had to balance the simplicity of the destination-passing API with the expressiveness we desired. In $\lambda_d$, we could have both—but only with the help of a tailored type system featuring ages.

In Chapters 3 and 4, we iterated through increasingly powerful destination-passing APIs in Haskell, each narrowing the gap with $\lambda_d$. By the end of Chapter 4, we recovered a system expressive enough to implement all motivating examples from $\lambda_d$, with no adaptations required further than mere translation of a syntax to another.

The last important contribution of our work is practical as well. In the course of developing DPS Haskell, we implemented the API within *Compact regions*—a separate memory space in the Haskell heap that is only loosely overseen by the garbage collector. Our initial motivation was to ensure that memory mutations caused by destination-filling primitives would not interfere with the invariants expected by the garbage collector and the runtime system. But this design choice turned out to bring additional benefits we had not anticipated. *Compact regions* are often used to store long-lived data structures more efficiently by avoiding repeated garbage collection passes over them. Our DPS Haskell API offers, in fact, a more direct and efficient way to allocate such structures in compact regions safely. The benchmarks in Section 3.6 support this claim: destination passing—implemented inside compact regions—proves to be a valuable tool in specific scenarios, either enabling more direct, imperative-like programming when desired (without compromising memory safety), or allowing more efficient compact region use, e.g. in the parser example we developed in Section 3.5.2.

[36] Haskell was also chosen due to the industrial context of this PhD, with funding from a company with industrial functional programming in its DNA, and due to my supervisor Arnaud Spiwack's involvement in Linear Haskell ⟦Bernardy et al., 2018; Spiwack, 2023a; Spiwack et al., 2022⟧.

Of all the approaches we explored in our journey, the one from Chapter 3 is likely the most viable for practical adoption, hitting a good balance between simplicity and usability. The vision behind $\lambda_d$—where every data structure is built through destination passing—diverges radically from conventional functional programming styles. This divergence is justified by the need to rigorously explore the theoretical foundations and safety considerations of destination passing, but it means $\lambda_d$ addresses concerns that are rather niche within the broader landscape of everyday programming. Meanwhile, the API developed in Chapter 4 is a significant effort to expand what is achievable relying solely on linearity as a safety mechanism; but its current form is too complex and cluttered to be made into an industrial-grade library. It showcases the extent of what we were able to achieve, yet there's likely room for simplification and refinement.

With these reflections in mind, we hope we have presented the reader with a clear and enjoyable overview of functional destination passing, through a journey rooted in type theory, language design, and practical implementation concerns.

## Future perspectives

There are still many interesting threads that can be pursued to improve this work. The most immediate is probably to further refine the existing DPS Haskell implementation. This involves, in particular, fine-tuning its performance, which remains one of the main selling points of the approach. This is even more important given its dual role as both a destination-passing interface and a better compact region allocation mechanism, as described above.

With a longer time frame, we can envision other improvements on both the practical and theoretical sides, as we detail below.

**Destination passing support for data structures with unpacked fields** A first direction for improvement would be to extend destination-passing support to data structures with unpacked fields—that is, fields that embed child elements directly instead of pointing to them.

Until now, we have restricted our attention to purely linked structures, and for good reason: they allow efficient composition (i.e., merging) of structures with holes.

Throughout both our theoretical work and implementations, our goal has been to perform destination filling as simple, single memory write. Also, since Minamide's foundational work, composability has been a key feature of structures with holes, crucial for building constructs like ampar-based difference lists.

When composing two ampars using `extendComp` (or filling a parent ampar's destination with a child ampar using `fillComp`), if the involved fields are indirected, then composition boils down to writing the address of the child into the hole of the parent. See below (we annotate memory locations with `@addr =` syntax, and denote holes by □):

```
extendComp :: ∀ s t u. Ampar t u ⊸ Ampar s (Dest 1 t) ⊸ Ampar s u

parentAmpar :: Ampar [t] (Dest 1 [t]) = newAmpar tok1 &extend @'(:) @1 1
-- in memory:  Ampar 0x05ba (Dest 0x05ca)
-- @0x05ba =   0xcb1e : □
-- @0xcb1e =   I# 1#

childAmpar :: Ampar [t] (Dest 1 [t]) = newAmpar tok2 &extend @'(:) @1 2
-- in memory:  Ampar 0x0840 (Dest 0x0850)
-- @0x0840 =   0xbc4e : □
-- @0xbc4e =   I# 2#

resAmpar :: Ampar [t] (Dest 1 [t]) = parentAmpar &extendComp childAmpar
-- in memory:  Ampar () (Dest 0x0850)
-- @0x05ba =   0xcb1e : 0x0840
-- @0xcb1e =   I# 1#
-- @0x0840 =   0xbc4e : □
-- @0xbc4e =   I# 2#
```

To perform the composition, only the second field of the *cons* constructor at `0x05ba` needs to be updated. The resulting ampar inherits the destination from `childAmpar`, which still correctly points to the hole at `0x0850`.

However, things become more complicated when unpacked fields are involved:

```
extendComp :: ∀ s t u. Ampar t u ⊸ Ampar s (Dest 1 t) ⊸ Ampar s u

-- GHC does not support UNPACK for polymorphic fields so we cannot do:
-- data Pair a b = Pair {-# UNPACK #-} !a {-# UNPACK #-} !b
data IntPair = IntPair {-# UNPACK #-} !Int {-# UNPACK #-} !Int
data NestedPair = NestedPair {-# UNPACK #-} !Int {-# UNPACK #-} !IntPair

parentAmpar :: Ampar NestedPair (Dest 1 IntPair)
  = newAmpar tok1 &extend @'NestedPair @1 1
-- in memory:  Ampar 0x05ba (Dest 0x05ca)
-- @0x05ba =   NestedPair 1# □

childAmpar :: Ampar IntPair (Dest 1 Int) = newAmpar tok2 &extend @'IntPair @1 2
-- in memory:  Ampar 0x0840 (Dest 0x0850)
-- @0x0840 =   IntPair 2# □

resAmpar :: Ampar NestedPair (Dest 1 Int) = parentAmpar &extendComp childAmpar
-- in memory:  Ampar 0x05ba (Dest 0x0850)
-- @0x05ba =   NestedPair 1# 2# □
-- @0x0840 =   IntPair 2# □
```

Now, to perform composition, we must copy the fields of the child structure into the hole of the parent, as the parent field is annotated as unpacked. For instance, here the payload `2#` □ is copied into the hole at `0x05ca`. But doing so invalidates the destination inherited from the child: it still points to `0x0850`, while the remaining hole in `resAmpar` now lives at `0x5d2`!

Addressing this would require tracking and updating destination pointers dynamically whenever composition happens throughout an unpacked field. But such a mechanism would impose significant overhead and undermine the simplicity and efficiency of our current model.

A more realistic path forward—one we are likely to adopt in [Bagrel, 2024b]—is to distinguish destinations that point to unpacked fields from those that point to indirected ones.[37] We would then restrict the use of `fillComp` and `extendComp` to cases where the destination targets an indirected field. This approach preserves much of the expressiveness of destination-passing and the composition of incomplete structures, while still supporting unpacked fields. The new restrictions would apply only to those unpacked fields (that were not supported before) and would not compromise the rest of the system, allowing destinations to remain simple pointers, with no need for complex update mechanisms.

**DPS Haskell, outside of compact regions** A second direction for future work is to implement destination-passing support directly within Haskell's standard garbage-collected heap, rather than relying on the protective boundary of compact regions.

The main challenge lies in the close interaction between memory safety and the internals of GHC's garbage collector, particularly regarding when and where immutability assumptions are relied upon. GHC uses a copying collector with a variety of performance-critical optimizations, and violating its expectations about immutability could lead to subtle and potentially catastrophic bugs. For instance, if structures with holes are moved by the GC, then any live destination referring to them must be updated accordingly to avoid dangling pointers.

[37] Whether a field is unpacked or indirected can be determined using `GHC.Generics`.

Another challenge is that GHC's runtime system is complex and actively evolving, with key implementation details often spread across various documents, so acquiring a complete and up-to-date understanding of this system would likely require input from experienced contributors to the compiler and runtime.

Still, adapting the DPS Haskell implementation to run directly on the standard GC heap could significantly lower the barrier to adoption. Programmers would be able to use destination-passing selectively, without the overhead and constraints associated with compact regions. It would likely bring substantial performance improvements as well, by avoiding the need to copy data into compact regions. For example, in our *map* function using destinations, every value in the input list must currently be copied into the region, even though destination-passing is only needed for the spine. This direction remains uncertain—the runtime implications are subtle and the engineering effort nontrivial—but if achievable, the payoff could be great.

**Proof of safety for the practical DPS Haskell APIs** By the end of this PhD, we achieved our two main goals: demonstrating that functional destination passing can be made safe, and providing evidence that practical implementations of destination passing in an industrial-grade functional language are feasible and can deliver performance benefits.

The last natural step is to reconcile these two strands of work, by using the theoretical framework developed in $\lambda_d$ (Chapter 2) to prove that the practical implementations introduced in Chapters 3 and 4 are indeed safe.

Such a proof would involve translating each operator from the DPS Haskell APIs into their counterparts in $\lambda_d$, and showing that for any well-typed DPS Haskell program, there exists a suitable age assignment that makes the translated $\lambda_d$ program typecheck. We recall here that despite Haskell being notoriously call-by-need, our DPS implementation rely on compact regions in which any data structure is forced to normal form, which matches the call-by-value semantics of $\lambda_d$. So there should not be a mismatch on that front.

The main obstacle though is that despite their common roots, there are subtle but key differences between $\lambda_d$ and the various DPS Haskell APIs. For example, in the version from Chapter 3, the absence of destinations for linear data has wide-reaching implications: it allows destinations of heterogeneous ages to be safely stored in usual Haskell data structures like standard library lists. This flexibility would be difficult to justify in $\lambda_d$, which lacks any notion of non-destination-based data structures. Similarly, in the API from Chapter 4, we rely both on a non-capturing `updWith` operator (unlike the original **upd**$\ltimes$), and on the `extend` functions, which would effectively translate to a controlled yet capturing version of **upd**$\ltimes$.

The key, then, may not be to reuse $\lambda_d$ exactly as it stands, but to design a slightly adapted formal language tailored to the specific API we aim to verify. Much of the existing formalization—especially the infrastructure for managing typing contexts, hole renaming, and substitution—could likely be reused with minimal changes. While the original proofs (Section 2.7, ⟦Bagrel and Spiwack, 2025b⟧) were not written with reusability in mind (as we expected $\lambda_d$ to be general enough in its current form), only a relatively small portion depends directly on the typing rules of individual operators. Adapting these rules to match the ones of the DPS Haskell API from Chapter 3 for example would be time-consuming, but probably tractable. In return, this path offers a promising opportunity to generalize our formal development and extend its impact.